\documentclass[12pt,superscriptaddress,letter]{revtex4}%
\usepackage{amssymb}
\usepackage{color}
\usepackage{graphicx}
\usepackage{dcolumn}
\usepackage{bm}
\usepackage[header,title,page,titletoc]{appendix}
\usepackage{amsmath}
\usepackage{subfigure}
\usepackage{amsfonts}%
\providecommand{\U}[1]{\protect \rule{.1in}{.1in}}

\begin{document}
\title{Theory for Quantum Spacetime }
\author{Su-Peng Kou}
\thanks{Corresponding author}
\email{spkou@bnu.edu.cn}
\affiliation{Center for Advanced Quantum Studies, School of Physics and Astronomy, Beijing
Normal University, Beijing 100875, China}

\pacs{04.20.Cv, 12.10.-g}

\begin{abstract}
Quantum gravity (or quantum spacetime) is to unify general relativity and
quantum mechanics into a single theoretical framework and presented as the
most important open puzzle in fundamental physics. The development of a
microscopic theory of quantum spacetime becomes the key problem about quantum
gravity. This paper is the solution to this problem. The starting point of
this paper is very simple -- physical variant with higher-order variability
(see the below discussion). Based on this simple starting point, a microscopic
theory for quantum spacetime is developed, including its matrix representation
for quantum states, its time evolution, its geometry quantization, its
generalized symmetry, its canonical quantization, and the uncertainty
principle, black hole, AdS/CFT correspondence, scattering amplitudes of
gravitons... The result leads to a great unification of matter and spacetime
--\ the particles constitute the basic blocks of spacetime and spacetime is
really a multi-particle system that is made of matter. As a result, this work
would help researchers to understand the mysteries in quantum gravity.

\end{abstract}
\maketitle
\newpage
\tableofcontents

\newpage

\newpage

\section{Introduction}

Gravity is a natural phenomenon by which all objects attract each other
including galaxies, stars, human-being and even elementary particles. Hundreds
of years ago, Newton discovered the inverse-square law of universal
gravitation, $F=G\frac{Mm}{r^{2}}$ where $G$ is the Newton constant, $r$ is
the distance, and $M$ and $m$ are the possess masses for two objects. In
Newton's theory for gravity, matter and spacetime are two different
fundamental objects. The spacetime is always regarded as a rigid background,
on which matter moves. The success of Newton's theory has led to the belief of
"\emph{mechanics principle of gravity}". One hundred years ago, the
establishment of general relativity by Einstein is a milestone to learn the
underlying physics of gravity that provides a unified description of gravity
as a geometric property of spacetime. From Einstein's equations, $R_{\mu \nu
}-\frac{1}{2}Rg_{\mu \nu}=\frac{8\pi G}{c^{4}}T_{\mu \nu},$ the gravitational
force is really an effect of curved spacetime\cite{ein}. Here $R_{\mu \nu}$ is
the 2nd rank Ricci tensor, $R$ is the curvature scalar, $g_{\mu \nu}$ is the
metric tensor, and $T_{\mu \nu}$ is the energy-momentum tensor of matter. $c$
is speed of light. The success of general relativity has led to the belief of
"\emph{geometry principle of gravity}". According to this belief, when the
spacetime becomes curved, the matter freely moves along the geodesic lines. On
the other hand, the matter curves the spacetime. John Archibald Wheeler had
said, "\textit{Spacetime tells matter how to move, and matter tells time and
space how to curve}."

Today, general relativity becomes a fundamental branch of physics that agrees
very well with experiments and provides an accurate description of the dynamic
behaviors of macroscopic objects. However, in microcosmic world, the objects
obey quantum mechanics (also known as quantum physics or quantum theory). The
development of new quantum foundation for gravity (or quantum gravity) becomes
one of the most important trouble in modern physics. I show five unsolve
problems for quantum gravity:

\begin{enumerate}
\item Our spacetime is still very mysterious and far from being well
understood. What's the exact \emph{microstructure} of spacetime near Planck
length $l_{p}\simeq1.6\times10^{-33}\mathrm{cm}$? Does \emph{geometric}
structure have quantization characteristics, and what are the quantization rules?

\item In 1997, Juan Maldacena proposed the Anti de Sitter - Conformal Field
theory (AdS/CFT) correspondence\cite{ma}. A few year ago, the AdS/CFT
correspondence has been extended to a generalized mapping between usual
quantum conformal field theories and gravity\cite{witten}. However, AdS/CFT
correspondence is still a conjecture and far from being well understood.
What's the \emph{exact} rule of AdS/CFT correspondence within the framework of
quantum gravity rather than just a conjecture?

\item Black hole is one of most mysterious object in our universe. For black
holes, the spacetime inside becomes too curved to be seen. What's the exact
\emph{microstructure} of spacetime around black hole near Planck length?
What's the exact \emph{microstructure} of spacetime inside black hole? And,
how to characterize it?

\item In the framework of quantum field theory, it is believed that the
gravitational interaction comes from exchanging virtual gravitons - spin-2
bosonic particles. The primary approach to quantization of gravitational
interaction leads to \emph{unsolvable divergences}. How \emph{quantize}
gravitational waves correctly?

\item Scattering amplitudes play a fundamental role in modern quantum physics.
By detecting scattering amplitudes, people could extract logical predictions
for particle scattering from the complex formalism of particle physics. In
2003, Witten developed the theory that provides a strikingly compact formula
for tree--level scattering amplitudes in four-dimensional (4D) Yang-Mills
theory in terms of an integral over the moduli space of maps from the
$n$-punctured sphere in momentum space\cite{Witten:2003nn}. Furthermore, it
was known that these representations are supported on solutions of the
scattering equations by using cohomology classes on ambitwistor
space\cite{am1}. What's the exact \emph{microstructure} of the scattering
amplitudes for different particles? How to calculate \emph{loop} amplitudes?
Why \emph{amplituhedron}?
\end{enumerate}

Based on different principles, to develop a new theory for quantum gravity
there are different candidates to solve the problem of quantum gravity,
including gauge theory for the Lorentz group\cite{Utiyama,mm}, superstring
theory\cite{ss} and quantum loop theory\cite{loop}, noncommutative
geometry\cite{con}, it from qubit, ... Unfortunately, all these theories
didn't solved above troubles. As a result, quantum gravity is still a big
challenge for physicists.

To develop a theory for quantum gravity satisfactorily, a complete theory
beyond both quantum mechanics and general relativity must be developed in
unison rather than only providing certain theory with quantized gravitational
waves. Then, we reexamine the entire foundation of modern physics and
find\emph{ three} \emph{hidden} assumptions. These assumptions are commonly
referred to as agreed upon by people and are deeply hidden.

\emph{One} hidden assumption is the \emph{separation of spacetime and matter}.
In modern physics, all physical objects belong to two different types --
matter and spacetime. People are familiar to spacetime as a "stage" and all
kinds of physical processes of matter (or elementary particles) on it, and
take it for granted. The situation looks like ants moving on the elastic
surface of a balloon. In general relativity, although there exists interaction
between matter and spacetime, we have a dualism of two different objects,
matter and spacetime.

The \emph{second} hidden assumption is the \emph{validity of quantum
mechanics}. People always assume that to develop a theory for quantum gravity,
the fundamental principle of quantum mechanics is correct. Therefore,
\textquotedblleft time\textquotedblright \ means the evolution of quantum
states that must satisfy the (generalized) Schrodinger equation. However, we
will point out that this hidden assumption leads people to the wrong fork in
the road towards quantum gravity.

The \emph{third} hidden assumption is about \emph{invariant/symmetry} in the
possible theory for quantum gravity. People always take it for granted that it
is invariant/symmetry that characterizes the quantum systems (including the
quantum spacetime). For example, the theory for quantum gravity based on
supersymmetry is developed. This belief of "\emph{symmetry induce
interaction}" in a certain sense prevents people from obtaining the correct theory.

In the following parts, we will point out that the \emph{three hidden
assumptions are all misleading}. In the paper, an inspiring idea is that\emph{
the particle is basic block of spacetime and the spacetime is made of matter}.
Therefore, according to this idea, the matter is really certain "changing" of
\textquotedblleft spacetime\textquotedblright \ itself rather than extra things
on it. This is the \emph{new idea} for the foundation of quantum gravity and
the development of a complete theory and then becomes starting point of this
paper. In the paper, we point out that all physical processes of our world be
intrinsically described by the processes of the changings of a physical
variant -- a system "\emph{uniform} \emph{changing}"\cite{kou1}. Another key
point of the new theory is \emph{higher-order variability} rather the
gauge/global symmetry. Now, the principle of "symmetry induce interaction" is
replaced by the principle of "variability induce interaction". We have
a\ "\emph{variability principle of gravity}". According to this principle, a
theory for quantum gravity is developed. Quantum mechanics and general
relativity are unified, i.e.,
\begin{align*}
&  \text{Quantum mechanics + general relativity}\\
&  \Longrightarrow \text{Theory of a physical variant.}%
\end{align*}

The paper is organized as below. In Sec. II, we develop a fundamental theory
for quantum spacetime. In Sec. III, we develop fundamental theory for AdS/CFT
correspondence. In Sec. IV, we develop the theory for black hole. In Sec. V,
we develop the theory for scattering amplitudes. In Sec. VI, we draw the conclusion.

\newpage

\section{Quantum Spacetime -- Unification of Matter and Spacetime}

\subsection{Fundamental mathematic theory for higher-dimensional variants}

Our classical world can be regarded as \emph{"non-changing"} structure that is
described by usual classical "field" on Cartesian space. In the paper of
\cite{kou1}, we generalize usual classical "field" to "variant". We call the
new mathematic structure to be \emph{variant theory}. As a result, usual
classical field (for example, $f(x)$) is suitable to characterize a system
with "\emph{non-changing}" structure, i.e.,
\[
\text{"Classical field on space": Non-changing structure;}%
\]
Variant theory is suitable to characterize a system with "\emph{changing}" or
"\emph{operating}" structure, i.e.,
\[
\text{"Space on space": Changing structure.}%
\]
In particular, for higher-dimensional variant, their longitudinal changings
and transverse changings interplay each other and the resulting rules help us
develop a theory for quantum gravity.

\subsubsection{Review on usual variant theory}

\paragraph{General variants}

A variant describes "changing" structure, of which the element object is
"group-changing elements" $\delta \phi^{a}$. So, it is quite different from
usual fields $g(x)$ that characterize "non-changing" structure, of which the
element object is "group element" $\phi^{a}$. Here, the word "changing"\ means
a space-like structure of a set of number's changing on Cartesian space.
Therefore, a variant is theory describing the space dynamics rather than field
dynamics on Cartesian space. In a word, it describes a "space" on the other.

A higher-dimensional variant $V_{\mathrm{\tilde{G},}d}[\Delta \phi^{\mu},\Delta
x^{\mu},k_{0}^{\mu}]$ ($d>1$)\ is defined by\ a mapping between a
d-dimensional group-changing space $\mathrm{C}_{\mathrm{\tilde{G},}d}$ with
total size $\Delta \phi^{\mu}$\ and Cartesian space $\mathrm{C}_{d}$\ with
total size $\Delta x^{\mu}$, i.e.,%
\begin{align}
V_{\mathrm{\tilde{G},}d}[\Delta \phi^{\mu},\Delta x^{\mu},k_{0}^{\mu}]  &
:\nonumber \\
\mathrm{C}_{\mathrm{\tilde{G},}d}  &  =\{ \delta \phi^{\mu}\}
\Longleftrightarrow \mathrm{C}_{d}=\{ \delta x^{\mu}\}
\end{align}
where \textit{\textrm{\~{G}} }is a non-compact Lie group with\textit{ }$N$
generator and $N<d.$ $\Longleftrightarrow$\ denotes an ordered mapping under
fixed changing rate of integer multiple $k_{0}^{\mu}$. Here, the
group-changing space $\mathrm{C}_{\mathrm{\tilde{G}},d}(\Delta \phi^{\mu}%
)$\textit{\ }is described by a series of numbers of group element $\phi^{\mu}$
of $\mu$-th generator independently in size order along $a$-th
direction.\ $\delta \phi^{\mu}$ denotes group-changing element along $\mu
$-direction rather than group element (or element of group). $\delta \phi^{\mu
}$ is defined by an infinitesimal group-changing operation with $d$
directions, $\tilde{U}(\delta \phi_{i})=(\prod_{a=1}^{d}(\tilde{U}(\delta
\phi_{i}^{a}))$ with $\tilde{U}(\delta \phi_{i}^{a})=e^{i((\delta \phi_{i}%
^{a}T^{a})\cdot \hat{K}_{a})}$, $\hat{K}_{a}=-i\frac{d}{d\phi^{a}}.$\

Now, we take a 1D variant $V_{\mathrm{\tilde{U}(1),}1}[\Delta \phi,\Delta
x,k_{0}]$ as an example to show the concept. $V_{\mathrm{\tilde{U}(1),}%
1}[\Delta \phi,\Delta x,k_{0}]$ is one dimensional (1D) group-changing space
$\mathrm{C}_{\mathrm{\tilde{U}(1)},1}(\Delta \phi)$\textit{ }on Cartesian space
$\mathrm{C}_{1}$, i.e.,
\begin{align}
V_{\mathrm{\tilde{U}(1),}1}[\Delta \phi,\Delta x,k_{0}]  &  :\\
\mathrm{C}_{\mathrm{\tilde{U}(1)},1}(\Delta \phi)  &  =\{ \delta \phi \}
\Longleftrightarrow \mathrm{C}_{1}=\{ \delta x\}.\nonumber
\end{align}
\textit{ }According to above definition,\ for a 1D variant $V_{\mathrm{\tilde
{U}(1),}1}[\Delta \phi,\Delta x,k_{0}],$ we have
\begin{equation}
\delta \phi_{i}=k_{0}n_{i}\delta x_{i}%
\end{equation}
where $k_{0}$ is a constant real number and $n_{i}$ is an integer number.
$k_{0}n_{i}$ is changing rate for $i$-th space element, i.e., $k_{0}%
n_{i}=\delta \phi_{i}/\delta x_{i}$. Therefore, for the 1D variant
$\mathrm{C}_{\mathrm{\tilde{U}(1)},1}(\Delta \phi)$, we have a series of
numbers of infinitesimal elements to record its information.\quad Different 1D
variants $V_{\mathrm{\tilde{U}(1),}1}[\Delta \phi,\Delta x,k_{0}]$ are
characterized by different distributions of $n_{i}$. As a result, in some
sense, a variant can be described by "\emph{function}" of $n_{i}$ under constraints.

For a higher-dimensional case $V_{\mathrm{\tilde{G},}d}[\Delta \phi^{\mu
},\Delta x^{\mu},k_{0}^{\mu}]$, along a given direction (for example, $\mu
$-direction), the situation is similar to the 1D case by considering the
corresponding distributions of $n_{i}^{\mu}.${}We then take $d$-dimensional
$\mathrm{\tilde{S}\tilde{O}}$\textrm{(d)} variant $V_{\mathrm{\tilde{S}%
\tilde{O}(d)},d}[\Delta \phi^{\mu},\Delta x^{\mu},k_{0}^{\mu}]$ as an example.
A $d$-dimensional $\mathrm{\tilde{S}\tilde{O}}$\textrm{(d)} variant is a
mapping between Clifford group-changing space $\mathrm{C}_{\mathrm{\tilde
{S}\tilde{O}(d)},d}$\ and a rigid spacetime $\mathrm{C}_{d},$ i.e.,\textit{ }%
\begin{align}
V_{\mathrm{\tilde{S}\tilde{O}(d)},d}[\Delta \phi^{\mu},\Delta x^{\mu}%
,k_{0}^{\mu}]  &  :\nonumber \\
\mathrm{C}_{\mathrm{\tilde{S}\tilde{O}(d)},d}(\Delta \phi^{\mu})  &  =\{
\delta \phi^{\mu}\} \Leftrightarrow \mathrm{C}_{d}=\{ \delta x^{\mu}\}
\end{align}
where a Clifford group-changing space\textit{ }$\mathrm{C}_{\mathrm{\tilde
{S}\tilde{O}(d)},d}(\Delta \phi^{\mu})$\textit{\ }is described by $d$ series of
numbers of group elements $\phi^{\mu}$ arranged in size order with unit
"vector" as Gamma matrices $\Gamma^{\mu}$ obeying Clifford algebra $\{
\Gamma^{i},\Gamma^{j}\}=2\delta^{ij}$. The total size along $\mu$-direction of
$\mathrm{C}_{\mathrm{\tilde{S}\tilde{O}(d)},d}(\Delta \phi^{\mu})$ is\textit{
}$\Delta \phi^{\mu}$\textit{. }$\mu$ labels the spatial direction.\textit{
}$\Leftrightarrow$\ denotes an ordered mapping with fixed changing rate of
integer multiple $k_{0}.$ The $d$-dimensional Clifford group-changing
space\textit{ }$\mathrm{C}_{\mathrm{\tilde{S}\tilde{O}(d}),d}(\Delta \phi^{\mu
})$ has orthogonality, i.e.,%
\begin{equation}
\left \vert \mathbf{\phi}_{\mathrm{A}}-\mathbf{\phi}_{\mathrm{B}}\right \vert
^{2}=%
{\displaystyle \sum \nolimits_{\mu}}
(\phi_{\mathrm{A,}\mu}e^{\mu}-\phi_{\mathrm{B},\mu}e^{\mu})^{2}%
\end{equation}
where $\mathbf{\phi}_{\mathrm{A}}=%
{\displaystyle \sum \nolimits_{\mu}}
\phi_{\mathrm{A,}\mu}e^{\mu}$ and $\mathbf{\phi}_{\mathrm{B}}=%
{\displaystyle \sum \nolimits_{\mu}}
\phi_{\mathrm{B,}\mu}e^{\mu}$.

In particular, we point out that $\mathrm{C}_{\mathrm{\tilde{S}\tilde{O}%
(d+1)},d+1}$ is noncommutative space obeying noncommutative geometry. Its
coordinates are phase angles $\delta \phi^{\mu}$ of non-compact $\mathrm{\tilde
{S}\tilde{O}}$\textrm{(d+1)} Lie group; the coordinate unit vectors
$\mathbf{e}^{\mu}$ (the fundamental vectors along $\phi^{\mu}$-direction)
becomes $\Gamma^{\mu},$ i.e., $\mathbf{e}^{\mu}=\Gamma^{\mu}.$ The
anti-commutation condition matrices $\Gamma^{\mu}$ of Clifford group-changing
space indicate a non-commutating character\cite{con}, i.e.,
\begin{equation}
\{ \mathbf{e}^{\mu},\mathbf{e}^{\nu}\}=\{ \Gamma^{\mu},\Gamma^{\nu}%
\}=2\delta_{\mu \nu}%
\end{equation}
and
\begin{equation}
\lbrack \mathbf{e}^{\mu},\mathbf{e}^{\nu}]=\left[  \Gamma^{\mu},\Gamma^{\nu
}\right]  \neq0.
\end{equation}

\paragraph{Uniform variants}

Uniform variant (U-variant) is an important variant. A d-dimensional U-variant
$V_{0,d}[\Delta \phi^{\mu},\Delta x^{\mu},k_{0}^{\mu}]$ for group-changing
space $C_{\mathrm{\tilde{G}},d}(\Delta \phi^{\mu})$ of non-compact Lie group
\textrm{\~{G}} is defined by a perfect, ordered mapping between a
d-dimensional Clifford group-changing space $\mathrm{C}_{\mathrm{\tilde{G}}%
,d}(\Delta \phi^{\mu})$ and the d-dimensional Cartesian space $\mathrm{C}_{d}$,
i.e.,
\begin{align*}
V_{\mathrm{\tilde{G},}d}[\Delta \phi^{\mu},\Delta x^{\mu},k_{0}^{\mu}]  &
:C_{\mathrm{\tilde{G}},d}(\Delta \phi^{\mu})=\{ \delta \phi^{\mu}\} \\
&  \Leftrightarrow \mathrm{C}_{d}=\{ \delta x^{\mu}\}
\end{align*}
where $\Leftrightarrow$\ denotes an ordered mapping under fixed changing rate
of integer multiple $k_{0}^{\mu},$\ and $\mu$ labels the spatial
direction\textit{.} In particular, for a U-variant, the total size $\Delta
\phi^{\mu}$ of $\mathrm{C}_{\mathrm{\tilde{G}},d}$\ exactly matches the total
size $\Delta x^{\mu}$ of $\mathrm{C}_{d}$, i.e., $\Delta \phi^{\mu}=k_{0}^{\mu
}\Delta x^{\mu}$. A U-variant with infinite size ($\Delta x\rightarrow \infty$)
has 1-th order variability, i.e.,%
\begin{equation}
\mathcal{T}(\delta x^{\mu})\leftrightarrow \hat{U}(\delta \phi^{\mu}%
)=e^{i\cdot \delta \phi^{\mu}T^{\mu}}%
\end{equation}
where $\mathcal{T}(\delta x^{\mu})$ is the spatial translation operation on
$\mathrm{C}_{d}$ along $x^{\mu}$-direction and $\hat{U}(\delta \phi^{\mu})$ is
shift operation on $\mathrm{C}_{\mathrm{\tilde{G}},d}(\Delta \phi^{\mu})$, and
$\delta \phi^{\mu}=k_{0}^{\mu}\delta x^{\mu}$. That means when one translate
along Cartesian space $\delta x^{\mu},$ the corresponding shifting of
group-changing space $\mathrm{C}_{\mathrm{\tilde{G}},d}$ along $\mu$-th
direction is $\delta \phi^{\mu}=k_{0}^{\mu}\delta x^{\mu}.$

For example, a 1D U-variant $V_{\mathrm{\tilde{U}(1),}1}$ is defined by a
perfect, ordered mapping between a 1D group-changing space $\mathrm{C}%
_{\mathrm{\tilde{U}(1),}1}(\Delta \phi)$ and the 1D Cartesian space\textit{
}$\mathrm{C}_{1}$. For a uniform variant with infinite size ($\Delta
x\rightarrow \infty$), to characterize 1-th order variability, we have the
following relationship,
\begin{equation}
\mathcal{T}(\delta x)\leftrightarrow \hat{U}(\delta \phi)=e^{i\cdot \delta \phi}%
\end{equation}
where $\delta \phi=k_{0}\delta x.$ $\mathcal{T}(\delta x)$ is the spatial
translation operation on $\mathrm{C}_{1}$ and $\hat{U}(\delta \phi)$ is shift
operation on $\mathrm{C}_{\mathrm{\tilde{U}(1),}1}(\Delta \phi)$. According to
the 1-th order variability, for the 1D U-variant $\mathrm{C}_{\mathrm{\tilde
{U}(1)},1}(\Delta \phi)$, we have an ordered series of numbers $n_{i}=1$ of
infinitesimal elements.

In addition, $V_{\mathrm{\tilde{U}(1),}1}$ is described by a complex field
\begin{equation}
\mathrm{z}_{u}(x)=\exp(i\phi(x))
\end{equation}
in Cartesian space where $\phi(x)=\phi_{0}+k_{0}x$ that corresponds to a
spiral line on a cylinder with fixed radius. We may regard a 1D U-variant to
be a knot/link structure between the curved line of $\mathrm{z}_{u}(x)$ and
the straight line at center of $\mathrm{z}(x)=0$.

People had known that a knot/link can be projected by counting the crossings
(or zeroes named in this paper) of the corresponding lines. With the help of
the knot projection (K-projection), people can locally obtain the property of
the variant. We then introduce the\emph{ }K-projection of the curved line of
1D U-variant along a given direction $\theta$ on the straight line at center
of $\mathrm{z}(x)=0$ in 2D space $\{ \xi(x),\eta(x)\}$. In mathematics, the
K-projection is defined by%
\begin{equation}
\hat{P}_{\theta}\left(
\begin{array}
[c]{c}%
\xi(x)\\
\eta(x)
\end{array}
\right)  =\left(
\begin{array}
[c]{c}%
\xi_{\theta}(x)\\
\left[  \eta_{\theta}(x)\right]  _{0}%
\end{array}
\right)
\end{equation}
where $\xi_{\theta}(x)$ is variable and $\left[  \eta_{\theta}(x)\right]
_{0}$ is constant. In the following parts we use $\hat{P}_{\theta}$ to denote
the projection operators. Because the projection direction out of the curved
line is characterized by an angle $\theta$ in $\{ \xi,\eta \}$ space, we have
\begin{equation}
\left(
\begin{array}
[c]{c}%
\xi_{\theta}\\
\eta_{\theta}%
\end{array}
\right)  =\left(
\begin{array}
[c]{cc}%
\cos \theta & \sin \theta \\
\sin \theta & -\cos \theta
\end{array}
\right)  \left(
\begin{array}
[c]{c}%
\xi \\
\eta
\end{array}
\right)
\end{equation}
where $\theta$ is angle \textrm{mod}($2\pi$), i.e. $\theta \operatorname{mod}%
2\pi=0.$ So the curved line of 1D variant is described by the function
\begin{equation}
\xi_{\theta}(x)=\xi(x)\cos \theta+\eta(x)\sin \theta.
\end{equation}
In the following parts, we call $\theta \in \lbrack0,2\pi)$ projection angle.
Under projection, each zero corresponds to a solution of the equation
\begin{equation}
\hat{P}_{\theta}[\mathrm{z}(x)]\equiv \xi_{\theta}(x)=0.
\end{equation}
We call the equation to be zero-equation and its solutions to be
zero-solution. For this 1D U-variant $V_{\mathrm{\tilde{U}(1)},1}(\Delta
\phi,\Delta x,k_{0})$, from the its analytics representation $\mathrm{z}%
_{u}(x)\sim e^{ik_{0}\cdot x}$, we get the zero-solutions to be%
\begin{equation}
x=l_{0}\cdot n/2+\frac{l_{0}}{2\pi}(\theta+\frac{\pi}{2})
\end{equation}
where $n$ is an integer number, and $l_{0}=2\pi/k_{0}$. This is called zero
lattice, of which each zero corresponds to a crossing. The zero lattice is a
lattice of "two-sublattice" with discrete spatial translation symmetry. In
other words, with total size $l_{0}$, a unit cell with $2\pi$ phase changing
has two zeroes. The original non-compact $\mathrm{\tilde{U}(1)}$ group turns
into a field of compact \textrm{U(1)} group on 1D uniform zero lattice of
"two-sublattice", i.e.,%
\begin{equation}
\phi(x)=2\pi N(x)+\varphi(x).
\end{equation}

For higher-dimensional $\mathrm{\tilde{S}\tilde{O}}$\textrm{(d)} U-variant
$V_{\mathrm{\tilde{S}\tilde{O}(d)},d}[\Delta \phi^{\mu},\Delta x^{\mu}%
,k_{0}^{\mu}],$ we have 1-th order variability along an arbitrary spatial
direction, i.e.,
\begin{align}
\mathcal{T}(\delta x^{i})  &  \leftrightarrow \hat{U}^{\mathrm{T}}(\delta
\phi^{i})=e^{i\delta \phi^{i}\Gamma^{i}},\nonumber \\
\text{ }i  &  =x_{1},x_{2},\text{...},x_{d},
\end{align}
where $\delta \phi^{i}=k_{0}\delta x^{i}$ and $\Gamma^{i}$ are the Gamma
matrices obeying Clifford algebra $\{ \Gamma^{i},\Gamma^{i}\}=2\delta^{ij}$.
Therefore, $\hat{U}^{\mathrm{T}}(\delta \phi^{i})$ is (spatial) translation
operation on Clifford group-changing space rather than the generator of a
(non-compact) $\mathrm{\tilde{S}\tilde{O}}$\textrm{(d)}\ group. For the
higher-dimensional $\mathrm{\tilde{S}\tilde{O}}$\textrm{(d)} U-variant
$V_{\mathrm{\tilde{S}\tilde{O}(d)},d}[\Delta \phi^{\mu},\Delta x^{\mu}%
,k_{0}^{\mu}]$, by generalizing to the K-projection to the $d$ 1D variants of
non-compact Abelian group $\mathrm{\tilde{S}\tilde{O}}$\textrm{(d)}, we have
$d$-dimensional zero lattice. The original non-compact $\mathrm{\tilde{G}}$
group turns into a field of compact \textrm{G} group on $d$-dimensional
uniform zero lattice of "two-sublattice", i.e.,
\begin{equation}
\phi^{\mu}(x)=2\pi N^{\mu}(x)+\varphi^{\mu}(x).
\end{equation}
Along $\mu$-th spatial direction of the zero lattice, the lattice site is
labeled by $N^{\mu}.$ Consequently, after doing D-projection together with
K-projection, we can also relabel the group-changing space $\mathrm{C}%
_{\mathrm{\tilde{S}\tilde{O}(d)},d}(\Delta \phi^{a})$ by $2d$ numbers ($N^{\mu
}(x),\varphi^{\mu}(x)$): $\varphi^{\mu}(x)$ is compact phase angle of $\mu$-th
group generator of the compact group, the other is the integer winding number
of unit cell of zero lattice $N^{\mu}(x)$. Although $\mathrm{C}%
_{\mathrm{\tilde{S}\tilde{O}(d+1)},d+1}$ is noncommutative space obeying
noncommutative geometry, the $d$-dimensional uniform zero lattice is
commutative space obeying commutative geometry, $\left[  \hat{U}^{\mathrm{T}%
}(\delta \phi^{\mu}(x),\hat{U}^{\mathrm{T}}(\delta \phi^{\nu}(x)\right]
=2\delta^{\mu \nu}$.

\paragraph{Perturbative uniform variants}

Perturbative uniform variant (P-variant) is another important type of variant
that can be generated by perturbatively changings on a uniform one. In
general, one may imagine that U-variants and P-variants correspond to ground
states and excited states in quantum many-body systems, respectively.

A d-dimensional P-variant $V_{d}[\Delta \phi^{\mu},\Delta x^{\mu},k_{0}^{\mu}]$
for group-changing space $\mathrm{C}_{\mathrm{\tilde{G}},d}(\Delta \phi^{\mu})$
of non-compact Lie group \textrm{\~{G}} is defined by a quasi-perfect, ordered
mapping between a d-dimensional Clifford group-changing space $\mathrm{C}%
_{\mathrm{\tilde{G}},d}(\Delta \phi^{\mu})$ and the d-dimensional Cartesian
space $\mathrm{C}_{d}$, i.e.,
\begin{align}
V_{\mathrm{\tilde{G},}d}[\Delta \phi^{\mu},\Delta x^{\mu},k_{0}^{\mu}]  &
:\nonumber \\
\mathrm{C}_{\mathrm{\tilde{G}},d}(\Delta \phi^{\mu})  &  =\{ \delta \phi^{\mu}\}
\Leftrightarrow \mathrm{C}_{d}=\{ \delta x^{\mu}\}.
\end{align}
where $\Leftrightarrow$\ denotes an ordered mapping under fixed changing rate
of integer multiple $k_{0}^{\mu},$\ and $\mu$ labels the spatial direction.
The adjective "quasi-perfect" means the total size $\Delta \phi^{\mu}$ of
$\mathrm{C}_{\mathrm{\tilde{G}},d}$\ doesn't exactly match the total size
$\Delta x^{\mu}$ of $\mathrm{C}_{d}$, i.e., $\Delta \phi^{\mu}\neq k_{0}^{\mu
}\Delta x^{\mu},$ and $\left \vert (\Delta \phi^{\mu}-k_{0}^{\mu}\Delta x^{\mu
})/\Delta \phi^{\mu}\right \vert \ll1$. According to above mismatch condition
$\Delta \phi^{\mu}\neq k_{0}^{\mu}\Delta x^{\mu},$ and $\left \vert (\Delta
\phi^{\mu}-k_{0}^{\mu}\Delta x^{\mu})/\Delta \phi^{\mu}\right \vert \ll1$, for a
P-variant, there must exist more than one type of group-changing elements on
it. Therefore, for a P-variant, there exist two kinds of group-changing
elements $\delta \phi^{A},$ $\delta \phi^{B}$ on d-dimensional Cartesian space
$\mathrm{C}_{d}$. The perturbative condition becomes
\begin{align}
\Delta \phi^{\mu}  &  =\sum \limits_{i}\delta \phi^{A}+\sum \limits_{j}\delta
\phi_{j}^{B},\\
\left \vert \sum \limits_{i}\delta \phi_{j}^{A}\right \vert  &  \gg \left \vert
\sum \limits_{j}\delta \phi_{j}^{B}\right \vert .\nonumber
\end{align}

In general, for P-variants, beside 1-th order representation without doing
K-projection and 0-th order representation under K-projection, there exists an
additional representation -- hybrid-order representation under partial
K-projection. By using hybrid-order representation under partial K-projection,
we have a usual quantum field description for a P-variant. The key point is to
consider the group-changing elements $\delta \phi^{B}$ to be extra objects on a
rigid uniform zero lattice that is partial K-projected from original U-variant.

We take 1D P-variant $V_{\mathrm{\tilde{U}(1),}1}[\Delta \phi,\Delta x,k_{0}]$
of non-compact $\mathrm{\tilde{U}(1)}$ Lie group as an example to show its
hybrid-order representation under partial K-projection.

Firstly, we do partial K-projection on the original U-variant
$V_{0,\mathrm{\tilde{U}(1),}1}[\Delta \phi^{A},\Delta x,k_{0}]$ and get a
compact group on zero lattice of "two-sublattice", i.e., $\phi(x)=2\pi
N(x)+\varphi(x).$ We then relabel the group-changing space $\mathrm{C}%
_{\mathrm{U(1)},1}(\Delta \phi)$ by two numbers ($N(x),\varphi(x)$):
$\varphi(x)$ is compact phase angle, the other is the integer winding number
of unit cell of zero lattice $N(x)$. $\varphi(x)$ can be canceled by choosing
a special projection angle $\theta$. Next, we do \emph{compactification }for
the extra group-changing elements $\delta \phi^{B}$. On the zero lattice
$N(x),$ to exact determine an extra group-changing element, one must know its
position of lattice site $N(x)$ together with its phase angle on this site
$\varphi(x).$ Due to the compactification, the non-compact phase angle $\phi$
turns into a compact one $\varphi.$ As a result, on zero lattice, the extra
group-changing elements $\delta \phi_{i}^{B}(x_{i})$ of $\hat{U}(\delta \phi
_{i}^{B}(x_{i}))$ is reduced into group operation $\hat{U}(\delta \varphi
_{i}(N_{i}(x_{i})))$. Here, $\hat{U}(\delta \varphi_{i}(N_{i}(x_{i})))$ is a
local phase operation that changing phase angle from $\varphi_{0}$ to
$\varphi_{0}+\delta \varphi_{i}(N_{i}(x_{i})).$ Therefore, we have a group of
local phase operations on zero lattice. By using the usual quantum field of
compact \textrm{U(1)} group, we can fully describe it.

For a higher-dimensional $\mathrm{\tilde{S}\tilde{O}}\mathrm{(d)}$ P-variant
$V_{\mathrm{\tilde{S}\tilde{O}(d),}d}[\Delta \phi^{\mu},\Delta x^{\mu}%
,k_{0}^{\mu}]$, we can use similar approach to represent the system. In
continuum limit, a higher-dimensional P-variant $V_{\mathrm{\tilde{S}\tilde
{O}(d)\tilde{G},}d}[\Delta \phi^{\mu},\Delta x^{\mu},k_{0}^{\mu}]$ is
characterized by a usual quantum field of compact \textrm{U(1)}$\times
$\textrm{SO(d)} group in quantum field theory.

\subsubsection{The changings of $\mathrm{\tilde{S}\tilde{O}}$\textrm{(d)}
variants}

The changings of $\mathrm{\tilde{S}\tilde{O}}$\textrm{(d)} variant
($V_{\mathrm{\tilde{S}\tilde{O}(d),}d}[\Delta \phi^{\mu},\Delta x^{\mu}%
,k_{0}^{\mu}]$) is prelude of our universe in physics. In this paper we focus
on its different types of changings.\ For $V_{\mathrm{\tilde{S}\tilde{O}%
(d),}d}[\Delta \phi^{\mu},\Delta x^{\mu},k_{0}^{\mu}]$, there are three types
of changings: global/local expand/contract, and local shape changings.

1) Globally \emph{expanding} or \emph{contracting} $\mathrm{C}_{\mathrm{\tilde
{S}\tilde{O}(d)},d}(\Delta \phi^{a})$\ with changing its corresponding size on
Cartesian space $\mathrm{C}_{d}$: The operation of contraction/expansion on
group-changing space is $\tilde{U}(\delta \phi^{a})=e^{i(\delta \phi^{a}%
T^{a})\cdot \hat{K}^{a}}$ where $\delta \phi^{a}=(\Delta \phi^{a})^{\prime
}-\Delta \phi^{a}$ and $\hat{K}^{a}=-i\frac{d}{d\phi^{a}}.$ In the following
part, we point out that globally expand/contract of group-changing space in a
variant corresponds to the generation/annihilate of particles in quantum mechanics;

2) Locally \emph{expanding} or \emph{contracting} $\mathrm{C}_{\mathrm{\tilde
{S}\tilde{O}(d)},d}(\Delta \phi^{a})$\ without changing its corresponding size
on Cartesian space $\mathrm{C}_{d}$: The operation of contraction/expansion on
group-changing space becomes local. In the following part, we point out that
this type of time-dependent changings of a variant corresponds to the motion
of particles in quantum mechanics with fixed particle's number;

3) Locally \emph{shape} changings\emph{ }on Cartesian space $\mathrm{C}_{d}$:
Locally shape changing of $\mathrm{C}_{\mathrm{\tilde{S}\tilde{O}(d)}%
,d}(\Delta \phi^{a})$ on Cartesian space $\mathrm{C}_{d}$\ ($d>1$) leads to
curved space and is relevant to the theory of quantum gravity.

In the earlier paper, we had give detailed discussion on the theory of
changings from global/local expand/contract. In this paper, we will focus on
the third type of changings (local shape changings) and the relationship
between three types of changings.

\subsubsection{Representations for shape changings of $\mathrm{\tilde{S}%
\tilde{O}}$\textrm{(d)} variant}

An $\mathrm{\tilde{S}\tilde{O}}$\textrm{(d)} variant is described by mappings
between the Clifford group-changing space and Cartesian space%
\begin{align}
V_{\mathrm{\tilde{S}\tilde{O}(d)},d}[\Delta \phi^{i},\Delta x^{i},k_{0}^{i}]
&  :\nonumber \\
\mathrm{C}_{\mathrm{\tilde{S}\tilde{O}(d)},d}(\Delta \phi^{i})  &  =\{
\delta \phi^{i}\} \Leftrightarrow \mathrm{C}_{d}=\{ \delta x^{i}\}
\end{align}
These mappings are characterized by
\begin{equation}
\mathcal{T}(\delta x^{i})\leftrightarrow \hat{U}^{\mathrm{T}}(\delta \phi
^{i})=e^{i\cdot \delta \phi^{i}\Gamma^{i}}%
\end{equation}
where $\delta \phi^{i}=k_{0}^{i}\cdot(\Delta x^{i}).$ Without considering the
total volume changing of the system and with the fixed changing rate
$k_{0}^{i}=k_{0}$, the local shape changings comes from local changings of the
$d-1$ compact phase angles $\delta \phi^{i}$. To characterize $\delta \phi^{i}$,
there are two kinds of representations -- geometry representation by fixing
$\Gamma^{i}$ and matrix representation by fixing $\Delta x^{i}$.

To derive the two representations, we do K-projection on $\mathrm{\tilde
{S}\tilde{O}}$\textrm{(d)} uniform variant $V_{\mathrm{\tilde{S}\tilde{O}%
(d),}d}[\Delta \phi^{i},\Delta x^{i},k_{0}]$ and get a uniform d-dimensional
zero lattice. Then, we consider the perturbation on it and get a perturbative
uniform variant. The extra changings of an original uniform variant can be
characterized either by a non-uniform zero lattice within geometric
representation or a deformed matrix network within matrix representation.

\paragraph{Geometric representation}

Firstly, we discuss the geometry representation for a perturbative uniform
variant by considering a non-uniform zero lattice.

Now, we begin with a uniform $\mathrm{\tilde{S}\tilde{O}}$\textrm{(d)} variant
by geometry representation via \textquotedblleft \emph{topological
lattice}\textquotedblright \ on Cartesian spacetime.

Along an arbitrary direction after shifting the distance $l_{0}$ (or $t_{0}$),
the phase angle of the ground state changes $2\pi.$ We then do
\emph{compactification} on the Clifford group-changing space $\mathrm{C}%
_{\mathrm{\tilde{S}\tilde{O}(d)}}$. After compactification, the coordinate of
$\mathrm{C}_{\mathrm{\tilde{S}\tilde{O}(d)}}$ along the given direction
$\mathbf{e}^{\mu}$ is reduced to a compact one, i.e., $\phi^{i}(x)=2\pi
N^{i}(x)+\varphi^{i}(x).$ We relabel a position in spacetime by two numbers
($\varphi^{i}(x),N^{i}(x)$): $\varphi^{i}(x)$ is a small phase angle
$\varphi^{\mu}(x)\in \lbrack0,2\pi)$, the other is a very large integer number
$N^{i}(x)$. Now, we have a theory of \emph{compact} \textrm{SO(d)} group on a
lattice labeled by $n^{i}(x)$ that make up a \emph{\textquotedblleft
topological\textquotedblright} version lattice. We call it\emph{ topological
spacetime}. It is obvious that for the unit cell of the topological lattice,
there are $2^{d}$ zeroes.

Then, the topological lattice of a uniform $\mathrm{\tilde{S}\tilde{O}}%
$\textrm{(d)} variant is defined as:

\textit{Definition: A topological lattice of a uniform $\mathrm{\tilde
{S}\tilde{O}}$\textrm{(d)} variant is defined by considering periodically
changing of phases of which the phase angle changes }$2\pi$\textit{ during
shifting a lattice distance. The lattice sites are denoted by }$N^{i}%
(x)=\frac{1}{2\pi}\phi^{i}(x)-\frac{1}{2\pi}\varphi^{i}(x).$

Now, we have a \emph{geometry representation} of a uniform $\mathrm{\tilde
{S}\tilde{O}}$\textrm{(d)} variant that is a uniform d-dimensional topological
lattice with fixed lattice sites $l_{0}\Delta N^{i}.$ In general, we may set
$l_{0}=t_{0}=1$.

From above discussion, according to the higher-order variability, the
perturbative uniform variant is characterized by the local spatial translation
operators $\mathcal{T}(\Delta x^{i})\leftrightarrow U^{\mathrm{T}}(\delta
\phi^{i}).$ On Cartesian space, the spatial coordinates locally change,
$(x^{i})_{\mathrm{curved}}=(x^{i})^{\prime}$. Correspondingly, the spatial
translation operators locally change, i.e.,
\begin{equation}
\mathcal{T}(\Delta x^{i})\rightarrow \mathcal{T}((\Delta x^{i})^{\prime
})\leftrightarrow \hat{U}^{\mathrm{T}}=e^{i\Gamma^{i}k_{0}(\Delta
x^{i})^{\prime}}.
\end{equation}
Now, the original uniform topological lattice with uniform lattice distances
$\Delta x^{\mu}$ slightly deviated from the original position: the distances
between two nearest-neighbor lattice sites deform, i.e., $(\Delta x^{\mu
}(N))^{\prime}-\Delta x^{i}=e_{i}(N),$ where $e_{i}(N)$ are vierbein fields
that are the difference between the geometric unit-vectors of the original
frame and the deformed frame.

Then, we discuss the theory in continuum limit.

In the continuum limit $\Delta x^{\mu}\gg1$, the spatial coordinates become
continuously changing
\begin{equation}
(\Delta x^{i}(N))^{\prime}\rightarrow \Delta x^{i}(x).
\end{equation}
Now, in geometry representation, the non-uniform topological lattice is
characterized by a curved space. The geometry fields (vierbein fields $e^{a}$
and spin connections $\omega^{ab}$) of the curved space are determined by the
non-uniform local coordinates, $(\Delta x^{i}(x))^{\prime}$. To characterize
the deformed topological lattice, with the help of the vierbein fields $e^{a}%
$, the space metric is defined by
\begin{equation}
e_{i}^{a}e_{b}^{i}=\delta_{b}^{a}\,,\quad e_{i}^{a}e_{a}^{j}=\delta_{i}^{j},
\end{equation}
and
\begin{equation}
e_{\alpha}^{a}e_{\beta}^{b}=g_{\alpha \beta}.
\end{equation}
The Riemann curvature 2-form is written as
\begin{equation}
R_{b}^{a}=d\omega_{b}^{a}+\omega_{c}^{a}\wedge \omega_{b}^{c},
\end{equation}
where $R_{b\mu \nu}^{a}\equiv e_{\alpha}^{a}e_{b}^{\beta}R_{\beta \mu \nu
}^{\alpha}$ are the components of the usual Riemann tensor projection on the
tangent space.

\paragraph{Matrix representation}

Next, we discuss the matrix representation for a perturbative uniform
$\mathrm{\tilde{S}\tilde{O}}$\textrm{(d)} variant by the changings of the
$\Gamma$-matrix on a uniform zero lattice. Within matrix representation, the
(perturbative) uniform $\mathrm{\tilde{S}\tilde{O}}$\textrm{(d)} variant is
characterized by a (deformed) \emph{matrix network}.

Then we define matrix network:

\textit{Definition: The matrix network of a perturbative uniform
$\mathrm{\tilde{S}\tilde{O}}$\textrm{(d)} variant is described by }%
$\Gamma^{\{n^{i},m^{j}\}}$\textit{ on the links between two nearest-neighbor
lattice sites }$n^{i}$\textit{ and }$m^{j}$\textit{ of the topological lattice
of spacetime. Or, }$\Gamma^{\{n^{i},m^{j}\}}$\textit{ on different paired
links of the topological lattice of spacetime constitute\ a matrix network. }

In the following parts, we will show that in continuum limit, the matrix
network turns into field for a special \textrm{SO(d)} rotor $\Gamma^{i}(x,t)$.
The matrix network $\Gamma^{\{n^{i},m^{j}\}}$ on links of the topological
lattice becomes indispensable to characterize different perturbative uniform
$\mathrm{\tilde{S}\tilde{O}}$\textrm{(d)} variant.

According above discussion, the deformation process of a uniform
$\mathrm{\tilde{S}\tilde{O}}$\textrm{(d)} variant\ can be representation by
local operations, $\hat{S}(x)$. We then use matrix representation to
characterize these shape changings via local operations, i.e.,
\begin{align}
\mathcal{T}((\Delta x^{i})^{\prime})  &  \leftrightarrow \hat{U}=e^{i\Gamma
^{i}k_{0}(\Delta x^{i})^{\prime}}\nonumber \\
&  =\hat{S}(x)\mathcal{T}(\Delta x^{\mu})(\hat{S}(x))^{-1},
\end{align}
\ where the operation $\hat{S}(x)=e^{i\phi_{i}(x)\Gamma^{i}}$ characterizes
the local changes.

Consequently, under the local operations $\hat{S}(x)$, the uniform matrix
network $\Gamma_{\mathrm{flat}}^{\{n^{i},m^{i}\}}$ on flat spacetime turns
into a non-uniform one $\Gamma_{\mathrm{curved}}^{\{n^{i},m^{i}\}}(x)$, i.e.,
\begin{equation}
\Gamma_{\mathrm{curved}}^{\{n^{i},m^{i}\}}(x)=\hat{S}(x)\Gamma_{\mathrm{flat}%
}^{\{n^{i},m^{i}\}}(\hat{S}(x))^{-1}.
\end{equation}
In particular, we emphasize that the coordinates do not change any more, i.e.,
$(x^{i}(x))_{\mathrm{curved}}=(x^{i}(x))_{\mathrm{flat}}.$

In continuum limit, the matrix network turns into field for a \textrm{SO(d)}
rotor $\Gamma^{i}(x)$, i.e.,
\[
\Gamma^{i}(x,t)=\hat{S}(x)\Gamma^{i}(\hat{S}(x))^{-1}.
\]
Now, the coordinate unit vectors $\mathbf{e}^{i}$ (the fundamental vectors
along $x^{i}$-direction of spacetime becomes $\Gamma^{\mu},$ i.e.,
$\mathbf{e}^{i}=\Gamma^{\mu}.$ The anti-commutation condition matrices
$\Gamma^{i}$ of Clifford group-changing space indicate a quantum character of
spacetime\cite{con}, i.e.,
\begin{equation}
\{ \mathbf{e}^{\mu},\mathbf{e}^{\nu}\}=\{ \Gamma^{\mu},\Gamma^{\nu}%
\}=2\delta_{\mu \nu}%
\end{equation}
and
\begin{equation}
\lbrack \mathbf{e}^{\mu},\mathbf{e}^{\nu}]=\left[  \Gamma^{\mu},\Gamma^{\nu
}\right]  \neq0.
\end{equation}

\subsection{Fundamental physics theory for quantum spacetime}

In this paper, we focus on the\textit{ }($d+1$)-dimensional $\mathrm{\tilde
{S}\tilde{O}}$\textrm{(d+1)} physical variant\textit{ }$V_{\mathrm{\tilde
{S}\tilde{O}(d+1)},d+1}(\Delta \phi^{\mu},\Delta x^{\mu},k_{0},\omega_{0})$
that plays the role of physical reality in our universe\cite{kou1}. Therefore,
our world is really a \emph{uniform}, \emph{holistic} changing structure with
1-th order spatial-tempo variability.

According to the 1-th order spatial-tempo variability, physical laws (special
relativity, general relativity and quantum mechanics) emerge. To make it
clear, we introduce the tower of changings.

Modern physics comes from the tower of changings with the changings in
different levels and different physical laws emerge from the changings in
different levels:

\begin{enumerate}
\item 0-th level physics structure is the uniform physical variant -- a
uniform \emph{changing} structure in Cartesian space named "\emph{vacuum}" or
"\emph{ground state}" in usual physics;

\item 1-th level physics structure is the global expansion and contraction
types of "\emph{changings}" of the physical variant named "\emph{matter}" in
usual physics. Now, the size of the group-changing space is changed;

\item 2-th level physics structure is the "\emph{changings}" of the physical
variant without size changings. There are two types of motions: one is local
expansion and contraction changings, which is named "\emph{quantum motion}" of
matter, the other is local shape changings, which is named "\emph{spacetime
curving}".
\end{enumerate}

\subsubsection{$\mathrm{\tilde{S}\tilde{O}}$\textrm{(d+1)} physical variants}

Firstly, we introduce the ($d+1$)-dimensional $\mathrm{\tilde{S}\tilde{O}}%
$\textrm{(d+1)} physical variant\textit{ }$V_{\mathrm{\tilde{S}\tilde{O}%
(d+1)},d+1}(\Delta \phi^{\mu},\Delta x^{\mu},k_{0},\omega_{0})$ that is the
physical reality in our world, a mapping between\textrm{ }$\mathrm{\tilde
{S}\tilde{O}}$\textrm{(d+1)} Clifford group-changing space\textit{
}$\mathrm{C}_{\mathrm{\tilde{S}\tilde{O}(d+1)},d+1}$\textit{\ }and a rigid
spacetime $\mathrm{C}_{d+1},$\textit{ }i.e.,\textit{ }%
\begin{align}
V_{\mathrm{\tilde{S}\tilde{O}(d+1)},d+1}[\Delta \phi^{\mu},\Delta x^{\mu}%
,k_{0}^{\mu}]  &  :\nonumber \\
\mathrm{C}_{\mathrm{\tilde{S}\tilde{O}(d+1)},d+1}  &  =\{ \delta \phi^{\mu
}\} \nonumber \\
&  \Leftrightarrow \mathrm{C}_{d+1}=\{ \delta x^{\mu}\}
\end{align}
where $\Leftrightarrow$\ denotes an ordered mapping with fixed changing rate
of integer multiple $k_{0}$ or $\omega_{0},$\ and $\mu$ labels the spatial
direction. A ($d+1$)-dimensional Clifford group-changing space\textit{
}$\mathrm{C}_{\mathrm{\tilde{S}\tilde{O}(d+1)},d+1}(\Delta \phi^{\mu})$\textit{
}is described by $d+1$ series of numbers of group elements $\phi^{\mu}$
arranged in size order with unit "vector" as Gamma matrices $\Gamma^{\mu}$
obeying Clifford algebra $\{ \Gamma^{i},\Gamma^{j}\}=2\delta^{ij}$. In
particular, we set light speed $c=1,$ and have $\omega_{0}=k_{0}$.

The ($d+1$)-dimensional Clifford group-changing space\textit{ }$\mathrm{C}%
_{\mathrm{\tilde{S}\tilde{O}(d+1)},d+1}(\Delta \phi^{\mu})$ has orthogonality.
A ($d+1$)-dimensional Clifford group-changing space $\mathrm{C}%
_{\mathrm{\tilde{S}\tilde{O}(d+1)},d+1}(\Delta \phi^{\mu})$ obeys
non-commutating geometry due to $\{ \Gamma^{\mu},\Gamma^{\nu}\}=2\delta
^{\mu \nu}$. For two vectors in $\mathrm{C}_{\mathrm{\tilde{S}\tilde{O}%
(d+1)},d+1}(\Delta \phi^{\mu})$, $\mathbf{\phi}_{\mathrm{A}}=\phi
_{\mathrm{A,}\mu}e^{\mu}$ and $\mathbf{\phi}_{\mathrm{B}}=\phi_{\mathrm{B}%
,\mu}e^{\mu},$ the add and subtract rules become $\mathbf{\phi}_{\mathrm{A}%
}\pm \mathbf{\phi}_{\mathrm{B}}=%
{\displaystyle \sum \nolimits_{\mu}}
(\phi_{\mathrm{A,}\mu}e^{\mu}+\phi_{\mathrm{B},\mu}e^{\mu}).$ The distance
between $\mathbf{\phi}_{\mathrm{A}}$ and $\mathbf{\phi}_{\mathrm{B}}$ becomes
$\left \vert \mathbf{\phi}_{\mathrm{A}}-\mathbf{\phi}_{\mathrm{B}}\right \vert
^{2}=%
{\displaystyle \sum \nolimits_{\mu}}
(\phi_{\mathrm{A,}\mu}e^{\mu}-\phi_{\mathrm{B},\mu}e^{\mu})^{2}.$

In the following parts, we develop a new, and complete theoretical framework
for quantum gravity based on the Variant hypothesis:

\textit{Variant hypothesis about physical reality -- Physical reality in our
universe is a (}$d+1$\textit{)-dimensional }$\mathrm{\tilde{S}\tilde{O}}%
$\textrm{(d+1)} \textit{physical variant }$V_{\mathrm{\tilde{S}\tilde{O}%
(d+1)},d+1}(\Delta \phi^{\mu},\Delta x^{\mu},k_{0},\omega_{0})$\textit{. In our
universe, we have }$d=3$\textit{.}

\subsubsection{Higher-order variability for physical variant -- 0-th level
physics structure}

As the base of the tower, the uniform $\mathrm{\tilde{S}\tilde{O}}%
$\textrm{(d+1)}\textit{ }physical variant becomes the 0-th level physics
structure. To accurately characterize the physical variant, we consider its
1-th order spatial-tempo variability, which corresponds to its
geometry/dynamic properties, respectively.

The 1th order spatial-tempo variability is determined by the following
equation,
\begin{equation}
\mathcal{T}(\delta x^{\mu})\leftrightarrow \hat{U}(\delta \phi^{\mu}),
\end{equation}
where $\hat{U}(\delta \phi^{\mu})=e^{i\cdot \delta \phi^{\mu}\Gamma^{\mu}},$
$\Gamma^{\mu}$ is Gamma generator $\{ \Gamma^{i},\Gamma^{i}\}=2\delta^{ij}$
and $\delta \phi^{\mu}=k_{0}\delta x^{\mu}$ is the corresponding phase
angle\textit{. }In particular, due to $c=1,$ we have the characterized
length/time $l_{0}=t_{0}=\frac{2\pi}{k_{0}}$\textbf{. }$l_{0}=2l_{p}$ is the
twice of Planck length (This fact will be proved in the following parts). For
simplicity, we can denote it by the following equation
\begin{align}
\mathcal{T}(\delta x)  &  \leftrightarrow \hat{U}(\phi),\text{ }\nonumber \\
\text{or }\mathcal{T}(\delta x)\cdot \hat{U}^{-1}(\phi)  &  =1.
\end{align}

We point out that quantum flat spacetime looks like a special \emph{spacetime
crystal with topological constraints.} In 2012, Frank Wilczek proposed the
idea of time crystal \cite{Wilczek2012}, of which a many-body system
self-organizes in time and starts spontaneously to undergo a periodic motion.
If there is an additional topological constraint on spacetime crystal, it
turns into a spacetime with 1-th order variability of tempo transformation.

On the other hand, 1-th order rotation variability is defined by
\begin{equation}
\hat{U}^{\mathrm{R}}\leftrightarrow \hat{R}_{\mathrm{space}}%
\end{equation}
where $\hat{U}^{\mathrm{R}}$ is (compact) \textrm{SO(d+1)} rotation operator
on Clifford group-changing space $\hat{U}^{\mathrm{R}}\Gamma^{I}\mathbf{(}%
\hat{U}^{\mathrm{R}})^{-1}=\Gamma^{I^{\prime}},$ and $\hat{R}_{\mathrm{space}%
}$ is \textrm{SO(d+1)} rotation operator on Cartesian space, $\hat
{R}_{\mathrm{space}}x^{I}\hat{R}_{\mathrm{space}}^{-1}=x^{I^{\prime}}.$ After
doing a global composite rotation operation $\hat{U}^{\mathrm{R}}\cdot \hat
{R}_{\mathrm{space}}$, the system is invariant. The 1-th order rotation
variability will play important role to determine scattering amplitude for
gravitational waves on twistor space.

Physical law always comes from linearization from "\emph{uniform}
\emph{changing}" of a system.

According to spatial variability $\mathcal{T}(\delta x^{i})\leftrightarrow
\hat{U}^{\mathrm{T}}(\delta \phi^{i})=e^{i\cdot k_{0}\delta x^{i}\Gamma^{i}},$
($i=x_{1},x_{2},$...$,x_{d}$), we have a fixed spatial changing rate for the
system, i.e., $k_{0}\neq0$. With linearization at $k=k_{0}$, we have
dispersion as
\begin{equation}
\omega=\omega_{0}+c(k-k_{0})
\end{equation}
where $c=\frac{\partial \omega}{\partial k}\mid_{k=k_{0}}$ becomes an effective
"light" velocity. Then\emph{ Lorentz invariant emerges}.

In addition, quantum mechanics emerges from 1-th order tempo variability (or a
uniform motion of the group-changing space along $\Gamma^{t}$ direction),
i.e., $\omega_{0}\neq0$. For uniform physical variant, the energy density
$\rho_{E}=\frac{\Delta E}{\Delta V}$ is constant. With linearization at
$\omega=\omega_{0}$, we have
\begin{align}
\rho_{E}(\omega_{0}+\delta \omega)  &  =\rho_{E}(\omega_{0})\nonumber \\
+\frac{\delta \rho_{E}}{\delta \omega}  &  \mid_{\omega=\omega_{0}}\delta
\omega+...
\end{align}
where $\frac{\delta \rho_{E}}{\delta \omega}\mid_{\omega=\omega_{0}}=\rho_{J}$
is called the density of (effective) "angular momentum". In the following
parts, we point out that the "angular momentum" $\rho_{J}$ of an elementary
particles is just Planck constant $\hbar$ and the quantization condition in
quantum mechanics comes from the linearization of energy density $\rho_{E}$
via $\omega$ near $\omega_{0}$.

\subsubsection{Matter -- size changings of group-changing space}

In this section, we discuss the 1-th level physics structure by defining matter.

Matter is defined as globally \emph{expanding} or \emph{contracting}
$\mathrm{C}_{\mathrm{\tilde{S}\tilde{O}(d+1)},d+1}$ group-changing space\ with
changing its corresponding size in rigid space $\mathrm{C}_{d+1}$. Globally
expand/contract of group-changing space corresponds to the
generation/annihilate of particles in quantum mechanics. The generation or
annihilation operation of matter is defined by the operator of
contraction/expansion of $\mathrm{C}_{\mathrm{\tilde{S}\tilde{O}(d+1)},d+1}$
group-changing space in Cartesian space\textit{ }$\mathrm{C}_{d}$, i.e.,
$\hat{U}(\delta \phi^{a})=e^{i(\delta \phi^{a})\cdot \hat{K}^{a}}$ where
$\delta \phi^{a}$ and $\hat{K}^{a}=-i\frac{d}{d\phi^{a}}$ ($a=x,y,z,t$).

When we consider matter on spacetime, the original uniform physical variant
turns into the perturbative uniform physical variant that is about expanding
or contracting $\mathrm{C}_{\mathrm{\tilde{S}\tilde{O}(d+1)},d+1}$
group-changing space in rigid/curved spacetime.

\subsubsection{Motions -- changings of mappings between $\mathrm{C}%
_{\mathrm{\tilde{S}\tilde{O}(d+1)},d+1}$ and $\mathrm{C}_{d+1}$}

In this part, we discuss the 2-th level physics structure by classifying the
types of motion that corresponds to different types of time-dependent
changings of $\mathrm{\tilde{S}\tilde{O}}$\textrm{(d+1)} physical variants
$V_{\mathrm{\tilde{S}\tilde{O}(d+1)},d+1}(\Delta \phi^{\mu},\Delta x^{\mu
},k_{0},\omega_{0})$ without size changings of group-changing space
$\mathrm{C}_{\mathrm{\tilde{S}\tilde{O}(d+1)},d+1}.$ There are two types of
motions, one is about motion of matter that corresponds to locally expanding
or contracting $\mathrm{C}_{\mathrm{\tilde{S}\tilde{O}(d+1)},d+1}(\Delta
\phi^{a})$\ without changing its corresponding size on Cartesian space
$\mathrm{C}_{d+1}$; The other is about curving of spacetime that corresponds
to locally shape changings\emph{ }on Cartesian space $\mathrm{C}_{d+1}$.

In earlier paper of \cite{kou1}, we had studied the motion of matter. Locally
expand/contract of group-changing space corresponds to the classical/quantum
motion of particles with fixed particle's number. Quantum motion describes the
ordered relative motion between group-changing elements of the elementary
particles that is characterized by Sch\"{o}rdinger equation. Classical motion
describes certain globally shift of a quantum/classical object with
ordered/disordered group-changing elements that is characterized by Newton equation.

Except for the motion for matter, there exists another type of motion --
curving spacetime that characterizes the\emph{ shape changings }of the
physical variant. The gravitational waves are collective modes curving
spacetime. In this paper, we focus on this type of motion.

\subsubsection{Invariant/symmetry}

In modern physics, it was known invariance/symmetry plays important role in
modern physics. In this section, we will show how invariance/symmetry emerge
from higher-order variability.

As shown in Fig.1, invariance/symmetry can be regarded as \emph{shadow} of
variability: 0-level invariance (or fixity) determines the invariance of
physical laws with fixed physical constants; 1-level invariance (or topology
stationarity) determines the invariance of the matter; 2-level invariance (or
symmetry) determines the invariance of motions.\begin{figure}[ptb]
\includegraphics[clip,width=0.63\textwidth]{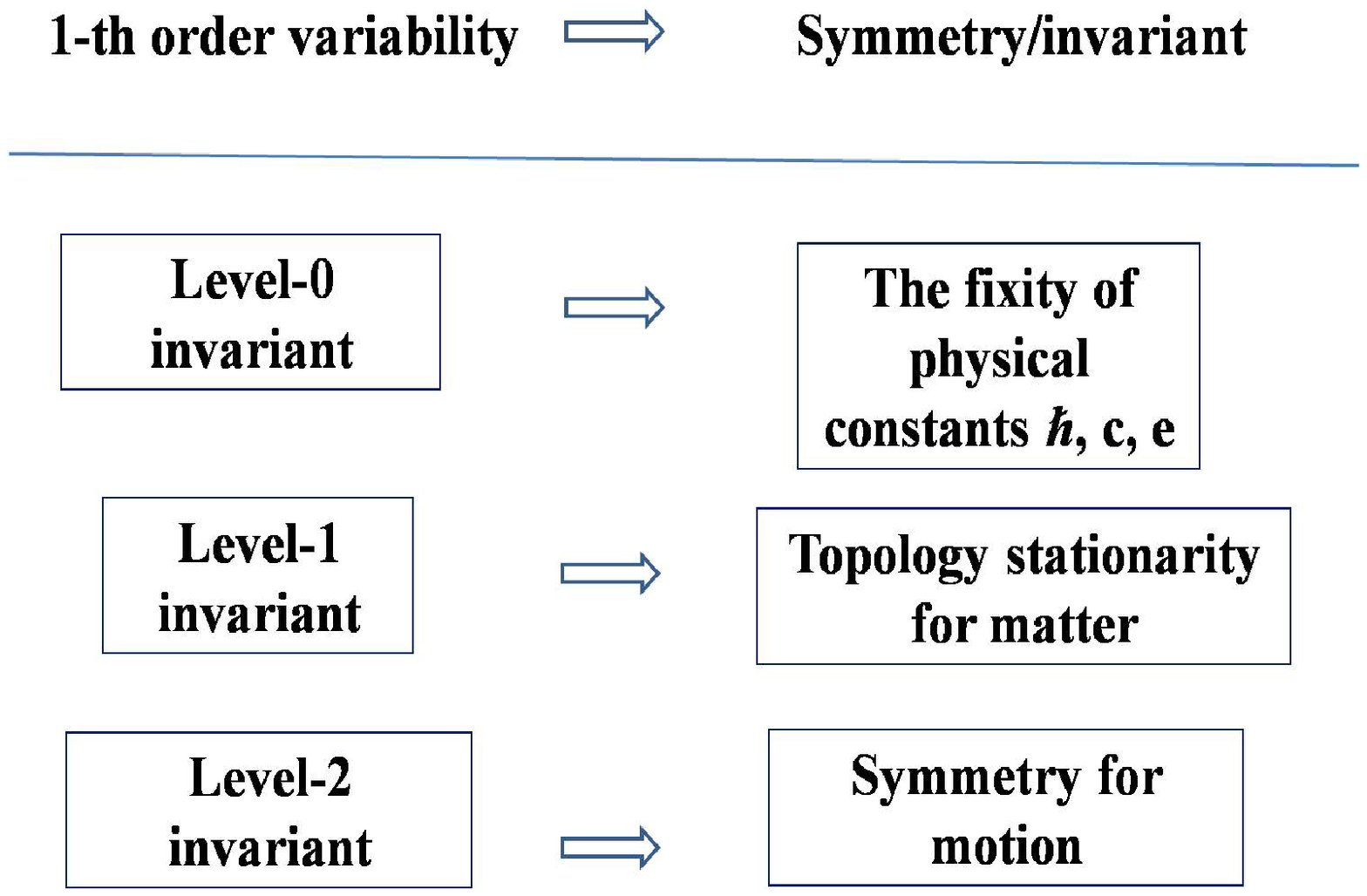}\caption{Invariance/symmetry
can be regarded as \emph{shadow} of variability}%
\end{figure}

\paragraph{Level-0 invariant: The fixity of physical constants}

Firstly, we discuss the invariant of 0-th level physics structure for physical reality.

For the level-0 physics, we have a uniform physical variant $V_{\mathrm{\tilde
{S}\tilde{O}(d+1)},d+1}(\Delta \phi^{\mu},\Delta x^{\mu},k_{0},\omega_{0})$
with 1-th order variability. The changing rates of group-changing spaces are
invariant that leads to \emph{fixity }of physical constants. Physical law
always comes from linearization from a system with "\emph{uniform}
\emph{changing}". The \emph{fixity} indicates an invariant of physical laws
(Lorentz invariant, and quantization condition, Schr\"{o}dinger equation,
...). The specific manifestation of invariance is the fixity of physical
constants, such as light speed $c$, Planck constant $\hslash$, ... All these
physical constants don't change with time and place. We point out that such an
invariance (or fixity) is protected by the 1-th order variability.

\paragraph{Level-1 invariant: Topology stationarity}

Next, we discuss the invariant of 1-th level physics structure for matter.

It was known that matter corresponds to globally expand or contract of the
group-changing space $\mathrm{C}_{\mathrm{\tilde{S}\tilde{O}(d+1)},d+1}$ with
changing the size of the system. Elementary particles are $\pi$ phase changing
along different directions.

There exists an invariant for matter, i.e., their sizes of group-changing
space can never be changed. Such a invariance is called \emph{topology
stationarity} of matter. During the processes of motion, the size of the given
elementary particle doesn't change any more. Therefore, the topological
properties of a moving elementary particle are invariant.

The invariant of matter leads to \emph{differential homeomorphism invariance.}
The differential homeomorphism invariance is not usual symmetry/invariant of
the system. Instead, it is symmetry/invariant for matter. The differential
homeomorphism invariance denotes \emph{synchronous variability} between
quantum spacetime and matter.

In addition, in the following part, to characterize the topology stationarity
and the unification of spacetime and matter, we introduce a new concept --
\emph{the charge of spacetime} or \emph{spacetime charge} that will plays
important role in the \emph{general symmetry} for quantum spacetime.

\paragraph{Level-2 invariant: Symmetry for motion}

Finally, we discuss the invariant of 2-th level physics structure for motions.

It was known that motion corresponds to locally expand or contract of the
group-changing space $\mathrm{C}_{\mathrm{\tilde{S}\tilde{O}(d+1)},d+1}$
without changing their corresponding sizes. Different states of motions
correspond to different mappings between $\mathrm{C}_{\mathrm{\tilde{S}%
\tilde{O}(d+1)},d+1}$ and $\mathrm{C}_{d+1}$. If two states (or different
mappings between $\mathrm{C}_{\mathrm{\tilde{S}\tilde{O}(d+1)},d+1}$, and
$\mathrm{C}_{d+1}$ have same energy, we call such an invariance to be
\emph{symmetry} of motions.

For uniform physical variant under compactification, there exist two kinds of
symmetries -- one is about (discrete) translation symmetry $T(\delta x^{\mu
}),$ the others is about global symmetry (compact \textrm{U(1) }rotation
symmetry and global compact \textrm{SO(d+1) }rotation symmetry). Let us show
the detail.

According to the 1-th order variability $\mathcal{T}\leftrightarrow \hat{U}$,
under compactification, the continuous translation operation $\mathcal{T}%
(\delta x^{\mu})$ of the U-variant is reduced into a discrete spatiotemporal
translation symmetry $T(\delta x^{\mu})$ on the zero lattice, i.e.,
\begin{equation}
\mathcal{T}(\delta x^{\mu})\rightarrow T(\delta N^{\mu}).
\end{equation}
For zero lattice, one lattice site is equivalence to another. Then, for the
uniform zero lattice, we have a reduced translation symmetry denoted by the
following equation
\begin{equation}
T(\delta N^{\mu})\rightarrow1.
\end{equation}

On the other hand, under compactification, the operation $\hat{U}^{\mu}$ of
non-compact $\mathrm{\tilde{S}\tilde{O}}\mathrm{(d+1)}$ group belongs to
compact \textrm{U(1)}$\times$\textrm{SO(d+1)} group. On each lattice site of
zero lattice, we have an invariant under the compact \textrm{U(1)}$\times
$\textrm{SO(d+1)} group, i.e.,
\begin{equation}
\hat{U}^{\mu}\rightarrow \hat{U}_{\mathrm{U(1)}}\otimes \hat{U}%
_{\mathrm{SO(d+1)}}.
\end{equation}
For simplicity, we can denote them by the following equations
\begin{equation}
\hat{U}_{\mathrm{SO(d+1)}}\rightarrow1,\text{ }\hat{U}_{\mathrm{U(1)}%
}\rightarrow1.\nonumber
\end{equation}

After compactization and continuum $l_{0}\rightarrow0$, the 1-th order
variability is reduced to continuous spatiotemporal translation invariance,
together with internal compact \textrm{U(1)}$\times$\textrm{SO(d+1)} symmetry.
Therefore, with considering the spatiotemporal translation symmetry (or
$T(\delta x)=1$), the momentum $p$ along given spatial direction, mass $m$,
and energy $E$\ become conserved quantities; with considering the internal
\textrm{U(1) }phase symmetry, the particle number $N$ becomes a conserved
quantity; with considering the internal \textrm{SO(d+1)} symmetry, the angular
momentum becomes a conserved quantity.

For curved spacetime (a perturbative uniform physical variant), the situation
becomes complex. We don't have spatiotemporal translation invariance and
internal \textrm{SO(d+1)} rotating symmetry. Momentum, energy and angular
momentum are no more conserved quantities. However, the internal compact
\textrm{U(1)} symmetry is not broken. As a result, the particle number is
still a conserved quantity that corresponds to the globally expand or contract
of the group-changing space $\mathrm{C}_{\mathrm{\tilde{S}\tilde{O}(d+1)}%
,d+1}$ with changing the corresponding size. This characterizes topology
stationarity of matter.

\paragraph{Summary}

In the end of this section, we give a summary.

For 0-th level physics structure for physical reality, we have level-0
invariant that is the fixity of physical constants; For 1-th level physics
structure for matter, we have level-1 invariant that is the topology
stationarity of matter; for 2-th level physics structure for motion, we have
level-2 invariant that is the symmetry of motion. For a uniform physical
variant under compactification, we have both translation symmetry $T(\delta
x^{\mu})\ $and global symmetry (compact \textrm{U(1) }rotation symmetry and
global compact \textrm{SO(d+1) }rotation symmetry).

In addition, we point out that there exist additional invariant --
\emph{emergent (local) \textrm{SO(3,1)} Lorentz invariance.} We point out that
the (local) \textrm{SO(3,1)} Lorentz invariant is not a usual
symmetry/invariant of the system but a constraint from linear dispersion, or
the invariance of dispersion. The emergent (local) Lorentz invariant makes the
situation much more complex.

To characterize the internal, compact \textrm{SO(3+1)} structure of an
elementary particles by the description with non-compact \textrm{SO(3,1)}
Lorentz invariance, the theory for quantum spacetime becomes a theory with
infinite gauge fields! Now, we have an \textrm{SO(3)}$^{\mathrm{SO(3+1)}}$
gauge structure,\ of which each group element of \textrm{SO(3+1)} group for a
3D sub-manifold \textrm{M}$_{3}^{\mu}$ corresponds to an \textrm{SO(3)} gauge
theory. For different 3D sub-manifold \textrm{M}$_{3}^{\mu}$, there exist
different gauge fields, $A_{\mu}(x)$. In the following parts, we will discuss
this issue in detail.

\subsection{Theory for quantum flat spacetime}

\subsubsection{Quantum flat spacetime -- 0-th level physics structure}

Firstly, we develop the theory of quantum flat spacetime.

The quantum flat spacetime is a uniform physical variant that is defined as a
perfect mapping between Clifford group-changing space\textit{ }$\mathrm{C}%
_{\mathrm{\tilde{S}\tilde{O}(3+1)}}$\ and Cartesian spacetime $\mathrm{C}%
_{3+1}$, i.e.,
\begin{align*}
&  \text{Quantum flat spacetime }\\
&  =\text{ Uniform $\mathrm{\tilde{S}\tilde{O}}$\textrm{(d+1)} physical
variant.}%
\end{align*}

In mathematics, a flat quantum spacetime is defined by ordered mapping, i.e.,
the mapping from usual Cartesian spacetime $\mathrm{C}_{d+1}$ to the Clifford
group-changing space \textrm{C}$_{\mathrm{\tilde{S}\tilde{O}(d+1)}}$, i.e.,
\begin{equation}
\{ \phi^{\mu}\} \in \text{\textrm{C}}_{\mathrm{\tilde{S}\tilde{O}(d+1)}%
}\Leftrightarrow \{x^{\mu}\} \in \mathrm{C}_{d+1},
\end{equation}
where $\Leftrightarrow$ denotes space-mapping. Now, the size of the Cartesian
spacetime $\mathrm{C}_{d+1}$ matches the Clifford group-changing space
\textrm{C}$_{\mathrm{\tilde{S}\tilde{O}(d+1)}}$ and the changing rates along
different directions are all constant.

From definition of quantum flat spacetime, there exists 1-th order variability
of both spatial-tempo transformation and rotation transformation, i.e.,%
\begin{equation}
\mathcal{T}(\delta x^{\mu})\leftrightarrow \hat{U}(\delta \phi^{\mu}),
\end{equation}
where $\hat{U}(\delta \phi^{\mu})=e^{i\Gamma^{\mu}k_{0}\delta x^{\mu}}$ are
group translation operations in non-compact $\mathrm{\tilde{S}\tilde{O}}%
$\textrm{(3+1)} Lie group. The wave vector $k_{0}=\omega^{0}=\frac{2\pi}%
{l_{0}}$ ($c=1$) and $l_{0}=t_{0}$ is the characterized length/time.
$\Gamma^{\mu}$ are the Gamma matrices in the massive Dirac model.

To characterize the quantum flat spacetime, there are two types of
representations -- geometry representation, or matrix representation. Due to
the ordered mapping, the two representations are equal and can be transformed
each other.

We firstly characterize a quantum flat spacetime by geometry representation
via \textquotedblleft \emph{topological lattice}\textquotedblright \ on
Cartesian spacetime. \begin{figure}[ptb]
\includegraphics[clip,width=0.67\textwidth]{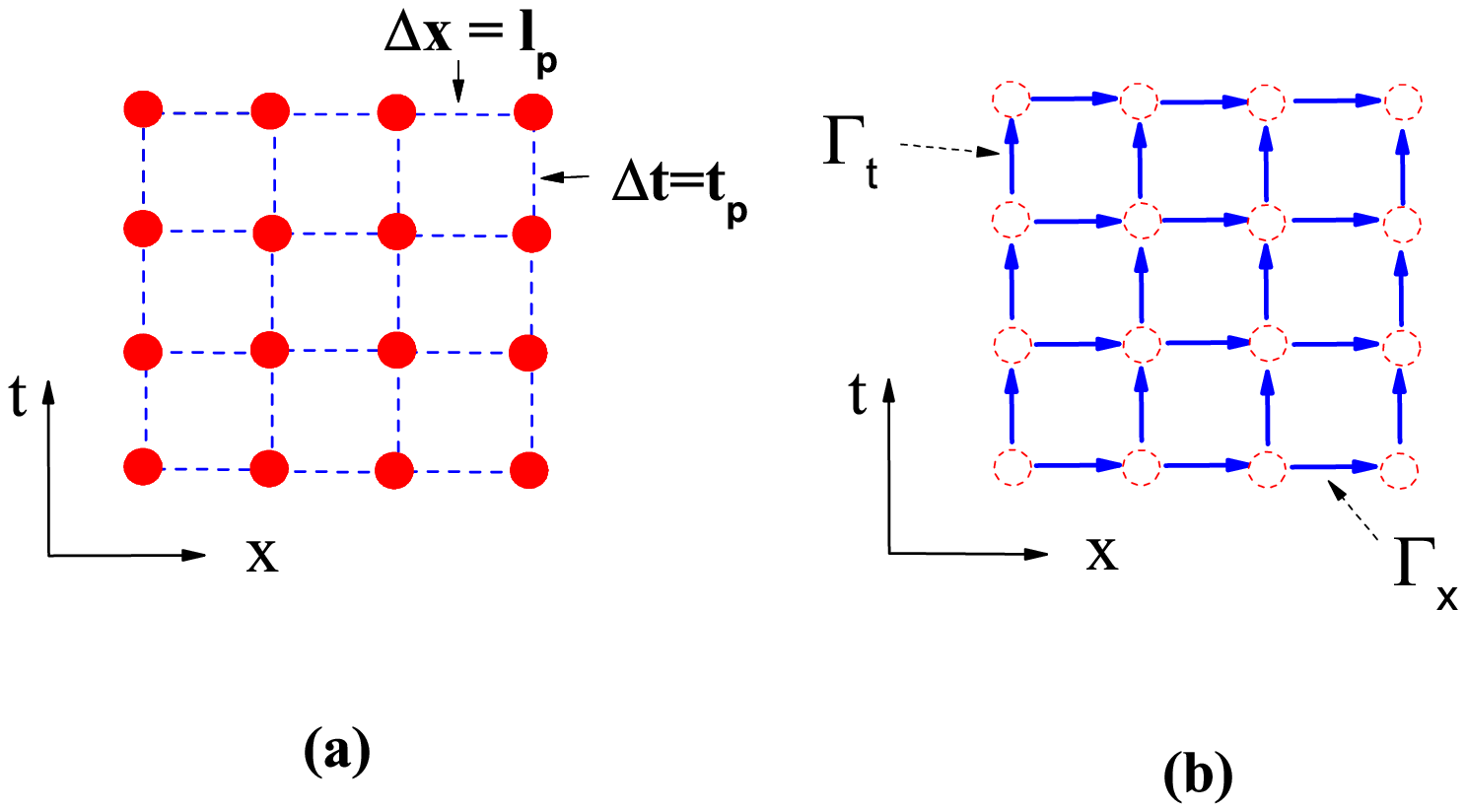}\caption{An illustration
for 1+1D flat quantum spacetime: (a) is geometry representation with 2D
uniform topological lattice that is denoted by solid red spots. The lattice
distance along spatial/tempo direction is Planck length/time ($l_{0}$/$t_{0}%
$). During an spatial/tempo shifting Planck length $l_{p}=l_{0}/2$ (or
$t_{0}/2$), the phase change of the vacuum is $\pi$; (b) is the matrix
representation with 2D uniform matrix network. The matrix network is described
by $\Gamma_{\mathrm{flat}}^{\{N^{\mu},M^{\mu}\}}$ (or $\Gamma_{x}$ and
$\Gamma_{t}$) on all links between two nearest-neighbor lattice sites (solid
blue arrows). }%
\end{figure}

According to the variability, the vacuum of quantum spacetime is defined by
the following relation,\quad \
\begin{equation}
\mathcal{T}(\delta x^{\mu})\leftrightarrow \hat{U}(\delta \phi^{\mu}%
)=e^{ik_{0}^{\mu}\cdot(\delta x^{\mu})\Gamma^{\mu}},
\end{equation}
where $k_{0}^{\mu}=k_{0}$. Along an arbitrary direction $e^{\mu}$
($\mu=x,y,z,t$) after shifting the distance $l_{0}$ (or $t_{0}$), the phase
angle of the ground state changes $2\pi.$ We then do \emph{compactification}
on the Clifford group-changing space $\mathrm{C}_{\mathrm{\tilde{S}\tilde
{O}(d+1)}}$. After compactification, the coordinate of $\mathrm{C}%
_{\mathrm{\tilde{S}\tilde{O}(d+1)}}$ along the given direction $\mathbf{e}%
^{\mu}$ is reduced to a compact one, i.e., $\phi^{\mu}(x)=2\pi N^{\mu
}(x)+\varphi^{\mu}(x).$ We relabel a position in spacetime by two numbers
($\varphi(x),N(x)$): $\varphi^{\mu}(x)$ is a phase angle $\varphi^{\mu}%
(x)\in \lbrack0,2\pi)$, $N^{\mu}(x)$ is winding number. Now, we have a theory
of \emph{compact} \textrm{SO(d+1)} group on a crystal labeled by $N^{\mu}(x)$
and get \emph{\textquotedblleft topological\textquotedblright} version lattice.

For quantum flat spacetime, the topological lattices along tempo direction and
those along spatial direction are symmetric and will be indistinguishable.
Now, we have two character lengths, the Planck length $l_{p}=G^{1/2}$ and
lattice unit of topological lattice $l_{0}$. What's the relationship between
them? In the following sections, we will answer this question and get
$l_{0}=2l_{p}$.

Fig.2(a) shows a 2D topological lattice of quantum flat spacetime. The sites
of the topological lattice of flat spacetime are $l_{0}N^{x}$ along an
arbitrary spatial direction and $t_{0}N^{t}$ along an arbitrary spatial/tempo
direction. Here, $N^{x}$ and $N^{t}$ are integer numbers. After shifting the
distance $\Delta x=l_{0}$,\ the phase angle of the system changes $2\pi,$
i.e., $T(l_{0})=e^{i\Gamma^{x}2\pi}=1;$ After shifting the time interval
$\Delta t=t_{0}$ along a tempo direction,\ the phase angle of the system
changes $2\pi,$ i.e., $T(t_{0})=e^{i\Gamma^{t}2\pi}=1.$ Therefore, the
periodic motion of vacuum indicates the existence of an internal
\textquotedblleft clock\textquotedblright \ of our spacetime with a period of
time $t_{0}$.

As illustrated in Fig.2(a), we have a \emph{geometry representation} of a
quantum flat spacetime that is a uniform (1+1)D topological lattice with fixed
lattice sites $l_{0}\Delta N^{\mu}.$ In general, we may set the lattice
distance to be unit $l_{0}=t_{0}=1$. In continuum limit, the quantum spacetime
is reduced to a usual, commutative Minkovski spacetime rather than
noncommutative spacetime.

Next, to characterize the quantum flat spacetime, we introduce matrix
representation via a \textquotedblleft \emph{matrix network}\textquotedblright.
The matrix network is described by $\Gamma^{\{N^{\mu},M^{\mu}\}}$ on the
\emph{links} between two nearest-neighbor lattice sites $N^{\mu}$ and $M^{\mu
}$ of the topological lattice. Or, $\Gamma^{\{N^{\mu},M^{\mu}\}}$ on different
links of the topological lattice of spacetime constitute\ a matrix network.
Fig.2(b) shows the matrix network $\Gamma^{\{N^{\mu},M^{\mu}\}}$ on links of
2D topological lattice that is indispensable to characterize different quantum spacetimes.

With the help of matrix representation, we can define "quantum states" of a
spacetime. A physical system in quantum mechanics is described by a Hilbert
space $\mathcal{E}$ that becomes the state space of the quantum system.

For the case of $d=3$, under matrix representation the Hilbert space
$\mathcal{E}$ of quantum spacetime consists of all four-by-four matrices on
links $\{N^{\mu},M^{\mu}\}$,
\begin{equation}
\mathcal{E}:\mathcal{H}_{QST}=\mathcal{H}_{\{(0,0,0,0),(1,0,0,0)\}}%
\otimes...\mathcal{H}_{\{N^{\mu},M^{\mu}\}}.
\end{equation}
The states of flat quantum spacetime are characterized by a constant matrix
network, $\{ \Gamma_{\mathrm{flat}}^{\{N^{\mu},M^{\mu}\}}(x),$ $\mu
=x,y,z,t\},$ i.e.,%
\begin{align}
\Gamma_{\mathrm{flat}}  &  =(\Gamma_{\mathrm{flat}}^{\{N^{x},M^{x}\}}%
,\Gamma_{\mathrm{flat}}^{\{N^{y},M^{y}\}}(x),\Gamma_{\mathrm{flat}}%
^{\{N^{z},M^{z}\}}(x),\Gamma_{\mathrm{flat}}^{\{N^{t},M^{t}\}}(x))\nonumber \\
&  =(\tau^{x}\otimes \sigma^{x},\tau^{x}\otimes \sigma^{y},\tau^{x}\otimes
\sigma^{z},\tau^{z}\otimes \vec{1}).
\end{align}

In the following parts, under matrix representation we may denote the ground
state of flat quantum spacetime in the Hilbert space $\mathcal{E}$ by vacuum
state $\left \vert \mathrm{vac}\right \rangle .$ Now, the corresponding
relationship $\mathcal{T}(\delta x^{\mu})\leftrightarrow \hat{U}(\delta
\phi^{\mu})=e^{ik_{0}^{\mu}\cdot(\delta x^{\mu})\Gamma^{\mu}}$ can be written
as an equation
\begin{align*}
\mathcal{T}(\delta x^{\mu})\left \vert \mathrm{vac}\right \rangle  &  =\hat
{U}(\delta \phi^{\mu})\left \vert \mathrm{vac}\right \rangle \\
&  =e^{ik_{0}^{\mu}\cdot(\delta x^{\mu})\Gamma^{\mu}}\left \vert \mathrm{vac}%
\right \rangle .
\end{align*}

In the continuum limit, the Gamma matrix of matrix network is reduced to the
usual Gamma matrix in the Dirac equation $\Gamma^{\mu}$. In particular, we
point out that the matrix network turns into an \textrm{SO(3+1)} rotor, i.e.,
\[
\Gamma_{\mathrm{flat}}^{\{N^{\mu},M^{\mu}\}}(x)\rightarrow \Gamma^{\mu}(x,t).
\]

In summary, for a (3+1)D quantum flat spacetime, we have a uniform topological
lattice. Under geometry representation, the uniform topological lattice has
fixed lattice distances $l_{0}\Delta N^{\mu};$ under matrix representation, a
the uniform topological lattice has\ uniform matrix network with fixed Gamma
matrix $\Gamma_{\mathrm{flat}}^{\{N^{\mu},M^{\mu}\}}$ on its links.

Therefore, geometry representation is a "classical" representation, under
which the uniform topological lattice indicates a flat commutative spacetime;
matrix representation is a "quantum" representation, under which the uniform
matrix network indicates a \textquotedblleft ground state\textquotedblright%
\ for quantum spacetime.

\subsubsection{Matter}

\paragraph{Zero Hypothesis of elementary particles: zero as elementary
particle}

By using geometry representation under D-projection and K-projection (not
compactification), a uniform physical variant is reduced into a uniform zero
lattice. According to earlier discussion, zero number is a \emph{topological}
invariable that characterizes different topological equivalence classes of the
system. Then, to develop 1-th level physics structure, we had given the
Hypothesis for elementary particles:\textit{ }

\textit{Elementary particle is zero of an }$\mathrm{\tilde{S}\tilde{O}}%
$\textrm{(d+1)} \textit{physical variant} $V_{\mathrm{\tilde{S}\tilde{O}%
(d+1)},d+1}(\Delta \phi^{\mu},\Delta x^{\mu},k_{0},\omega_{0})$ \textit{under
D-projection and K-projection. }

As a result, a uniform physical variant is mapped onto a many-particle system,
i.e.,
\[
\text{Uniform physical variant}\Longleftrightarrow \text{Many-particle
system}.
\]
An elementary particle is a zero that is the information unit of the system,
i.e.,%
\[
\text{Information unit }\Longleftrightarrow \text{Zero}\Longleftrightarrow
\text{Elementary particle. }%
\]
This fact also means that the spacetime is composed of elementary particles
and the block of space (or strictly speaking, spacetime) is an elementary particle.

\paragraph{Topological property of elementary particles}

Each elementary particle corresponds to a zero with $\pi$-phase changing along
arbitrary direction. Therefore, these elementary particles become topological
defects of quantum spacetime and play the role of "magnetic monopole" on
matrix network. Let us give a brief discussion on this fact.

Along arbitrary spatial direction of the physical variant, the local Gamma
matrices around a topological defect at center are switched on the tangentia
sub-manifold. When there exists a topological defect, the periodic boundary
condition of the system along an arbitrary direction is changed into
anti-periodic boundary condition, $\Delta \phi_{x}=\pi,$ $\Delta \phi_{y}=\pi,$
$\Delta \phi_{z}=\pi.$ A topological defect not only phase switching along a
spatial direction, but also becomes topological defect along tempo direction,
i.e., along $t$-direction, a fermionic topological defect is also an
anti-phase changing denoted by $e^{i\Gamma^{5}\cdot \Delta \phi_{t}},$
$\Delta \phi_{t}=\pi.$

In the following parts, we will provide detailed discussion on the topological
structure of elementary particles and introduce \emph{spacetime charge} (or
\emph{charge of spacetime}) to characterize the topological property of
elementary particles.

\paragraph{Geometric property of elementary particles}

It was known that an elementary particle is information unit (or a zero) of
the physical variants\textit{ }$V_{\mathrm{\tilde{S}\tilde{O}(d+1)}%
,d+1}(\Delta \phi^{\mu},\Delta x^{\mu},k_{0},\omega_{0})$. For a uniform
physical variant, the zeroes have uniform distribution. Therefore, along
arbitrary direction, the size of an elementary particle is $\pi/k_{0}%
=\frac{l_{0}}{2}$ where $l_{0}$\ is the minimum distance between two zeroes.
As a result, in $d$-dimensional space, the volume of an elementary particle is
finite, $\Delta V=(\frac{l_{0}}{2})^{d}$. The exact volume $\Delta V$ of an
elementary particle is given by $\pm l_{0}^{3}=\pm(2l_{p})^{3}$. In the
following parts, we will provide detailed calculation on this result.

\paragraph{Dynamic property of elementary particles}

It was known that an elementary particle has fixed "angular momentum".

The angular momentum of a uniform physical variant has a uniformly
distribution, or the angular momentum density $\rho_{J}$ is constant. Then,
for an elementary particle with fixed volume, the "angular momentum" is given
by
\[
J_{F}=\rho_{J}\Delta V.
\]
$J_{F}$ plays the role of Planck constant $\hbar$ in quantum mechanics.
Because Planck constant $\hbar$ characterizes the constant motion on Clifford
group-changing space, the changings of the distribution of group-changing
elements on Cartesian space\textit{ }$\mathrm{C}_{d+1}$ will never change its
value, i.e.,%
\[
\hbar=\mathrm{constant}.
\]

\subsubsection{Motion}

In this section, we discuss the motion of physical variant. Without
considering curving spacetime, the motion comes from globally shifting of
elementary particles on spacetime.

\paragraph{Effective Dirac model for elementary particles}

In this section, we derive the effective Hamiltonian for elementary particles.

We firstly define generation operator of elementary particle $c_{i}^{\dagger
}\left \vert 0\right \rangle =\left \vert i\right \rangle ,$ on uniform zero
lattice. We write down the hopping Hamiltonian. The hopping term between two
nearest neighbor sites $i$ and $j$ on uniform zero lattice becomes
\begin{equation}
\mathcal{H}_{\left \{  i,j\right \}  }=Jc_{i}^{\dagger}(t)\mathbf{T}_{\left \{
i,j\right \}  }c_{j}(t)
\end{equation}
where $\mathbf{T}_{\left \{  i,j\right \}  }$ is the transfer matrix between two
nearest neighbor sites $i$ and $j$ and $c_{i}(t)$ is the annihilation operator
of elementary particle at the site $i$. $J=\frac{c}{l_{0}}$ is an effective
coupling constant between two nearest-neighbor sites that fits light speed $c$
in low energy limit. According to variability, $\left \vert i\right \rangle
=e^{il_{0}(\hat{k}^{\mu}\cdot \Gamma^{\mu})/2}\left \vert j\right \rangle ,$ the
transfer matrix $\mathbf{T}_{\left \{  i,j\right \}  }$ between $\left \vert
i\right \rangle $ and $\left \vert j\right \rangle $ is defined by $\mathbf{T}%
_{\left \{  i,j\right \}  }=\left \langle i\mid j\right \rangle =e^{il_{0}(\hat
{k}^{\mu}\cdot \Gamma^{\mu})/2}.$ After considering the contribution of the
terms from all sites, the effective Hamiltonian is obtained as%
\begin{equation}
\mathcal{H}=%
{\displaystyle \sum \limits_{\{i,j\}}}
\mathcal{H}_{\left \{  i,j\right \}  }=J%
{\displaystyle \sum \limits_{\{i,j\}}}
c_{i}^{\dagger}\mathbf{T}_{\left \{  i,j\right \}  }c_{i+e^{I}}.
\end{equation}

In continuum limit, we have%
\begin{align}
\mathcal{H}  &  =J%
{\displaystyle \sum \limits_{\mu}}
{\displaystyle \sum \limits_{i}}
c_{i}^{\dagger}(e^{il_{0}(\hat{k}^{\mu}\cdot \Gamma^{\mu})/2})c_{i+\mathbf{e}%
_{\mu}}\\
&  =l_{0}J%
{\displaystyle \sum \limits_{\mu}}
{\displaystyle \sum \limits_{k^{\mu}}}
c_{k^{\mu}}^{\dagger}[\cos(k^{\mu}\cdot \Gamma^{\mu})]c_{k^{\mu}}%
\end{align}
where the dispersion in continuum limit is
\begin{equation}
E_{k}\simeq \pm c\sqrt{[(\vec{k}-\vec{k}_{0})\cdot \vec{\Gamma}]^{2}%
+((\omega-\omega_{0})\cdot \Gamma^{t})^{2}},
\end{equation}
where $\vec{k}_{0}=\frac{2}{l_{0}}(\frac{\pi}{2},\frac{\pi}{2},\frac{\pi}%
{2}),$ $\omega_{0}=\frac{\pi}{2}\frac{2}{l_{0}}c$.

We then re-write the effective Hamiltonian to be
\begin{equation}
\mathcal{H}=\int(\Psi^{\dagger}\hat{H}\Psi)d^{3}x
\end{equation}
where $\hat{H}=\vec{\Gamma}\cdot \Delta \vec{p}$ with $\vec{\Gamma}=(\Gamma
^{x},\Gamma^{y},\Gamma^{z})$ and
\begin{align}
\Gamma^{t}  &  =\tau^{z}\otimes \vec{1}\mathbf{,}\text{ }\Gamma^{x}=\tau
^{x}\otimes \sigma^{x},\\
\Gamma^{y}  &  =\tau^{x}\otimes \sigma^{y},\text{ }\Gamma^{z}=\tau^{x}%
\otimes \sigma^{z}.\nonumber
\end{align}
$\vec{p}=\hbar \Delta \vec{k}$ is the momentum operator. This is a model for
massless Dirac fermions.

To obtain the particle's mass, we must tune $\omega_{0}.$ If $\omega_{0}\neq
ck_{0}$ the Dirac fermion have mass, i.e., $m=\hbar(\omega_{0}-ck_{0})/c^{2}$.
We then re-write the effective Hamiltonian to be\cite{dirac}
\begin{equation}
\mathcal{H}=\int(\Psi^{\dagger}\hat{H}\Psi)d^{3}x
\end{equation}
where
\begin{equation}
\hat{H}=\vec{\Gamma}\cdot \Delta \vec{p}+m\Gamma^{t}%
\end{equation}
with $\vec{\Gamma}=(\Gamma^{x},\Gamma^{y},\Gamma^{z})$ and
\begin{align}
\Gamma^{t}  &  =\tau^{z}\otimes \vec{1}\mathbf{,}\text{ }\Gamma^{x}=\tau
^{x}\otimes \sigma^{x},\\
\Gamma^{y}  &  =\tau^{x}\otimes \sigma^{y},\text{ }\Gamma^{z}=\tau^{x}%
\otimes \sigma^{z}.\nonumber
\end{align}
$\vec{p}=\hbar \Delta \vec{k}$ is the momentum operator. This is a massive Dirac model.

The Lagrangian $L$\ of fermionic particles becomes
\begin{equation}
L=\bar{\Psi}(i\gamma^{\mu}\hat{\partial}_{\mu}-m)\Psi
\end{equation}
where $\gamma^{\mu}$ are the Gamma matrices defined as $\gamma^{1}=\gamma
^{0}\Gamma^{x}$, $\gamma^{2}=\gamma^{0}\Gamma^{y},$ $\gamma^{3}=\gamma
^{0}\Gamma^{z}$, $\gamma^{0}=\Gamma^{t}.$ The Gamma matrices $\Gamma^{I}$
($I=x,y,z$) and $\Gamma^{t}$ obey Clifford algebra, i.e., $\{ \Gamma
^{I},\Gamma^{t}\}=0$, and $\{ \Gamma^{I},\Gamma^{J}\}=0.$

\paragraph{Geometry property of moving elementary particles}

Based on the theory of physical variant, quantum motion describes locally
expanding or contracting group-changing space. In addition, it characterizes
the ordered relative motion between group-changing elements of the elementary
particles. In this part, we discuss the physical picture for quantum motion
from point view of geometry.

Firstly, we give a geometric picture for quantum motion of plane waves along
certain direction, $\psi(x,t)=Ce^{-i\Delta \omega \cdot t+i\Delta k\cdot x}$.

In 1-th order representation, quantum motion describes an extra uniformly
shifting of extra group-changing elements on group-changing space $\phi
=t\cdot \Delta \omega$, of which the "velocity" is just $\Delta \omega$. On
Cartesian space, this is spiral motion by combining rotating in phase angle
$\varphi(t)=(t\cdot \Delta \omega)\operatorname{mod}(2\pi)$ and translating on
Cartesian space synchronously. The pitch on Cartesian space is $\frac{2\pi
}{\Delta k}.$ The period of rotation motion of phase angle is $\frac{2\pi
}{\Delta \omega}$. This result indicates the existence of different between a
static particle with $\Delta k=0$ and a moving one with $\Delta k\neq0$. And,
from it, one can see that the absolute change for a moving particle.

Next, we define \emph{motion charge }(or charge of motion).

For a moving elementary particle described by $\psi(x,t)=Ce^{-i\Delta
\omega \cdot t+i\Delta \vec{k}\cdot \vec{x}},$ the changing rate $\vec{k}_{0}$
turns into $\vec{k}_{0}+\Delta \vec{k}.$ Due to the topology stationarity of
elementary particle, the size in group-changing space is fixed to be $\pi
$\ along arbitrary direction. Therefore, the size of the elementary particle
on Cartesian spacetime \textrm{C}$_{d+1}$ changes from $\pi/k_{0}=\frac{l_{0}%
}{2}$ ($k_{0}=\left \vert \vec{k}_{0}\right \vert $) to $\pi/\left \vert \vec
{k}_{0}+\Delta \vec{k}\right \vert \simeq \frac{l_{0}}{2}-\frac{l_{0}}{2}%
(\frac{\Delta \vec{k}}{k_{0}}).$ We call $\vec{Q}=\frac{\Delta \vec{k}}{k_{0}}$
to be motion charge (or charge of motion) for a moving elementary particle.
See the illustration in Fig.6(a).

In addition, there exists motion charge (or charge of motion) $\frac
{\Delta \omega}{\omega_{0}}$ along tempo direction. $\frac{\Delta \omega}%
{\omega_{0}}$ characterizes the size changing of a moving elementary particle
in Cartesian spacetime \textrm{C}$_{d+1}$ along tempo direction. For a massive
elementary particle, the motion charge along tempo direction is
\[
Q_{t}=\frac{\Delta \omega}{\omega_{0}}=\frac{mc^{2}}{\omega_{0}\hbar}.
\]

On flat spacetime, according to Noether's theorem, with the spatial/tempo
translation symmetry, we have conservation rule for energy-momentum tensor.
The energy-momentum tensor for elementary particles is defined by
\[
T_{\mu \nu}=\bar{\psi}\gamma^{\nu}\partial_{\mu}\psi=\psi^{\dagger}\gamma
^{0}\gamma^{\nu}\partial_{\mu}\psi.
\]
For the case $\nu=0,$ we have $T_{\mu0}=\psi^{\dagger}\partial_{\mu}\psi$ that
are just the energy and momentum. The momentum is proportional to the motion
charge along given direction,
\[
\Delta \vec{p}=\hbar k_{0}\vec{Q}.
\]
However, the energy of an elementary particle isn't proportional to the motion
charge along tempo direction. Instead, it characterizes the global effect from
both motion charge from spatial direction and that from tempo direction,
\begin{align*}
\Delta E  &  =\sqrt{(c\Delta \vec{k}_{0})^{2}+m^{2}c^{4}}\\
&  =\sqrt{(ck_{0})^{2}\vec{Q}^{2}+\hbar^{2}\omega_{0}^{2}Q_{t}^{2}}.
\end{align*}
\emph{This is the key point that leads to a complex theory for quantum
gravity. }

Finally, we point out that for an elementary particle with finite motion
charge, the spacetime becomes disturbed. As a result, other elementary
particles feel the effect of gravitational force and the "charge" for
gravitational interaction is just the charge of motion.

\paragraph{Motion: absoluteness and relativity?}

\subparagraph{Emergent $\mathrm{SO(1,3)}$ Lorentz invariant and special
relativity}

According to above discussion, the dispersion of the elementary particles in
continuum limit is described by%
\begin{equation}
\Delta \omega=\pm \sqrt{(c\Delta \vec{k})^{2}+m^{2}}.
\end{equation}
Then, we have $(\Delta \omega)^{2}-(c\Delta \vec{k})^{2}=m^{2}=$ constant that
becomes a constraint on the changing of wave vector $\Delta \vec{k}$ and that
of angular frequency $\Delta \omega.$ The constraint from dispersion on
$\Delta \vec{k}$ and $\Delta \omega$ results another constraint on the spacetime
interval $\Delta s^{2}$ between two events
\[
-(c\Delta t)^{2}+(\Delta \vec{x})^{2}=\Delta s^{2}%
\]
where $\Delta \vec{x}$ is the distance between the space coordinates and
$\Delta t$ is the distance between the time coordinate. Hence,
$\mathrm{SO(1,3)}$ Lorentz invariant emerges. To keep the invariant of $\Delta
s^{2},$ the $\mathrm{SO(1,3)}$ Lorentz transformation is obtained as
\begin{equation}
\left(
\begin{matrix}
ct^{\prime}\\
x^{\prime}\\
y^{\prime}\\
z^{\prime}%
\end{matrix}
\right)  =\left(
\begin{matrix}
\gamma & -\gamma v/c & 0 & 0\\
-\gamma v/c & \gamma & 0 & 0\\
0 & 0 & 1 & 0\\
0 & 0 & 0 & 1
\end{matrix}
\right)  \left(
\begin{matrix}
ct\\
x\\
y\\
z
\end{matrix}
\right)
\end{equation}
where $\gamma=\frac{1}{(1-v^{2}/c^{2})^{1/2}}.$

Based on above equation, we can develop special relativity as Einstein had
done. Now, due to the linear dispersion, the speed of light has the same value
$c$ in any inertial frame. On the other hand, due to Lorentz invariant, all
inertial frames are equivalent.

Now, we consider the physical processes with two classical objects A and B on
spacetime. Strictly speaking, we consider two classical/quantum objects that
undergo classical motion. In general, we may assume that A is the object being
measured with velocity $v$ and B is the rest measuring instrument with zero
velocity (clock or ruler). The results are well known.

During measurement, according to special relativity, the simultaneity
disappears, and the inference of time depends on one's frame of reference.
Clocks at different points can only be synchronized in the given frame. For A
object moving with velocity $v$ along the $x$-axis of a rest frame $S$, we
have a clock at rest in the system $S$. Two consecutive ticks of this clock
are then characterized by $\Delta x=0$. If we want to know the relation
between the times between these ticks as measured in both objects, we have
$\Delta t^{\prime}=\gamma \Delta t$ (for events in which $\Delta x=0)$ that is
larger than the time $\Delta t$ between these ticks as measured in the rest
frame of the clock. This phenomenon is called \emph{time dilation.}

Similarly, suppose we have a measuring rod at rest in the unprimed system $S$.
In this system, the length of this rod is written as $\Delta x$. If we want to
find the length of this rod as measured in the `moving' system $S^{\prime}$,
we must make sure to measure the distances $x^{\prime}$ to the end points of
the rod simultaneously in the primed frame $S^{\prime}$. In other words, the
measurement is characterized by $\Delta t^{\prime}=0$, which we can combine
with the fourth equation to find the relation between the lengths $\Delta x$
and $\Delta x^{\prime}$: $\Delta x^{\prime}=(1/\gamma)\Delta x$, $\Delta
t^{\prime}=0.$ This shows that the length $\Delta x^{\prime}$ of the rod as
measured in the 'moving' frame $S^{\prime}$ is shorter than the length $\Delta
x$ in its own rest frame. This phenomenon is called \emph{length contraction}
or \emph{Lorentz contraction.}

\subparagraph{The relativity for absolute motion}

According to above discussion, one can see that the motion has both relativity
and absoluteness.

On the one hand, special relativity describes the measurement of two classical objects.

Absolute coordinate system\emph{ }had played important role in classical
mechanics. From Galileo, people found that objects free from external
influence would either remain at rest or move in a straight line at a constant
speed. This is Galileo's Principle of Inertia and was popularized as Newton's
first law. Now, the object moving in a straight line at constant speed
$\vec{v}$ is described by $\vec{x}(t)=\vec{x}_{0}+\vec{v}t$. However, due to
the relativity for motion, one must define inertial frame that is a simply
frame as a coordinatization of spacetime. Under the transformation of frame
(basic Galilean transformation), $\vec{x}^{\prime}=\vec{x}-\vec{v}t,$
$t^{\prime}=t$, the motion becomes relative. However, people assumed that
there may still exist an absolute coordinate system called ether. Matter and
light move inside ether. As a result, by considering ether to be the inertial
frame, the state of moving or rest can be distinguished. Everyone is familiar
with the later stories. Einstein developed the theory of special relativity.
Then, the ether does not exist. There doesn't exist absolute coordinate system
and different inertial frames are equivalence. As a result, for the two
classical objects A and B on spacetime, the inertial frames for A and B are
equivalent. That means the existence of relativity of motion.

On the other hand, according to above discussion, our spacetime is a physical
variant that plays the role of an absolute coordinate system.

For a flat spacetime that is characterized by a uniform physical variant
$V_{\mathrm{\tilde{S}\tilde{O}(d+1)},d+1},$ the matter (or elementary
particles) comes from size changing of group-changing space and the particle's
motion is characterized by finite wave vector $\Delta \vec{k}\neq0$ or the
finite motion charge $\vec{Q}=\frac{\Delta \vec{k}}{k_{0}}$. For a particle,
$\Delta \vec{k}$\ determines the group velocity (or the absolute velocity)
$\vec{v}=\frac{c^{2}}{E(\Delta \vec{k})}\Delta \vec{k}$ where $E(\Delta \vec{k})$
is its energy. As a result, we had provided a hidden assumption -- \emph{the
uniform physical variant is the absolute coordinate system or the inertial
frame}. Then, we point out that a moving particle is different from a rest one
by comparing their motion charge.

In addition, quantum flat spacetime is known to be a special spacetime crystal
with topological constraints. The 1-th order tempo variability implies a
regular motion of the group-changing space along $\Gamma^{t}$ direction, i.e.,
$\omega_{0}\neq0$. Therefore, the regular motion of the group-changing space
along $\Gamma^{t}$ direction with $\omega_{0}\neq0$ plays the role of an
immanent clock. This clock loks like\emph{ }the existence of a universal time
from Newton: \textquotedblleft \textit{Absolute, true, and mathematical time,
of itself, and from its own nature, flows equably without relation to anything
external}.\textquotedblright \ The 1-th order spatial variability implies an
immanent ruler. The motion leads to the changings of the immanent clock/ruler
that is characterized by motion charge. That means the existence of
absoluteness of motion.

\subparagraph{How to resolve this contradiction?}

The answer is "\emph{The absoluteness and relativity describe different
aspects of motion}".

Firstly, we consider the difference between the absoluteness of motion and the
relativity of motion.

Now, motion is absoluteness by considering the uniform physical variant to be
absolute coordinate. In particular, the absolute changings from motion is
characterized by the mapping between group-changing space and Cartesian space
and becomes the changings of a physical variant. Next, when we consider the
processes for spacetime and matter during measurement, the situation changes.
The relativity from motion is characterized by the mapping between the motion
state of A and that of B. Here, A and B are different mappings between
group-changing space and Cartesian space.

Next, we consider the unification of the absoluteness of motion and the
relativity of motion.

Now, we can consider the physical variant (quantum spacetime) itself to be A
object. For a moving clock or ruler relative to spacetime, we can also set B
to be the inertial frame. Under measurement, the spacetime becomes moving
object and obeys special relativity, i.e., the immanent clock/ruler (or
changing rate) of physical variant changes by Lorentz transformation, i.e.,
$\omega_{0}\rightarrow \omega_{0}^{\prime}=\gamma^{-1}\omega_{0}$ and
$k_{0}\rightarrow k_{0}^{\prime}=\gamma k_{0}$.

In summary, although we have absolute coordinate (the quantum spacetime
itself), during classical measurement, the special relativity still holds. We
say that \emph{absolute motion becomes relative during classical measurement}.

\subsection{Theory for quantum curved spacetime}

In above section we developed the theory for quantum flat spacetime.

For quantum flat spacetime, the vacuum (or ground state) obeys 1-th order
variability of both spatial-tempo transformation and rotation transformation,
i.e.,%
\begin{equation}
\mathcal{T}(\Delta x)\leftrightarrow \hat{U}(\delta \phi^{\mu})=e^{i\Gamma^{\mu
}\delta \phi^{\mu}},
\end{equation}
or%
\[
\mathcal{T}(\delta x^{\mu})\left \vert \mathrm{vac}\right \rangle =\hat
{U}(\delta \phi^{\mu})\left \vert \mathrm{vac}\right \rangle =e^{i\Gamma^{\mu
}\delta \phi^{\mu}}\left \vert \mathrm{vac}\right \rangle .
\]
where $\delta \phi^{\mu}=k_{0}\delta x^{\mu}$ are group translation operations
in non-compact $\mathrm{\tilde{S}\tilde{O}}$\textrm{(3+1)} Lie group. The wave
vector $k_{0}=\omega^{0}=\frac{2\pi}{l_{0}}$ and $l_{0}=t_{0}$ is the
characterized length/time. $\Gamma^{\mu}$ are the Gamma matrices in the
massive Dirac model.

In this section, we develop the theory of quantum curved spacetime.

Now, a quantum curved spacetime is an $\mathrm{\tilde{S}\tilde{O}}(d+1)$
perturbative physical variant that is described by \emph{inhomogeneous
space-mapping},
\begin{equation}
\{ \phi^{\mu}\} \in \text{\textrm{C}}_{\mathrm{\tilde{S}\tilde{O}(3+1)}%
}\Leftrightarrow \{x^{\mu}\} \in \mathrm{C}_{3+1},
\end{equation}
where $\Leftrightarrow$ denotes inhomogeneous\emph{ }space-mapping.

\subsubsection{Geometric/matrix representation for quantum curved spacetime}

To characterize the quantum curved spacetime, there are two types of
representations -- geometry representation and matrix representation. In the
following parts, we provide the detailed discussion on two different
representations one by one.

\paragraph{Geometry representation}

Firstly, we discuss the geometry representation for quantum curved spacetime.

From the above discussion, it was known that a quantum flat spacetime is
uniquely characterized by the spatial/tempo translation operators
\begin{equation}
\mathcal{T}(\Delta x^{\mu})\leftrightarrow \hat{U}=e^{i\Gamma^{\mu}k_{0}\Delta
x^{\mu}}.
\end{equation}
The situation doesn't change for the case of quantum curved spacetime. On
curved spacetime, spatiotemporal coordinates locally change,
\begin{equation}
(x^{\mu})_{\mathrm{curved}}=(x^{\mu})^{\prime}.
\end{equation}
This leads to geometry representation for the shape changings of quantum
curved spacetime (or the physical variant). Correspondingly, the spatial/tempo
translation operators locally change, i.e.,
\begin{equation}
\mathcal{T}(\Delta x^{\mu})\rightarrow \mathcal{T}((\Delta x^{\mu})^{\prime
})\leftrightarrow \hat{U}=e^{i\Gamma^{\mu}k_{0}\cdot(\Delta x^{\mu})^{\prime}}%
\end{equation}
or
\begin{align*}
\mathcal{T}(\Delta x^{\mu})\left \vert \mathrm{vac}\right \rangle  &  =\hat
{U}(\Delta \phi^{\mu})\left \vert \mathrm{vac}\right \rangle \\
&  =e^{i\Gamma^{\mu}k_{0}\cdot(\Delta x^{\mu})^{\prime}}\left \vert
\mathrm{vac}\right \rangle .
\end{align*}

\begin{figure}[ptb]
\includegraphics[clip,width=0.67\textwidth]{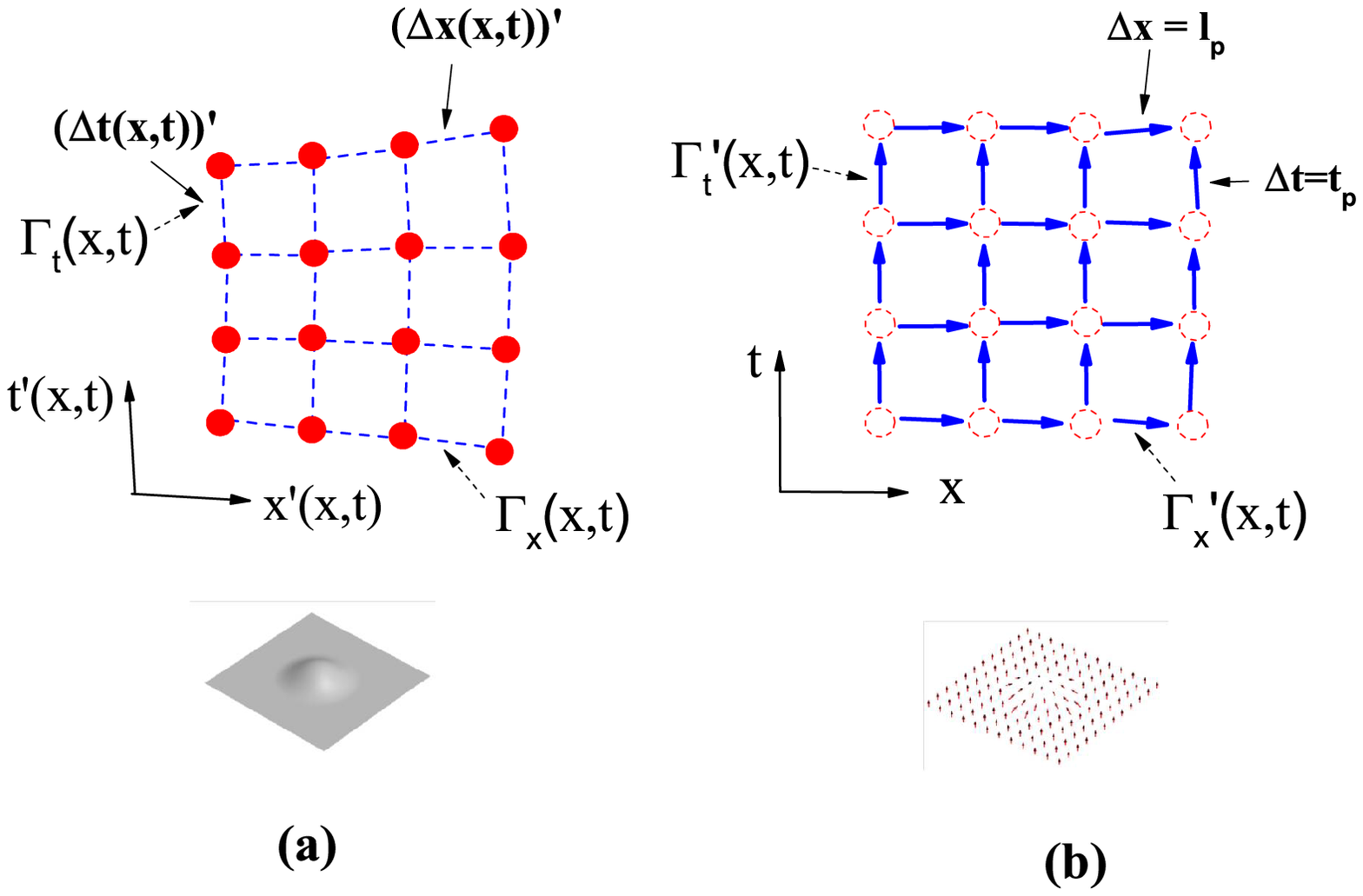}\caption{An illustration
for 1+1D curved spacetime: (a) is the geometry representation with 2D deformed
topological lattice that is denoted by solid red spots; (b) is the matrix
representation with 2D deformed matrix network that are described by
$\Gamma_{\mathrm{curved}}^{\{N^{\mu},M^{\mu}\}}$ (or $\Gamma_{x}^{\prime
}(x,t)$ and $\Gamma_{t}^{\prime}(x,t)$) on all links between two
nearest-neighbor lattice sites (solid blue arrows). }%
\end{figure}

As illustrated in Fig.3(a), we have a geometry representation of a quantum
curved spacetime\ -- (3+1)D deformed topological lattice. Now, the original
uniform topological lattice with uniform lattice distances $\Delta x^{\mu}$
slightly deviated from the original position: the distances between two
nearest-neighbor lattice sites on virtual spacetime lattice deform, i.e.,
$(\Delta x^{\mu}(N))^{\prime}-\Delta x^{\mu}=e_{\mu}(N),$ where $e_{\mu}(N)$
are vierbein fields that are the difference between the geometric unit-vectors
of the original frame and the deformed frame.

In particular, we emphasize that under geometry representation, the matrix
network $\Gamma^{\{N^{\mu},M^{\mu}\}}$ on links of the deformed topological
lattice is invariant, i.e.,
\begin{equation}
\Gamma_{\mathrm{curved}}^{\{N^{\mu},M^{\mu}\}}=\Gamma_{\mathrm{flat}%
}^{\{N^{\mu},M^{\mu}\}}.
\end{equation}
In general, we may set $l_{0}=t_{0}=1$.

Then, we discuss the theory in continuum limit.

In the continuum limit $\Delta x^{\mu}\gg1$, the spatiotemporal coordinates
become continuously changing
\begin{equation}
(\Delta x^{\mu}(N))^{\prime}\rightarrow \Delta x^{\mu}(x).
\end{equation}
Now, in geometry representation, with fixed Gamma matrix $\Gamma
_{\mathrm{curved}}^{\{N^{\mu},M^{\mu}\}}=\Gamma_{\mathrm{flat}}^{\{N^{\mu
},M^{\mu}\}}$, quantum spacetime turns into a classical curved one. The
geometry fields (vierbein fields $e^{a}$ and spin connections $\omega^{ab}$)
are determined by the non-uniform local coordinates, $(\Delta x^{\mu
}(x))^{\prime}$. With the help of the vierbein fields $e^{a}$, the space
metric is defined by
\begin{equation}
e_{\mu}^{a}e_{b}^{\mu}=\delta_{b}^{a}\,,\quad e_{\mu}^{a}e_{a}^{\nu}%
=\delta_{\mu}^{\nu},
\end{equation}
and
\begin{equation}
\eta_{ab}e_{\alpha}^{a}e_{\beta}^{b}=g_{\alpha \beta},
\end{equation}
where $\eta_{ab}$ is the Minkowskian matrix $\eta_{ab}=$ diag$\left(
-1,1,1,1\right)  .$ The Riemann curvature 2-form is written as
\begin{equation}
R_{b}^{a}=d\omega_{b}^{a}+\omega_{c}^{a}\wedge \omega_{b}^{c},
\end{equation}
where $R_{b\mu \nu}^{a}\equiv e_{\alpha}^{a}e_{b}^{\beta}R_{\beta \mu \nu
}^{\alpha}$ are the components of the usual Riemann tensor projection on the
tangent space.

In the continuum limit, the Lagrangian for particles on curved spacetime turns
into
\begin{equation}
L=\sqrt{-g}\bar{\Psi}(e_{a}^{\mu}\gamma^{a}(i\hat{\partial}_{\mu}+i\omega
_{\mu})-m)\Psi,
\end{equation}
where $e_{a}^{\mu}$ denotes the vierbein fields, $\omega_{\mu}=(\omega_{\mu
}^{0I}\gamma^{0I}/2,\omega_{\mu}^{IJ}\gamma^{IJ}/2)$ ($I,J=1,2,3$) are spin
connections and $\gamma^{ab}=-\frac{1}{4}[\gamma^{a},\gamma^{b}]$
($a,b=0,1,2,3$). In particular, the gamma matrices $\gamma^{\mu}=\gamma
^{0}\Gamma^{\mu}$ are all fixed as the flat ones.

The theory is invariant under all possible coordinate transformations
\begin{equation}
x^{\mu}\rightarrow(x^{\mu}(x))^{\prime}\,,
\end{equation}
where $(x^{\mu}(x))^{\prime}$ is invertible, differentiable and with a
differentiable inverse. Under the above transformation, the metric transforms
as
\begin{equation}
g_{\mu \nu}(x)\rightarrow g_{\mu \nu}^{\prime}(x^{\prime})=\frac{\partial
x^{\rho}}{\partial x^{\prime \mu}}\, \frac{\partial x^{\sigma}}{\partial
x^{\prime \nu}}\,g_{\rho \sigma}(x)\,.
\end{equation}
For physical variant, the coordinate transformations come from the
transformation of mappings between \textrm{C}$_{\mathrm{\tilde{S}\tilde
{O}(3+1)}}\ $and $\mathrm{C}_{3+1}$ without changing \textrm{C}%
$_{\mathrm{\tilde{S}\tilde{O}(3+1)}}$.

In addition, this model described by $\mathcal{S}$ is also invariant under
local \textrm{SO(3,1)} Lorentz transformation $L(x)=e^{\theta_{ab}%
(x)\gamma^{ab}}$ as
\begin{align}
\Psi(x)  &  \rightarrow \Psi^{\prime}(x)=L(x)\Psi(x),\nonumber \\
\gamma^{\mu}  &  \rightarrow(\gamma^{\mu}(x))^{\prime}=L(x)\gamma^{\mu
}(S(x))^{-1},\nonumber \\
\omega_{\mu}  &  \rightarrow \omega_{\mu}^{\prime}(x)=L(x)\omega_{\mu
}(x)(S(x))^{-1}\nonumber \\
&  +S(x)\partial_{\mu}(S(x))^{-1}.
\end{align}
$\gamma^{5}$ is invariant under local \textrm{SO(3,1)} Lorentz symmetry as
\begin{equation}
\gamma^{5}\rightarrow(\gamma^{5})^{\prime}=L(x)\gamma^{5}(L(x))^{-1}%
=\gamma^{5}.
\end{equation}

In particular, we point out that such a local \textrm{SO(3,1)} Lorentz
symmetry is an emergent symmetry rather than the original one.

\paragraph{Matrix representation}

\subparagraph{$\Gamma$-matrix representation}

Next, we discuss the matrix representation for quantum curved spacetime.

It was known that one can record its information of curving process\ by local
spatiotemporal operations, $\hat{S}(x)=e^{i\phi_{\mu}(x)\Gamma^{\mu}}$. Then,
by using $\hat{S}(x)$, we introduce a special matrix representation --
$\Gamma$-\emph{matrix representation} to characterize the shape changings of
spacetime, i.e.,
\begin{align}
\mathcal{T}((\Delta x^{\mu})^{\prime})  &  \leftrightarrow \hat{U}%
=e^{i\Gamma^{\mu}k_{0}(\Delta x^{\mu})^{\prime}}\nonumber \\
&  =\hat{S}(x)\mathcal{T}(\Delta x^{\mu})(\hat{S}(x))^{-1}.
\end{align}
\ Under the operation $\hat{S}(x),$ the ground state of spacetime $\left \vert
\mathrm{vac}(x)\right \rangle $ turns into
\begin{equation}
\left \vert \mathrm{vac}(x)\right \rangle \rightarrow \left \vert \mathrm{vac}%
(x)\right \rangle ^{\prime}=\hat{S}(x)\left \vert \mathrm{vac}(x)\right \rangle .
\end{equation}
As a result, the changes of quantum states of spacetime are characterized by
the changings of $\hat{S}(x)$!

Consequently, under the local operations $\hat{S}(x)$, the uniform matrix
network $\Gamma_{\mathrm{flat}}^{\{N^{\mu},M^{\mu}\}}$ on flat spacetime turns
into a non-uniform one $\Gamma_{\mathrm{curved}}^{\{N^{\mu},M^{\mu}\}}(x)$,
i.e.,
\begin{equation}
\Gamma_{\mathrm{curved}}^{\{N^{\mu},M^{\mu}\}}(x)=\hat{S}(x)\Gamma
_{\mathrm{flat}}^{\{N^{\mu},M^{\mu}\}}(\hat{S}(x))^{-1}.
\end{equation}
In particular, we emphasize that the spatiotemporal coordinates do not change
any more, i.e., $(x^{\mu}(x))_{\mathrm{curved}}=(x^{\mu}(x))_{\mathrm{flat}}.$
Now, we have%
\[
\mathcal{T}(\Delta x^{\mu})\left \vert \mathrm{vac}\right \rangle =\hat
{U}(\Delta \phi^{\mu})\left \vert \mathrm{vac}\right \rangle =e^{i(\Gamma^{\mu
})^{\prime}k_{0}\cdot \Delta x^{\mu}}\left \vert \mathrm{vac}\right \rangle
\]
where $(\Gamma^{\mu})^{\prime}=\Gamma_{\mathrm{curved}}^{\{N^{\mu},M^{\mu}%
\}}(x)$. See the illustration of a curved 2D spacetime described by a deformed
matrix network in Fig.3(b).

In summary, we have a language of quantum mechanics for spacetime.

The Hilbert space $\mathcal{E}$ of quantum spacetime consists of all
four-by-four matrices on links $\{N^{\mu},M^{\mu}\}$,
\begin{align}
\mathcal{E}  &  :\mathcal{H}_{QST}=\mathcal{H}_{\{(0,0,0,0),(1,0,0,0)\}}%
\otimes...\nonumber \\
&  \otimes \mathcal{H}_{\{N^{\mu},M^{\mu}\}}.
\end{align}
The states of quantum spacetime are characterized by different matrix network
\begin{equation}
\{ \Gamma_{\mathrm{curved}}^{\{N^{\mu},M^{\mu}\}}(x),\mu=x,y,z,t\}.
\end{equation}
We call this representation of quantum spacetime $\hat{S}(x)=e^{i\phi_{\mu
}(x)\Gamma^{\mu}}$ to be $\Gamma$\emph{-matrix representation. }In continuum
limit, the matrix network turns into field for a \textrm{SO(4)} rotor
$\Gamma^{\mu}(x,t)$, i.e.,
\[
\Gamma^{\mu}\rightarrow(\Gamma^{\mu})^{\prime}(x,t)=\hat{S}(x)\Gamma^{\mu
}(\hat{S}(x))^{-1}.
\]
Within matrix representation, the \emph{parallel transport} is defined by a
special motion along fixed $(\Gamma^{\mu})^{\prime}(x,t)).$

\subparagraph{$\gamma$-matrix representation}

Within $\Gamma$-matrix representation,\emph{ }the quantum spacetime is
described by a matrix network, a field for a \textrm{SO(4)} rotor $\Gamma
^{\mu}(x,t)$, i.e., $\Gamma^{\mu}(x,t)=\hat{S}(x)\Gamma^{\mu}(\hat{S}%
(x))^{-1}.$ However, we cannot directly use $\Gamma$-matrix representation to
characterize a quantum spacetime and its dynamics. To clearly keep Lorentz
covariance, an equivalent, better representation of quantum spacetime is
$\gamma$\emph{-matrix representation,}
\begin{equation}
\hat{S}(x)=e^{\phi_{ab}(x)\gamma^{ab}}(\gamma^{ab}=-\frac{1}{4}[\gamma
^{a},\gamma^{b}]).
\end{equation}
Now, we can use $\gamma$-matrix representation to characterize the changings
of \textrm{SO(4)} rotor $\Gamma^{\mu}(x,t)$ by the representation of local
\textrm{SO(3,1)} Lorentz\ group. However, due to the mismatch of the
operations on $\gamma^{\mu}$ and those on $\Gamma^{\mu}$ (or matrix network
$\Gamma_{\mathrm{curved}}^{\{N^{\mu},M^{\mu}\}}$), we have big \emph{trouble}.
Let provide a detailed discussion on the trouble.

By defining $\gamma^{0}=\Gamma^{5}$, the small deformation on (3+1)D
topological lattice along the i-th spatial direction from $\hat{S}(x)$ is
given by $e^{i\Gamma^{i}\cdot \delta \phi_{i}}$ in $\Gamma$-matrix
representation, or, $e^{i\gamma^{0a}\cdot \delta \phi_{i}\delta_{ia}}$ in
$\gamma$-matrix representation. Under $e^{i\Gamma^{i}\cdot \delta \phi_{i}%
}=e^{i\gamma^{0a}\cdot \delta \phi_{i}\delta_{ia}}$, the lattice distance along
the i-th spatial direction correspondingly changes, i.e.,
\begin{equation}
\Delta x^{i}\rightarrow(\Delta x^{i})^{\prime}=\Delta x^{i}+\frac{l_{0}}{2\pi
}\delta \phi_{i}.
\end{equation}
However, without $e^{i\delta \phi_{t}\cdot \Gamma^{5}}$ in $\hat{S}%
(x)=e^{\phi_{ab}(x)\gamma^{ab}}$, no operation in $\gamma$-matrix
representation leads to $\Delta t\rightarrow(\Delta t)^{\prime}=\Delta
t+\frac{t_{0}}{2\pi}\delta \phi_{t}.$ Or, the small change of lattice distance
along tempo direction \emph{cannot} be well defined in $\gamma$-matrix
representation. That means using $\gamma$-matrix representation, we can only
characterize the changes of a 3D subspace $(x,y,z)$ in (3+1)D topological lattice.

To completely characterize the deformation of the (3+1)D matrix network in
$\gamma$-matrix representation, we introduce two new concepts --
\emph{generalized gamma matrices and their round-robin}.

\textit{Definition -- generalized gamma matrices and their round-robin: The
gamma matrices }$\tilde{\gamma}^{\mu}$\textit{ are defined as }$\tilde{\gamma
}^{1}=\tilde{\gamma}^{0}\Gamma^{x}$\textit{, }$\tilde{\gamma}^{2}%
=\tilde{\gamma}^{0}\Gamma^{y},$\textit{ }$\tilde{\gamma}^{3}=\tilde{\gamma
}^{0}\Gamma^{z}$\textit{, }$\tilde{\gamma}^{0}=\Gamma^{t}$\textit{ where }%
\begin{equation}
\tilde{\gamma}^{0}=\alpha \Gamma^{x}+\beta \Gamma^{y}+\gamma \Gamma^{z}%
+\delta \Gamma^{t}%
\end{equation}
\textit{ with }$\alpha^{2}+\beta^{2}+\gamma^{2}+\delta^{2}=1$\textit{. Here,
}$\alpha,$\textit{ }$\beta,$\textit{ }$\gamma,$\textit{ }$\delta$\textit{ are
real number. The changes of}$\  \tilde{\gamma}^{\mu}$\textit{ by tuning the
values of }$\alpha,$\textit{ }$\beta,$\textit{ }$\gamma,$\textit{ }$\delta
$\textit{ is called round-robin of generalized gamma matrices, i.e., }%
$\tilde{\gamma}^{\mu}\rightarrow \tilde{\gamma}^{\mu^{\prime}}$\textit{. Now,
local transformation turns into }$\hat{S}(x)\Longrightarrow \tilde
{S}(x)=e^{\phi_{ab}(x)\tilde{\gamma}^{ab}}$\textit{ }$(\tilde{\gamma}%
^{ab}=-\frac{1}{4}[\tilde{\gamma}^{a},\tilde{\gamma}^{b}])$\textit{. }

Then, with the help of the generalized gamma matrices $\tilde{\gamma}^{0}$ and
their round-robin, we develop the $\gamma$-matrix representation to
characterize the deformation of the spacetime.

For curved spacetime, under a theory with fixed generalized gamma matrices
$\tilde{\gamma}^{0}$, we can only describe a corresponding 3D sub-manifold in
(3+1)D curved spacetime that is denoted by $\mathrm{M}_{3}^{\mu}$
perpendicular to $\mathbf{e}^{\mu}=\alpha \mathbf{e}^{x}+\beta \mathbf{e}%
^{y}+\gamma \mathbf{e}^{z}+\delta \mathbf{e}^{t}$ with $\alpha^{2}+\beta
^{2}+\gamma^{2}+\delta^{2}=1$. The usual 3D space is thus denoted by
\textrm{M}$_{3}^{\mu=t}$ perpendicular to time direction $\mathbf{e}^{t}$.
Under a given round-robin of generalized gamma matrices, the theory of a 3D
sub-manifold denoted by \textrm{M}$_{3}^{\mu}$ perpendicular to $\mathbf{e}%
^{\mu}$ is changed to the theory of another denoted by \textrm{M}$_{3}%
^{\mu^{\prime}}$ perpendicular to $\mathbf{e}^{\mu^{\prime}}$.

In addition, we point out that under an arbitrary round-robin of generalized
gamma matrices, although the mass term for fermionic particles in Lagrangian
changes its formula from $m\bar{\Psi}\Psi$ to $m\bar{\Psi}\Gamma^{i^{\prime}%
}\Gamma^{5}\Psi,$ the Hamiltonian does not change any more!

\subparagraph{$g$-matrix representation (or gauge representation)}

In the continuum limit, we upgrade the $\gamma$-matrix representation of
quantum spacetime to a $g$\emph{-matrix representation}, by which we can
easily characterize topological structures of quantum spacetime. This is also
called \emph{gauge representation}.

Firstly, we consider $\gamma$-matrix representation with the general gamma
matrices defined by $\gamma^{0}=\Gamma^{5}$. Now, the local transformation of
spacetime $\hat{S}(x)=e^{\phi_{ab}(x)\gamma^{ab}}$ $(\gamma^{ab}=-\frac{1}%
{4}[\gamma^{a},\gamma^{b}])$ is a combination of spin rotation transformation
$\hat{R}(x)$ and spatial transformation along $i$-direction ($i=x,y,z$)
$\hat{S}^{i}(x)=e^{i\delta \phi^{i}(x)\cdot \Gamma^{i}},$ i.e.,
\begin{equation}
\hat{S}(x)=\hat{R}(x)\oplus \hat{S}^{i}(x).
\end{equation}
Here, $\oplus$ denotes operation combination. $a$, $b$ denote internal
indices. Under a non-uniform \textrm{SO(4)} transformation $\hat{S}(x)$, we
have\textrm{ }%
\begin{align}
\gamma^{0}  &  \rightarrow \hat{S}(x)\gamma^{0}(\hat{S}(x))^{-1}\nonumber \\
&  =(\gamma^{0}(x))^{\prime}=%
{\displaystyle \sum \nolimits_{a}}
\gamma^{a}n^{a}(x),
\end{align}
where $n^{a}(x)=(n^{1},n^{2},n^{3},n^{0})=(\vec{n},n^{0})$ is a unit
\textrm{SO(4)} vector-field.

To characterize the curved spacetime, we introduce an auxiliary gauge field
$A_{\mu}^{ab}(x)$ that is written into two parts: \textrm{SO(3)} parts
\begin{equation}
A^{ab}(x)=\mathrm{tr}(\gamma^{ab}(\hat{S}(x))d(\hat{S}(x))^{-1})
\end{equation}
and \textrm{SO(4)/SO(3)} parts
\begin{align}
A^{a0}(x)  &  =\mathrm{tr}(\gamma^{a0}\hat{S}(x))d(\hat{S}(x))^{-1}%
)\nonumber \\
&  =\gamma^{0}d(\gamma^{a}(x)).
\end{align}
The total field strength $\mathcal{F}^{IJ}(x)$ of $a,b=1,2,3$ components can
be divided into two parts $\mathcal{F}^{ab}(x)=F^{ab}+A^{a0}\wedge A^{b0}.$
According to pure gauge condition, we have the Maurer-Cartan equation,
\begin{equation}
\mathcal{F}^{ab}(x)=F^{ab}+A^{a0}\wedge A^{b0}\equiv0
\end{equation}
or
\begin{equation}
F^{ab}=dA^{ab}+A^{ac}\wedge A^{cb}\equiv-A^{a0}\wedge A^{b0}.
\end{equation}
Here, $a$, $b$, $c$ all denote internal indices.

Now, in continuum limit, we have a strange quantum field theory on flat
spacetime. The Lagrangian for particles on curved spacetime turns into
\begin{equation}
L=\bar{\Psi}(\gamma^{\mu}(x)(i\hat{\partial}_{\mu}+i\omega_{\mu}(x))-m)\Psi,
\end{equation}
where $\gamma^{x,y,z}(x)$ is not constant Gamma matrix, but a rotor field.

We then do the transformation of round-robin. To do the transformation of
round-robin, we consider the quantum states of another 3D sub-manifold
\textrm{M}$_{3}^{\mu^{\prime}}$ ($\mu^{\prime}\neq t,$ for example,
$\mu^{\prime}=y$) in (3+1)D spacetime.

For \textrm{M}$_{3}^{\mu^{\prime}}$, under the round-robin of generalized
gamma matrices, we can define
\begin{equation}
\tilde{\gamma}^{0}=\alpha \Gamma^{1}+\beta \Gamma^{2}+\gamma \Gamma^{3}%
+\delta \Gamma^{5}%
\end{equation}
with $\alpha^{2}+\beta^{2}+\gamma^{2}+\delta^{2}=1$. The local transformation
turns into $\tilde{U}(x)=e^{\tilde{\phi}_{ab}(x)\tilde{\gamma}^{ab}}$. The
auxiliary gauge field $\tilde{A}^{ab}(x)$ and the gauge field strength turn
into
\begin{equation}
\tilde{A}^{ab}(x)=\mathrm{tr}(\tilde{\gamma}^{ab}(\tilde{S}(x))d(\tilde
{S}(x))^{-1})
\end{equation}
and
\begin{align}
\tilde{F}^{ab}  &  =d\tilde{A}^{ab}+\tilde{A}^{ac}\wedge \tilde{A}%
^{cb}\nonumber \\
&  \equiv-\tilde{A}^{a0}\wedge \tilde{A}^{b0},
\end{align}
respectively.\ After considering the mathematical set of all gauge fields
$\tilde{A}^{ab}(x)$ from generalized gamma matrices, we have an equivalent
description of the quantum states of curved spacetime through these gauge
fields. This is a new type of gauge structure -- an \textrm{SO(3)}%
$^{\mathrm{SO(4)}}$ gauge structure,\ of which each group element of
$\mathrm{SO(4)}$ group for a 3D sub-manifold \textrm{M}$_{3}^{\mu}$
corresponds to an \textrm{SO(3)} gauge theory. For different 3D sub-manifolds
\textrm{M}$_{3}^{\mu}$, there exist different gauge fields, $A_{\mu}(x)$.
Therefore, there are infinite gauge fields for the \textrm{SO(3)}%
$^{\mathrm{SO(4)}}$ gauge structure.

In summary, we have a correspondence between curved spacetime and
\textrm{SO(3)}$^{\mathrm{SO(4)}}$ gauge fields. If we insist on using flat
spacetime to represent the quantum theory of curved spacetime, we have an
\textrm{SO(3)}$^{\mathrm{SO(4)}}$ gauge fields. For an arbitrary 3D
sub-manifold \textrm{M}$_{3}^{I},$ the Lagrangian for particles becomes
\begin{equation}
L=\bar{\Psi}(\gamma^{J}(x,t)(i\hat{\partial}_{\mu}+i\omega_{\mu}%
(x,t))-m\Gamma^{I}(x,t)\Gamma^{5})\Psi,
\end{equation}
where $\gamma^{J}(x)$ for $J\neq I$ is not constant Gamma matrix.

\paragraph{Intrinsic relationship between geometry representation and matrix
(or gauge) description}

Because the matrix representation (including gauge representation)\ and the
geometric representation characterize the same quantum curved spacetime, there
must exist an inevitable connection between them. Let us show it.

We firstly show the relationship between gauge fields $A^{ab}(x)$\ in gauge
representation\ and vierbein fields $e^{a}(x)$ in geometric representation.

On the one hand, to characterize the changes of a topological lattice, we
consider a curved spacetime by using a geometry representation. On the
deformed topological lattice, the \textquotedblleft lattice
distances\textquotedblright \ become dynamic vector fields. We define the
vierbein fields $e^{a}(x)$ that are supposed to transform homogeneously under
the local symmetry, and behave as ordinary vectors under local transformation
along $x^{a}$-direction,
\begin{equation}
e^{a}(x)=dx^{a}(x)\text{ and }e_{\mu}^{a}(x)=\frac{\partial x^{a}(x)}%
{\partial \xi_{\mu}},
\end{equation}
where $\xi_{\mu}$ denotes the coordinate variable of the flat topological lattice.

On the other hand, within the representation of $\Gamma^{5}=\gamma^{0}$, we
consider a varied vector-field
\begin{align}
(\gamma^{0}(x))^{\prime}  &  =\hat{S}(x)\gamma^{0}(\hat{S}(x))^{-1}\nonumber \\
&  =%
{\displaystyle \sum \nolimits_{a}}
\gamma^{a}n^{a}(x),
\end{align}
where $n^{a}(x)=(n^{1},n^{2},n^{3},n^{0})$ is a unit \textrm{SO(4)}
vector-field in $\gamma$-matrix representation.

For the smoothly deformed vector-fields $n^{a}(x)\ll1$, we have
\begin{align}
n^{a}(x)  &  =\frac{dx^{a}(x)}{l_{0}}=\frac{d\phi^{a}(x)}{2\pi}\nonumber \\
&  =\mathrm{tr}[\gamma^{0}d\gamma^{a}(x)]=A^{a0}(x),\text{ }a=1,2,3,
\end{align}
where $N^{a}(x)$ denotes the numbers of a topological lattice. Thus, the
relationship between $e^{a}(x)$ and $A^{a0}(x)$ is obtained as
\begin{equation}
e^{a}(x)\equiv l_{0}A^{a0}(x),\text{ }a=1,2,3.
\end{equation}

Under round-robin of generalized gamma matrices, for another 3D subspace
\textrm{M}$_{3}^{\mu^{\prime}}$ ($\mu^{\prime}\neq t$) within another
representation of $\Gamma^{a}=\tilde{\gamma}^{0},$\ we have
\begin{equation}
e^{0}(x)=l_{0}\tilde{A}^{a0}(x).
\end{equation}

After considering these relationships, the correspondence between geometry
representation for topological lattice and $\Gamma$/$\gamma$/$g$-matrix
representation for matrix network constitutes an important clue of the article.

\paragraph{Summary}

For a (3+1)D quantum curved spacetime, we have a deformed (3+1)D topological
lattice with fluctuated lattices in geometry representation and a non-uniform
(3+1)D matrix network with fluctuated Gamma matrix on its links in matrix
representation. Under Lorentz covariance, we use $\gamma$-matrix/gauge
representation to characterize the changings of \textrm{SO(4)} matrix network
$\Gamma^{\mu}(x,t)$. This leads to an \textrm{SO(3)}$^{\mathrm{SO(4)}}$ gauge
structure,\ of which each group element of $\mathrm{SO(4)}$ group for a 3D
sub-manifold \textrm{M}$_{3}^{\mu}$ corresponds to an \textrm{SO(3)} gauge
theory. By using the \textrm{SO(3)}$^{\mathrm{SO(4)}}$ gauge theory, we have a
local field description for curved spacetime. This will play important role in
the unification of matter and spacetime.

\paragraph{Quantized geometry for quantum curved spacetime}

In general, the curved spacetime is described by non-Euclidean geometry.
\emph{What do the traditional geometric quantities (for example, volume) mean
in quantum spacetime?} In this section, we will discuss geometric quantities
of topological defect for quantum spacetime by using matrix (gauge)
representation. We focus on the 3D space \textrm{M}$_{3}^{\mu=t}$ in (3+1)D
quantum spacetime by fixing $\Gamma^{t}=\gamma_{0}.$

Firstly, we can show the quantized geometry of quantum flat spacetime.

Now, the \textquotedblleft unit\textquotedblright \ of 3D bulk is that with
smallest 3-volume $\Delta V_{0}$ for a unit sell with $2^{3}$ zeroes (a block
of quantum spacetime). An arbitrary 3D bulk can be regarded as a system with a
lot of bulk \textquotedblleft unit\textquotedblright. This fact leads to the
volume quantization of a 3D bulk of a quantum flat spacetime, i.e., $\Delta
V=N\cdot \Delta V_{0}$ where $N$ is an position integer number about uni cells.

Next, we provide a detailed calculation on the 3-volume in 3D curved space
with topological defects.

In Riemannian geometry, the 3-volume for $\mathcal{M}$ in 3D curved space is
defined by
\begin{equation}
\Delta V=\frac{1}{3!}%
{\displaystyle \int \limits_{\mathcal{M}}}
\epsilon_{abc}e_{\mathcal{M}}^{a}\wedge e_{\mathcal{M}}^{b}\wedge
e_{\mathcal{M}}^{c},
\end{equation}
where $e_{\mathcal{M}}^{a,b,c}$ denote the local frame of $\mathcal{M}$ in 3D
curved space. According to the above section, for quantum spacetime there
exists a correspondence between the geometry representation of topological
lattice and $\Gamma$/$\gamma$/$g$-matrix representation of matrix network. We
transform the geometric value to topological value in gauge representation.

By using following equation, $e_{\mathcal{S}}^{a}\wedge e_{\mathcal{S}}%
^{b}=(l_{0})^{2}A_{\mathcal{S}}^{a0}\wedge A_{\mathcal{S}}^{b0}$, the 3-volume
$\Delta V$ becomes
\begin{align}
\Delta V  &  =\frac{1}{3!}%
{\displaystyle \int \limits_{\mathcal{M}}}
\epsilon_{abc}e_{\mathcal{M}}^{a}\wedge e_{\mathcal{M}}^{b}\wedge
e_{\mathcal{M}}^{c}\nonumber \\
&  =\frac{1}{3!}l_{0}^{3}%
{\displaystyle \int \limits_{\mathcal{M}}}
\epsilon_{abc}\mathrm{tr}(A_{\mathcal{M}}^{a0}\wedge A_{\mathcal{M}}%
^{b0}\wedge A_{\mathcal{M}}^{c0}),
\end{align}
where
\begin{align}
A^{a0}(x)  &  =\mathrm{tr}(\gamma^{a0}\hat{S}(x))d(\hat{S}(x))^{-1}%
)\nonumber \\
&  =\mathrm{tr}(\gamma^{0}(x)d(\gamma^{a}(x))).
\end{align}
Then, from the equations, $(\gamma^{0}(x))^{\prime}=\hat{S}(x)\gamma^{0}%
(\hat{S}(x))^{-1}=%
{\displaystyle \sum \nolimits_{a}}
\gamma^{a}n^{a}(x)$ and $n^{a}(x)=\mathrm{tr}[\gamma^{0}d\gamma^{a}(x)],$ we
get%
\begin{align}
\Delta V  &  =\frac{1}{3!}l_{0}^{3}%
{\displaystyle \int \limits_{\mathcal{M}}}
\mathrm{tr}[\epsilon_{abc}\gamma^{0}(x)\wedge d(\gamma^{a}(x))\nonumber \\
&  \wedge d(\gamma^{b}(x))\wedge d(\gamma^{c}(x))]\nonumber \\
&  =\frac{1}{3!}(\frac{l_{0}}{2\pi})^{3}%
{\displaystyle \int \limits_{\mathcal{M}}}
\epsilon_{abc}n^{0}(x)\wedge d(n^{a}(x))\nonumber \\
&  \wedge d(n^{b}(x))\wedge d(n^{c}(x)),
\end{align}
where $n^{a}(x)=(n^{1}(x),n^{2}(x),n^{3}(x),n^{0}(x))$ is a unit
\textrm{SO(4)} vector-field in $\gamma$-matrix representation.

We consider 3-volume of topological defects in 3D space \textrm{M}$_{3}%
^{\mu=t}$ that is related to the issue of the size of particle in usual x/y/z space.

Now, the gamma matrix $\gamma^{0}$ is fixed to $\Gamma^{t}=\tau^{z}\otimes
\vec{1}$ and the other four-by-four matrices are reduced to three two-by-two
Pauli matrices, i.e., $\gamma^{a}(x)\rightarrow \sigma^{I}$ ($I=x,y,z$).
Correspondingly, the \textrm{SO(4)} vector-field $n^{a}(x)=(n^{1}%
(x),n^{2}(x),n^{3}(x),n^{0}(x))$ is reduced to an \textrm{SO(3)} vector-field
$n^{I}(x)=(n^{x}(x),n^{y}(x),n^{z}(x))$. The definition of 3-volume turns
into
\begin{align}
\Delta V  &  =\frac{1}{3!}(l_{0}^{3}%
{\displaystyle \int \limits_{\mathcal{M}}}
\mathrm{tr}[\epsilon_{IJK}\wedge d(\mathcal{N}^{I}(x))\nonumber \\
&  \wedge d(\mathcal{N}^{J}(x))\wedge d(\mathcal{N}^{K}(x))],
\end{align}
where $\mathcal{N}^{I}(x)=s(x)\sigma^{I}(s(x))^{-1}$ and $s(x)$ is a
two-by-two matrix reduced from the four-by-four matrix $\hat{S}(x).$ As a
result, we have
\begin{align}
\Delta V  &  =\frac{1}{3!}l_{0}^{3}%
{\displaystyle \int \limits_{\mathcal{M}}}
\epsilon_{IJK}d(n^{I}(x))d(n^{J}(x))\wedge d(n^{K}(x))\nonumber \\
&  =\frac{1}{3!}l_{0}^{3}%
{\displaystyle \int \limits_{\mathcal{M}}}
\epsilon_{IJK}d[n^{I}(x)d(n^{J}(x))\wedge d(n^{K}(x))]\nonumber \\
&  =\frac{1}{3!}l_{0}^{3}%
{\displaystyle \oint \nolimits_{\mathcal{S}}}
\epsilon_{IJK}[n^{I}(x)d(n^{J}(x))\wedge d(n^{K}(x))]\nonumber \\
&  =4\pi l_{0}^{3}q_{m},
\end{align}
where $q_{m}=\frac{1}{3!4\pi}%
{\displaystyle \oint \nolimits_{\mathcal{S}}}
\epsilon_{IJK}[n^{I}(x)d(n^{J}(x))\wedge d(n^{K}(x))]$ is the Pontriagin
number and $\mathcal{S}$ is the closed surface enclosing $\mathcal{M}$ in 3D
space. Therefore, we also have
\begin{equation}
q_{m}=\frac{1}{4\pi}%
{\displaystyle \oint \nolimits_{\mathcal{S}}}
F_{\mathcal{S}}^{IJ},
\end{equation}
where $F_{\mathcal{S}}^{IJ}=dA_{\mathcal{S}}^{IJ}+A_{\mathcal{S}}^{Ic}\wedge
A_{\mathcal{S}}^{cJ}\equiv-A_{\mathcal{S}}^{I0}\wedge A_{\mathcal{S}}^{J0}$ is
the strength of gauge fields on $\mathcal{S}$.

This result indicates that for quantum spacetime an object with topological
property has fixed 3-volume. The corresponding 3-volume is determined by a
topological invariant, or the \textquotedblleft magnetic
charge\textquotedblright \ of \textrm{SO(3)}$^{\mathrm{SO(4)}}$ gauge fields.
Above result also means when one locally change the 3-volume $\Delta V$, the
quantum spacetime changes highly non-locally with changing the number of
\textquotedblleft magnetic monopole\textquotedblright \
\begin{equation}
q_{m}=\frac{\Delta V}{4\pi l_{0}^{3}}.
\end{equation}

Using similar approach, we can find that in quantum spacetime, the volume
$\Delta V$ for $\mathcal{M}$ in an arbitrary 3D subspace \textrm{M}$_{3}^{\mu
}$ ($\mu=t$ or $\mu \neq t$) of topological defect with the \textquotedblleft
magnetic charge\textquotedblright \ of \textrm{SO(3)}$^{\mathrm{SO(4)}}$ gauge
fields on flat spacetime becomes,
\[
\text{3-volume }\Delta V=4\pi l_{0}^{3}q_{m}\text{ in }\mathrm{M}_{3}^{\mu},
\]
where $q_{m}$ is the number of \textquotedblleft magnetic
monopole\textquotedblright.

Let us give a simple argument on the geometry quantization for curved
spacetime. At the micro level, quantum flat spacetime is reduced to the
topological lattice. The changing of 3-volume for a given geometric object
must be quantized, of which the value is topological invariable.

\subsubsection{Theory for matter in quantum curved spacetime}

In this part, we discuss the property of matter (elementary particles) in
(3+1)D quantum curved spacetime.

\paragraph{Topological property of matter}

According to above discussion, it was known that an elementary particle is
information unit of Clifford group-changing space. The generation/annihilation
of an elementary particle leads to \emph{contraction/expansion} $\pi$-phase
changing of Clifford group-changing space along an arbitrary direction. As a
result, when there exists an excited elementary particle, the periodic
boundary condition of systems along arbitrary direction is changed into
anti-periodic boundary condition. Therefore, an elementary particle plays the
role of topological defect on quantum spacetime.

It was known that an elementary particle is $\pi$-phase changing along
different direction in quantum spacetime. When there exists an elementary
particle, the periodic boundary condition of the ground state along an
arbitrary direction is changed into anti-periodic boundary condition,
$\Delta \phi^{i}=\pi.$ Along arbitrary direction $\phi^{i}$, the local Gamma
matrices around an elementary particle at center are switched on the tangentia
sub-spacetime. Consequently, along given direction (for example $x^{i}%
$-direction), the local Gamma matrices on the tangential sub-space are
switched by $e^{i\Gamma^{i}\cdot \Delta \phi_{x^{i}}}$ ($\Delta \phi_{x^{i}}=\pi
$): Along $x^{i}$-direction, in the limit of $x^{i}\rightarrow-\infty$, we
have the local Gamma matrices on the tangential sub-space as $\Gamma^{j}$ and
$\Gamma^{k}$; in the limit of $x^{i}\rightarrow \infty$, we have the local
Gamma matrices on the tangential sub-space as
\begin{equation}
e^{i\Gamma^{i}\cdot \Delta \phi_{x^{i}}}(\Gamma^{j})e^{-i\Gamma^{i}\cdot
\Delta \phi_{x^{i}}}=-\Gamma^{j}%
\end{equation}
and
\begin{equation}
e^{i\Gamma^{i}\cdot \Delta \phi_{x^{i}}}(\Gamma^{k})e^{-i\Gamma^{i}\cdot
\Delta \phi_{x^{i}}}=-\Gamma^{k}.
\end{equation}

Due to the rotation symmetry in (3+1)D quantum spacetime, a topological defect
becomes monopole on arbitrary 3D sub-manifold. Along $t$-direction, the
generation of an elementary particle leads to an anti-phase changing
$\Delta \phi_{t}=\pi.$ The local Gamma matrices around a topological defect at
center are switched on the tangentia sub-spacetime along arbitrary
direction.\ That means the elementary particle becomes a "magnetic monopole"
for Gamma matrices $\Gamma^{\mu}(x,t)$.

Then, we use $g$-matrix representation to characterize the topological
property of elementary particles. With help of $g$-matrix representation, we
point out that each elementary particle traps unit \textquotedblleft magnetic
charge\textquotedblright \ of quantum spacetime.

Firstly, we set $\Gamma^{5}=\gamma_{0}.$ By using $g$-matrix representation,
an elementary particle traps a "magnetic charge" of the auxiliary gauge field,
i.e.,
\begin{equation}
N_{F}=\int \sqrt{-g}\Psi^{\dagger}\Psi dV=-q_{m}%
\end{equation}
where $q_{m}=\frac{1}{4\pi}%
{\displaystyle \int}
\epsilon_{jk}\epsilon_{ijk}F_{jk}^{jk}\cdot dS_{i}$ is the "magnetic" charge
of auxiliary gauge field $A^{jk}$. For single particle $N_{F}=1$, the
"magnetic" charge is $q_{m}=1$. Then, we write down the following constraint%
\begin{equation}%
{\displaystyle \int}
\rho_{F}dV=-\frac{1}{4\pi}%
{\displaystyle \int}
\epsilon_{jk}\epsilon_{ijk}F_{jk}^{jk}\cdot dS_{i}%
\end{equation}
where
\begin{align}
F^{ij}  &  =dA^{ij}+A^{ik}\wedge A^{kj}\\
&  \equiv-A^{i0}\wedge A^{j0}\nonumber
\end{align}
and $\rho_{F}=\sqrt{-g}\Psi^{\dagger}\Psi.$ Here, $dV$ and $dS$ are
infinitesimal volume and infinitesimal area on 3D space, respectively. The
upper indices of $F_{jk}^{jk}$ label the local Gamma matrices on the
tangential sub-space and the lower indices of $F_{jk}^{jk}$ denote the spatial
direction. The non-zero Gaussian integrate $\frac{1}{4\pi}%
{\displaystyle \int}
\epsilon_{jk}\epsilon_{ijk}F_{jk}^{jk}\cdot dS_{i}$ just indicates the local
Gamma matrices on the tangential sub-space $A^{i0}\wedge A^{j0}$ to be the
local frame of an orientable sphere with fixed chirality.

We call the equation ($N_{F}=q_{m}$) to be \emph{spacetime Gaussian theorem}
that determines the time evolution of quantum spacetime, i.e.,%
\[
\text{Einstein' equations}\Longleftrightarrow \text{ Spacetime Gaussian
theorem.}%
\]
That means an elementary particle becomes a topological defect of gauge field
in 3D sub-manifold \textrm{M}$_{3}^{\mu=0}$. This leads to an equivalence
principle between matter and topological defect of spacetime in the 3D
sub-manifolds \textrm{M}$_{3}^{\mu=0}$.

In general, under round-robin of generalized gamma matrices $\tilde{\gamma
}^{0}=\alpha \Gamma^{1}+\beta \Gamma^{2}+\gamma \Gamma^{3}+\delta \Gamma^{5}$, an
elementary particle becomes a topological defect of gauge field in arbitrary
3D sub-manifold \textrm{M}$_{3}^{\mu}$! This leads to an equivalence principle
between matter and topological defect of spacetime in arbitrary 3D
sub-manifolds \textrm{M}$_{3}^{\mu}$.

In addition, we give an additional comment on the fermionic statistics of the
elementary particles. On the one hand, because a fermionic particle $\Psi$ as
a spinor in the defining representation of \textrm{SU(2) }group, each particle
has $\frac{1}{2}$ \textquotedblleft electrical charge\textquotedblright \ by
coupling $\omega^{0b}.$ On the other hand, each particle has unit
\textquotedblleft magnetic charge\textquotedblright. Therefore, the fermionic
statistics of elementary particles is obtained \cite{monopole}.

\paragraph{Geometric property of matter}

In traditional quantum mechanics (or quantum field theory), an elementary
particle (for example, an electron) is considered as an infinitesimal point.
Accurately predicting electron's size (or its volume) is an important puzzle.
The importance of predicting elementary particle's size is the same as
predicting the size of the Earth. In this part, we calculate the size of
elementary particles and give an accurate result.

According to above discussion, it was known that 3-volume of spacetime $\Delta
V$ of topological defects of spacetime $q_{m}$ is determined by,
\[
\text{3-volume }\Delta V=4\pi l_{0}^{3}q_{m},
\]
where $q_{m}$ is the number of \textquotedblleft magnetic
monopole\textquotedblright \ and $l_{0}=2l_{p}$ is the twice of Planck length
(This fact will be proved in the following parts).

On the other hand, an elementary particle plays the role of a topological
defect of spacetime on\textit{ }\textrm{M}$_{3}^{\mu}$ ($\mu=x,y,z,t$), i.e.,%
\begin{equation}
N_{F}=-q_{m}.
\end{equation}
where $N_{F}$ denotes the number of particles. Thus, we have
\begin{equation}
N_{F}=-(4\pi l_{0}^{3})^{-1}\Delta V,
\end{equation}
That means the particles have finite 3-volume, or the changing of 3-volume for
spacetime is really determined by the changing of particle number!

Finally, an elementary particle with $N_{F}=\pm1$ has a fixed 3-volume as%
\begin{align}
\Delta V  &  =\pm4\pi l_{0}^{3}=\pm4\pi(2l_{p})^{3}\nonumber \\
&  \sim \pm4.1\times10^{-97}\mathrm{cm}^{3}.
\end{align}
An elementary particle is not only the block of 3D space \textrm{M}$_{3}%
^{\mu=t}$ but also the block of arbitrary 3D-sub-manifold \textrm{M}$_{3}%
^{\mu \neq t}.$ For example, in (2+1)D spacetime \textrm{M}$_{3}^{\mu=z},$ a
Dirac particle has fixed 3-volume as
\[
\Delta V=\pm4\pi l_{0}^{2}t_{0}.
\]

As a result, the distribution of the geometric object from single elementary
particle is obviously described by particle's wave function $\psi
(x)=\sqrt{\Omega(x)}e^{i\varphi(x)}$. Its time evolution obeys Schrodinger's
equation $i\hbar \frac{d\psi(x,t)}{dt}=\hat{H}\psi(x,t)$ where $\hat{H}$ is the
Hamiltonian of elementary particles. Therefore, the density of elementary
particle $\Omega(x)=\int \psi^{\ast}(x)\psi(x)dV$ denotes the distribution of
the changings of 3-volume in space. Then, we have the changings of 3-volume in
given region $\mathcal{M}$ is
\begin{align*}
\Delta V  &  =4\pi(l_{0}^{3})\Delta N_{F}\\
&  =4\pi(l_{0}^{3})\int_{\mathcal{M}}\psi^{\ast}(x)\psi(x)dV.
\end{align*}

In the end, we call the result to be \emph{the principle of equivalence
between matter and spacetime} i.e.,%
\[
\text{Particle}\Longleftrightarrow \text{Block of spacetime.}%
\]

In addition, we obtain a \emph{triangular equivalence principle} about matter
in quantum spacetime. See Fig.4 that shows the intrinsic relationship between
\textquotedblleft Dirac elementary particle\textquotedblright \ (or the
matter), \textquotedblleft3-volume\textquotedblright \ (or the quantum
spacetime itself) and \textquotedblleft magnetic monopole\textquotedblright%
\ (or the topological defect of quantum spacetime). This figure can be
considered as a quantum generalization of the equivalence principle in
classical gravity to triangular equivalence principle about matter in quantum spacetime.

This result also indicates that a particle has a finite size along tempo
direction. Or, \textquotedblleft time\textquotedblright \ is also reality! To
make it clear, we classify the types of changings along tempo direction for an
elementary particle: one is $\pi$-phase changing along tempo direction that is
about its geometry property (or particle's structure), the other is extra
phase changings along tempo direction that is about its dynamic property (or
usual motion).\begin{figure}[ptb]
\includegraphics[clip,width=0.67\textwidth]{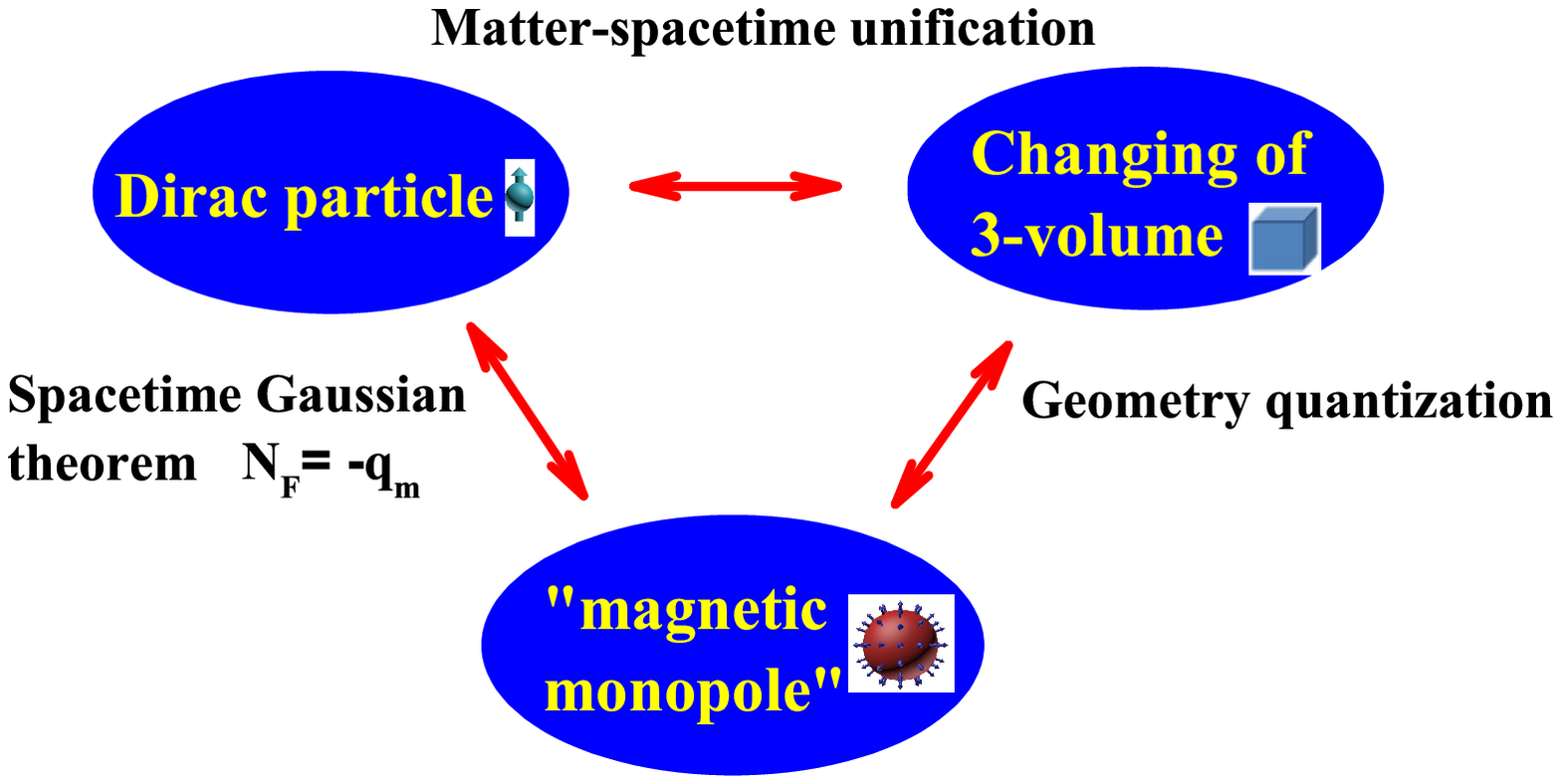}\caption{An illustration
of the triangular equivalence principle in quantum spacetime. This is an
intrinsic relationship between \textquotedblleft Dirac (elementary particle)
particle\textquotedblright \ (or the matter), \textquotedblleft changing\quad
of\quad3-volume\textquotedblright \ (or the quantum spacetime itself) and
\textquotedblleft magnetic monopole\textquotedblright \ (or the topological
defect of quantum spacetime). Here, $N_{F}$ denotes the number of particles,
$q_{m}$ denotes the \textquotedblleft magnetic charge\textquotedblright \ in
gauge representation of quantum spacetime, $\Delta V$ denotes the
changing\quad of\quad3-volume in 3D space of a quantum spacetime. $l_{0}$ is
the lattice constant of the topological lattice with $l_{0}=2l_{p}$ where
$l_{p}$ is Planck length.}%
\end{figure}

\paragraph{Unification of matter and spacetime}

In the first section, we have pointed out that there exists a hidden
assumption -- the separation of spacetime and matter. In general relativity,
although there exists interaction between matter and spacetime, there is a
dualism of two different objects, matter and spacetime and matter may move in
(flat or curved) spacetime. In above section, we found that a particle has a
fixed size rather than a point in spacetime. In this section, we point out the
particles constitute the basic blocks of quantum spacetime and the quantum
spacetime is really a multi-particle system and made of matter.

We then discuss the relationship between different changings of quantum
flat/curved spacetime.

The quantum flat/curved spacetime is uniquely characterized by the coordinates
$\Delta x^{\mu}$ and the local vector's unit $\Gamma^{\mu}(x)$. So, the
changes of a quantum spacetime can be divided into two types, one is
longitudinal about $\Delta x^{\mu}$ (or the contraction/expansion processes
with finite volume changing), and the other is transverse changings about
$\Gamma^{\mu}(x)$ (or shape changings without 3-volume changing). Then,
\emph{what's the intrinsic relationship between longitudinal changings and
transverse changings}? Let us give an answer.

For the case of longitudinal changings of quantum spacetime along $\mu$-th
direction, we have
\begin{equation}
\Delta x^{\mu}\rightarrow(\Delta x^{^{\mu}})^{\prime}=\lambda^{^{\mu}}\Delta
x^{^{\mu}}%
\end{equation}
and
\begin{equation}
\mathbf{e}^{\mu}=\Gamma^{\mu}\rightarrow(\Gamma^{\mu})^{\prime}=\Gamma^{\mu},
\end{equation}
where $\lambda^{\mu}$ is a constant value. Under the longitudinal (or size)
changings of quantum spacetime, the total volume will increase or decrease,
$\Delta V\rightarrow(\Delta V)^{\prime}\neq \Delta V$.

For the case of transverse changings of quantum spacetime, the
\textquotedblleft shape\textquotedblright \ of the system is deformed. Now, we
have
\begin{equation}
\mathbf{e}^{\mu}=\Gamma^{\mu}\rightarrow(\Gamma^{\mu}(x))^{\prime}\neq
\Gamma^{\mu}.
\end{equation}
Under the geometry representation, we have a curved spacetime
\begin{equation}
\Delta x^{\mu}\rightarrow(\Delta x^{^{\mu}}(x))^{\prime}.
\end{equation}
that is characterized by a matrix network $\{ \Gamma_{\mathrm{curved}%
}^{\{N^{\mu},M^{\mu}\}}(x),\mu=x,y,z,t\}$ or the auxiliary gauge fields
$A^{ab}(x)$.\begin{figure}[ptb]
\includegraphics[clip,width=0.67\textwidth]{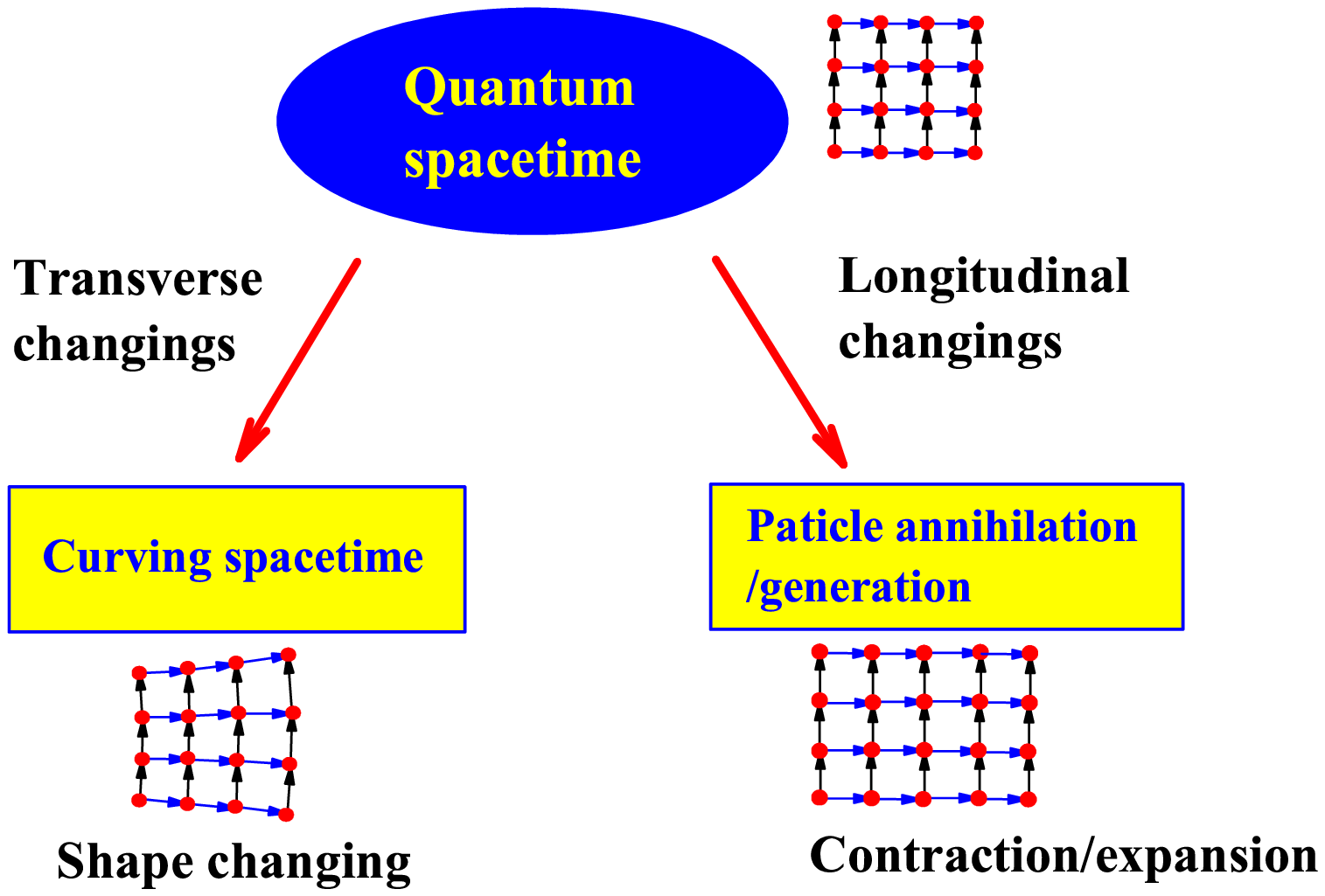}\caption{Classification
of changes of a quantum spacetime -- shape changes (or the processes of
curving spacetime) and contraction/expansion changes (or the processes of
particle annihilation/generation).}%
\end{figure}

Fig.5 shows the two classes of changings of a quantum spacetime: one is about
transverse changings -- shape changings (or the processes for curving
spacetime) that is characterized by a matrix network $\{ \Gamma
_{\mathrm{curved}}^{\{N^{\mu},M^{\mu}\}}(x),\mu=x,y,z,t\}$ and the other is
longitudinal changings -- contraction/expansion changings (or the processes
for single particle annihilation/generation). This result indicates the
unification of quantum mechanics and gravity.

\subsubsection{Theory for Motion and gravity on quantum spacetime}

In this part, we study the motion and gravity of quantum spacetime. We point
out that when additional local longitudinal changings occur (a locally
contraction/expansion changing from motion), transverse changings (or shape
changings) occur. Quantum spacetime becomes globally curved, like a bent
plastic cloth. This gives the mechanism of gravitational force.

\paragraph{Einstein-Hilbert action as topological BF term for $\mathrm{SO(3)}%
^{\mathrm{SO(4)}}$ gauge fields}

Elementary particles play the role of topological defects of quantum
spacetime. To characterize the topological constraint, we introduce
topological BF term. The situation is similar to the Chern-Simons terms in
(2+1)D topological field theory. Under the Chern-Simons term, the local
constraint from flux-charge binding is guaranteed. However, according to the
existence of \textrm{SO(3)}$^{\mathrm{SO(4)}}$ gauge structure, the situation
here is more complex than that for (2+1)D Chern-Simons theory. For different
3D sub-manifolds of the 4D topological lattice, we must define different gauge
fields. It is round-robin of generalized gamma matrices that changes one gauge
class to another. Let us show the details.

We firstly study the local topological constraint on 3D sub-manifold by
setting $\Gamma^{5}=\gamma_{0}.$

Now, an elementary particle traps a "magnetic charge", i.e.,
\begin{equation}
N_{F}=\int \sqrt{-g}\Psi^{\dagger}\Psi dV=-q_{m} \label{monopole}%
\end{equation}
where $q_{m}=\frac{1}{4\pi}%
{\displaystyle \int}
\epsilon_{jk}\epsilon_{ijk}F_{jk}^{jk}\cdot dS_{i}$ is the "magnetic" charge
of auxiliary gauge field $A^{jk}$. For single particle $N_{F}=1$, the
"magnetic" charge is $q_{m}=-1$. We have
\begin{equation}
\rho_{F}=-\epsilon_{0bcd}\epsilon_{0ijk}\frac{1}{4\pi}\hat{D}_{i}F_{jk}^{cd},
\label{con}%
\end{equation}
where $\rho_{F}$ is the density of elementary particles.

We next use Lagrangian approach to characterize the local topological
constraint, $N_{F}=-q_{m}$.

The local topological constraint in Eq.(\ref{con}) can be re-written into
\begin{equation}
\frac{i}{4}\text{\textrm{tr}}\sqrt{-g}\bar{\Psi}\gamma^{i}(\gamma^{0i}%
/2)\Psi=-\epsilon_{jk}\epsilon_{ijk}\frac{1}{4\pi}\hat{D}_{i}F_{jk}^{jk}
\label{12}%
\end{equation}
where $\hat{D}_{i}=i\hat{\partial}_{i}+i\omega_{i}$ is covariant derivative in
(3+1)D spacetime. In the path-integral formulation, to enforce such local
topological constraint, we may add a topological BF term $S_{\mathrm{MBF}}$ in
the action that is
\[
S_{\mathrm{BF1}}=-\frac{i}{4}\text{\textrm{tr}}\sqrt{-g}\bar{\Psi}\varpi
_{0}^{0i}\gamma^{i}(\gamma^{0i}/2)\Psi+\epsilon_{0ijk}\epsilon_{0ijk}%
\varpi_{0}^{0i}\frac{1}{4\pi}\hat{D}_{i}F_{jk}^{jk}%
\]
where $\varpi^{0i}$ is a field that plays the role of Lagrangian multiplier.
The upper index $i$ of $\varpi^{0i}$ denotes the local radial Gamma matrix
around a topological defect, along which the Gamma matrix doesn't change.
Thus, we use the dual field $\varpi^{0i}$ to enforce the topological
constraint in Eq.(\ref{monopole}). That is, to denote the upper index of
$F^{jk}$ that is\ the local tangential Gamma matrices,\ we set antisymmetric
property of upper index of $\varpi^{0i}$ and that of $F^{jk}$.

On the other hand, because $\varpi^{0i}$ and $\omega^{0i}$ have the same
\textrm{SO(3,1)} generator $(\gamma^{0i}/2)$, due to \textrm{SO(3,1)} Lorentz
invariance we can do Lorentz transformation and absorb the dual field
$\varpi^{0i}$ into $\omega^{0i}$, i.e., $\omega^{0i}\rightarrow(\omega
^{0i})^{\prime}=\omega^{0i}-\varpi^{0i}$. As a result, the dual field
$\varpi^{0i}$ is replaced by $\omega^{0i}$ and the first term $-\frac{i}{4}%
$\textrm{tr}$\sqrt{-g}\bar{\Psi}\varpi_{0}^{0i}\gamma^{i}(\gamma^{0i}/2)\Psi$
in $S_{\mathrm{BF1}}$ is absorbed into the Lagrangian of Dirac fermions.

Then, we have%
\begin{align}
S_{\mathrm{BF1}}  &  =\epsilon_{0ijk}\epsilon_{0ijk}\varpi_{0}^{0i}\frac
{1}{4\pi}\hat{D}_{i}F_{jk}^{jk}\\
&  =\text{A total differential term}-\frac{1}{4\pi}\int \epsilon_{0ijk}\,
\epsilon_{0\nu \lambda \kappa}\,R_{0\nu}^{0i}F_{\lambda \kappa}^{jk}\text{ }%
d^{4}x\\
&  =-\frac{1}{4\pi}\int \epsilon_{0ijk}\,R^{0i}\wedge F^{jk}\nonumber
\end{align}
where
\begin{equation}
R^{0i}=d\omega^{0i}+\omega^{0j}\wedge \omega^{ji}.
\end{equation}
From $F^{jk}\equiv-A^{j0}\wedge A^{k0}$ and $e^{i}\wedge e^{j}=(l_{0}%
)^{2}A^{j0}\wedge A^{k0}.$ The induced topological BF term $S_{\mathrm{MBF1}}$
is linear in the conventional strength in $R^{0i}$ and $F^{jk}$. This term is
becomes
\begin{equation}
S_{\mathrm{BF1}}=\frac{1}{4\pi l_{0}^{2}}\int \epsilon_{0ijk}R^{0i}\wedge
e^{j}\wedge e^{k}.
\end{equation}

Furthermore, we use Lagrangian approach to characterize the deformation from a
topological defect on other 3D sub-manifold on (3+1)D spacetime. In general,
for other operation descriptions $\tilde{\gamma}^{0}=\alpha \Gamma^{1}%
+\beta \Gamma^{2}+\gamma \Gamma^{3}+\delta \Gamma^{5}$, a topological defect also
play the role of magnetic monopole and traps a "magnetic charge" of the
corresponding auxiliary gauge fields.

Using the similar approach, we derive another topological BF term
$S_{\mathrm{BF2}}$ in the action that is
\begin{equation}
S_{\mathrm{BF2}}=-\frac{1}{4\pi}\int \epsilon_{0ijk}\,R^{0i}\wedge \tilde
{F}^{jk}\nonumber
\end{equation}
where $R^{0i}=d\omega^{0i}+\omega^{0j}\wedge \omega^{ji}$. From $\tilde{F}%
^{k0}\equiv-\tilde{A}^{kj}\wedge \tilde{A}^{j0}$ and $e^{i}\wedge e^{j}%
=(l_{0})^{2}\tilde{A}^{i0}\wedge \tilde{A}^{j0},$ this term becomes
\begin{equation}
S_{\mathrm{BF2}}=\frac{1}{4\pi l_{0}^{2}}\int \epsilon_{ijk0}R^{0i}\wedge
e^{j}\wedge e^{k}.
\end{equation}
This topological BF term enforces another local topological constraint for
topological defect on $\{x_{i},x_{j},t\}$-sub-manifold. The topological BF
term becomes
\begin{equation}
\frac{1}{4\pi l_{0}^{2}}\int \epsilon_{ijk0}R^{ij}\wedge e^{k}\wedge e^{0}.
\end{equation}

Finally, with the help of a complete set of reduced Gamma matrices
$\gamma^{\mu},$ the total topological BF term is obtained as%
\begin{equation}
S_{\mathrm{BF}}=%
{\displaystyle \sum \limits_{i}}
S_{\mathrm{BF}i}.
\end{equation}
Now, the upper index of the topological BF term $R^{ij}\wedge e^{k}\wedge
e^{l}$ becomes symmetric, i.e., $i,j,k,l=1,2,3,0$.

The full topological BF term $S_{\mathrm{BF}}$ that enforces local topological
constraints for topological defect on all 3D sub-manifold in (3+1)D spacetime,
turns into the Einstein-Hilbert action $S_{\mathrm{EH}}$ as
\begin{align}
S_{\mathrm{BF}}  &  =S_{\mathrm{EH}}=\frac{1}{4\pi l_{0}^{2}}\int
\epsilon_{ijkl}R^{ij}\wedge e^{k}\wedge e^{l}\nonumber \\
&  =\frac{1}{4\pi l_{0}^{2}}\int \sqrt{-g}Rd^{4}x\nonumber \\
&  =\frac{1}{16\pi l_{p}^{2}}\int \sqrt{-g}Rd^{4}x.
\end{align}
This equation indicates that $l_{0}$ is the twice of Planck length,
$l_{0}=2l_{p}$. As a result, we have
\begin{align}
&  \text{The Einstein-Hilbert action}\\
&  =\text{ Topological BF terms for }\mathrm{SO(3)}^{\mathrm{SO(4)}}\text{
gauge fields.}\nonumber
\end{align}

Finally, from above discussion, under geometry representation, we derived an
effective theory of quantum spacetime as%
\begin{align}
S  &  =\mathcal{S}_{\mathrm{4D}}+S_{\mathrm{EH}}\\
&  =\int \sqrt{-g(x)}\bar{\Psi}(e_{a}^{\mu}\gamma^{a}\hat{D}_{\mu}-m)\Psi \text{
}d^{4}x\nonumber \\
&  +\frac{1}{16\pi G}\int \sqrt{-g}R\text{ }d^{4}x\nonumber
\end{align}
where $\mathcal{S}_{\mathrm{4D}}$ characterizes the action for elementary
particles and $G=l_{p}^{2}$. In Einstein-Hilbert action $S_{\mathrm{EH}}$, the
scalar tensor ${R}$ is obtained from the curvature tensor as%
\begin{align}
{R}  &  =g^{\mu \nu}{R}_{\mu \nu},\quad{R}_{\mu \nu}=g^{\rho \sigma}R_{\rho
\mu \sigma \nu},\nonumber \\
R_{\mu \rho \sigma}^{\nu}  &  =\frac{\partial \Gamma_{\mu \sigma}^{\nu}}{\partial
x^{\rho}}-\frac{\partial \Gamma_{\mu \rho}^{\nu}}{\partial x^{\sigma}}%
+\Gamma_{\lambda \rho}^{\nu}\Gamma_{\mu \sigma}^{\lambda}-\Gamma_{\lambda \sigma
}^{\nu}\Gamma_{\mu \rho}^{\lambda}\,,
\end{align}
where $\Gamma_{\mu \sigma}^{\nu}$ are the affine connections
\begin{equation}
\Gamma_{\nu \rho}^{\mu}=\frac{1}{2}g^{\mu \lambda}(\frac{\partial g_{\lambda \nu
}}{\partial x^{\rho}}+\frac{\partial g_{\lambda \rho}}{\partial x^{\nu}}%
-\frac{\partial g_{\nu \rho}}{\partial x^{\lambda}}).
\end{equation}

\paragraph{Time evolution of quantum spacetime and Einstein equations}

According to above discussion, the total action of quantum spacetime is
obtained as%
\begin{align}
S  &  =\mathcal{S}_{\mathrm{4D}}+S_{\mathrm{EH}}\\
&  =\int \sqrt{-g(x)}\bar{\Psi}(e_{a}^{\mu}\gamma^{a}\hat{D}_{\mu}-m)\Psi \text{
}d^{4}x\nonumber \\
&  +\frac{1}{16\pi l_{p}^{2}}\int \sqrt{-g}R\text{ }d^{4}x.\nonumber
\end{align}
After considering the energy-momentum tensor $T_{\mu \nu}=\bar{\psi}\gamma
^{\nu}\partial_{\mu}\psi$, the variation of the total action with respect to
$g_{\mu \nu}$ leads to the traditional Einstein equations,
\begin{align}
G_{\mu \nu}  &  ={R}_{\mu \nu}-\frac{1}{2}g_{\mu \nu}{R}\nonumber \\
&  =\frac{8\pi \,G}{c^{4}}T_{\mu \nu}.
\end{align}
This classical equation describes the evolution of spacetime.

\emph{How about the evolution of quantum spacetime?}

Because the Einstein-Hilbert action $S_{\mathrm{EH}}$ is only a pure
topological constraint term, the Hamiltonian for quantum spacetime themselves
(without considering matter) becomes zero, i.e.,
\begin{equation}
\hat{H}\equiv0!
\end{equation}
Therefore, the evolution of quantum spacetime can not satisfy Schrodinger
equation! Instead, the time evolution in quantum spacetime is determined
spacetime Gaussian theorem. Therefore, the evolution of quantum spacetime is
\emph{self-induced} and does not satisfy the Schrodinger equation. This leads
to time evolution in quantum spacetime itself.

\paragraph{Gravitational waves on quantum spacetime}

Gravitational wave comes from the fluctuating of spacetime \cite{gra}, i.e.,%
\begin{align}
g_{\mu \nu}(x)  &  =\eta_{ab}[e_{\mu}^{a}(x)\cdot e_{\nu}^{b}(x)]\nonumber \\
&  =\eta_{\mu \nu}+h_{\mu \nu}(x),\quad|h_{\mu \nu}(x)|\ll1\,,
\end{align}
where the perturbative field $h_{\mu \nu}$ is a tensor under Lorentz
transformations and coordinate transformations.

At linear order in $h_{\mu \nu}$ the affine connections and curvature tensor
read
\begin{align}
\Gamma_{\mu \rho}^{\nu}  &  =\frac{1}{2}\eta^{\nu \lambda}\,(\partial_{\rho
}h_{\lambda \mu}+\partial_{\mu}h_{\lambda \rho}-\partial_{\lambda}h_{\mu \rho
})\,,\nonumber \\
{R}_{\mu \rho \sigma}^{\nu}  &  \simeq \partial_{\rho}\Gamma_{\mu \sigma}^{\nu
}-\partial_{\sigma}\Gamma_{\mu \rho}^{\nu}.
\end{align}
By introducing the so-called trace-reverse tensor
\begin{equation}
\overline{h}^{\mu \nu}=h^{\mu \nu}-\frac{1}{2}\eta^{\mu \nu}h,
\end{equation}
where $h=\eta_{\alpha \beta}h^{\alpha \beta}$ and $\overline{h}=-h$, the
equation of motion in vacuum turns into
\begin{equation}
\eta_{\rho \sigma}\, \partial^{\rho}\partial^{\sigma}\overline{h}_{\nu \sigma
}=0.
\end{equation}
Gravitational waves propagate at the speed of light. We denote the field
$h_{ij}$ which satisfies the following transverse and traceless gauge
conditions,
\begin{align}
h^{00}  &  =0,\quad h^{0i}=0,\quad \nonumber \\
\partial_{i}h^{ij}  &  =0,\quad h^{ii}=0.
\end{align}

For the case of $+$ polarization described by $h_{ij}^{\mathrm{TT}}%
=h_{+}\left(
\begin{array}
[c]{cc}%
1 & 0\\
0 & -1
\end{array}
\right)  \sin(\omega t-kz)$, we have $\xi_{i}=[x_{0}+\delta x(t),y_{0}+\delta
y(t)],$ where
\begin{align}
\delta x(t)  &  =\frac{h_{+}}{2}\,x_{0}\, \sin(\omega t-kz),\nonumber \\
\delta y(t)  &  =-\frac{h_{+}}{2}\,y_{0}\, \sin(\omega t-kz).
\end{align}
For the case of $\times$ polarization, we have
\begin{align}
\delta x(t)  &  =\frac{h_{\times}}{2}\,y_{0}\, \sin(\omega t-kz),\nonumber \\
\delta y(t)  &  =\frac{h_{\times}}{2}\,x_{0}\, \sin(\omega t-kz)t.
\end{align}

We then take gravitational wave with $+$ polarization along z-direction as an
example to show its quantum spacetime.

Under geometry representation, a gravitational wave with $\times$ polarization
along the z-direction is defined by periodically\ oscillating of lattice
distances on (3+1)D topological lattice, i.e.,
\begin{equation}
(\Delta x^{\mu}(x))_{\mathrm{curved}}=(\Delta x^{\mu}(x))^{\prime},
\end{equation}
where
\begin{align}
\delta x(x)  &  =\varepsilon \cdot x_{0}\sin(\omega t-kz),\nonumber \\
\delta y(x)  &  =\varepsilon \cdot y_{0}\sin(\omega t-kz),\nonumber \\
\delta z(x)  &  =0,\text{ }\delta t(x)=0.
\end{align}
Here, $\varepsilon$ is very tiny, $\phi_{0}\rightarrow0$. We then derive the
local operations
\begin{equation}
\hat{S}(x)=\exp \{i\frac{\varepsilon}{2}(x^{2}\Gamma^{x}+y^{2}\Gamma^{y}%
)\sin(\omega t-kz)\}.
\end{equation}
Now, the ground state turns into $\left \vert \mathrm{vac}(x)\right \rangle
^{\prime}=\hat{S}(x)\left \vert \mathrm{vac}(x)\right \rangle .$ Under spatial
transformation $\mathcal{T}(\delta x)$, we have
\begin{equation}
\mathcal{T}(\delta x)\left \vert \mathrm{vac}(x)\right \rangle ^{\prime
}=e^{i\Gamma^{x}k_{0}\delta x\{1+x\varepsilon \sin(\omega t-kz)\}}\left \vert
\mathrm{vac}(x)\right \rangle ;
\end{equation}
Under spatial transformation $\mathcal{T}(\delta y)$, we have
\begin{equation}
\mathcal{T}(\delta y)\left \vert \mathrm{vac}(x)\right \rangle ^{\prime
}=e^{i\Gamma^{y}k_{0}\delta y\{1+y\varepsilon \sin(\omega t-kz)\}}\left \vert
\mathrm{vac}(x)\right \rangle ;
\end{equation}
Under spatial transformation $\mathcal{T}(\delta z)$, we have
\begin{equation}
\mathcal{T}(\delta z)\left \vert \mathrm{vac}(x)\right \rangle ^{\prime
}=e^{i\Gamma^{z}k_{0}\delta z}\left \vert \mathrm{vac}(x)\right \rangle ;
\end{equation}
Under spatial transformation $\mathcal{T}(\delta t)$, we have
\begin{equation}
\mathcal{T}(\delta t)\left \vert \mathrm{vac}(x)\right \rangle ^{\prime
}=e^{i\Gamma^{t}\omega_{0}\delta t}\left \vert \mathrm{vac}(x)\right \rangle .
\end{equation}

Under $\Gamma$-matrix representation, the perturbation of the spacetime comes
from fluctuating matrix network, i.e.,
\begin{equation}
\Gamma_{\mathrm{curved}}^{\{N^{\mu},M^{\mu}\}}(x)=\hat{S}(x)\Gamma
_{\mathrm{flat}}^{\{N^{\mu},M^{\mu}\}}\hat{S}(x))^{-1}.
\end{equation}
As a result, we have the periodically\ oscillating Gamma matrices
$\Gamma_{\mathrm{curved}}^{\{N^{\mu},M^{\mu}\}}(x)$. To locally derive the
matrix network $\Gamma_{\mathrm{curved}}^{\{N^{\mu},M^{\mu}\}}(x)$ around the
point $(0,0,0,0)$, we have $x_{0}=l_{0}=1$ or $y_{0}=l_{0}=1.$ Then, the local
operations on the links $\{N^{\mu},M^{\mu}\}=\{(0,0,0,0),(1,0,0,0)\}$ turn
into
\begin{equation}
\hat{S}(x)=\exp \{ \frac{i\varepsilon}{2}\Gamma^{x}\sin(\omega t-kz)\};
\end{equation}
the local operations on the links $\{N^{\mu},M^{\mu}%
\}=\{(0,0,0,0),(0,1,0,0)\}$ turn into
\begin{equation}
\hat{S}(x)=\exp \{ \frac{i\varepsilon}{2}\Gamma^{y}\sin(\omega t-kz)\};
\end{equation}
the local operations on the links $\{N^{\mu},M^{\mu}%
\}=\{(0,0,0,0),(1,1,0,0)\}$ turn into
\begin{equation}
\hat{S}(x)=\exp \{i\frac{\varepsilon}{2}(\Gamma^{x}+\Gamma^{y})\sin(\omega
t-kz)\};...
\end{equation}
So, different gravitational waves are described by different matrix networks.

After obtaining the $\Gamma$-matrix representation, we get the $\gamma$-matrix
representation, i.e.,
\begin{align}
\hat{S}(x)  &  =\exp \{i\frac{\varepsilon}{2}(x^{2}\Gamma^{x}+y^{2}\Gamma
^{y})\sin(\omega t-kz)\} \nonumber \\
&  =\exp \{i\frac{\varepsilon}{2}(-x^{2}\gamma^{12}+y^{2}\gamma^{23}%
)\sin(\omega t-kz)\}.
\end{align}
Because there is no change of lattice distance along tempo direction, there is
no necessity to do a round-robin.\emph{ }Under the definition of $\gamma
^{0}=\Gamma^{5}$, the gauge representation can also be derived as
\begin{equation}
A_{\mu}^{ab}(x)=\mathrm{tr}(\gamma^{ab}(\hat{S}(x))\frac{d}{\partial x_{\mu}%
}(\hat{S}(x))^{-1})
\end{equation}
and
\begin{align}
A_{\mu}^{a0}(x)  &  =\mathrm{tr}(\gamma^{a0}\hat{S}(x))d(\hat{S}%
(x))^{-1})\nonumber \\
&  =\gamma^{0}\frac{d}{\partial x_{\mu}}(\gamma^{a}(x))^{-1},
\end{align}
where $a,b=1,2,3$.

By using similar approach, we can obtain quantum representation for other
curved spacetimes.

Another important problem is \emph{scattering amplitude }for gravitons. It was
known that this problem is relevant to `type II' ambitwistor
superstrings\cite{am1}. In the following parts, we will separately show the
calculations on scattering amplitude for gravitons.

\paragraph{Gravitational force and "weak" equivalent principle}

Gravitational force leads to attraction effect on massive objects. As a
result, gravitational force is responsible for keeping the planets in motion
around the Sun and the Moon around the Earth. Newton was the first to discover
the laws of gravitational force,
\[
F=G\frac{m_{A}m_{B}}{r^{2}}%
\]
where $G=\frac{c^{3}}{\hbar}l_{p}^{2}=\frac{c^{5}}{\hbar}t_{p}^{2}$ is the
Newton constant, $r$ is the distance, and $m_{A}$ and $m_{B}$ are the possess
masses for two objects A and B. By setting $c=1$ and $\hbar=1,$ we have
$G=l_{p}^{2}=t_{p}^{2}.$ For elementary particles, the gravitational force
between them is very tiny.

In this part, we discuss the gravitational interaction between two massive
elementary particles.

According to above discussion, there exists motion charge (or charge of
motion) $\frac{\Delta \omega}{\omega_{0}}$ along tempo direction for massive
elementary particles,
\[
Q_{t}=\frac{\Delta \omega}{\omega_{0}}=\frac{mc^{2}}{\omega_{0}\hbar}%
\]
with $m=\hbar(\omega_{0}-ck_{0})/c^{2}.$ The motion charge along tempo
direction characterizes the size changing of a moving elementary particle in
Cartesian spacetime \textrm{C}$_{3+1}$ along tempo direction. By using the
motion charge $Q_{t},$ we can rewrite the gravitational force
\begin{align}
F  &  =G\frac{m_{A}m_{B}}{r^{2}}\label{f}\\
&  =\frac{c^{5}}{\hbar}t_{p}^{2}m_{A}m_{B}\frac{1}{r^{2}}\nonumber \\
&  =\frac{c^{5}}{\hbar}(\frac{2\pi}{\omega_{0}})^{2}m_{A}m_{B}\frac{1}{r^{2}%
}\nonumber \\
&  =\kappa \frac{Q_{t}^{A}Q_{t}^{B}}{r^{2}}\nonumber
\end{align}
where $Q_{t}^{A}=\frac{m_{A}c^{2}}{\omega_{0}\hbar},$ $Q_{t}^{B}=\frac
{m_{B}c^{2}}{\omega_{0}\hbar},$ and $\kappa=2\pi c\hbar.$

From above equation of gravitational force, we find that the motion charge
becomes dimensionless parameter characterizes gravitational interaction. The
smaller the motion charge (or mass), the smaller the gravity. Based on
Eq.\ref{f}, we give an explanation on the microscopic physical mechanism for
gravitational force from the motion charge.

When a massive elementary particle is generated onto a quantum flat spacetime,
the 3-volume locally changes anisotropically along tempo direction (due to
existence of motion charge along tempo direction). As a result, the spacetime
will be globally deformed due to the existence of local anisotropy induced by
particle's mass. Due to symmetry of different direction in 3D space, the
changings of shape anisotropy along tempo direction has inverse square law.
When the quantum spacetime is curved by the local shape anisotropy induced by
a massive elementary particle, the shape of other elementary particles becomes
changed. As a result, gravitational force appears and the motion charge can be
regarded as the charge of gravitational interaction. The larger of motion
charge (particle's mass), the larger anisotropy of the shape of an elementary
particle, then the larger of the gravitational interaction.

In addition, we discuss \emph{"weak" equivalent principle} between inertial
mass and gravitational mass.

Einstein had proposed this equivalent principle, i.e., \textit{inertial mass
about dispersion }$(\Delta \omega)=(c\Delta \vec{k})^{2}+m^{2}$\textit{ and
gravitational mass about interaction }$F=G\frac{m_{A}m_{B}}{r^{2}}$\textit{
are the same thing. }To explain the "weak" equivalent principle, the key point
is motion charge $Q_{t}=\frac{mc^{2}}{\omega_{0}\hbar}$.

On the one hand, the motion charge $Q_{t}$ is proportional to the inertial
mass $m=\frac{\omega_{0}\hbar}{c^{2}}Q_{t}$ that characterizes the deviation
of periodic motion from $ck_{0},$ i.e., $(\omega_{0}-ck_{0})=Q_{t}\omega_{0}.$
Then, the motion charge plays important role in dispersion $(\Delta
\omega)=(c\Delta \vec{k})^{2}+m^{2}=(c\Delta \vec{k})^{2}+(\frac{\omega_{0}%
\hbar}{c^{2}}Q_{t})^{2};$ On the other hand, the motion charge $Q_{t}$
characterizes the local anisotropy of spacetime induced by the extra massive
elementary particles. Then, the local anisotropy curves the spacetime. There
exists gravitational interaction $F=\kappa \frac{Q_{t}^{A}Q_{t}^{B}}{r^{2}}$
(or $F=G\frac{m_{A}m_{B}}{r^{2}}$) between two massive elementary particles.

In summary, we say that the "weak" equivalent principle between inertial mass
and gravitational mass comes from the equivalent between the deviation of
dispersion $ck_{0}$ for the elementary particle and the deviation of isotropy
of quantum spacetime.

\paragraph{Quantum motion on curved spacetime and "strong" equivalence
principle}

In this part, we discuss the motion of elementary particles in a curved
spacetime and provide an explanation on "strong" equivalence principle.

According to general relativity, the motion of elementary particles is
described by the well known geodesic equation
\begin{equation}
\frac{d^{2}x^{\mu}}{d\tau^{2}}+\Gamma_{\lambda \sigma}^{\mu}\frac{dx^{\lambda}%
}{d\tau}\frac{dx^{\sigma}}{d\tau}=0,\text{ }\mu=0,1,2,3
\end{equation}
where $\Gamma_{\lambda \sigma}^{\mu}$ is the Christoffel connection for a
Riemannian metric. The derivative of the four-position $x^{\mu}$ with respect
to an affine parameter $\tau$ is the contravariant four-velocity,
$\frac{dx^{\mu}}{d\tau}=u^{\mu},$ in units where $c=1$. In our theory, we have
the same geodesic equation. In matrix representation, the geodesic lines
correspond to the lines with same Gamma matrix $\Gamma^{a}$.

\begin{figure}[ptb]
\includegraphics[clip,width=0.47\textwidth]{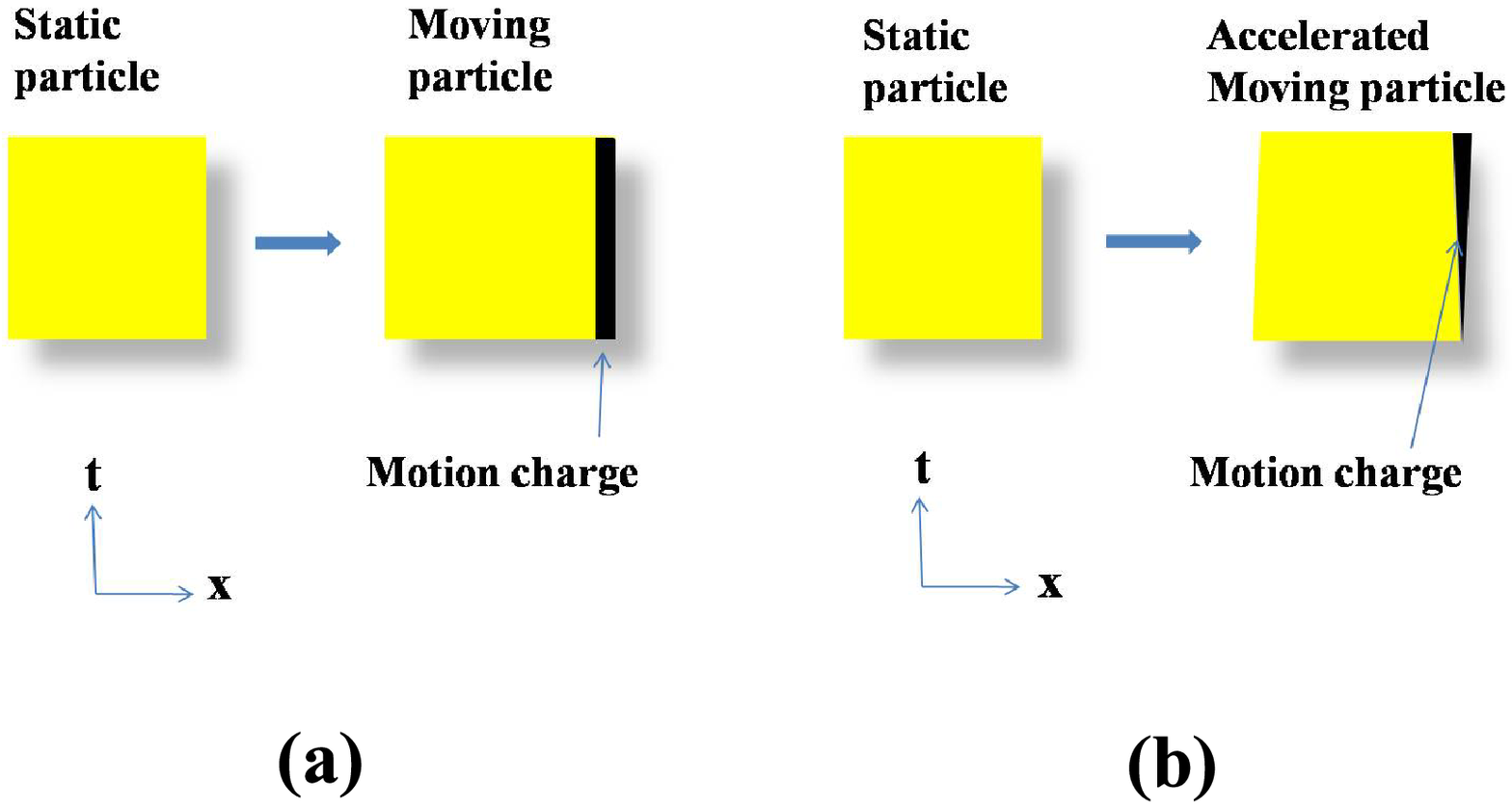}\caption{(a) An
illustration of shape changing and motion charge for a moving elementary
particle. Now, the motion charge is constant and the shape of moving
elementary particle is rectangle; (b) The shape changing for an accelerated
elementary particle. Now, the motion charge becomes time-dependent. $x$
denotes the coordinate along moving direction. Now, the shape of moving
elementary particle is trapezoid.}%
\end{figure}

On the one hand, we discuss the accelerated elementary particle.

For an accelerated elementary particle, the velocity $\vec{v}$ is no more
constant, $\vec{v}=\vec{v}(t).$ Now, wave vector becomes time-dependent
\[
\Delta \vec{k}(t)=\frac{E(\Delta \vec{k})}{c^{2}}\vec{v}(t)
\]
\ where $E(\Delta \vec{k})$ is its energy. So, the instantaneous motion charge
along motion direction is also time-dependent
\[
\vec{Q}(t)=\frac{\Delta \vec{k}(t)}{k_{0}}.
\]
The time-dependent\ motion charge indicates that the anisotropy of the
elementary particle becomes time-dependent. At two ends of an elementary
particles along tempo direction, due to different motion charges, the
particle's shape changes from \emph{rectangle} to \emph{trapezoid}. See the
illustration in Fig.6(b). The situation is same to that for an elementary
particle in curved spacetime. Now, the particle's shape in curved spacetime is
also trapezoid that corresponds to certain accelerated elementary particles.
This is just the mechanism of "strong" equivalence principle.

In summary, from the point view of particle's geometry, "strong" equivalence
principle indicates the equivalence between the trapezoid-like geometric
structure for the accelerated elementary particle on flat spacetime and that
for the elementary particle on curved spacetime.

\subsection{Generalized symmetry for quantum spacetime}

To define a quantum spacetime, a key point is to generalize "symmetry" or
"invariant" of usual field to (higher-order) variability. \emph{What's the
invariant/symmetry of quantum spacetime}? In this section, we develop the
theory about generalized symmetry for quantum spacetime.

\subsubsection{Review on generalized symmetry for quantum fields}

Firstly, we review generalized symmetry for quantum field theories (or quantum
many-body systems)\cite{s}. Generalized symmetry plays important role to unify
different physical phenomena in quantum field theory, condensed matter theory,
and particle physics. There exist different types of generalized symmetries,
such as higher-form symmetries, higher-group symmetries, non-invertible symmetries.

Generally, with the help of Noether's theorem, for a quantum field or a
quantum many-body system, a global continuous symmetry \textrm{G} is known to
guarantee \emph{conservation current} $J_{\mu}$ satisfying
\begin{equation}
\partial^{\mu}J_{\mu}=0.
\end{equation}
As a result, conservation current and symmetry become two sides of a coin.
Conservation indicates that for a moving object, its charge $Q$ does not
change over time. For example, for a quantum system with (0-form) generalized
symmetry \textrm{U(1)}, we have
\begin{equation}
\lbrack \hat{Q},\hat{H}]=0
\end{equation}
where $\hat{H}$ is the Hamiltonian of the system. Due to $[\hat{Q},\hat
{H}]=0,$ we then define a time-independent unitary operator -- \emph{symmetry
operator} (or \emph{topological operator})
\begin{equation}
U=e^{i\alpha \hat{Q}}%
\end{equation}
that denotes a family of operators within a limited region of spacetime. For
an object with charge $\hat{Q}$ under a (0-form) generalized symmetry created
by local operators $\psi^{\dagger}(x)$, we have
\begin{equation}
\psi^{\dagger}(x)\rightarrow U\psi^{\dagger}(x)U^{-1}=e^{iQ\alpha}%
\psi^{\dagger}(x).
\end{equation}
This describes the \emph{changing structure} for particle that is generated by
$\psi^{\dagger}(x)$ under $U.$

In addition, for a charge operator $\hat{Q},$ we have a canonical quantization
condition of charge as $[\varphi,\hat{Q}]=i.$ According to the canonical
quantization condition, $U=e^{i\alpha \hat{Q}}$ is an operator that changing
the phase angle $\alpha.$

This formalism is naturally generalized to the cases of \emph{extended}
operators. $p$-form generalized global symmetries act on $p$-dimensional
charged operators and are implemented by $(d-p-1)$-dimensional surface
operators. For general integer $p\geq-1$, a $p$-form symmetry means the
existence of topological operators $U_{\alpha}(\Sigma_{D-p-1})$ labeled by a
group element $\alpha$ and a closed codimension-$(p+1)$ submanifold of
spacetime. Here, $\Sigma$ is a closed $d$-dimensional surface, of codimension
one in spacetime. For coincident submanifolds, these operators satisfy the
\textquotedblleft fusion rule" $U_{\alpha}(\Sigma)U_{\beta}(\Sigma
)=U_{\alpha+\beta}(\Sigma)$. The operators charged under a $p$-form symmetry
are supported on $p$-dimensional loci, and create $p$-brane excitations. The
conservation law asserts that the $(p+1)$-dimensional world-volume of these
excitations will not have boundaries.

In summary, generalized symmetry is a concept that characterizes both
\emph{variability} and \emph{invariant/symmetry} of quantum systems: On one
hand, it characterizes local variability via $\psi^{\dagger}(x)\rightarrow
U\psi^{\dagger}(x)U^{-1}=e^{iQ\alpha}\psi^{\dagger}(x).$ This equation means
that local field induces phase changing; on the other hand, it characterizes
global Invariant/symmetry via $UQU^{-1}=Q$ and $\frac{dQ}{dt}=0,$ or
$[Q,\hat{H}]=0$. This equation means that the charge $Q$ of the local field
$\psi^{\dagger}(x)$ is topological and invariant under time evolution.

Therefore, to define a generalized symmetry, one need to follow the following
research steps,
\begin{align*}
\text{A quantum system}  &  \rightarrow \text{ conservation current }J_{\mu}\\
&  \rightarrow \  \text{charge }\hat{Q}\\
&  \rightarrow \text{\ symmetry operator }U=e^{i\alpha \hat{Q}}\\
&  \rightarrow U\psi^{\dagger}(x)U^{-1}=e^{iQ\alpha}\psi^{\dagger}(x).
\end{align*}

\subsubsection{Generalized symmetry for quantum spacetime}

In above section, we discuss the generalized symmetry for quantum fields.
Generalized symmetry is a concept that characterize both \emph{variability}
and \emph{invariant/symmetry} of quantum systems. In this section, we turn to
study generalized symmetry for quantum spacetime.

Generalized symmetry for quantum spacetime is really a generalized
differential homeomorphism invariance that characterizes local variability.

Firstly, we define \emph{"topological" charge} of quantum spacetime.

According to above discuss, we have a spacetime Gaussian theorem $Q^{\mu
}=q_{m}^{\mu}.$ Then, for an arbitrary 3D subspace \textrm{M} of quantum
spacetime, the "topological" charge is just the number of \textquotedblleft
magnetic monopole\textquotedblright \
\begin{equation}
Q^{\mu}\rightarrow q_{m}^{\mu}=\frac{1}{4\pi}%
{\displaystyle \oint \nolimits_{\mathcal{S}}}
\epsilon_{cd}\epsilon_{ijk}F_{jk}^{cd}\cdot dS_{i}.
\end{equation}
We may call $q_{m}^{\mu}$ to be \emph{spacetime charge }(\emph{charge of
spacetime}). So, we have infinite "topological" charges, each of which
corresponds to an element of compact $\mathrm{SO(4)}$ group. Or, on each point
of compact $\mathrm{SO(4)}$ group space, we have a topological
spacetime-charge. So, to characterize a quantum spacetime, we must define
infinite topological spacetime-charges. In addition, we can change
"topological" charge of a 3D subspace \textrm{M}$_{3}^{\mu}$ to another
\textrm{M}$_{3}^{\mu^{\prime}}$ by doing round-robin $\mathrm{R}^{\mu
\mu^{\prime}}$ that corresponds to a global $\mathrm{SO(4)}$ rotation
operation, i.e.,%
\[
\mathrm{R}^{\mu \mu^{\prime}}Q^{\mu}(\mathrm{R}^{\mu \mu^{\prime}})^{-1}%
=Q^{\mu^{\prime}}.
\]
We call it a \emph{tribe} of spacetime charges. Hence, the situation is quite
different from that of generalized symmetry in quantum fields.

On the other hand, according to above discussion, for geometric objects in
quantum spacetime, the changing of \textquotedblleft magnetic
monopole\textquotedblright \ $q_{m}^{\mu}$ leads to the changing of the
3-volume, i.e.,
\begin{equation}
\Delta V^{\mu}=4\pi l_{0}^{3}q_{m}^{\mu}.
\end{equation}
That means topological "object" of quantum spacetime is also the geometric
object with finite 3-volume. The contraction/expansion of quantum spacetime
leads to changing of topological spacetime-charges.

Secondly, we define \emph{the invariant/symmetry }of quantum spacetime.\emph{
}

The "object" or "local operation" of quantum spacetime is local
contraction/expansion of spacetime that is also a cluster of elementary
particles with finite 3-volume. Especially, its shape can be deformed
arbitrarily on a curved spacetime. \emph{What's the invariant/symmetry?} To
answer the question, we check the types of operations (or changings), under
which the "object" (or "local operation") doesn't change.

It was known that the "object" (or "local operation") here is local
contraction/expansion of spacetime. The operations without changing 3-volume
of the quantum spacetime belong to the operations for invariant/symmetry.
Therefore, the invariant/symmetry is \emph{differential homeomorphism
invariance} and the operations for differential homeomorphism
invariance\ comes from \emph{local coordinate transformations,} including
spatial/tempo translation operation, space rotation, i.e., $x^{\mu}%
\rightarrow(x^{\mu}(x))^{\prime},$ where $(x^{\mu}(x))^{\prime}$ is
invertible, differentiable and with a differentiable inverse. For an arbitrary
3D subspace \textrm{M}$_{3}^{\mu},$ the topological charge of quantum
spacetime $Q^{\mu}$ (the 3-volume $\Delta V^{\mu}$, or particle's number
$N_{F}$) will never be changed under local coordinate transformation on 3D
subspace \textrm{M}$_{3}^{\mu}$. For another 3D subspaces \textrm{M}$_{3}%
^{\mu}$ by round-robin, we have same results. This fact had been called
topology stationarity of matter and can be easily understood by considering
the invariant of the size of Clifford group-changing space under different
mapping to Cartesian space.

Thirdly, we define \emph{symmetry operator}.

In above part, we point out that the invariant/symmetry is differential
homeomorphism invariance and the operations for differential homeomorphism
invariance\ comes from local coordinate transformations. However, due to the
mismatch of the operations on $\gamma^{\mu}$ and those on $\Gamma^{\mu}$ (or
matrix network $\Gamma_{\mathrm{curved}}^{\{N^{\mu},M^{\mu}\}}$), the symmetry
operator is unusual. We must split the four dimensional quantum spacetime into
3+1 where "3" represents 3D subspace \textrm{M}$_{3}^{\mu}$ and "1" the
residue 1D subspace $x^{\mu}$. As a result, the operations for
invariant/symmetry belong to two classes: one is about rotation/translation
operation $V_{\mathrm{M}_{3}^{\mu}}^{\mu}$ in 3D subspace \textrm{M}$_{3}%
^{\mu},$ the other is about translation operation $U^{\mu}(\phi^{\mu})$ in the
residue 1D subspace $x^{\mu}.$ To characterize the translation symmetry along
other direction, we do round-robin.

In addition, for a quantum spacetime, the Hamiltonian is zero, $\hat{H}%
\equiv0$. Therefore, we don't worry about the the condition of $[\hat{Q}^{\mu
},\hat{H}]=0$. For each topological operator $\hat{Q}^{\mu},$ we have a
canonical quantization condition of charge as $[\varphi^{\mu},\hat{Q}^{\mu
}]=i.$ According to the canonical quantization condition, $U^{\mu}%
=e^{i\varphi^{\mu}Q^{\mu}}$ is an operator that changing the phase angle
$\varphi^{\mu}$ along $\mu$-direction.

We then define the symmetry operator (or topological operator)
\begin{equation}
U^{\mu}=e^{i\phi^{\mu}\hat{Q}^{\mu}}=\exp(\frac{i\Delta \hat{V}^{\mu}\phi^{\mu
}}{4\pi l_{0}^{3}})
\end{equation}
that denotes a (translation) operator along $\mu$-direction within a limited
region of spacetime on 3D subspace \textrm{M}$_{3}^{\mu}$. $\Delta \hat{V}%
^{\mu}$ denotes an operator of local contraction/expansion of quantum
spacetime. Under the rotation/translation transformation $V_{\mathrm{M}%
_{3}^{\mu}}^{\mu}$, the topological operator $\hat{Q}^{\mu}$ is obvious
invariant, i.e.,
\[
V_{\mathrm{M}_{3}^{\mu}}^{\mu}Q^{\mu}(V_{\mathrm{M}_{3}^{\mu}}^{\mu}%
)^{-1}=Q^{\mu}.
\]

\subsubsection{Summary}

Generalized symmetry is a suitable way to characterize quantum spacetime. The
topological charge for quantum spacetime is topological spacetime-charges
$Q^{\mu}$ that is number of \textquotedblleft magnetic
monopole\textquotedblright \ $Q^{\mu}=q_{m}^{\mu}=\frac{1}{4\pi}%
{\displaystyle \oint \nolimits_{\mathcal{S}}}
\epsilon_{cd}\epsilon_{ijk}F_{jk}^{cd}\cdot dS_{i}$ in an arbitrary 3D
subspace \textrm{M.} Therefore, generalized symmetry for quantum spacetime
represents the invariance of coordinate transformations (that don't change
3-volume) for locally contracts/expands of quantum spacetime (that change
3-volume). In particular, for generalized symmetry of quantum spacetime, there
exist a tribe of symmetry operations $U^{\mu}$ for corresponding topological
spacetime-charges $Q^{\mu}$ and non-topological operation $V_{\mathrm{M}%
_{3}^{\mu}}^{\mu}.$

In the end, we compare the generalized symmetries for quantum fields and those
for quantum spacetime.

One is about \emph{Noether's theorem} and \emph{conservation current}. For
moving quantum object, its wave function varies and obeys Schrodinger
equation. However, for a quantum spacetime, its motion comes from shape
changing that curves the spacetime. It is Einstein equation rather than
Schrodinger equation that describes its states under time evolution.
Therefore, Noether's theorem can be applied to quantum field with global
continuous symmetry, but cannot be applied to a quantum spacetime with 1-th
order variability.

The second difference comes from \emph{topological charge }and\emph{ symmetry
operator}. For a usual quantum field with given symmetry, its topological
charge and symmetry operator is unique. However, for a quantum spacetime, we
have a tribe of topological charges and symmetry operators rather than a
single one. Or, there exist\ topological spacetime charges and corresponding
symmetry operators on each 3D subspace \textrm{M}$_{3}^{\mu}.$

The third difference is about \emph{"form"} of objects. In quantum system, we
have extended operators with different dimensions, such as 0D point-like
objects, 1D line-like objects, 2D surface-like objects,... However, for
quantum spacetime, all object have finite 3-volume in (3+1)D quantum spacetime
(or finite (d)-volume in (d+1)D quantum spacetime).

\subsection{Other issues relevant to quantum spacetime}

\subsubsection{Canonical quantization for quantum spacetime and spacetime
uncertainty}

In canonical quantization, if the action is written as $S=\int(\frac{dA}%
{dt}\cdot B)dt,$ where $A$ and $B$ are considered to be a pair of canonical
coordinate and canonical momentum. As a result, in quantum mechanics, we have
\begin{equation}
A\rightarrow \hat{A},\text{ }B\rightarrow \hat{B},
\end{equation}
and
\begin{equation}
\left[  \hat{A},\hat{B}\right]  =i.
\end{equation}
This leads to the uncertainty principle,
\begin{equation}
\Delta A\cdot \Delta B\geq \frac{1}{2}.
\end{equation}

Therefore, to derive the canonical quantization in quantum spacetime, the
action must be written as a standard form $S=\int(\frac{dA}{dt}\cdot B)dt$ and
check what are $A$ and $B$.

Firstly, we transform\ the Einstein-Hilbert action $S_{\mathrm{EH}}$ into
exterior derivative form,
\begin{align}
S_{\mathrm{EH}}  &  =\frac{1}{16\pi G}\int \sqrt{-g}Rd^{4}x\nonumber \\
&  =\frac{1}{16\pi G}\int \epsilon_{abcd}R^{ab}\wedge e^{c}\wedge e^{d}.
\end{align}
From the relationship between gauge fields $A^{ab}(x)$\ in gauge
representation\ or vierbein fields $e^{a}(x)$ in geometric representation
$e^{a}\wedge e^{b}=(l_{0})^{2}A^{a0}\wedge A^{b0},$ we have%
\begin{align}
&  \frac{1}{16\pi G}\int \epsilon_{0ijk}R^{0b}\wedge e^{c}\wedge e^{d}%
\nonumber \\
&  =\frac{1}{16\pi G}(l_{0})^{2}\int \epsilon_{0bcd}R^{0b}\wedge A^{c0}\wedge
A^{d0}.
\end{align}
With the help of the Maurer-Cartan equation $F^{ab}\equiv-A^{a0}\wedge
A^{b0},$ these terms turn into a topological one,
\begin{equation}
-\frac{1}{16\pi G}(l_{0})^{2}\int \epsilon_{0bcd}\,R^{0b}\wedge F^{cd}%
\end{equation}
where $R^{0a}=d\omega^{0a}+\omega^{0b}\wedge \omega^{ba}.$ After doing a
partial integral, we have
\begin{equation}
\frac{1}{16\pi G}(l_{0})^{2}\int \epsilon_{0bcd}\, \omega^{0b}\wedge(DF^{cd}),
\end{equation}
where $(DF^{cd})$ is proportional to the density of magnetic monopoles.

Then, we rewrite above action as canonical quantization formula. For the case
of a uniform $\omega_{0}^{0a}=\frac{de^{a}}{(l_{0})dt}$, we have
\begin{align}
\mathcal{L}  &  =\frac{1}{16\pi G}\frac{de^{a}}{(l_{0})dt}\epsilon_{acd}%
(l_{0})^{2}(\int DF^{cd})\nonumber \\
&  =\frac{1}{16\pi G(l_{0})^{2}}\frac{de^{a}}{dt}\cdot V_{\mathrm{total}}^{a},
\end{align}
where $V_{\mathrm{total}}^{a}=(l_{0})^{3}%
{\displaystyle \oint \nolimits_{\mathcal{S}}}
F_{\mathcal{S}}^{IJ}$ is the total 3-volume that is perpendicular to the
direction $e^{a}$. For example, when $a=t,$ $V_{\mathrm{total}}^{a}$ denotes
the 3-volume of 3D space and $e^{a}$ is the uniform vierbein field along tempo direction.

As a result, the vierbein fields $e^{a}$ (that is proportional to the total
size $L^{a}=\Delta x^{a}$ of coordinates $x^{a}$) and the total 3-volume
$V_{\mathrm{total}}^{a}$ perpendicular to this direction\ become a pair of
canonical coordinate and canonical momentum operators. As a result, in quantum
mechanics, we have $e^{a}\rightarrow \hat{e}^{a}$ $V_{\mathrm{total}}%
^{a}\rightarrow \hat{V}_{\mathrm{total}}^{a}$, and
\begin{equation}
\lbrack e^{a},\frac{1}{16\pi G(l_{0})^{2}}\hat{V}_{\mathrm{total}}^{a}]=i,
\end{equation}
or
\begin{equation}
\lbrack e^{a},\hat{V}_{\mathrm{total}}^{a}]=i16\pi G(l_{0})^{2}.
\end{equation}
That means $L^{a}$ and total 3-volume $V_{\mathrm{total}}^{a}$ perpendicular
to this direction do not commutate,
\begin{equation}
\lbrack L^{a},\hat{V}_{\mathrm{total}}^{a}]=i16\pi G(l_{0})^{2}.
\end{equation}
This leads to an uncertainty principle of quantum spacetime
\begin{equation}
\Delta V_{\mathrm{total}}^{a}\cdot \Delta e^{a}>8\pi Gl_{0}^{2}%
\end{equation}
or
\begin{equation}
\Delta V_{\mathrm{total}}^{a}\cdot L^{a}>8\pi Gl_{0}^{2}.
\end{equation}

From the relationship between particles, and 3-volume of them, $N_{F}=(4\pi
l_{0}^{3})^{-1}\Delta V,$ we have $\hat{N}_{F}=(4\pi l_{0}^{3})^{-1}\Delta
\hat{V}$ where $\hat{N}_{F}$ is the operator of particle number. On the other
hand, the canonical quantization condition of quantum spacetime is obtained
as
\begin{align}
\lbrack \hat{L}^{a},\hat{V}_{\mathrm{total}}^{a}]  &  =[\hat{L}^{a},4\pi
(l_{0})^{3}\hat{N}_{F}]\nonumber \\
&  =i16\pi(l_{p})^{2}(l_{0})^{2}=i4\pi(l_{0})^{4},
\end{align}
where $L^{a}=\Delta x^{a}$ is total size of coordinates $x^{a}$ and
$V_{\mathrm{total}}^{a}$ is the total 3-volume perpendicular to this
direction. As a result, the canonical quantization condition of quantum
spacetime becomes
\begin{equation}
\lbrack \hat{L}^{a},\hat{N}_{F}]=il_{0}%
\end{equation}
or
\begin{equation}
\lbrack \hat{N}^{a},\hat{N}_{F}]=i,
\end{equation}
where $\hat{N}^{a}=\frac{\hat{L}^{a}}{l_{0}}$ denotes the operator of lattice
sites of a (3+1)D topological lattice.\

This canonical quantization condition means particle number on space does not
commutate the lattice number along tempo direction! \emph{Why?} The reason
comes from the fact that elementary particle as changing unit in Clifford
group-changing space. It was known that, the generation/annihilation of an
elementary particle leads to $\pi$-phase changing of Clifford space along an
arbitrary direction, including both spatial direction and tempo direction.
Therefore, many elementary particles have finite 3-volume $\Delta V=4\pi
(l_{0})^{3}\hat{N}_{F}$ in \textrm{M}$_{3}^{\mu}$ and $\hat{N}_{F}\pi$-phase
changing along $\phi^{\mu}$ direction. The synchronous changings of elementary
particles in different directions in quantum spacetime naturally leads to a
canonical quantization $[\hat{N}^{\mu},\hat{N}_{F}]\neq0$. We may call it
\emph{spacetime duality} (\textrm{M}$_{3}^{\mu}$ and its complementary space
$\phi^{\mu}$) for canonical quantization condition in quantum spacetime.

\subsubsection{It from qubit and "Whole wave functions"}

"\emph{It from Qubit}" is a belief to understand the origins of spacetime from
quantum entanglement. To follow this idea, there are two different
methodologies: One is \emph{Reductionism} from top to down, the other is
\emph{Emergence} from down to up.

Following the methodology of Reductionism, people try to understand the nature
of spacetime by studying the quantum entanglement of spacetime. An example is
about the conjecture of $\mathrm{ER=EPR}$\cite{ER}; Following the methodology
of Emergence, people try to understand the nature of spacetime by constructing
certain many-body models and studying its ground states and excitations. An
attempt is from certain local (bosonic) models (or a qubit model)\cite{wen}.
The goal is to find the emergence of gravitational waves and gravitons
(helicity $\pm2$ excitations) with a linear dispersion as the low energy excitations.

In this paper, we only focus on the issue about the methodology of Emergence.

According to above discussion, spacetime is really a many-body system of
matter and elementary particle is block unit of spacetime. So, we consider
spacetime as many-body systems and try to write down its "\emph{Whole wave
function}". Here, the "Whole wave function" is a representation for physical
variant rather than the wave function from solving Schrodinger's equation. In
a word, it is beyond quantum mechanics.

To obtain the "Whole wave function", there are four steps.

Step 1: Obtain the "Whole wave function" of a simple 1D space.

The simplest 1D space is a uniform variant $V_{\mathrm{\tilde{U}(1),}1}%
[\Delta \phi,\Delta x,k_{0}]$ that is 1D group-changing space $\mathrm{C}%
_{\mathrm{\tilde{U}(1)},1}(\Delta \phi)$\textit{\ }on Cartesian space
$\mathrm{C}_{1}$, i.e.,
\begin{align}
V_{\mathrm{\tilde{U}(1),}1}[\Delta \phi,\Delta x,k_{0}]  &  :\mathrm{C}%
_{\mathrm{\tilde{U}(1)},1}(\Delta \phi)=\{ \delta \phi \} \nonumber \\
&  \Longleftrightarrow \mathrm{C}_{1}(\Delta x)=\{ \delta x\}
\end{align}
where $\Longleftrightarrow$\ denotes an ordered mapping under fixed changing
rate of integer multiple $k_{0}$.\ For this 1D uniform variant
$V_{\mathrm{\tilde{U}(1),}1}[\Delta \phi,\Delta x,k_{0}],$ the size $\Delta
\phi$ of the non-compact $\mathrm{\tilde{U}(1)}$ group is $N\pi$, and the size
$\Delta x$ of the Cartesian space $\mathrm{C}_{1}$\ is $L$.

Under K-projection on this uniform variant $V_{\mathrm{\tilde{U}(1),}1}%
[\Delta \phi,\Delta x,k_{0}]$, we have a uniform zero lattice, of which each
zero is just a fermionic elementary particle. We\ may regard the 1D space as
1D "fermionic" system in terms of a Slater determinant,
\begin{equation}
\Psi_{N}(\{x_{j}\})=\det \left(
\begin{array}
[c]{cccc}%
x_{1}^{0} & x_{1}^{1} & ... & x_{1}^{N-1}\\
x_{2}^{0} & x_{2}^{1} & ... & x_{2}^{N-1}\\
\vdots & \vdots & ... & \vdots \\
x_{N}^{0} & x_{N}^{1} & ... & x_{N}^{N-1}%
\end{array}
\right)  . \label{eqSlater}%
\end{equation}
The determinant takes into account all permutations of the $N$ particles
(zeroes) over the $N$ particle positions, $x_{1},...,x_{N}$, and may be
rewritten by Vandermonde determinant,
\begin{equation}
\Psi_{N}(\{x_{j}\})=\prod_{i<j}\left(  x_{i}-x_{j}\right)  .
\label{eqVandermonde}%
\end{equation}

One can check the 1-th order variability
\begin{equation}
\mathcal{T}(\delta x)\rightarrow \hat{U}(\delta \phi)=e^{i\cdot \delta \phi}%
\end{equation}
where $\delta \phi=k_{0}\delta x$. Without 1-th order variability along tempo
direction, this "Whole wave function" is not a physical variant. So, it has
trivial physical consequences.

Step 2: Obtain the "Whole wave function" of a simple (1+1)D quantum spacetime.

Now, we have an $\mathrm{\tilde{S}\tilde{O}}$\textrm{(1+1)} physical variants
that is mapping between (1+1)D $\mathrm{\tilde{S}\tilde{O}}$\textrm{(1+1)}
Clifford group-changing space $\mathrm{C}_{\mathrm{\tilde{S}\tilde{O}%
(1+1)},1+1}(\Delta \phi^{\mu})$ and a rigid spacetime $\mathrm{C}_{1+1}(\Delta
x^{\mu}).$ Here, $\mathrm{\tilde{S}\tilde{O}}$\textrm{(1+1}$\mathrm{)}$
denotes an $\mathrm{\tilde{S}\tilde{O}}$\textrm{(1+1) }non-compact group and
$\mu$ denotes an index for arbitrary orthogonal direction of spacetime.

Under K-projection on the (1+1)D $\mathrm{\tilde{S}\tilde{O}}$\textrm{(1+1)}
physical variants, we have a (1+1)D uniform zero lattice, of which each zero
is a fermionic elementary particle. We\ may also regard the (1+1)D spacetime
as (1+1)D fermionic system in terms of a Vandermonde determinant,
\begin{equation}
\Psi(\{ \hat{x}_{j}\})=\prod_{i<j}\left(  \hat{x}_{i}-\hat{x}_{j}\right)  .
\end{equation}
In particular, for uniform case, $\hat{x}$ is
\[
\hat{x}=x\sigma_{x}+t\sigma_{t},
\]
where $\{ \sigma_{x},\sigma_{t}\}=0.$

One can check the 1-th order variability
\begin{equation}
\mathcal{T}(\delta x^{\mu})\leftrightarrow \hat{U}(\delta \phi^{\mu}),
\end{equation}
where $\hat{U}(\delta \phi^{\mu})=e^{i\cdot \delta \phi^{\mu}\sigma^{\mu}}$
($\mu=x,t$). With 1-th order variability along tempo direction, this "Whole
wave function" is a physical variant with non-trivial physical consequences.

Step 3: Obtain the wave function of (3+1)D spacetime.

Now, we have an $\mathrm{\tilde{S}\tilde{O}}$\textrm{(3+1)} physical variants
that is mapping between (3+1)D $\mathrm{\tilde{S}\tilde{O}}$\textrm{(3+1)}
Clifford group-changing space $\mathrm{C}_{\mathrm{\tilde{S}\tilde{O}%
(3+1)},3+1}(\Delta \phi^{\mu})$ and a rigid spacetime $\mathrm{C}_{1+1}(\Delta
x^{\mu}).$

Under K-projection on the (3+1)D $\mathrm{\tilde{S}\tilde{O}}$\textrm{(3+1)}
physical variants, we have a (3+1)D uniform zero lattice, of which each zero
is also a fermionic elementary particle. The lattice distances along spatial
and tempo directions determine light speed and Planck constant. We\ then
consider the (3+1)D spacetime as (3+1)D fermionic system in terms of a
Vandermonde determinant,
\begin{equation}
\Psi_{N}(\{ \hat{x}_{j}\})=\prod_{i<j}\left(  \hat{x}_{i}-\hat{x}_{j}\right)
.
\end{equation}
For uniform case, we have
\[
\hat{x}=x\Gamma^{x}+y\Gamma^{y}+z\Gamma^{z}+t\Gamma^{t},\text{ }%
\]
where $(d+1)$-by-$(d+1)$ Gamma matrices $\Gamma^{\mu}$ obeying Clifford
algebra $\{ \Gamma^{i},\Gamma^{j}\}=2\delta^{ij}.$

This is just the conjecture about "Whole wave function" for our universe!

Step 4: Developing the theory from the "Whole wave function".

The approach to developing the theory from the "Whole wave function" had been
given in above sections yet. In particular, the quantum mechanics and gravity
emerge simultaneously. In other words, the approach is beyond quantum
mechanics. Therefore, the evolution of quantum spacetime is self-induced
without "Hamiltonian". This leads to the rule of general relativity.

In addition, we point out that we have the ability to construct the "Whole
wave function" for curved spacetime, AdS, even the spacetime with black holes.

For the case of different curved spacetimes, we replace a uniform "Whole wave
function" by non-uniform ones, i.e.,
\begin{equation}
\Psi(\{ \hat{x}_{j}\})=\prod_{i<j}\left(  \hat{x}_{i}^{\prime}-\hat{x}%
_{j}^{\prime}\right)  .
\end{equation}
where
\[
\hat{x}^{\prime}=x^{\prime}\Gamma^{x}+y^{\prime}\Gamma^{y}+z^{\prime}%
\Gamma^{z}+t^{\prime}\Gamma^{t}.
\]
The coordinates $(x^{\mu})_{\mathrm{curved}}=(x^{\mu})^{\prime}$ become
non-uniform and the Gamma matrices are still fixed. An example is AdS. By
replacing $z$ by $iz,$ the whole wave function of a typical (uniform) AdS is
written as
\begin{equation}
\Psi(\{ \hat{x}_{j}\})=\prod_{i<j}\left(  \hat{x}_{i}-\hat{x}_{j}\right)  .
\end{equation}
where
\[
\hat{x}=x\Gamma^{x}+y\Gamma^{y}+iz\Gamma^{z}+t\Gamma^{t}.
\]
This provides an opportunity to check the validity of AdS/CFT correspondence.

\subsubsection{Quantum spacetime -- noncommutative or commutative?}

\paragraph{Review on the theory for noncommutative geometry}

Space (or spacetime) is always considered to have smooth manifold structure
with the commutative algebra of functions generated by coordinates $x^{\mu},$
i.e., $[\hat{x}^{\mu},\hat{x}^{\nu}]=0.$ A. Connes developed an alternative
theory for space that is represented by a noncommutative algebra through
noncommutative geometry\cite{con}. For noncommutative geometry, there exists a
duality between algebra and geometry. More precisely there is a duality
between certain categories of geometric spaces and categories of algebras
representing those spaces. It was believed to relevant to quantum gravity.

The noncommutative space of a noncommutative geometry is a kind of
quantization, analogous to canonical quantization in physics, which replaces
an algebra of functions on a phase space with a Heisenberg (Weyl) algebra of
operators on a Hilbert space, i.e.,%
\[
\lbrack \hat{x}^{\mu},\hat{x}^{\nu}]=i\theta^{\mu \nu}%
\]
where in the canonical case $\theta^{\mu \nu}$ is an antisymmetric constant
matrix of dimension length-squared, and by letting the fields on
noncommutative spacetime be functions of the noncommutative coordinate
operators. In physics, the first application is Snyder's \textquotedblleft
quantized spacetime\textquotedblright \ which originates from the 5D de Sitter
space\cite{Snyd47}. It preserves Lorentz invariance, but breaks translational
invariance\cite{Yang47}.

There exists a standard procedure to consider a quantum field on
noncommutative spacetime that is function of the noncommutative coordinate
operators. In general, through Weyl quantization the noncommutative algebra of
operators can be represented on the algebra of ordinary functions on classical
spacetime by using the noncommutative Moyal $\star$-product. By the
noncommutative Moyal $\star$-product, a usual function $f(x)$ is replaced by
Weyl operator
\[
f(x)\rightarrow \hat{W}[f]=%
{\displaystyle \int}
d^{D}x[f(x)\hat{\Delta}(x)]
\]
where $\hat{\Delta}(x)=%
{\displaystyle \int}
\frac{d^{D}x}{(2\pi)^{D}}e^{-ik_{\mu}\hat{x}^{\mu}}e^{ik_{\nu}x^{\nu}}.$ So,
we have%
\[
f(x)=\mathrm{Tr}[\hat{W}[f]\hat{\Delta}(x)].
\]

Now, one replaces the usual point-wise product of functions, $f(x)$ and
$g(x)$, by the noncommutative Moyal $\star$-product,
\begin{equation}
(f\star g)(x)=f(x)\exp(\frac{i}{2}\overleftarrow{\partial}_{\mu}\theta^{\mu
\nu}\overrightarrow{\partial}_{\nu})g(x)
\end{equation}
where
\begin{multline}
\exp(\frac{i}{2}\overleftarrow{\partial}_{\mu}\theta^{\mu \nu}\overrightarrow
{\partial}_{\nu})g(x)=\sum_{n=1}^{\infty}\frac{1}{n!}\left(  \frac{i}%
{2}\right)  ^{n}\nonumber \\
\times \theta^{\mu_{1}\nu_{1}}\cdots \theta^{\mu_{n}\nu_{n}}\partial_{\mu_{1}%
}\cdots \partial_{\mu_{n}}f(x)\partial_{\nu_{1}}\cdots \partial_{\nu_{n}}g(x).
\end{multline}
Then, the commutator of field operators, $\hat{\phi}(x)$ and $\hat{\psi}(x)$,
is represented on the algebra of functions by the Moyal bracket:
\[
\lbrack \phi(x),\psi(x)]_{\star}=\phi(x)\star \psi(x)-\phi(x)\star \psi(x).
\]
So, we have
\[
\mathrm{Tr}[\hat{W}[f_{1}]\hat{W}[f_{2}]]=%
{\displaystyle \int}
d^{D}x[f(_{1}x)f_{2}(x)].
\]

In summary, by using above standard procedure, we "put" different types of
quantum fields on noncommutative space-time.

However, the situation becomes complex after considering Lorentz invariance
(or usual Poincar\'{e} symmetry). On noncommutative space, the usual Lorentz
symmetry disappears. Instead, one has an invariant under the twisted
Poincar\'{e} algebra, deformed by the Abelian twist element $\mathcal{F}%
=e^{\frac{i}{2}\theta^{\mu \nu}P_{\mu}\otimes P_{\nu}},$ where $P_{\mu
}=-i\partial_{\mu}$ are the generators of translations for spacetime. Or, on
the noncommutative spacetime, relativistic invariance means invariance under
twisted Poincar\'{e} transformations rather than a usual one\cite{2005,2009}.

\paragraph{Clifford group-changing space as noncommutative space}

We point out that the Clifford group-changing space $\mathrm{C}%
_{\mathrm{\tilde{S}\tilde{O}(d+1)},d+1}$ is really noncommutative space
obeying noncommutative geometry. Its coordinates are phase angles $\delta
\phi^{\mu}$ of non-compact $\mathrm{\tilde{S}\tilde{O}}$\textrm{(d+1)} Lie
group; the coordinate unit vectors $\mathbf{e}^{\mu}$ (the fundamental vectors
along $\phi^{\mu}$-direction) becomes $\Gamma^{\mu},$ i.e., $\mathbf{e}^{\mu
}=\Gamma^{\mu}.$ The anti-commutation condition matrices $\Gamma^{\mu}$ of
Clifford group-changing space indicate a non-commutating character, i.e.,
\begin{equation}
\{ \mathbf{e}^{\mu},\mathbf{e}^{\nu}\}=\{ \Gamma^{\mu},\Gamma^{\nu}%
\}=2\delta_{\mu \nu}%
\end{equation}
and
\begin{equation}
\lbrack \mathbf{e}^{\mu},\mathbf{e}^{\nu}]=\left[  \Gamma^{\mu},\Gamma^{\nu
}\right]  \neq0.
\end{equation}
For such a noncommutative space with anti-commutation condition $\{
\Gamma^{\mu},\Gamma^{\nu}\}=2\delta_{\mu \nu}$, the \emph{parallelogram rule}
for vectors is similar to usual space.

Based on such a noncommutative space $\mathrm{C}_{\mathrm{\tilde{S}\tilde
{O}(d+1)},d+1}$, there are two approaches to develop quantum theories. See Fig.7.

Approach I is to consider the rigid spacetime $\mathrm{C}_{d+1}$ as base space
and the physical processes come from different mappings between the
noncommutative space $\mathrm{C}_{\mathrm{\tilde{S}\tilde{O}(d+1)},d+1}$ and
commutative base space $\mathrm{C}_{d+1}.$ For this case, the noncommutative
space (or group-changing space $\mathrm{C}_{\mathrm{\tilde{S}\tilde{O}%
(d+1)},d+1}$) becomes a physical object rather than a statics rigid space.
Now, we have a theory for dynamical noncommutative space (or group-changing
space). This is what I do.

Approach II is to consider the noncommutative space $\mathrm{C}%
_{\mathrm{\tilde{S}\tilde{O}(d+1)},d+1}$ as base space and physical processes
come from different mappings between one noncommutative space $\mathrm{C}%
_{\mathrm{\tilde{S}\tilde{O}(d+1)},d+1}$ and another noncommutative space
$\mathrm{C}_{\mathrm{\tilde{S}\tilde{O}(d+1)},d+1}^{\prime}.$ Now, we have a
theory with a background of noncommutative space (or group-changing space
$\mathrm{C}_{\mathrm{\tilde{S}\tilde{O}(d+1)},d+1}$). This is what others had
done based on noncommutative geometry.

\begin{figure}[ptb]
\includegraphics[clip,width=0.47\textwidth]{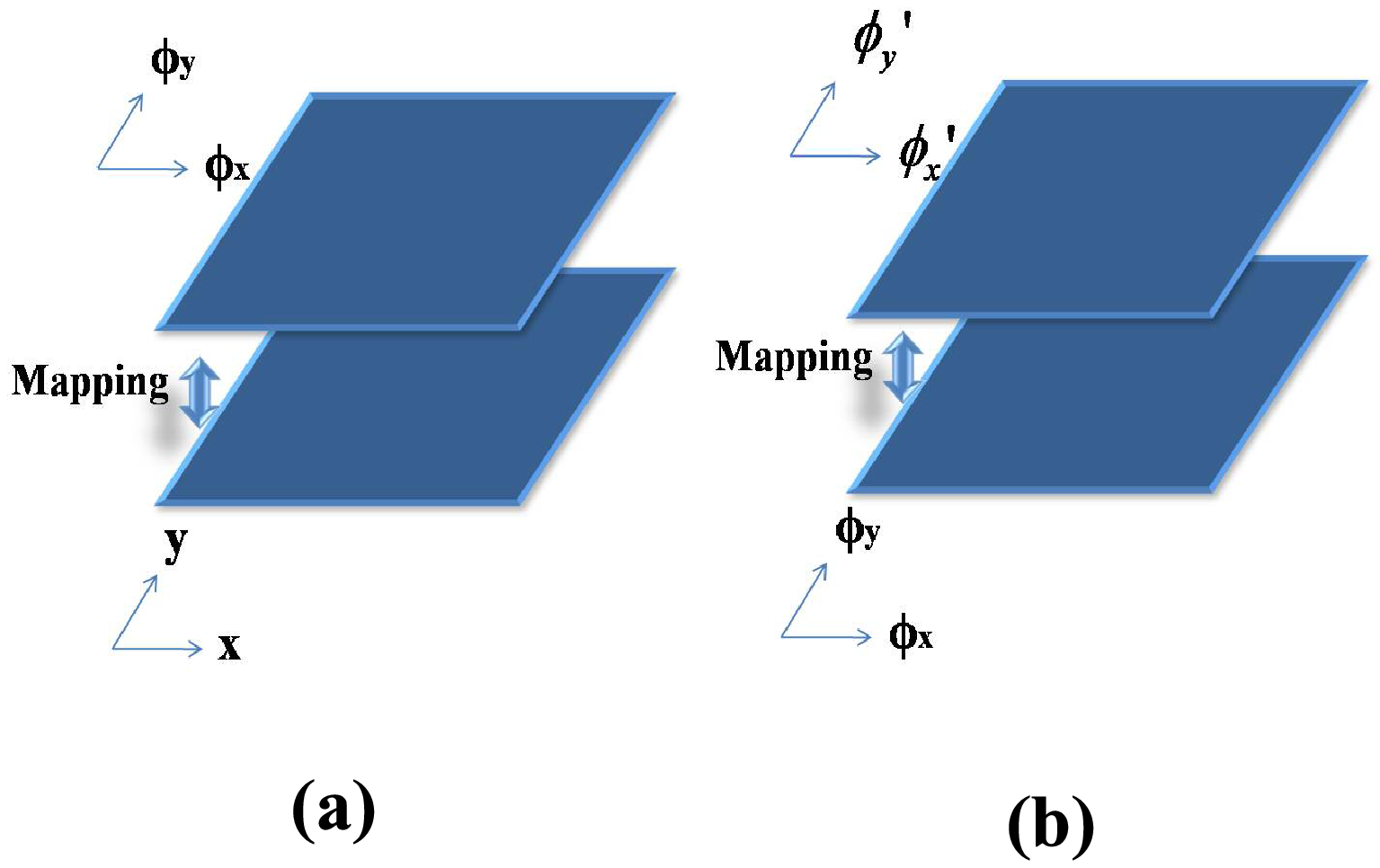}\caption{The difference
between the two approaches: (a) our case is to consider the rigid spacetime
$\mathrm{C}_{d+1}$ as base space and the physical processes come from
different mappings between the noncommutative space $\mathrm{C}_{\mathrm{\bar
{S}\bar{O}(d+1)},d+1}$ and base space $\mathrm{C}_{d+1}$, (b) for
noncommutative geometry, one considers the noncommutative space $\mathrm{C}%
_{\mathrm{\bar{S}\bar{O}(d+1)},d+1}$ as base space and physical processes come
from different mappings between this noncommutative space $\mathrm{C}%
_{\mathrm{\bar{S}\bar{O}(d+1)},d+1}$ and another noncommutative space
$\mathrm{C}_{\mathrm{\bar{S}\bar{O}(d+1)},d+1}^{\prime}$.}%
\end{figure}

\subparagraph{Approach I}

Firstly, we consider approach I.

Now, the noncommutative space $\mathrm{C}_{\mathrm{\tilde{S}\tilde{O}%
(d+1)},d+1}$ becomes dynamical object on base space $\mathrm{C}_{d+1}.$ To
characterize the dynamical processes from different mappings between the
noncommutative space $\mathrm{C}_{\mathrm{\tilde{S}\tilde{O}(d+1)},d+1}$ and
base space $\mathrm{C}_{d+1},$ the key point is to consider the noncommutative
space $\mathrm{C}_{\mathrm{\tilde{S}\tilde{O}(d+1)},d+1}$ as a physical object.

Our universe is really an $\mathrm{\tilde{S}\tilde{O}}$\textrm{(d+1)} physical
variant\textit{\ }$V_{\mathrm{\tilde{S}\tilde{O}(d+1)},d+1}(\Delta \phi^{\mu
},\Delta x^{\mu},k_{0},\omega_{0})$ that is a mapping between\textrm{\ }%
$\mathrm{\tilde{S}\tilde{O}}$\textrm{(d+1)} Clifford group-changing
space\textit{\ }$\mathrm{C}_{\mathrm{\tilde{S}\tilde{O}(d+1)},d+1}$%
\textit{\ }and a rigid spacetime $\mathrm{C}_{d+1}$. There are two types of
physical processes: one is about transverse changings or shape changings that
correspond to the processes for curving spacetime, the other is about
longitudinal changings -- contraction/expansion changings that correspond to
the processes for annihilating/generating matter. Then, there are two types of
"motion" (time-dependent "changings") in quantum spacetime -- one is about
motion of quantum spacetime itself, that is about transverse changings, the
other is about motion of matter, that is about longitudinal changings.

In the continuum limit, we derive an effective model for longitudinal
changings of the noncommutative space $\mathrm{C}_{\mathrm{\tilde{S}\tilde
{O}(d+1)},d+1}$ that is just Dirac model for elementary particles, $\bar{\Psi
}(ie_{a}^{\mu}\gamma^{a}\hat{\partial}_{\mu}-m)\Psi$ where $m$ is mass.
$\gamma^{\mu}$ are the Gamma matrices defined as $\gamma^{1}=\gamma^{0}%
\Gamma^{x}$, $\gamma^{2}=\gamma^{0}\Gamma^{y},$ $\gamma^{3}=\gamma^{0}%
\Gamma^{z}$, $\gamma^{0}=\Gamma^{t}.$ The Gamma matrices $\Gamma^{I}$
($I=x,y,z$) and $\Gamma^{t}$ obey Clifford algebra, i.e., $\{ \Gamma
^{I},\Gamma^{t}\}=0$, and $\{ \Gamma^{I},\Gamma^{J}\}=0.$ That means Lorentz
invariance is really an emergent phenomenon. The transverse changings of the
noncommutative space $\mathrm{C}_{\mathrm{\tilde{S}\tilde{O}(d+1)},d+1}$ is
just the curving of the spacetime. The effective action is the
Einstein-Hilbert action $S_{\mathrm{EH}}=\frac{1}{16\pi G}\int \sqrt{-g}%
Rd^{4}x.$ Finally, the total action $S$ is described by
\begin{align}
S  &  =\mathcal{S}_{\mathrm{4D}}+S_{\mathrm{EH}}\\
&  =\int \sqrt{-g(x)}\bar{\Psi}(e_{a}^{\mu}\gamma^{a}\hat{D}_{\mu}-m)\Psi \text{
}d^{4}x\nonumber \\
&  +\frac{1}{16\pi G}\int \sqrt{-g}R\text{ }d^{4}x\nonumber
\end{align}

In summary, the theory of dynamical noncommutative space becomes a consistent
theory for unifying quantum mechanics and gravity and thus becomes the hopeful
candidate about quantum gravity. In addition, to consider gauge fields (such
as Maxwell fields and Yang-Mills fields), we must introduce 2-th order
physical variants. This issue will be discussed in other paper.

\subparagraph{Approach II}

Next, we consider the approach II.

Now, a rigid noncommutative space $\mathrm{C}_{\mathrm{\tilde{S}\tilde
{O}(d+1)},d+1}$ becomes a background (or base space) for other noncommutative
space $\mathrm{C}_{\mathrm{\tilde{S}\tilde{O}(d+1)},d+1}^{\prime}$. The
dynamical processes come from different mappings between the noncommutative
space $\mathrm{C}_{\mathrm{\tilde{S}\tilde{O}(d+1)},d+1}$ and the other
$\mathrm{C}_{\mathrm{\tilde{S}\tilde{O}(d+1)},d+1}^{\prime}\ $that is a
physical object. The situation leads to a theory about noncommutative geometry.

Now, the physical system becomes a mapping between a dynamical $\mathrm{\tilde
{S}\tilde{O}}$\textrm{(d+1)} Clifford group-changing space\textit{\ }%
$\mathrm{C}_{\mathrm{\tilde{S}\tilde{O}(d+1)},d+1}^{\prime}$\textit{\ }and a
rigid one $\mathrm{C}_{\mathrm{\tilde{S}\tilde{O}(d+1)},d+1},$\textit{\ }%
i.e.,\textit{\ }%
\begin{equation}
\mathrm{C}_{\mathrm{\tilde{S}\tilde{O}(d+1)},d+1}^{\prime}=\{ \delta \phi_{\mu
}^{\prime}\} \Longleftrightarrow \mathrm{C}_{\mathrm{\tilde{S}\tilde{O}%
(d+1)},d+1}=\{ \delta \phi_{\mu}\}
\end{equation}
where $\Leftrightarrow$\ denotes an ordered mapping with fixed changing rate
of integer multiple $\lambda_{0},$\ and $\mu$ labels the spatial direction.
Both Clifford group-changing spaces\textit{\ }$\mathrm{C}_{\mathrm{\tilde
{S}\tilde{O}(d+1)},d+1}(\Delta \phi_{\mu})$\ and\textit{ }$\mathrm{C}%
_{\mathrm{\tilde{S}\tilde{O}(d+1)},d+1}^{\prime}$\textit{ }are described by
$d+1$ series of numbers of group elements $\phi_{\mu}$ arranged in size order.
Gamma matrices $\Gamma^{\mu}$ obey Clifford algebra $\{ \Gamma^{i},\Gamma
^{j}\}=2\delta^{ij}$.

We then consider the noncommutative space $\mathrm{C}_{\mathrm{\tilde{S}%
\tilde{O}(d+1)},d+1}^{\prime}$ as a many-body system with higher-order
variability,
\begin{equation}
\mathcal{T}(\delta \phi_{\mu})\leftrightarrow e^{i\delta \phi_{\mu}^{\prime
}\Gamma_{\mu}^{\prime}}%
\end{equation}
where $\delta \phi_{\mu}^{\prime}=\lambda_{0}^{\mu}\delta \phi_{\mu}$. In
particular, to get noncommutative geometry, the zero lattice of $\mathrm{C}%
_{\mathrm{\tilde{S}\tilde{O}(d+1)},d+1}^{\prime}$ on $\mathrm{C}%
_{\mathrm{\tilde{S}\tilde{O}(d+1)},d+1}$ cannot coincide that of
$\mathrm{C}_{\mathrm{\tilde{S}\tilde{O}(d+1)},d+1}$ on $\mathrm{C}_{d+1}.$
Hence, we have $\lambda_{0}^{\mu}\neq1$.

We can use the similar approach to do compactification of $\mathrm{C}%
_{\mathrm{\tilde{S}\tilde{O}(d+1)},d+1}^{\prime}$ on $\mathrm{C}%
_{\mathrm{\tilde{S}\tilde{O}(d+1)},d+1}$ and get \textquotedblleft
topological\textquotedblright \ version lattice together with a matrix
network.\textit{\ }As a result, to characterize physical processes for the
noncommutative space $\mathrm{C}_{\mathrm{\tilde{S}\tilde{O}(d+1)}%
,d+1}^{\prime}$ on $\mathrm{C}_{\mathrm{\tilde{S}\tilde{O}(d+1)},d+1}$, we
have a Dirac model on noncommutative spacetime,%
\begin{equation}
L=\bar{\not \Psi }\star(iD_{\mu}-m)\Psi.
\end{equation}
Now, the emergent Lorentz invariance becomes twisted. The transverse changings
of the noncommutative space $\mathrm{C}_{\mathrm{\tilde{S}\tilde{O}(d+1)}%
,d+1}^{\prime}$ become the curving of the noncommutative spacetime.

The situation is quite different from the physical variants for gauge fields.
Now, we have a mapping between their group-changing subspaces\textit{
}$\mathrm{C}_{1,\mathrm{\tilde{U}}_{1}\mathrm{(1)\in \tilde{G}}_{1},1}%
(\Delta \phi_{1,\mathrm{global}})$ and \textit{ }$\mathrm{C}_{2,\mathrm{\tilde
{U}}_{2}\mathrm{(1)\in \tilde{G}}_{2},2}(\Delta \phi_{2,\mathrm{global}}%
)$\textit{,} i.e.,%
\begin{align*}
\mathrm{C}_{1,\mathrm{\tilde{G}}_{1},d_{1}}(\Delta \phi_{1}^{\mu})  &
\Longleftrightarrow \mathrm{C}_{2,\mathrm{\tilde{G}}_{2},d_{2}}(\Delta \phi
_{2}^{\mu})\equiv \mathrm{C}_{1,\mathrm{\tilde{U}}_{1}\mathrm{(1)\in \tilde{G}%
}_{1},1}(\Delta \phi_{1,\mathrm{global}})\\
&  \Longleftrightarrow \mathrm{C}_{2,\mathrm{\tilde{U}}_{2}\mathrm{(1)\in
\tilde{G}}_{2},2}(\Delta \phi_{2,\mathrm{global}}):\{ \delta \phi
_{1,\mathrm{global}}\} \Leftrightarrow \{ \delta \phi_{2,\mathrm{global}}\}
\end{align*}
\textit{ }with the changing ratio $\lambda^{\lbrack12]}$. Here the elements of
two subgroup-changing spaces are $\delta \phi_{1,\mathrm{global}}=\left \vert
\delta \phi_{1}^{\mu}(x)\right \vert =\sqrt{%
{\displaystyle \sum \limits_{\mu}}
(\delta \phi_{1}^{\mu}(x))^{2}}$ and $\delta \phi_{2,\mathrm{global}}=\left \vert
\delta \phi_{2}^{\mu}(x)\right \vert =\sqrt{%
{\displaystyle \sum \limits_{\mu}}
(\delta \phi_{2}^{\mu}(x))^{2}},$ respectively. Here, if the changing ratio
$\lambda^{\lbrack12]}=1,$ we have an effective \textrm{U(1)} gauge field. See
the detailed discussion in Ref.\cite{kou1}.

\paragraph{Summary}

In the end, we give a summary.

For our universe (a physical variant), the matter and spacetime are unified
into single noncommutative space (or group-changing space $\mathrm{C}%
_{\mathrm{\tilde{S}\tilde{O}(d+1)},d+1}$). The dynamical physical processes of
the noncommutative space (or group-changing space $\mathrm{C}_{\mathrm{\tilde
{S}\tilde{O}(d+1)},d+1}$) are described by general relativity and quantum
mechanics. This is the approach I rather approach II. So, the key mistake of
noncommutative geometry by Connes and others comes from the separation of
matter and spacetime\cite{con}.

\subsubsection{Geometric Witten effect and spin geometry -- the road to loop
quantum gravity}

\paragraph{The action with Holst term for quantum spacetime}

In this part, we study a special quantum spacetime, of which there exists a
Holst term in the action\cite{hos}. So, our starting point is
\begin{align}
S  &  =\int \sqrt{-g}\bar{\Psi}(e_{a}^{\mu}\gamma^{a}\hat{D}_{\mu}-m)\Psi \text{
}d^{4}x\nonumber \\
&  +\frac{1}{16\pi G}\int \sqrt{-g}R\text{ }d^{4}x+\mathcal{S}_{\theta
,\mathrm{G}}%
\end{align}
where $\mathcal{S}_{\theta,\mathrm{G}}=-\frac{1}{16\pi G\beta}\int e_{a}\wedge
e_{b}\wedge R^{ab}$ is the Holst term. Here, $\beta$ is the Barbero-Immirzi
parameter\cite{bi}.

In general, this Holst term plays no role in the classical dynamics and only
has in non-perturbative quantum effects.

\paragraph{Geometric Witten effect and spin geometry}

In this part, we explore geometric Witten effect and discuss spin geometry by
considering an extra Holst term.

Firstly, we review Witten effect \cite{witten1} in usual quantum field theory.

We consider a magnetic monopole of gauge fields with finite magnetic charge,
\begin{equation}
q_{m}=\frac{1}{4\pi}%
{\displaystyle \oint \nolimits_{\mathcal{S}}}
F_{\mathcal{S}}^{IJ}\neq0.
\end{equation}
If we add a topological theta term $\Delta \mathcal{L}$ to the original
Lagrangian of the gauge fields,
\begin{equation}
\Delta \mathcal{L}=\theta \frac{e^{2}}{32\pi^{2}}\varepsilon^{\mu \nu \alpha \beta
}\mathrm{Tr}\left(  F_{\mu \nu}F_{\alpha \beta}\right)  , \label{theta}%
\end{equation}
there exists an induced electric charge $q_{e}$ of this magnetic monopole,
\begin{equation}
q_{e}=\frac{\theta}{2\pi}e. \label{effect}%
\end{equation}

According to earlier discussion, each elementary particle carries a unit
magnetic monopole with $q_{m}=\pm1.$ \emph{What's the corresponding Witten
effect?} In this part, we study this problem and explore the geometric Witten effect.

In exterior derivative form, the Einstein-Hilbert action $S_{\mathrm{EH}}$ was
transformed into
\begin{align*}
\frac{1}{16\pi G}\epsilon_{0bcd}\,R^{0b}\wedge F^{cd}  &  =\frac{1}{16\pi
G}\epsilon_{0bcd}\, \hat{D}\omega^{0b}\wedge F^{cd}\\
&  \rightarrow-\frac{1}{16\pi G}\epsilon_{0bcd}\, \omega^{0b}\wedge \hat
{D}F^{cd}.
\end{align*}
Here, we have used the following equation, $F^{jk}\equiv-A^{j0}\wedge A^{k0}$
and $e^{i}\wedge e^{j}=(2l_{P})^{2}A^{j0}\wedge A^{k0}$. By using similar
approach, the Holst term was transformed into
\begin{align*}
\frac{1}{16\pi G\beta}\,R^{ab}\wedge F^{ab}  &  =\frac{1}{16\pi G\beta}\,
\hat{D}\omega^{ab}\wedge F^{ab}\\
&  \rightarrow-\frac{1}{16\pi G\beta}\, \omega^{ab}\wedge \hat{D}F^{ab}.
\end{align*}
\

Then, the variation of the total action with respect to $\omega^{0b}$ leads to
the following equations
\begin{align}
\rho_{F}  &  =\sqrt{-g}\Psi^{\dagger}\Psi=(-\epsilon_{0bcd}\epsilon
_{0ijk}\frac{1}{16\pi G}\hat{D}_{i}F_{jk}^{cd})\nonumber \\
&  +(-\epsilon_{0ijk}\frac{1}{16\pi G\beta}\hat{D}_{i}F_{jk}^{ob}),
\end{align}
where $\rho_{F}$ is the density of fermions. After doing integral in 3D
subspace, we obtain
\begin{align}
N_{F}  &  =-\frac{1}{16\pi G}(l_{0})^{2}%
{\displaystyle \oint \nolimits_{\mathcal{S}}}
\epsilon_{cd}\epsilon_{ijk}F_{jk}^{cd}\cdot dS_{i}\nonumber \\
&  -\frac{1}{16\pi G\beta}(l_{0})^{2}%
{\displaystyle \oint \nolimits_{\mathcal{S}}}
\epsilon_{0b}\epsilon_{ijk}F_{jk}^{ob}\cdot dS_{i}\nonumber \\
&  =-\frac{1}{4G}(l_{0})^{2}q_{m}-\frac{1}{4G\beta}(l_{0})^{2}q_{s}.
\end{align}
where $N_{F}$ denotes the number of (fermionic) particles. Finally, after
considering the right dimension, we have%
\begin{equation}
N_{F}=-q_{m}-\frac{1}{\beta}q_{s}.
\end{equation}

Above equation indicates the \emph{geometric Witten effect}. For an elementary
particle with magnetic monopole $q_{m},$ there exists a new contribution
$-\frac{1}{\beta}q_{s}$ that is dependent on $\beta$.

In addition, we give a physical explanation on the geometric Witten effect
from the Holst term.

For the terms with $a,b\neq0,$ the Holst term was transformed into
\[
-\frac{1}{16\pi G\beta}\, \omega^{ab}\wedge \hat{D}F^{ab},\text{ }a,b\neq0.
\]
Then, the variation of the total Lagrangian with respect to $\omega^{ab}$
leads to the following equations
\begin{align}
j_{F}^{cd}  &  =\sqrt{-g}\Psi^{\dagger}\gamma^{0}\gamma^{c}\gamma^{d}%
\Psi \nonumber \\
&  =-\epsilon_{0bcd}\epsilon_{0ijk}\frac{1}{16\pi G}\hat{D}_{i}F_{jk}%
^{0c}\nonumber \\
&  -\epsilon_{0ijk}\frac{1}{16\pi G\beta}\hat{D}_{i}F_{jk}^{cd}.
\end{align}
This second term $-\epsilon_{0ijk}\frac{1}{16\pi G\beta}\hat{D}_{i}F_{jk}%
^{cd}$ indicates that the spin current/density of elementary particles traps
magnetic monopoles of spacetime. We call the geometry from spin ($N_{F}%
=-\frac{1}{\beta}q_{s}$) to be \emph{spin geometry.}

As a result, a quantum spacetime with an extra Holst term provides an
opportunity to display the existence of spin geometry. So, we call the usual
geometry from particle number ($N_{F}=-q_{m}=-\frac{\Delta V}{(l_{0})^{3}4\pi
}$) to be called \emph{charge geometry}.

\paragraph{Quantum loop description for spin geometry}

Loop quantum gravity is assumed to be a non-perturbative approach to the
quantum theory of gravity, in which no classical background metric is
used\cite{loop}\cite{as}. It has considerable successes to its quantum theory
of spatial geometry in which quantities such as area and volume are quantized
in units of the Planck length, and a calculation of black hole entropy. In
addition, to study the dynamics of spacetime, people developed spin foam
approach by attempting the construction of the path integral representation of
the theory.

Finally, we give a comment on quantum loop description for spin geometry.

Spin geometry is determined by the Holst term $-\frac{1}{16\pi G\beta}\int
e_{a}\wedge e_{b}\wedge R^{ab}$ and always proportional to the Immirzi
parameter $\beta.$\ Now, the spin changings lead to the changings of geometry.

In LQG, the Holst term $-\frac{1}{16\pi G\beta}\int e_{a}\wedge e_{b}\wedge
R^{ab}$ plays more important role than usual Einstein-Hilbert term $\frac
{1}{16\pi G}\int \epsilon^{abcd}e_{a}\wedge e_{b}\wedge R^{cd}.$ For example,
the quantum non-commuting relation is determined by $\beta$\cite{as}. In
addition, all physical results (the area, the volume, ...) are proportional to
the Immirzi parameter $\beta$. That means all these physical quantities come
from the Holst term $-\frac{1}{16\pi G\beta}\int e_{a}\wedge e_{b}\wedge
R^{ab}$ rather\ than Einstein-Hilbert term $\frac{1}{16\pi G}\int
\epsilon^{abcd}e_{a}\wedge e_{b}\wedge R^{cd}.$

\textit{A summary: LQG is a correct theory that characterize kinetic processes
for spin geometry, rather than a complete theory for quantum gravity including
both spin geometry and charge geometry. }

\subsection{Discussion and conclusion}

In the end of this paper, we draw the conclusion. The\emph{ }starting point of
this theory is very simple -- $\mathrm{\tilde{S}\tilde{O}}$\textrm{(d+1)}
physical variant\textit{ }$V_{\mathrm{\tilde{S}\tilde{O}(d+1)},d+1}(\Delta
\phi^{\mu},\Delta x^{\mu},k_{0},\omega_{0})$ with 1-th order variability,
\begin{equation}
\mathcal{T}(\delta x^{\mu})\leftrightarrow \hat{U}(\delta \phi^{\mu})=e^{i\cdot
k_{0}\delta x^{\mu}\Gamma^{\mu}}.
\end{equation}
Based on the simple starting point, we develop a complete theory for quantum
spacetime. In this part, we unified spacetime and matter into an
$\mathrm{\tilde{S}\tilde{O}}$\textrm{(d+1)} physical variant\textit{
}$V_{\mathrm{\tilde{S}\tilde{O}(d+1)},d+1}(\Delta \phi^{\mu},\Delta x^{\mu
},k_{0},\omega_{0})$. See the logical structure of the paper in Fig.8.

\begin{figure}[ptb]
\includegraphics[clip,width=0.9\textwidth]{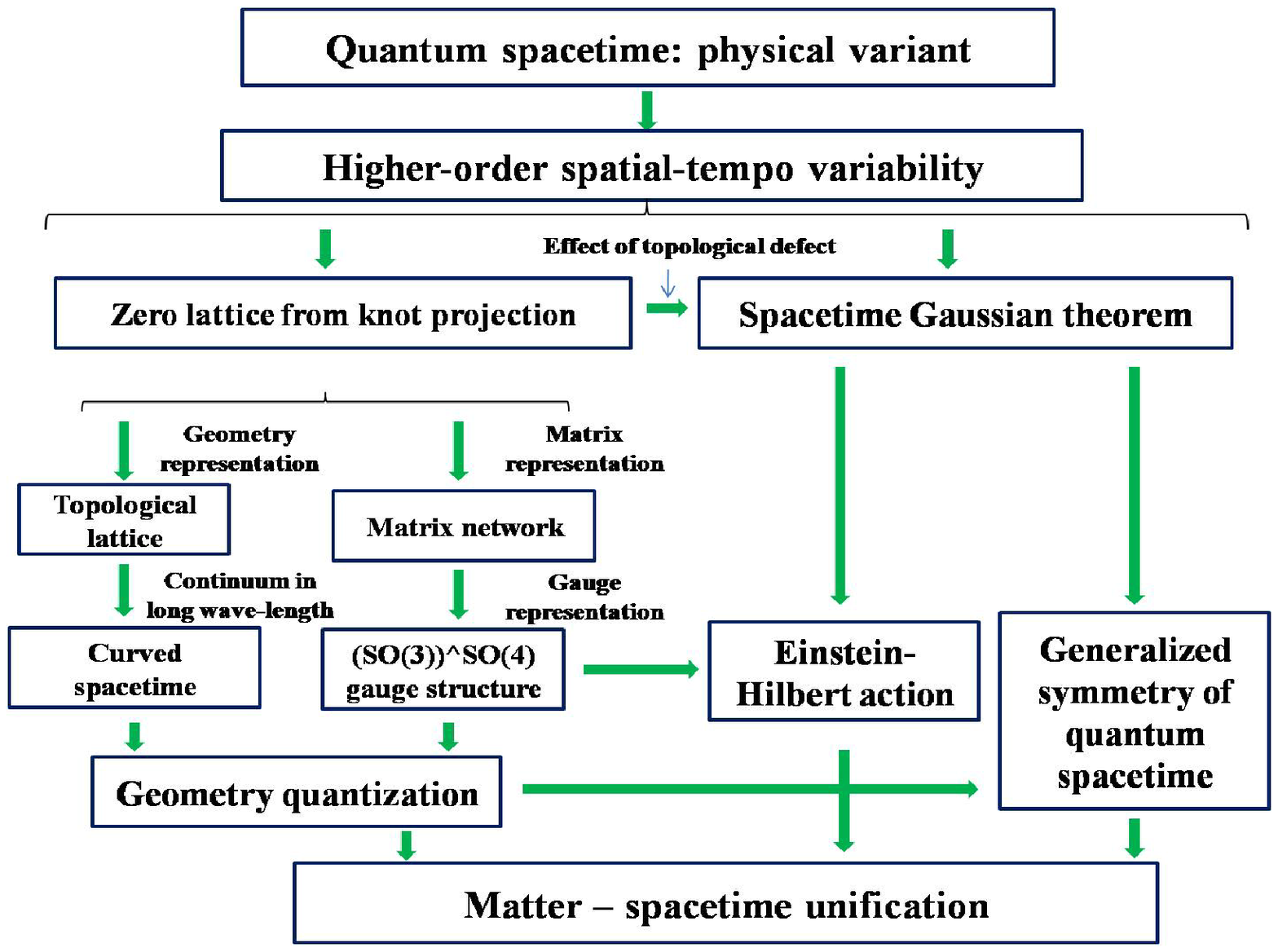}\caption{The logical
structure of the theory of quantum spacetime}%
\end{figure}

However, there are unsolved problems for quantum gravity, such as the
underlying mechanism of AdS/CFT correspondence\cite{ma}, quantum information
problem of black holes\cite{haw}, underlying mechanism of scattering amplitude
for tree Feymann diagram of gravitational waves (for example, why ambitwistor
superstring?\cite{am1})... In next parts, we will apply this theory to solve
above questions one by one.

\newpage

\section{Theory for Non-unitary Physical Variant: from AdS/CFT Correspondence
to AdS/NGT Equivalence}

\subsection{Introduction}

An important progress in modern physics is the Anti de Sitter - Conformal
Field theory (AdS/CFT) correspondence that was proposed by Juan Maldacena in
1997\cite{ma}. It characterizes the relationship between a quantum field
theory with conformal invariance on flat four dimensional (4D) spacetime, and
gravity theory for $AdS_{5}\times S^{5}$ (five dimensional Anti de Sitter
space times a 5-sphere). The flat 4D spacetime is the boundary (at infinity)
of the $AdS_{5}\times S^{5}$. Now, AdS/CFT correspondence between a creative
research field. The correspondence has been extended to a generalized mapping
between usual quantum conformal field theories beyond $N=4$ Super Yang-Mills
field theory and AdS\cite{witten}. It leads to the holographic nature of
gravity manifest, i.e., the perturbative metric fluctuations $g_{\mu \nu}$ of
AdS corresponds to stress tensor of CFT $T_{\mu \nu}$; a vector field (gauge
field) $A_{\mu}$ corresponds to a current $J^{\mu}$. In addition, the
holographic entangled entropy $S$ indicates the match between the scaling of
the CFT entropy density and the Bekenstein-Hawking entropy of minimum surface
in AdS\cite{haw}. As a result, the AdS/CFT correspondence gives us a geometric
description of QFT phenomena and may provide, understanding quantum field
theories at strong coupling (for example, QCD).

However, AdS/CFT correspondence is still a conjecture and far from being well
understood. We show following unsolve problems for fully understanding it:

\begin{enumerate}
\item What's the \emph{exact} rule of AdS/CFT correspondence within the
framework of quantum gravity rather than just a conjecture?

\item Why the perturbative metric fluctuations $g_{\mu \nu}$ of AdS correspond
to a boundary stress tensor $T_{\mu \nu}$ in CFT within the framework of
quantum gravity?

\item According to the dictionary from AdS/CFT correspondence, the particle's
mass $m$ in AdS plays the role of anomalous dimension $\nu$ in correlation
functions. Is it correct within the framework of quantum gravity? Why?

\item According to AdS/CFT correspondence, the gauge fields $A_{\mu}$ in AdS
correspond to usual current in CFT $J^{\mu}$. What does it mean within the
framework of quantum gravity?

\item According to AdS/CFT correspondence, there exists Ryu-Takayanagi's
formula of the holographic entangled entropy\cite{rt}. Is it correct within
the framework of quantum gravity? What's underlying mechanism of
Ryu-Takayanagi's formula?

\item How to characterize quantum fluctuations from gravitational waves in the
bulk of AdS by CFT beyond the boundary formula?
\end{enumerate}

According to above discussion, an inspiring idea is that\emph{ the particle is
basic block of spacetime and the spacetime is made of matter}. Therefore,
according to this idea, the matter is really certain "changing" of
\textquotedblleft spacetime\textquotedblright \ itself rather than extra things
on it. This is the \emph{new idea} for the foundation of quantum gravity and
the development of a complete theory and then becomes starting point of this
part\cite{kou1}. Another key point of the new theory is \emph{non-unitary
physical} with \emph{non-unitary higher-order variability}. In the following
parts, based on the theory of non-unitary physical variants, we provide a
fully understanding on AdS/CFT correspondence within the framework of quantum
gravity and answer above six questions.

We point out that all physical processes of system be intrinsically described
by the processes of the changings of a physical variant. In particular, the
elementary particles in AdS and those in CFT may have different structures.

The first theory about AdS comes from the d+1 dimensional complex zero
lattice,%
\begin{align*}
\text{AdS }  &  =\text{ A geometric representation }\\
&  \text{for complex zero lattice.}%
\end{align*}
The information unit (or elementary particle) is just the zero of the complex
zero lattice. Under the geometry representation of complex coordinates, the
theory is same to that for unitary physical variant. However, we point out
that the quantum mechanics in AdS is not Hermitian.

The second theory about CFT comes from (d-1)+1 dimensional real zero lattice,%
\begin{align*}
\text{CFT}  &  =\text{A kinetic representation }\\
&  \text{for real zero lattice.}%
\end{align*}
Now, there doesn't exist the zero solution along d-th direction without phase
changing. The information unit (or the elementary particle) changes.

\subsection{Fundamental mathematic theory for non-unitary variants}

Firstly, we develop the theory for non-unitary variants. Usual unitary variant
characterizes a system with "\emph{phase changing}" structure\cite{kou1},
i.e.,
\[
\text{Unitary variant: changing structure for phases;}%
\]
The non-unitary variant characterizes a system with "\emph{amplitude
operating}" structure, i.e.,%
\[
\text{Unitary variant: changing structure for amplitude.}%
\]
In particular, for non-unitary variants, their phase changings and amplitude
changings along different dimensions interplay each other and the resulting
rule helps us develop a theory for AdS/CFT.

\subsubsection{Non-unitary variant theory}

\paragraph{Non-unitary group-changing space}

In general, in quantum physics, the object of study is described by unitary
group $\mathrm{G}$ on Cartesian space $\mathrm{C}_{d}$, of which the operation
$U(g)$ obeys unitary condition, $\det(U(g))=1.$ The unitary condition
indicates that the group operation describes the (relative) phase change
between several modes. For example, for (non-Abelian) \textrm{SO(N)} group,
the group operation is $U(g)=e^{i\Theta}$ where $\Theta=\sum_{a=1}%
^{(n-1)n/2}\theta^{a}T^{a}$ and $\theta^{a}$ are a set of $(n-1)n/2$ constant
parameters, and $T^{a}$ are Hermitian $(n-1)n/2$ matrices representing the
generators of the Lie algebra of $\mathrm{SO(N)}$. In general, we have spinor
representation for \textrm{SO(N)} group. By introducing Gamma matrices obeying
Clifford Algebra $\Gamma^{a}$, $\{ \Gamma^{a},\Gamma^{b}\}=2\delta^{ab}$, the
generators of the Lie algebra of $\mathrm{SO(N)}$ become $-\frac{i}{4}%
[\Gamma^{a},\Gamma^{b}].$ For the case of $N=3$, both Gamma matrices and the
generators for $\mathrm{SO(3)}$ Lie group are Pauli matrices $\sigma^{x},$
$\sigma^{y},$ $\sigma^{z}$.

However, by generalizing usual Hermitian quantum mechanics to a non-Hermitian
one, we have non-unitary operation, of which $U(g)$ doesn't obey unitary
condition, $\det(U(g))\neq1.$ The non-unitary condition indicates that the
group operation describes the relative amplitude change between several modes.
For example, for (non-Abelian) \textrm{SO(N)} group, the group operation is
$U(g)=e^{i\Theta}$ where $\Theta=\sum_{a=1}^{(n-1)n/2}\theta^{a}T^{a}$ and
$\theta^{a}=e^{i\varphi^{a}}\left \vert \theta^{a}\right \vert $ are a set of
complex $(n-1)n/2$ constant parameters, and $T^{a}$ are Hermitian $(n-1)n/2$
matrices representing the generators of the Lie algebra of $\mathrm{SO(N)}$.
Here, we have $\varphi^{a}\neq0,$ $\pi.$

To define a non-unitary variant, we introduce non-unitary group-changing space
$\mathrm{C}_{\mathrm{\tilde{G}},d}(\Delta \phi^{a})$ for non-compact Lie group
\textrm{\~{G}}$_{(N,M)}$. Here \textrm{G} with "$\sim$" above means a
non-compact Lie group.

\textit{Definition}:\textit{\ The non-unitary }$d$\textit{-dimensional
group-changing space} $\mathrm{C}_{\mathrm{\tilde{G}},d}(\Delta \phi^{a}%
)$\textit{\ of non-compact \textrm{\~{G}} Lie group is described by }%
$N$\textit{ series of numbers of complex group element }$e^{i\varphi^{a}%
}\left \vert \delta \phi^{a}\right \vert $\textit{ of a-th generator
independently in size order. }$\Delta \phi^{a}$\textit{ denotes the size of the
group-changing space along a direction, a complex topological number. Here, at
least one of }$\varphi^{a}$\textit{ is not zero, i.e., }$\varphi^{a}\neq0,$
$\pi.$\textit{ }For a non-compact \~{G} Lie group, it has $N$ generators and
$N<d$.

For example, one dimensional (1D) non-unitary group-changing space
$C_{\mathrm{\tilde{U}(1)},1}(\Delta \phi)$\ of non-compact \textrm{\~{U}(1)}
group is described by a series of numbers of non-unitary group element
$e^{i\varphi}\left \vert \delta \phi \right \vert $ arranged in size order.
$\Delta \phi=\left \vert
{\displaystyle \sum}
\delta \phi^{a}\right \vert $ denotes the total size of the changing space that
turns to infinite, i.e., $\Delta \phi \rightarrow \infty$. For 1D non-unitary
group-changing space $\mathrm{C}_{\mathrm{\tilde{U}(1)},1}(\Delta \phi)$, we
have a series of infinitesimal non-unitary group-changing operations,
\begin{equation}
\prod_{i}(\tilde{U}(\delta \phi_{i}))
\end{equation}
where $\tilde{U}(\delta \phi_{i})=e^{i((\delta \phi_{i})\cdot \hat{K})}$,
$\hat{K}=-i\frac{d}{d\phi}.$\ Here, the i-th non-unitary operation $\hat
{U}(\delta \phi_{i})$ ($\delta \phi_{i}=e^{i\varphi}\left \vert \delta \phi
_{i}\right \vert $) generates an element of non-unitary group-changing that is
infinitesimal non-unitary group-changing operation.

For a $d$-dimensional non-unitary group-changing space $\mathrm{C}%
_{\mathrm{\tilde{G}},d}(\Delta \phi^{a})$, the element is an infinitesimal
$d$-dimensional non-unitary group-changing operation $\delta \phi
^{a}=e^{i\varphi^{a}}\left \vert \delta \phi^{a}\right \vert $ ($\delta \phi
^{a}\rightarrow0,$ $a=1,...,d$). We can also denote a $d$-dimensional
group-changing space $\mathrm{C}_{\mathrm{\tilde{G}},d}(\Delta \phi^{a})$ for
non-compact group \textrm{\~{G}} by a series of infinitesimal operations of
non-unitary group-changing,
\begin{equation}
\prod_{i}(\tilde{U}(\delta \phi_{i}))=\prod_{i}(\prod_{a=1}^{d}(\tilde
{U}(\delta \phi_{i}^{a})))
\end{equation}
where $\tilde{U}(\delta \phi_{i})=\prod_{a=1}^{d}(\tilde{U}(\delta \phi_{i}%
^{a}))$ and $\tilde{U}(\delta \phi_{i}^{a})=e^{i((\delta \phi_{i}^{a}T^{a}%
)\cdot \hat{K}_{a})}$, $\hat{K}_{a}=-i\frac{d}{d\phi^{a}}.$\ Here, the i-th
non-unitary operation $\hat{U}(\delta \phi_{i})$ ($\delta \phi_{i}=e^{i\varphi
}\left \vert \delta \phi_{i}\right \vert $) generates an element of non-unitary
group-changing that is infinitesimal non-unitary group-changing operation with
$d$ directions.

In particular, the operation $\tilde{U}(\delta \phi_{i})$ is a
"\emph{non-local}" operation that will change the size the group-changing
space $\mathrm{C}_{\mathrm{\tilde{G}},d}(\Delta \phi^{a})$, i.e., $\Delta
\phi^{a}\rightarrow \Delta \phi^{a}\pm \delta \phi_{i}^{a}$. On the contrary, the
local unitary/non-unitary group operation $\hat{U}(x_{i})=e^{\pm i\delta
\phi_{i}^{a}T^{a}}$ will never change the size of group space. In the
following part, we call\ $\delta \phi^{a}=e^{i\varphi^{a}}\left \vert \delta
\phi^{a}\right \vert $ that corresponds to $\tilde{U}(\delta \phi_{i}%
^{a})=e^{\pm i((\delta \phi_{i}^{a}T^{a})\cdot \hat{K}_{a})}$ ($\delta \phi
^{a}\rightarrow0$) to be non-unitary group-changing element for group-changing
space $\mathrm{C}_{\mathrm{\tilde{G}},d}(\Delta \phi^{a})$.

\paragraph{Non-unitary variant theory}

\subparagraph{Definition}

Non-unitary variant describes a structure of amplitude changings. We give a
definition about a general non-unitary variant.

\textit{Definition: A non-unitary variant }$V_{\mathrm{\tilde{G},}d}%
[\Delta \phi^{\mu},\Delta x^{\mu},k_{0}^{\mu}]$\textit{ is denoted by\ a
mapping between a d-dimensional non-unitary group-changing space }%
$\mathrm{C}_{\mathrm{\tilde{G},}d}$\textit{ with total size }$\Delta \phi^{\mu
}$\textit{\ and \textit{Cartesian }space }$\mathrm{C}_{d}$\textit{\ with total
size }$\Delta x^{\mu}$\textit{, i.e.,}%
\begin{align}
V_{\mathrm{\tilde{G},}d}[\Delta \phi^{\mu},\Delta x^{\mu},k_{0}^{\mu}]  &
:\mathrm{C}_{\mathrm{\tilde{G},}d}=\{ \delta \phi^{\mu}\} \nonumber \\
&  \Longleftrightarrow \mathrm{C}_{d}=\{ \delta x^{\mu}\}
\end{align}
\textit{where }$\Longleftrightarrow$\textit{\ denotes an ordered unitary
mapping under fixed changing rate of integer multiple }$k_{0}^{\mu}$\textit{.
}$k_{0}^{\mu}$\textit{\ is a real number. In particular, }$\delta \phi^{\mu
}=e^{i\varphi^{\mu}}\left \vert \delta \phi^{\mu}\right \vert $\textit{ denotes
non-unitary group-changing element along }$\mu$\textit{-direction (or element
of non-unitary group-changing space along }$\mu$-direction\textit{). }

Now, we take a 1D non-unitary variant $V_{\mathrm{\tilde{U}(1),}1}[\Delta
\phi,\Delta x,k_{0}]$ as an example to show the concept. $V_{\mathrm{\tilde
{U}(1),}1}[\Delta \phi,\Delta x,k_{0}]$ describes the mapping between 1D
non-unitary group-changing space $\mathrm{C}_{\mathrm{\tilde{U}(1)},1}%
(\Delta \phi)$\textit{ }and Cartesian space $\mathrm{C}_{1}$, i.e.,
\begin{align*}
V_{\mathrm{\tilde{U}(1),}1}[\Delta \phi,\Delta x,k_{0}]  &  :\\
\mathrm{C}_{\mathrm{\tilde{U}(1)},1}(\Delta \phi)  &  =\{ \delta \phi
=e^{i\varphi}\left \vert \delta \phi \right \vert \} \\
&  \Longleftrightarrow \mathrm{C}_{1}=\{ \delta x\}.
\end{align*}
According to above definition,\ for a 1D variant $V_{\mathrm{\tilde{U}(1),}%
1}[\Delta \phi,\Delta x,i\left \vert k_{0}\right \vert ],$ we have $\delta
\phi_{i}=e^{i\varphi}k_{0}n_{i}\delta x_{i}$ where $k_{0}$ is a constant real
number and $n_{i}$ is an integer number.

For a higher-dimensional case $V_{\mathrm{\tilde{G},}d}[\Delta \phi^{\mu
},\Delta x^{\mu},k_{0}^{\mu}]$, along different directions (for example, $\mu
$-direction), the situation is similar to the 1D case by considering the
corresponding distributions of $n_{i}^{\mu}.${}

We then take $d$-dimensional \textrm{\~{S}\~{O}(d)} non-unitary variant
$V_{\mathrm{\tilde{S}\tilde{O}(d)},d}[\Delta \phi^{\mu},\Delta x^{\mu}%
,k_{0}^{\mu}]$\ as an example, that is a prelude of AdS in physics. A
$d$-dimensional \textrm{\~{S}\~{O}(d)} non-unitary variant is a mapping
between non-unitary Clifford group-changing space $\mathrm{C}_{\mathrm{\tilde
{S}\tilde{O}(d)},d}$\ and a rigid spacetime $\mathrm{C}_{d},$ i.e.,\textit{ }%
\begin{equation}
V_{\mathrm{\tilde{S}\tilde{O}(d)},d}[\Delta \phi^{\mu},\Delta x^{\mu}%
,k_{0}^{\mu}]:\{ \delta \phi^{\mu}\} \Leftrightarrow \{ \delta x^{\mu}\}
\end{equation}
where a non-unitary Clifford group-changing space\textit{ }$\mathrm{C}%
_{\mathrm{\tilde{S}\tilde{O}(d)},d}(\Delta \phi^{\mu})$\textit{\ }is described
by $d$ series of numbers of complex group elements $\delta \phi^{\mu
}=e^{i\varphi^{\mu}}\left \vert \delta \phi^{\mu}\right \vert $ arranged in size
order with unit "vector" as $d$-by-$d$ Gamma matrices $\Gamma^{\mu}$ obeying
Clifford algebra $\{ \Gamma^{i},\Gamma^{j}\}=2\delta^{ij}$. The $d$%
-dimensional non-unitary Clifford group-changing space\textit{ }%
$\mathrm{C}_{\mathrm{\tilde{S}\tilde{O}(d}),d}(\Delta \phi^{\mu})$ has
orthogonality, i.e., $\left \vert \mathbf{\phi}_{\mathrm{A}}-\mathbf{\phi
}_{\mathrm{B}}\right \vert ^{2}=%
{\displaystyle \sum \nolimits_{\mu}}
\left \vert \phi_{\mathrm{A,}\mu}e^{\mu}-\phi_{\mathrm{B},\mu}e^{\mu
}\right \vert ^{2}$ where $\mathbf{\phi}_{\mathrm{A}}=%
{\displaystyle \sum \nolimits_{\mu}}
\phi_{\mathrm{A,}\mu}e^{\mu}$ and $\mathbf{\phi}_{\mathrm{B}}=%
{\displaystyle \sum \nolimits_{\mu}}
\phi_{\mathrm{B,}\mu}e^{\mu}$.

\subparagraph{Uniform non-unitary variant}

A d-dimensional uniform non-unitary variant (U-N-variant) $V_{0,d}[\Delta
\phi^{\mu},\Delta x^{\mu},k_{0}^{\mu}]$ for non-unitary group-changing space
$C_{\mathrm{\tilde{G}},d}(\Delta \phi^{\mu})$ of non-compact Lie group
\textrm{\~{G}} is defined by a perfect, ordered mapping between a
d-dimensional non-unitary group-changing space $\mathrm{C}_{\mathrm{\tilde{G}%
},d}(\Delta \phi^{\mu})$ and the d-dimensional Cartesian space $\mathrm{C}_{d}%
$, i.e.,
\begin{align}
V_{\mathrm{\tilde{G},}d}[\Delta \phi^{\mu},\Delta x^{\mu},k_{0}^{\mu}]  &  :\{
\delta \phi^{\mu}=e^{i\varphi^{\mu}}\left \vert \delta \phi^{\mu}\right \vert
\} \nonumber \\
&  \Leftrightarrow \{ \delta x^{\mu}\}
\end{align}
where $\Leftrightarrow$\ denotes an ordered mapping under fixed changing rate
of integer multiple $k_{0}^{\mu},$\ and $\mu$ labels the spatial
direction\textit{.} For a U-N-variant, the total size $\Delta \phi^{\mu}$ of
$\mathrm{C}_{\mathrm{\tilde{G}},d}$\ exactly matches the total size $\Delta
x^{\mu}$ of $\mathrm{C}_{d}$, i.e., $\left \vert \Delta \phi^{\mu}\right \vert
=\left \vert k_{0}^{\mu}\Delta x^{\mu}\right \vert $.

In particular, a U-N-variant with infinite size ($\Delta x\rightarrow \infty$)
has 1-th order unitary/non-unitary variability, i.e.,%
\begin{equation}
\mathcal{T}(\delta x^{\mu})\leftrightarrow \hat{U}(\delta \phi^{\mu}%
)=e^{i\cdot \delta \phi^{\mu}T^{\mu}}%
\end{equation}
where $\mathcal{T}(\delta x^{\mu})$ is the spatial translation operation on
$\mathrm{C}_{d}$ along $x^{\mu}$-direction and $\hat{U}(\delta \phi^{\mu})$ is
usual group operation on $\mathrm{C}_{\mathrm{\tilde{G}},d}(\Delta \phi^{\mu}%
)$, and $\delta \phi^{\mu}=e^{i\varphi^{\mu}}\left \vert \delta \phi^{\mu
}\right \vert $. That means when one translates along Cartesian space $\delta
x^{\mu},$ the corresponding amplitude along group-changing space
$\mathrm{C}_{\mathrm{\tilde{G}},d}$ is changing as $e^{i\cdot \delta \phi^{\mu
}T^{\mu}}=e^{i\cdot e^{i\varphi^{\mu}}\left \vert \delta \phi^{\mu}\right \vert
T^{\mu}}.$

\begin{figure}[ptb]
\includegraphics[clip,width=0.8\textwidth]{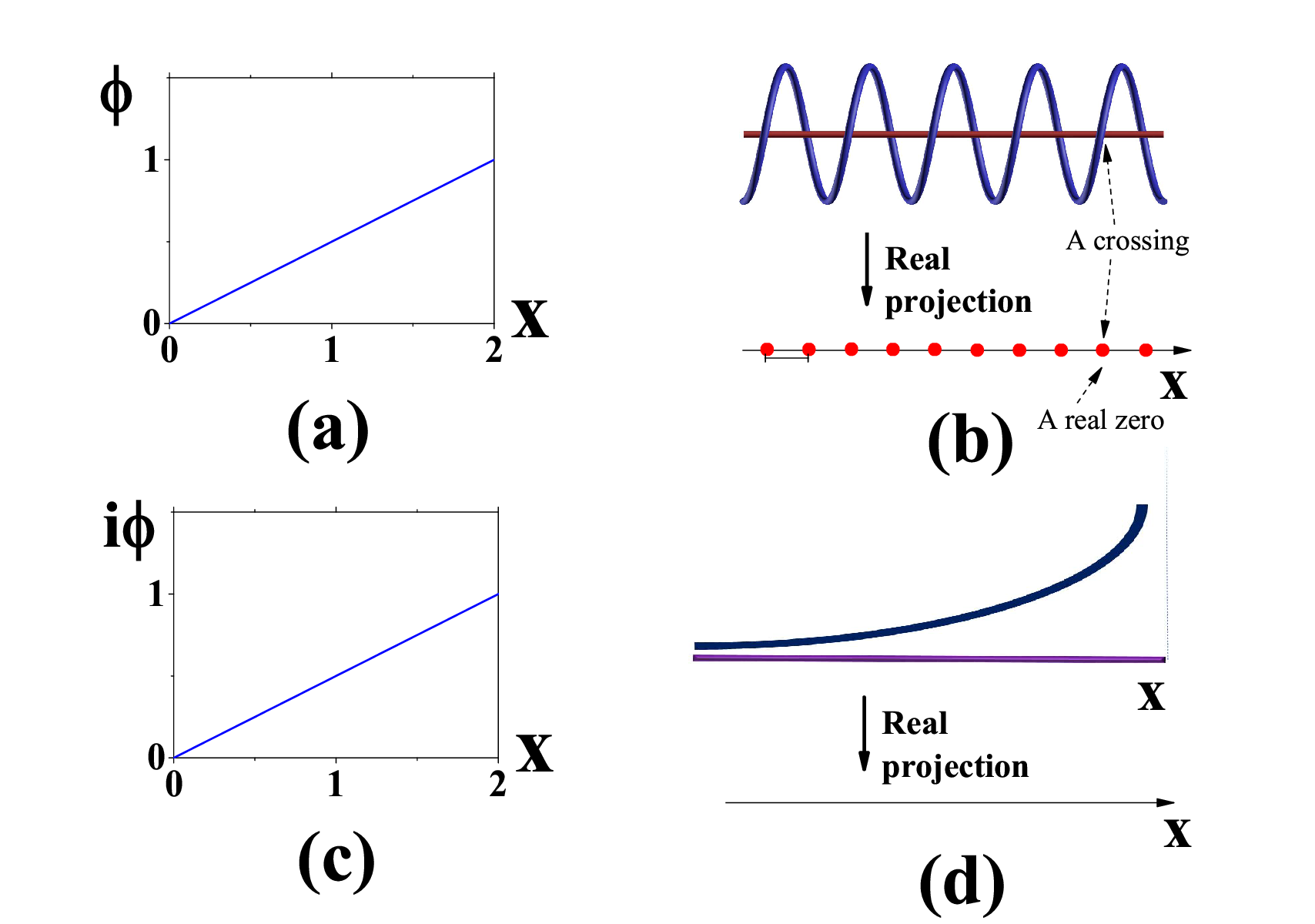}\caption{(Color online)
(a) Algebraic representation of 1D uniform unitary variant; (b) An
illustration of a 1D uniform unitary variant under geometry representation.
Phase changes along x-direction. Under knot-projection, we have a 1D crystal
of zeros (or zero lattice). Each crossing corresponds to a zero; (c) Algebraic
representation of 1D uniform non-unitary variant; (d) An illustration of a 1D
uniform non-unitary variant under geometry representation. The amplitude
rather than phase changes. Under real knot-projection, we don't have a zero
lattice. }%
\end{figure}

Next, we do knot projection (K-projection) on the U-N-variant and get the
corresponding zero lattice. See the illustration in Fig.9.

We take 1D U-N-variant $V_{0,\mathrm{\tilde{U}(1),}1}$ as an example. It is
described by a complex field $\mathrm{z}_{u}(x)=\exp(i\phi(x))$ in Cartesian
space where $\phi(x)=ie^{i\varphi}k_{0}x$. However, by taking $\tilde
{x}=e^{i\varphi}x$, the complex field $\mathrm{z}_{u}(x)=\exp(i\phi(x))$ in
Cartesian space becomes usual, i.e., $\phi(x)=ik_{0}\tilde{x}$. Now, in the
space denoted by the complex coordinates $\tilde{x}=e^{i\varphi}x,$ we have
knot like structure again and use the knot projection (K-projection) by
consider zero solution.

With the help of K-projection, people can locally obtain the property of the
variant. We introduce the\emph{ }K-projection of the curved line of 1D
U-N-variant along a given direction $\theta$ on the straight line at center of
$\mathrm{z}(\tilde{x})=0$ in 2D space $\{ \xi(\tilde{x}),\eta(\tilde{x})\}$.
In mathematics, the K-projection is defined by $\hat{P}_{\theta}\left(
\begin{array}
[c]{c}%
\xi(\tilde{x})\\
\eta(\tilde{x})
\end{array}
\right)  =\left(
\begin{array}
[c]{c}%
\xi_{\theta}(\tilde{x})\\
\left[  \eta_{\theta}(\tilde{x})\right]  _{0}%
\end{array}
\right)  $ where $\xi_{\theta}(\tilde{x})$ is variable and $\left[
\eta_{\theta}(\tilde{x})\right]  _{0}$ is constant. In the following parts we
use $\hat{P}_{\theta}$ to denote the projection operators. Under projection,
each zero corresponds to a solution of the equation $\hat{P}_{\theta
}[\mathrm{z}(\tilde{x})]\equiv \xi_{\theta}(\tilde{x})=0.$ For a 1D U-variant
$V_{\mathrm{\tilde{U}(1)},1}(\Delta \phi,\Delta x,k_{0})$, from the its
analytics representation $\mathrm{z}_{u}(\tilde{x})\sim e^{ik_{0}\cdot
\tilde{x}}$, we get the zero-solutions to be%
\begin{equation}
\tilde{x}=l_{0}\cdot n/2+\frac{l_{0}}{2\pi}(\theta+\frac{\pi}{2})
\end{equation}
or%
\begin{equation}
x=[l_{0}\cdot n/2+\frac{l_{0}}{2\pi}(\theta+\frac{\pi}{2})]e^{-i\varphi}%
\end{equation}
where $n$ is an integer number, and $l_{0}=2\pi/k_{0}$.

Because the zero solution is complex, we call the approach complex
K-projection and the corresponding zero lattice to be \emph{complex zero
lattice} that characterizes both phase changings and amplitude changings of
the system. As a result, the original 1D U-N-variant is reduced into a 1D
uniform complex zero lattice, of which each lattice site is characterized by
complex integer number. See the illustration of zero lattice under real knot
projection in Fig.9(b) and Fig.1(d).

For higher-dimensional \textrm{\~{S}\~{O}(d)} U-N-variant $V_{\mathrm{\tilde
{S}\tilde{O}(d)},d}[\Delta \phi^{\mu},\Delta x^{\mu},k_{0}^{\mu}],$ we have
1-th order unitary/non-unitary variability along different spatial directions,
i.e.,
\begin{equation}
\mathcal{T}(\delta x^{i})\leftrightarrow \hat{U}^{\mathrm{T}}(\delta \phi
^{i})=e^{i\cdot \delta \phi^{i}\Gamma^{i}},\text{ }i=x_{1},x_{2},\text{...}%
,x_{d},
\end{equation}
where $\delta \phi^{i}=\left \vert \delta \phi^{i}\right \vert e^{i\varphi_{i}%
}=k_{0}\delta x^{i}$ and $\Gamma^{i}$ are the Gamma matrices obeying Clifford
algebra $\{ \Gamma^{i},\Gamma^{i}\}=2\delta^{ij}$. Under K-projection, the
non-unitary variant turns into a $d$-dimensional uniform complex zero lattice,
$x^{i}=[l_{0}\cdot n^{i}+\frac{l_{0}^{i}}{\pi}(\theta+\frac{\pi}%
{2})]e^{-i\varphi^{i}}$.

In addition to complex K-projection, there exist other two different
K-projections -- real K-projection and imaginary K-projection.

For the representation under real K-projection, according to the zero equation
$\hat{P}_{\theta}[\mathrm{z}(\tilde{x}^{i})]\equiv \xi_{\theta}(\tilde{x}%
^{i})=\cos(k_{0}^{i}\cdot \tilde{x}^{i})=0$, we consider its real solutions.
Now, we have
\begin{align*}
\cos(k_{0}^{i}e^{i\varphi^{i}}\cdot x^{i})  &  =\cos(\cos(\varphi^{i}%
)k_{0}^{i}x^{i}+i\sin(\varphi^{i})k_{0}^{i}x^{i})\\
&  =\cos(\cos(\varphi^{i})k_{0}^{i}x^{i})\cosh(\sin(\varphi^{i})k_{0}^{i}%
x^{i})\\
&  -\sin(\cos(\varphi^{i})k_{0}^{i}x^{i})\sinh(\sin(\varphi^{i})k_{0}^{i}%
x^{i})\\
&  =0.
\end{align*}
We call it real zero lattice that characterizes the phase changings of the
system. For example, for the case of $\varphi^{i}=0,$ we have
\[
\cos(k_{0}^{i}e^{i\varphi^{i}}\cdot x^{i})=\cos(k_{0}^{i}x^{i})=0,
\]
of which the zero lattice is usual; for the case of $\varphi^{i}=\pm \frac{\pi
}{2},$ we have
\[
\cos(k_{0}^{i}e^{i\varphi^{i}}\cdot x^{i})=\cosh(k_{0}^{i}x^{i})=0.
\]
Now, there doesn't exist real zero solutions at all.

For the representation under imaginary K-projection, according to the zero
equation $\hat{P}_{\theta}[\mathrm{z}(\tilde{x}^{i})]\equiv \xi_{\theta}%
(\tilde{x}^{i})=\cos(k_{0}^{i}\cdot \tilde{x}^{i})=0$, we consider its
imaginary solutions where $\tilde{x}^{i}=ix^{i}$. Now, we have
\begin{align*}
\cos(k_{0}^{i}e^{i(\varphi^{i}-\frac{\pi}{2})}\cdot ix^{i})  &  =\cos
(k_{0}^{i}e^{i(\varphi^{i}-\frac{\pi}{2})}\cdot \tilde{x}^{i})\\
&  =\cos(\cos(\varphi^{i}-\frac{\pi}{2})k_{0}^{i}\tilde{x}^{i}\\
&  +i\sin(\varphi^{i}-\frac{\pi}{2})k_{0}^{i}\tilde{x}^{i})\\
&  =\cos(-\sin \varphi^{i}k_{0}^{i}\tilde{x}^{i}+i\cos \varphi^{i}k_{0}%
^{i}\tilde{x}^{i}).
\end{align*}
We call it imaginary zero lattice that characterizes the amplitude changings
of the system. For example, for the case of $\varphi^{i}=0,$ we have
\[
\cos(ik_{0}^{i}\tilde{x}^{i})=\cosh(k_{0}^{i}x^{i})=0.
\]
Now, there doesn't exist imaginary zero solutions at all. For the case of
$\varphi^{i}=\pm \frac{\pi}{2},$ we have
\[
\cos(k_{0}^{i}\cdot \tilde{x}^{i})=0.
\]

In summary, by the representation of complex K-projection, we can characterize
both phase changings and amplitude changings for a non-unitary variant; by the
representation of real K-projection, we can only characterize phase changings
that corresponds to the unitary physical processes; by the representation of
imaginary K-projection, we can only characterize amplitude changings of the
system. In the following part, we point out that based on the representation
of complex zero lattice we have a theory of AdS, while based on the
representation of real zero lattice, we have a theory of CFT.

\subparagraph{Perturbative non-unitary variant}

A d-dimensional perturbative non-unitary variant (P-N-variant) $V_{d}%
[\Delta \phi^{\mu},\Delta x^{\mu},k_{0}^{\mu}]$ for group-changing space
$\mathrm{C}_{\mathrm{\tilde{G}},d}(\Delta \phi^{\mu})$ of non-compact Lie group
\textrm{\~{G}} is defined by a quasi-perfect, ordered mapping between a
d-dimensional non-unitary group-changing space $\mathrm{C}_{\mathrm{\tilde{G}%
},d}(\Delta \phi^{\mu})$ and the d-dimensional Cartesian space $\mathrm{C}_{d}%
$, i.e.,
\begin{align}
V_{\mathrm{\tilde{G},}d}[\Delta \phi^{\mu},\Delta x^{\mu},k_{0}^{\mu}]  &  :\{
\delta \phi^{\mu}=e^{i\varphi^{\mu}}\left \vert \delta \phi^{\mu}\right \vert
\} \nonumber \\
&  \Leftrightarrow \{ \delta x^{\mu}\}.
\end{align}
where $\Leftrightarrow$\ denotes an ordered mapping under fixed changing rate
of integer multiple $k_{0}^{\mu},$\ and $\mu$ labels the spatial direction.
The adjective "quasi-perfect" means the total size $\Delta \phi^{\mu}$ of
$\mathrm{C}_{\mathrm{\tilde{G}},d}$\ doesn't exactly match the total size
$\Delta x^{\mu}$ of $\mathrm{C}_{d}$, i.e., $\left \vert \Delta \phi^{\mu
}\right \vert \neq \left \vert k_{0}^{\mu}\Delta x^{\mu}\right \vert .$

Under hybrid-order representation of partial K-projection, we have a usual
quantum field description for a P-N-variant. When we do partial K-projection
on the original U-N-variant $V_{0,\mathrm{\tilde{U}(1),}1}[\Delta \phi
^{A},\Delta x,k_{0}],$ we get a theory for AdS. On the contrary, if we use the
real K-projection, we get a quantum field theory on real zero lattice. This
leads to the CFT.

\subsubsection{Representations for shape changings of non-unitary
\textrm{\~{S}\~{O}(d)} variant}

In this part, we focus on non-unitary \textrm{\~{S}\~{O}(d)} variant that is
prelude of AdS in physics.

A non-unitary \textrm{\~{S}\~{O}(d)} variant is described by a mapping between
the non-unitary group-changing space and Cartesian space%
\begin{align}
V_{\mathrm{\tilde{S}\tilde{O}(d)},d}[\Delta \phi^{i},\Delta x^{i},k_{0}^{i}]
&  :\{ \delta \phi^{i}=e^{i\varphi^{i}}\left \vert \delta \phi^{i}\right \vert
\} \nonumber \\
&  \Leftrightarrow \{ \delta x^{i}\}.
\end{align}
These mappings are characterized by the local operations, $\mathcal{T}(\delta
x^{i})\leftrightarrow \hat{U}^{\mathrm{T}}(\delta \phi^{i})=e^{i\cdot \delta
\phi^{i}\Gamma^{i}}$ where $\delta \phi^{i}=k_{0}^{i}\cdot(\Delta x^{i}).$

There are two types of changings -- expand/contract, or shape changings. To
characterize the shape changings of \textrm{\~{S}\~{O}(d)} non-unitary variant
($V_{\mathrm{\tilde{S}\tilde{O}(d),}d}[\Delta \phi^{\mu},\Delta x^{\mu}%
,k_{0}^{\mu}]$), there are four representations -- geometry representations by
fixing Gamma matrices and matrix representation by fixing space coordinates on
complex zero lattice; geometry representations by fixing Gamma matrices and
matrix representation by fixing space coordinates on real zero lattice.

According to above discussions, we firstly introduce the complex zero lattice
by considering K-projection by replacing coordinates $\Delta x^{\mu}$ by
complex ones $\Delta \tilde{x}^{\mu}=\Delta x^{\mu}e^{\varphi^{\mu}}$. The
perturbative uniform variant can be characterized by a non-uniform complex
zero lattice within geometric representation by fixing Hermitian $\Gamma^{\mu
}$. The situation is same to that for unitary one. According to the
higher-order variability, the purterbative uniform variant is characterized by
the local spatial translation operators $\mathcal{T}(\Delta \tilde{x}^{\mu
})\rightarrow \hat{U}^{\mathrm{T}}(\delta \phi^{\mu}).$ On curved spacetime,
spatiotemporal coordinates locally change, $\tilde{x}^{\mu}\rightarrow
(\tilde{x}^{\mu})_{\mathrm{curved}}=(\tilde{x}^{\mu})^{\prime}$.
Correspondingly, under the geometric representation, the spatial translation
operators locally change, i.e.,
\begin{equation}
\mathcal{T}(\Delta \tilde{x}^{\mu})\rightarrow \mathcal{T}((\Delta \tilde{x}%
^{\mu})^{\prime})\leftrightarrow e^{i\Gamma^{\mu}k_{0}(\Delta \tilde{x}^{\mu
})^{\prime}}%
\end{equation}
Now, the distances between two nearest-neighbor lattice sites of complex zero
lattice deform, i.e., $(\Delta \tilde{x}^{\mu}(N^{\mu}))^{\prime}-\Delta
\tilde{x}^{\mu}=e^{\mu}(N^{\mu}),$ where $e^{\mu}(N^{\mu})$ are vierbein
fields that are the difference between the geometric unit-vectors of the
original frame and the deformed frame.

In the continuum limit $\Delta \tilde{x}^{\mu}\gg1$, the spatial coordinates
become continuous. Now, in geometry representation, the non-uniform complex
zero lattice is characterized by a curved space. The geometry fields (vierbein
fields $\tilde{e}^{a}$ and spin connections $\tilde{\omega}^{ab}$) of the
curved space are determined by the non-uniform local coordinates,
$(\Delta \tilde{x}^{\mu}(\tilde{x}))^{\prime}$. To characterize the deformed
complex zero lattice, with the help of the vierbein fields $\tilde{e}^{a}$,
the space metric is defined by $\tilde{e}_{i}^{a}\tilde{e}_{b}^{i}=\delta
_{b}^{a}\,,\quad \tilde{e}_{i}^{a}\tilde{e}_{a}^{j}=\delta_{i}^{j},$ and
$\tilde{e}_{\alpha}^{a}\tilde{e}_{\beta}^{b}=\tilde{g}_{\alpha \beta}.$ The
Riemann curvature 2-form is written as $\tilde{R}_{b}^{a}=d\tilde{\omega}%
_{b}^{a}+\tilde{\omega}_{c}^{a}\wedge \tilde{\omega}_{b}^{c},$ where $\tilde
{R}_{b\mu \nu}^{a}\equiv \tilde{e}_{\alpha}^{a}\tilde{e}_{b}^{\beta}\tilde
{R}_{\beta \mu \nu}^{\alpha}$ are the components of the usual Riemann tensor
projection on the tangent space.

In addition, we have another geometry representation by considering a real
zero lattice, i.e., $\Delta \tilde{x}^{\mu}=\Delta x^{\mu}e^{i\varphi^{\mu}%
}\rightarrow \Delta x^{\mu}.$ Now, $\Gamma^{\mu}$ becomes non-Hermitian,
constant matrices, i.e., $\Gamma^{\mu}\rightarrow \tilde{\Gamma}^{\mu}%
=\Gamma^{\mu}e^{i\varphi^{\mu}}.$ This leads to a theory of non-unitary
\textrm{SO(3)}$^{\mathrm{SO(4)}}$ gauge structure for the non-unitary variant.

Next, we discuss the matrix representations for a perturbative uniform
\textrm{\~{S}\~{O}(d)} non-unitary variant.

The information of the perturbative uniform \textrm{\~{S}\~{O}(d)} non-unitary
variant is recorded by the information of matrix network that is described by
$\Gamma^{\{N^{ii},M^{j}\}}$ on the links between two nearest-neighbor lattice
sites $N^{i}$ and $M^{j}$ of the zero lattices. For the matrix representation
on complex zero lattice, $\Gamma^{\{N^{ii},M^{j}\}}$ are Hermitian; while for
the matrix representation on real zero lattice, $\tilde{\Gamma}^{\{N^{ii}%
,M^{j}\}}$ become non-Hermitian.

Under matrix representations, the (perturbative) uniform \textrm{\~{S}%
\~{O}(d)} variant is characterized by a (deformed) \emph{matrix network}.
There are two types of matrix representations: One is about a non-Hermitian
matrix representation with non-Hermitian Gamma matrices $\Gamma^{\mu}$. Now,
the space coordinates are real constant, $\Delta x^{\mu}=\Delta \tilde{x}^{\mu
}e^{-i\varphi^{\mu}}$; The other is about a Hermitian matrix representation
with Hermitian, variable $\Gamma^{\mu}$. Now, the space coordinates are
complex, constant, $\Delta \tilde{x}^{\mu}=\Delta x^{\mu}e^{i\varphi^{\mu}}$.

In the end of this section, we point out that except for above four different
representations, there exist additional two\emph{ kinetic representations} by
fixing both Gamma matrices and space coordinates on complex (or real) zero
lattice. Now, the changing rate $k_{0}$ become fluctuated, i.e.,
\[
k_{0}\rightarrow k_{0}^{\mu}(x,t).
\]
Then, we have
\begin{equation}
\mathcal{T}(k_{0})\rightarrow \mathcal{T}(k_{0}^{\mu}(x,t))\leftrightarrow
e^{i\Gamma^{\mu}k_{0}^{\mu}(x,t)\Delta \tilde{x}^{\mu}}.
\end{equation}

\subsection{Theory for AdS}

\subsubsection{AdS as a special \textrm{\~{S}\~{O}(d+1)} non-unitary physical
variant}

Firstly, we introduce a special (d+1)-dimensional \textrm{\~{S}\~{O}(d+1)}
non-unitary physical variant\textit{ }$V_{\mathrm{\tilde{S}\tilde{O}%
(d+1)},d+1}(\Delta \phi^{\mu},\Delta x^{\mu},k_{0},\omega_{0})$ that is mapping
between\textrm{ \~{S}\~{O}(d+1)} non-unitary Clifford group-changing
space\textit{ }$\mathrm{C}_{\mathrm{\tilde{S}\tilde{O}(d+1)},d+1}$%
\textit{\ }and a rigid spacetime $\mathrm{C}_{d+1},$\textit{ }i.e.,\textit{ }%
\begin{align}
V_{\mathrm{\tilde{S}\tilde{O}(d+1)},d+1}[\Delta \phi^{\mu},\Delta x^{\mu}%
,k_{0}^{\mu}]  &  :\{ \delta \phi^{\mu}=\left \vert \Delta \phi^{\mu}\right \vert
e^{i\varphi^{\mu}}\} \nonumber \\
&  \Leftrightarrow \{ \delta x^{\mu}\}
\end{align}
where $\Leftrightarrow$\ denotes an ordered mapping with fixed changing rate
of integer multiple $k_{0}$ or $\omega_{0},$\ and $\mu$ labels the spatial
direction. In particular, we have
\[
\varphi^{\mu \neq d}=0,\text{ }\varphi^{\mu=d}=\pm \frac{\pi}{2}.
\]
Or, we have $\delta \phi^{\mu \neq d}=\pm \left \vert \Delta \phi^{d}\right \vert $
and $\delta \phi^{\mu=d}=\pm i\left \vert \Delta \phi^{d}\right \vert .$ This
\textrm{\~{S}\~{O}(d+1)} non-unitary physical variant is just that for AdS,
i.e.,
\[
\text{Flat AdS = Uniform \textrm{\~{S}\~{O}(d+1)} non-unitary physical
variant.}%
\]
In this part, we will develop a complete theoretical framework for AdS based
on the Variant hypothesis.

To accurately characterize the physical variant, we consider its 1-th order
spatial-tempo variability, which corresponds to its geometry/dynamic
properties, respectively.

The 1-th order spatial-tempo variability is determined by the following
equation,
\begin{equation}
\mathcal{T}(\delta x^{\mu})\leftrightarrow \hat{U}(\delta \phi^{\mu}),
\end{equation}
where $\hat{U}(\delta \phi^{\mu})=e^{i\cdot \delta \phi^{\mu}\Gamma^{\mu}}%
$\textbf{.} Along the d-th direction, we have a 1-th order non-unitary spatial
variability
\[
\mathcal{T}(\delta x^{d})\leftrightarrow \hat{U}(\delta \phi^{\mu}%
)=e^{i\cdot \delta \phi^{d}\Gamma^{d}}=e^{k_{0}x^{d}\Gamma^{d}}.
\]
In addition, there exists 1-th order rotation variability
\begin{equation}
\hat{U}^{\mathrm{R}}\leftrightarrow \hat{R}_{\mathrm{space}}%
\end{equation}
where $\hat{U}^{\mathrm{R}}$ is (non-compact), non-unitary \textrm{SO(d,1)}
rotation operator on Clifford group-changing space $\hat{U}^{\mathrm{R}}%
\Gamma^{I}\mathbf{(}\hat{U}^{\mathrm{R}})^{-1}=\Gamma^{I^{\prime}},$ and
$\hat{R}_{\mathrm{space}}$ is \textrm{SO(d,1)} rotation operator on Cartesian
space, $\hat{R}_{\mathrm{space}}x^{I}\hat{R}_{\mathrm{space}}^{-1}%
=x^{I^{\prime}}.$ After doing a global composite rotation operation $\hat
{U}^{\mathrm{R}}\cdot \hat{R}_{\mathrm{space}}$, the system is invariant.

\subsubsection{Theory for spacetime}

In this part, we develop the theory for curved AdS by on complex zero lattice.

Curved AdS is an $\mathrm{\tilde{S}\tilde{O}(d+1)}$ perturbative non-unitary
physical variant that is described by inhomogeneous space-mapping between
non-unitary Clifford group-changing space\textit{ }$\mathrm{C}_{\mathrm{\tilde
{S}\tilde{O}(3+1)}}$\textit{\ }and Cartesian spacetime $\mathrm{C}_{3+1}$. To
characterize the curved AdS, we do complex K-projection and get a complex zero
lattice, of which the lattice number becomes complex number. See the
illustration in Fig.10(a). In continuum limit, we have complex coordinates
$x^{\mu}\rightarrow \tilde{x}^{\mu}=e^{i\varphi^{\mu}}\cdot x^{\mu}$.

Fortunately, except for the coordinates become complex numbers, the geometry
representation and matrix representation for quantum curved AdS are same to
those for the unitary one (de Sitter space (dS)).

\begin{figure}[ptb]
\includegraphics[clip,width=0.8\textwidth]{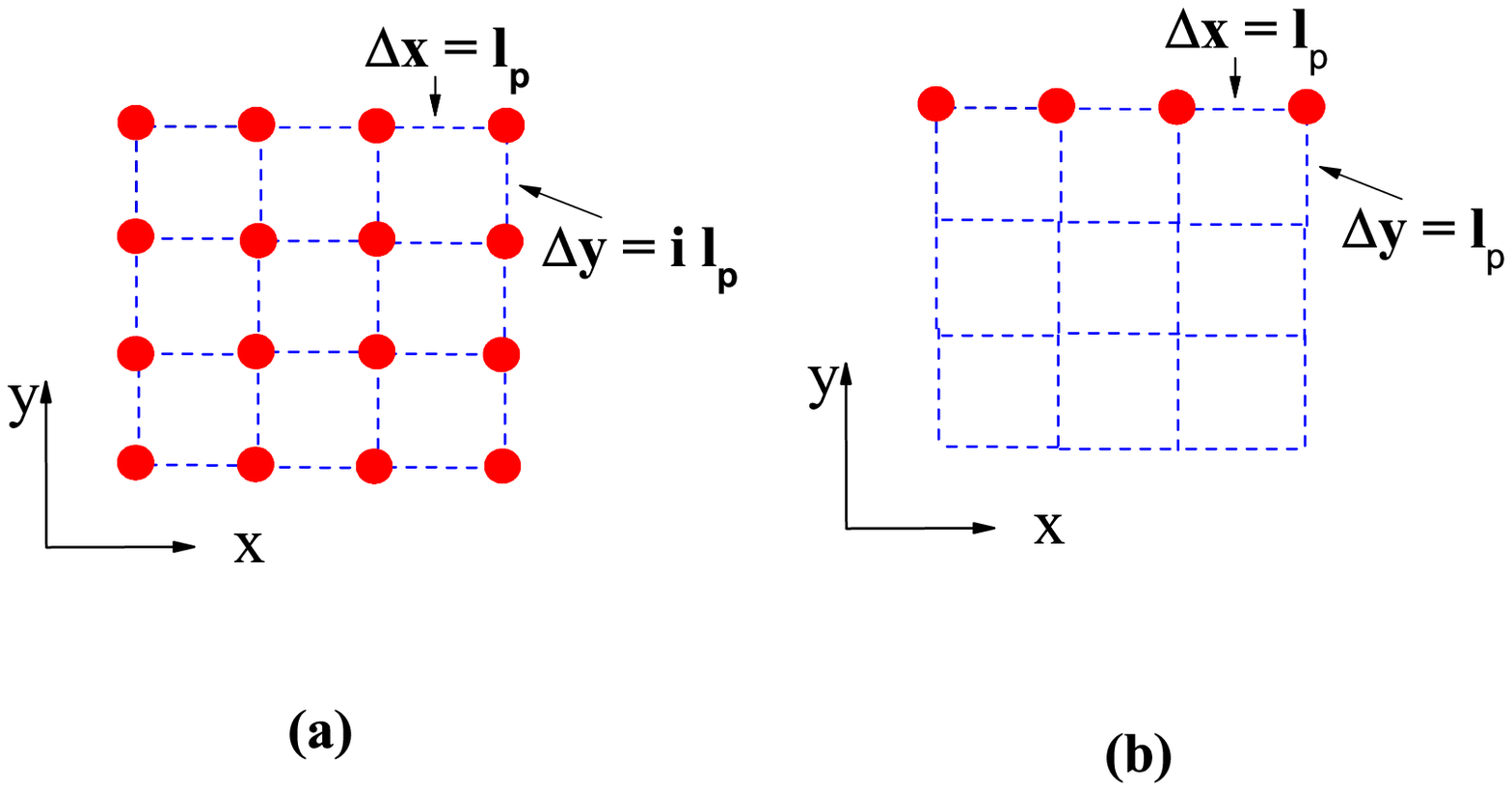}\caption{(Color online)
(a) A 2D uniform non-unitary variant under complex knot-projection. We have a
2D zero lattice. Along x-th direction, the lattice constant is real, along
y-th direction, the lattice distance is imaginary; (b) A 2D uniform
non-unitary variant under real knot-projection. We have a 1D zero lattice.
along x-th direction. The lattice constant is real.}%
\end{figure}

\subsubsection{Theory for matter}

Based on Geometry representation under D-projection and K-projection, a
uniform non-unitary physical variant is reduced into a uniform complex zero
lattice. We also assume that each zero corresponds to an elementary particle.
As a result, an elementary particle can be defined by a group of unitary
group-changing elements on complex coordinates,
\begin{equation}
\prod_{i}(\hat{U}(\delta \phi_{i}))=\prod_{i}(\prod_{\mu=1}^{d+1}(\hat
{U}(\delta \phi_{i}^{\mu})))
\end{equation}
where $\hat{U}(\delta \phi_{i})=\prod_{\mu=1}^{d+1}(\hat{U}(\delta \phi_{i}%
^{\mu}))$ and $\hat{U}(\delta \phi_{i}^{\mu})=e^{i((\delta \phi_{i}^{\mu}T^{\mu
})\cdot \hat{K}_{\mu})}$, $\hat{K}_{\mu}=-i\frac{d}{d\phi^{\mu}}.$\ Here, the
i-th unitary operation $\hat{U}(\delta \phi_{i})$ generates an element of
unitary group-changing that is infinitesimal unitary group-changing
operations. For an elementary, along an arbitrary direction, the total size of
group-changing elements is $%
{\displaystyle \sum \limits_{i}}
\delta \phi_{i}^{\mu}=\pi$.

According to above definition, one can see that the elementary particle is
same to that on dS. Therefore, the elementary particle on AdS becomes
topological defect of quantum spacetime and obey fermionic statistics. In
particular, we have
\[
\tilde{N}_{F}=N_{F}=-\tilde{q}_{m},
\]
where the number of particles $\tilde{N}_{F}$ is an integer, real number.
$\tilde{q}_{m}$ is the number of magnetic monopole of quantum spacetime,
\[
\tilde{q}_{m}=\frac{1}{4\pi}%
{\displaystyle \oint \nolimits_{\mathcal{\tilde{S}}}}
\tilde{F}_{\mathcal{\tilde{S}}}^{IJ}=\frac{1}{3!4\pi}%
{\displaystyle \oint \nolimits_{\mathcal{\tilde{S}}}}
\epsilon_{IJK}[\tilde{n}^{I}(x)d(\tilde{n}^{J}(x))\wedge d(\tilde{n}%
^{K}(x))].
\]
Here, $\mathcal{\tilde{S}}$ is the closed surface enclosing $\mathcal{\tilde
{M}}$ in 3D space. See the detailed definition of above equation in
Ref.\cite{kou1}. Because $\tilde{q}_{m}$ is defined on group-changing space,
it is real and same to that in unitary physical variant, i.e., $\tilde{q}%
_{m}=q_{m}$.

Furthermore, we point out that the geometry quantization for curved AdS is
similar to that for curved dS.

The lattice constant for the complex zero lattice of AdS is $\tilde{l}%
_{0}^{\mu}=e^{i\varphi^{\mu}}l_{0}.$ So, $\tilde{l}_{0}^{\mu}$ is $l_{0}$
along the directions with real coordinates; $\tilde{l}_{0}^{\mu}$ is $il_{0}$
along the direction with imaginary coordinate. The 3-volume $\Delta \tilde{V}$
of AdS is given by%
\[
\Delta \tilde{V}=(\tilde{l}_{0}^{\mu})^{3}4\pi \tilde{q}_{m}.
\]
Finally, with help of $\Delta \tilde{V}=(\tilde{l}_{0}^{\mu})^{3}4\pi \tilde
{q}_{m}$ and $\tilde{N}_{F}=N_{F}=-\tilde{q}_{m},$\ we have
\[
\tilde{N}_{F}=(4\pi(\tilde{l}_{0}^{\mu})^{3})^{-1}\Delta \tilde{V}.
\]
This equation that unifies spacetime and matter is also same to that for the
unitary case.

\subsubsection{Theory for motion}

Motion comes from different types of time-dependent changings of
\textrm{\~{S}\~{O}(d+1)} non-unitary physical variants $V_{\mathrm{\tilde
{S}\tilde{O}(d+1)},d+1}(\Delta \phi^{\mu},\Delta x^{\mu},k_{0},\omega_{0})$
without size changings of group-changing space $\mathrm{C}_{\mathrm{\tilde
{S}\tilde{O}(d+1)},d+1}.$

There are two types of motions, one is about motion of matter that corresponds
to locally expanding or contracting $\mathrm{C}_{\mathrm{\tilde{S}\tilde
{O}(d+1)},d+1}(\Delta \phi^{a})$\ without changing its corresponding size on
Cartesian space $\mathrm{C}_{d+1}$; The other is about curving of spacetime
that corresponds to locally shape changings\emph{ }on Cartesian space
$\mathrm{C}_{d+1}$. This is usually called gravitational waves. In this part,
due to different energy scales we call motion of matter to be \emph{fast
motion} and motion of gravitational waves to be \emph{slow motion}.

Firstly, we consider the motion of matter.

Using the earlier approach \cite{kou1}, the effective Hamiltonian for
elementary particles on complex spacetime is
\[
\mathcal{H}=\int(\Psi^{\dagger}(\mathbf{\tilde{x}})\hat{H}\Psi(\mathbf{\tilde
{x}}))d^{3}\tilde{x}%
\]
where $\hat{H}=\vec{\Gamma}\cdot \Delta \tilde{p}+m\Gamma^{t}$ with $\vec
{\Gamma}=(\Gamma^{x},\Gamma^{y},\Gamma^{z})$. Here, we have $\tilde{x}=x,$
$\tilde{y}=y,\  \tilde{z}=iz,$ $\tilde{t}=t.$\ This is a massive Dirac model on
spacetime with complex coordinates.\ We can also use $L_{\mathrm{particle}%
}=\bar{\Psi}(ie_{a}^{\mu}\gamma^{a}\hat{\partial}_{\mu}-m)\Psi$\ to describe
dynamics of elementary particles. $\gamma^{\mu}$ are the Gamma matrices
defined as $\gamma^{1}=\gamma^{0}\Gamma^{x}$, $\gamma^{2}=\gamma^{0}\Gamma
^{y},$ $\gamma^{3}=\gamma^{0}\Gamma^{z}$, $\gamma^{0}=\Gamma^{t}.$ With finite
mass $m$, the motion of elementary particles is always fast.

Secondly, we consider the motion of spacetime.

Using approach as in unitary physical variant, the action is obtained
\begin{equation}
S_{\mathrm{EH}}=\frac{1}{16\pi G}\int \sqrt{-g}\tilde{R}\text{ }d^{4}\tilde
{x}.\nonumber
\end{equation}
This action describes the dynamic of spacetime with complex coordinates itself.\

Finally, the total action is obtained as%
\begin{align}
S  &  =\mathcal{S}_{\mathrm{4D}}+S_{\mathrm{EH}}\\
&  =\int \sqrt{-g(\tilde{x})}\bar{\Psi}(e_{a}^{\mu}\gamma^{a}\hat{D}_{\mu
}-m)\Psi \text{ }d^{4}\tilde{x}\nonumber \\
&  +\frac{1}{16\pi G}\int \sqrt{-g}\tilde{R}\text{ }d^{4}\tilde{x}.\nonumber
\end{align}

However, the AdS has a special global shape of the Cartesian space
$\mathrm{C}_{d+1}$. This leads to additional term on the effective action,
i.e.,
\begin{align}
S  &  =\int \sqrt{-g(\tilde{x})}\bar{\Psi}(e_{a}^{\mu}\gamma^{a}\hat{D}_{\mu
}-m)\Psi \text{ }d^{4}\tilde{x}\nonumber \\
&  +\frac{1}{16\pi G}\int \sqrt{-g}\tilde{R}\text{ }d^{4}\tilde{x}+\int
\sqrt{-g}\Lambda \text{ }d^{4}\tilde{x}.\nonumber
\end{align}
Here, $\Lambda \,={\frac{d(d+1)}{L^{2}}}$ is a cosmological constant. The
constant $L$ is AdS radius. By using Poincare coordinates, we have
\begin{align}
\tilde{t}  &  =L\frac{1+x^{2}+z^{2}}{2z}\nonumber \\
\tilde{x}^{\mu}  &  =L\frac{x^{\mu}}{z}\label{eq:AdSPoincareEmb}\\
\tilde{x}^{d}  &  =L\frac{1-x^{2}-z^{2}}{2z}\nonumber
\end{align}
where $z>0$. According above discussion, the metric in $(d+1)$-dimensions for
flat AdS can be described by the so-called Poincare patch
\begin{equation}
d\tilde{s}^{2}\,=\,({\frac{L}{z})}^{2}(-dt^{2}+d\vec{x}^{2}+dz^{2}).
\label{AdS_metric}%
\end{equation}
The (conformal) boundary of the AdS space is located at AdS boundary of $z=0$.

In addition, we point out that $N_{d}l_{p}=2\pi L$ where $N_{d}$ is number of
complex zeroes long $x^{d}$-th direction and $l_{p}$ is Planck length.

Under the matrix representation, the spacetime becomes flat. However, the slow
motion of quantum spacetime (or fluctuating gravitational waves) leads to the
quantum fluctuations of the Gamma matrices in Dirac model, i.e.,
\[
\hat{H}=\vec{\Gamma}\cdot \Delta \tilde{p}+m\Gamma^{t}\rightarrow \hat{H}%
^{\prime}=\vec{\Gamma}(x,t)\cdot \Delta \tilde{p}+m\Gamma^{t}(x,t).
\]
This contributes an additional energy-momentum tensor. In general, we can use
the gauge field to characterize the of quantum fluctuations of the Gamma
matrices. See the detailed discussion in Ref.\cite{kou1}.

\subsubsection{Non-Hermitian quantum mechanics and spacetime skin effect}

In above section, we show that on AdS, the coordinates along d-th direction
becomes complex number.\emph{ What does it mean in our real world?} To
characterize the observables in quantum physics on AdS, we use kinetic representation.

We use kinetic representation with real coordinates and replace the complex
coordinates $\tilde{x}^{\mu}=e^{i\varphi^{\mu}}\cdot x^{\mu}$ by the real
coordinates $x,$\
\[
\tilde{x}^{\mu}\rightarrow x^{\mu}=e^{-i\varphi^{\mu}}\cdot \tilde{x}^{\mu}.
\]
\ Correspondingly, the changing rate becomes complex, i.e.,
\[
k_{0}\rightarrow \tilde{k}_{0}^{\mu}=e^{i\varphi^{\mu}}\cdot k_{0}.
\]
Now, Gamma matrices $\Gamma^{\mu}$ are still Hermitian.

Firstly, we discuss motion of matter.

Matter is defined by globally expanding or contracting $\mathrm{C}%
_{\mathrm{\tilde{S}\tilde{O}(d+1)},d+1}$ group-changing space on rigid space
$\mathrm{C}_{d+1}$. Along $\mu$-th ($\mu \neq d$) direction, the matter comes
from the phase changings; while along $\mu$-th ($\mu=d$) direction, the matter
comes from amplitude changings.

Globally expand/contract of group-changing space corresponds to the
generation/annihilate of elementary particles in quantum mechanics. Each
elementary particle corresponds to an zero with $\pi$-phase changing along the
direction; along $\mu$-th ($\mu=d$) direction, the elementary particle becomes
a "non-unitary" zero with $i\pi$-phase ($\pi$ amplitude) changing changing.
The total size of the group-changing elements for an elementary particle to be
$\pi$\ along $\mu$-th ($\mu \neq d$) direction and $i\pi$ along $\mu$-th
($\mu=d$) direction. This leads to \emph{non-Hermitian fermionic statistics}.
We call the elementary particles to be \emph{non-Hermitian elementary
particles.}

To describe the motion for non-Hermitian elementary particles, we replace
$\tilde{x}$ by $x$ and $k^{\mu}$ by $\tilde{k}^{\mu},$ i.e.,%
\[
\tilde{x}\rightarrow x,\text{ }\tilde{y}\rightarrow y,\text{ }\tilde
{z}\rightarrow z\text{ }%
\]
and
\[
\tilde{p}_{x}\rightarrow p_{x},\text{ }\tilde{p}_{y}\rightarrow p_{y},\text{
}\tilde{p}_{z}\rightarrow ip_{z}\text{.}%
\]
The effective Hamiltonian for non-Hermitian elementary particles is obtained
as
\[
\mathcal{H}=\int(\bar{\Psi}^{\dagger}(\mathbf{x})\hat{H}\Psi(\mathbf{x}%
))d^{3}x
\]
where $\hat{H}=\Gamma \cdot \Delta \tilde{p}+m\Gamma^{t}$ with $\Delta \tilde
{p}^{\mu}=\hbar \Delta \tilde{k}^{\mu}=(\hbar k^{x},\hbar k^{y},i\hbar k^{z})$.
Here, $\Psi^{\dagger}(\mathbf{x})$ denotes the generalized creation operation
for non-Hermitian elementary particles, of which the amplitude changes
$e^{\pi}$ along z-direction and phase changes $e^{i\pi}$ along other
directions. The corresponding Lagrangian is obtained as $L_{\mathrm{particle}%
}=\bar{\Psi}(i\gamma^{\mu}\tilde{\partial}_{\mu}-m)\Psi$.

Next, based on above effective Hamiltonian $\hat{H}$, we discuss the physical
properties of non-Hermitian elementary particles.

A key point is \emph{spacetime skin effect}.

According to non-unitary variability along d-th direction, non-unitary
operation $\hat{U}(\delta \phi^{d})=e^{k_{0}x^{d}\Gamma^{d}}$ on AdS changes
the relative weight of the eigenstates of $\Gamma^{d}$. Therefore, the
relative weight of $\Gamma^{d}$ exponentially grows/decreases towards the
boundary $x^{d}\rightarrow \pm \infty$ along d-th direction: in the limit of
$x^{d}\rightarrow \infty,$ the amplitude of eigenstates with positive
elgenvalues diverge while the amplitude of eigenstates with negative
elgenvalues turns to zero; the amplitude of eigenstates with negative
elgenvalues diverge in the limit of $x^{d}\rightarrow-\infty$ while the
amplitude of eigenstates with positive elgenvalues turns to zero. This
indicates the existence of spacetime skin effect.

According to spacetime skin effect, the main degrees of freedom for elementary
particles will concentrate on the boundary rather than in bulk! When particles
move along d-th direction, the quantum states are characterized by
$\Delta \tilde{k}^{d}=i\Delta k^{d}$ that is an imaginary value! Now, we have
the particle's amplitude rather than phase changes. For elementary particles
along d-th direction, the wave function is solved to be $\Psi(x^{d})\sim
e^{i(x^{d}\cdot \Delta \tilde{k}^{d})\Gamma^{z}}=e^{-(x^{d}\cdot \Delta
k)\Gamma^{d}}$. Due to $\Psi(x^{d})\sim e^{-(x^{d}\cdot \Delta k)\Gamma^{z}},$
we find that elementary particles gather at the boundary of the system,
$x^{d}\rightarrow \pm \infty.$

An additional representation is about complex matrix network. Now, we have the
real coordinates $x^{\mu}$ and wave vectors $k^{\mu}.$ As a result, the Gamma
matrices $\tilde{\Gamma}^{\mu}=e^{i\varphi^{\mu}}\Gamma^{\mu}$ become
non-Hermitian, i.e., $\tilde{\Gamma}^{\mu}\neq(\tilde{\Gamma}^{\mu})^{\dagger
}$. The non-Hermitian Gamma matrices $\tilde{\Gamma}^{\mu}$ leads to a
non-Hermitian quantum mechanics. The Hamiltonian becomes non-Hermitian, i.e.,
\[
\mathcal{H}=\int(\bar{\Psi}^{\dagger}(\mathbf{x})\hat{H}\Psi(\mathbf{x}%
))d^{3}x
\]
where $\hat{H}=\tilde{\Gamma}\cdot \Delta p+m\Gamma^{t}$ with $\tilde{\Gamma
}=(\tilde{\Gamma}^{x},\tilde{\Gamma}^{y},\tilde{\Gamma}^{z})=(\Gamma
^{x},\Gamma^{y},i\Gamma^{z})$. By using non-Hermitian Gamma matrices
$\tilde{\Gamma}^{\mu},$ we can also characterize the spacetime skin effect.
The result is consistent to above.

\subsection{Theory for CFT}

Curved AdS is an $\mathrm{\tilde{S}\tilde{O}(d+1)}$ non-unitary physical
variant described by an inhomogeneous space-mapping by a mapping between
non-unitary group-changing Clifford group-changing space\textit{ }%
$\mathrm{C}_{\mathrm{\tilde{S}\tilde{O}(3+1)}}$\textit{\ }and Cartesian
spacetime $\mathrm{C}_{3+1}$. Under real K-projection, we have (d-1)+1
dimensional real zero lattice. The theory turns into CFT on the boundary of
the system.

\subsubsection{Theory for spacetime}

Firstly, we focus on theory of $\mathrm{\tilde{S}\tilde{O}(d+1)}$ non-unitary
physical variant on real zero lattice.

Under real K-projection, the original non-unitary physical variant
$V_{\mathrm{\tilde{S}\tilde{O}(d+1)},d}[\Delta \phi^{\mu},\Delta x^{\mu}%
,k_{0}^{\mu}]$ is reduced into a (d-1)+1 dimensional uniform real zero
lattice: Along $\mu$-th ($\mu \neq d$) direction, there exists zero lattice, of
which the lattice site is denoted by $N^{\mu};$ Along $\mu$-th ($\mu=d$)
direction, there doesn't exist zero lattice. As a result, we have a (d-1)+1
dimensional zero lattice with real lattice number. The Gamma matrices
$\Gamma^{\mu}$ are Hermitian.

In particular, we point out that the (d-1)+1 dimensional zero lattice is the
sub-spacetime of the whole system, of which the normal lines are fixed to be
$\Gamma^{d}.$ Hence, for the real zero lattice, the corresponding spacetime in
continuum limit has uniform direction of normal lines $\Gamma^{d}.$ By setting
$\Gamma^{d}$ to a constant Gamma matrix, the spacetime must be flat and cannot
be curved!

\subsubsection{Theory for matter}

In CFT, we assume that each zero of the real zero lattice corresponds to an
elementary particle.

An elementary particle is a group of unitary group-changing elements on real
coordinates,
\begin{equation}
\prod_{i}(\hat{U}(\delta \phi_{i}))=\prod_{i}(\prod_{\mu=1}^{(d-1)+1}(\hat
{U}(\delta \phi_{i}^{\mu})))
\end{equation}
where $\hat{U}(\delta \phi_{i})=\prod_{\mu=1}^{d-1)}(\hat{U}(\delta \phi
_{i}^{\mu}))$ and $\hat{U}(\delta \phi_{i}^{\mu})=e^{i((\delta \phi_{i}^{\mu
}T^{\mu})\cdot \hat{K}_{\mu})}$, $\hat{K}_{\mu}=-i\frac{d}{d\phi^{\mu}}%
.$\ Here, the i-th unitary operation $\hat{U}(\delta \phi_{i})$ generates an
element of unitary group-changing that is infinitesimal unitary group-changing
operations. For an elementary, along an arbitrary direction ($\mu \neq d$), the
total size of group-changing elements is $%
{\displaystyle \sum \limits_{i}}
\delta \phi_{i}^{\mu \neq d}=\pi$. Therefore, these elementary particles obey
fermionic statistics.

However, along $\mu$-th ($\mu=d$) direction, the total size of non-unitary
group-changing space about the elementary particle is same to the size of the
system $L_{d}$. Now, each zero of real zero lattice corresponds to
$L_{d}/l_{0}$ zeroes of complex zero lattice. $L_{d}/l_{0}$ is the total
lattice number along $\mu$-th ($\mu=d$) direction of complex zero lattice.
That means, each elementary particle on real zero lattice becomes a composite
zero with $L_{d}/l_{0}$ zeroes of complex zero lattice.

\subsubsection{Theory for motion}

\paragraph{Classification of motions}

Firstly, we classify the types of motions on real zero lattice.

There are two motions -- one is fast motion about expanding and contracting
the real zero lattice; the other is slow motion about the "shape" changing of
the zero lattice.

The fast motion comes from the motion of the elementary particle (or a real
zero). Because the mass $m_{R}=mL_{d}/l_{0}$ (see below discussion) of
elementary particle diverges, the motion is very fast.

The slow motion comes from the fluctuations of gravitational waves along the
boundary of the system. Now, we may consider the ground state to be a
many-body system of real zeroes (or elementary particles with mass $m_{R}$).
Without considering curving spacetime from real zero lattice, the fluctuations
of gravitational waves lead to fluctuations of Gamma matrices.

\paragraph{Theory for fast motion}

Firstly, we consider the theory for fast motion.

According to above discussion, each elementary particle of real zero
corresponds to $L_{d}/l_{0}$ zeroes of complex zero lattice. As a result, in
the thermodynamic limit, the mass $m_{R}$ of the elementary particle on real
zero lattice diverges
\[
m_{R}=L_{d}/l_{0}m\rightarrow \infty,\text{ }L_{d}\rightarrow \infty.
\]
The reselection of information unit of the system leads to the changing of
effective Hamiltonian. Now, the effective Hamiltonian for elementary particles
on (d-1)+1 dimensional zero lattice is obtained by
\[
\mathcal{H}_{(d-1)+1}^{\mathrm{fast}}=\int(\Psi^{\dagger}(\mathbf{x})\hat
{H}_{(d-1)+1}^{\mathrm{fast}}\Psi(\mathbf{x}))d^{3}x
\]
where $\hat{H}_{(d-1)+1}^{\mathrm{fast}}=\vec{\Gamma}\cdot \Delta \vec{p}%
+m_{R}\Gamma^{t}$ ($m_{R}=L_{d}/l_{0}m$). According to above Hamiltonian, the
energy $\Delta E$ for fast motion is $\pm \sqrt{\left \vert \Delta \vec
{p}\right \vert ^{2}+m_{R}^{2}}$ and wave function is plane waves
$\psi(x,t)=Ce^{-i\Delta \omega \cdot t+i\Delta \vec{k}\cdot \vec{x}}=Ce^{-i\Delta
E\cdot t/\hbar+i\Delta \vec{p}\cdot \vec{x}/\hbar}.$ This Hamiltonian describes
fast motion with very high energy and is irrelevant to low energy physics. The
fast motion can also be characterized by motion charge $Q^{\mu}=(\vec{Q}%
,Q_{t})=(\frac{\Delta \vec{k}}{k_{0}},\frac{\Delta \omega}{\omega_{0}})$.

On real zero lattice, there exists non-Hermitian polarization effect that
corresponds to the spacetime skin effect on complex zero lattice.

It was known that the spacetime skin effect comes from the non-unitary
variability along $\mu$-th ($\mu=d$) direction $\hat{U}(\delta \phi
^{d})=e^{k_{0}x^{d}\Gamma^{d}}$ that can be considered as a non-unitary
operation on elementary particles in AdS (or complex zero lattice).

On real zero lattice, the corresponding non-unitary operation also leads to
non-Hermitian polarization effect. Now, the non-unitary variability along
$\mu$-th ($\mu=d$) direction $\hat{U}(\delta \phi_{I}^{d})=e^{k_{0}x^{d}%
\Gamma^{d}}$ becomes a global non-unitary operation on a real zero (for
example, $I$-th), i.e.,%
\begin{align*}
\hat{U}_{\mathrm{global}}(\delta \phi_{I}^{d})  &  =%
{\displaystyle \prod \limits_{x^{d}}}
\hat{U}(\delta \phi_{I}^{d})\\
&  =e^{\frac{1}{l_{0}}\int k_{0}x^{d}\Gamma^{d}dx}\sim e^{\frac{L_{d}^{2}%
}{2l_{0}^{2}}\Gamma^{d}}.
\end{align*}
Under the global non-unitary operation $\hat{U}_{\mathrm{global}}\sim
e^{\frac{L_{d}^{2}}{2l_{0}^{2}}\Gamma^{d}}$, the relative weight of the
elementary particle of real zero exponentially grows/decreases towards the
boundary $z\rightarrow \pm \infty.$ For example, in the limit of $x^{d}%
\rightarrow \infty,$ the amplitude of eigenstates with positive elgenvalues
diverge while the amplitude of eigenstates with negative elgenvalues turns to
zero. As a result, the degrees of freedom for the real zero becomes fully
polarized on the boundary. For each real, its quantum states are at
exceptional points (EPs)\cite{kato}. We call it \emph{non-Hermitian
polarization effect}.

We point out that the non-Hermitian polarization effect is \emph{robust. }When
we consider the dynamical processes in bulk, there may exist slightly
changings of the amplitude for different eigenstates of $\Gamma^{d}$. As a
result, the global non-unitary operation $\hat{U}_{\mathrm{global}}$ becomes
slightly changes, i.e.,
\begin{align*}
\hat{U}_{\mathrm{global}}  &  =%
{\displaystyle \prod \limits_{x^{d}}}
\hat{U}(\delta \phi^{d})\\
&  \rightarrow \hat{U}_{\mathrm{global}}^{\prime}=%
{\displaystyle \prod \limits_{x^{d}}^{\prime}}
\hat{U}(\delta \phi^{d}).
\end{align*}
Because $\hat{U}_{\mathrm{global}}$ comes from integrating all imaginary
zeroes along $\mu$-th ($\mu=d$) direction, the perturbative changings cannot
eliminate the non-Hermitian polarization effect. The quantum states for the
real zero are always at EPs and the degrees of freedom for fast motion are
frozen. This result again indicates the observable physical processes are
irrelevant to fast motion.

\paragraph{Theory for slow motion}

Next, we consider the physical processes from slow motion that describes the
shape changings of boundary of the system. The gravitational waves moving
along certain direction parallel to the boundary lead to fluctuations of Gamma
matrices. See the illustration in Fig.11.

\begin{figure}[ptb]
\includegraphics[clip,width=0.67\textwidth]{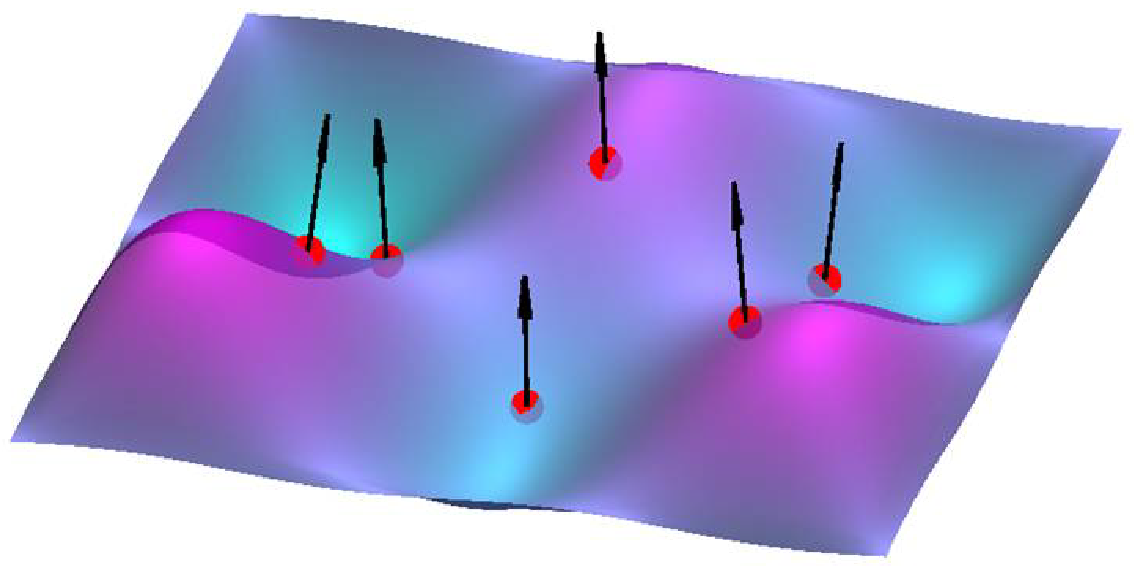}\caption{(Color online)
An illustration of the relationship of the fluctuations of external normal
directions (Gamma matrix along $x^{d}$-th direction $\Gamma^{d}(x)$) and the
shape fluctuations of the boundary of the system}%
\end{figure}

Now, we consider the real zero lattice to be a many-body system at half
filling, of which the elementary particle is a real zero that is a composite
zero with $L_{d}/l_{0}$ zeroes of complex zero lattice, i.e.,
\[
\text{Quantum spacetime (AdS) }\rightarrow \text{ Many-body system (CFT).}%
\]
The changing of physical picture from a quantum spacetime to a many-body
system leads to the changing of whole story!

Firstly, we consider the 1-th order variability for slow motion.

Along the spatial direction except for the d-th direction, i.e.,
\begin{align}
\mathcal{T}(\delta x^{i})  &  \leftrightarrow \hat{U}^{\mathrm{T}}(\delta
\phi^{i})=e^{i\cdot \delta \phi^{i}\Gamma^{i}},\text{ }\nonumber \\
i  &  =x_{1},x_{2},\text{...},x_{d-1},
\end{align}
where $\delta \phi^{i}=k_{0}\delta x^{i}$ and $\Gamma^{i}$ are the Gamma
matrices obeying Clifford algebra $\{ \Gamma^{i},\Gamma^{i}\}=2\delta^{ij}$.
The result doesn't change.

The system with 1-th order variability along tempo direction indicates a
uniform motion of the group-changing space along $\Gamma^{t}$ direction. After
considering the contribution from mass $m_{R},$ the original "angular
velocity" of the system $\omega_{0}$ turns into
\[
\omega_{0}\rightarrow \omega_{0}^{R}=\omega_{0}+\Delta \omega.
\]
where $\Delta \omega=\frac{m_{R}c^{2}}{\hbar}=\frac{mc^{2}L_{d}}{\hbar l_{0}}$.
Then, we have a \emph{renormalized} 1-th order variability along tempo
direction, i.e.,%
\begin{equation}
\mathcal{T}(\delta t)\leftrightarrow \hat{U}_{R}^{\mathrm{T}}(\delta \phi
^{t})=e^{i\cdot \delta \phi^{t}\Gamma^{t}},
\end{equation}
where $\hat{U}_{R}^{\mathrm{T}}(\delta \phi^{t})$ is renormalized (tempo)
translation operation on Clifford group-changing space and $\delta \phi
^{t}=(\omega_{0}+\Delta \omega)\delta t$.

In addition,\emph{ }1-th order rotation variability becomes
\emph{renormalized},
\begin{equation}
\hat{U}^{\mathrm{R}}\leftrightarrow \hat{R}_{\mathrm{space}}%
\end{equation}
where $\hat{U}^{\mathrm{R}}$ is \textrm{\~{S}\~{O}((d-1)+1)} rotation operator
on Clifford group-changing space $\hat{U}^{\mathrm{R}}\Gamma^{I}\mathbf{(}%
\hat{U}^{\mathrm{R}})^{-1}=\Gamma^{I^{\prime}},$ and $\hat{R}_{\mathrm{space}%
}$ is \textrm{\~{S}\~{O}((d-1)+1)} rotation operator on Cartesian space,
$\hat{R}_{\mathrm{space}}x^{I}\hat{R}_{\mathrm{space}}^{-1}=x^{I^{\prime}}.$
After doing a global composite rotation operation $\hat{U}^{\mathrm{R}}%
\cdot \hat{R}_{\mathrm{space}}$, the uniform ((d-1)+1)-dimensional
\textrm{\~{S}\~{O}((d-1)+1)} physical variants is invariant.

Secondly, we consider the size of elementary particles for slow motion.

Under real K-projection, the zero lattice along tempo direction becomes
renormalized, i.e., the size is changed from Planck time $c/l_{0}=l_{t}$ to
$c/l_{0}\lambda^{-1}=\lambda^{-1}l_{t}$ where the scaling coefficient
$\lambda$ is%
\[
\lambda=1+N_{d}\frac{mc^{2}}{\omega_{0}\hbar}=1+N_{d}Q_{t}.
\]
Consequently, the size of the elementary particle is renormalized, of which
the operators $\Psi^{\dagger}$ or $\Psi$ are replaced by $\Psi_{R}^{\dagger}$
or $\Psi_{R}$. Now, after considering the size renormalization along tempo
direction, the motion charge for the elementary particle of real zero is
forced to be zero, or $m_{R}=0$!

Thirdly, we consider the effective Hamiltonian for slow motion.

The slow motion from boundary fluctuations of the system can be characterized
by fluctuations of the normal direction of the boundary (or $\Gamma^{d}$).
Now, we have a model of \textrm{\~{S}\~{O}((d-1)+1)} quantum rotor field. The
fluctuations of $\Gamma^{d}$\ leads to the fluctuations of Gamma matrices
$\vec{\Gamma}$. As a result, the effective model becomes
\[
\hat{H}_{(d-1)+1}^{\mathrm{slow}}=c\vec{\Gamma}\cdot \vec{p}^{\mathrm{slow}},
\]
of which the fields are Gamma matrices rather than Dirac fermions. For excited
modes, the energy is given by $\Delta E^{\mathrm{slow}}=\pm c\left \vert
\Delta \vec{p}^{\mathrm{slow}}\right \vert .$ The motion charge along given
spatial direction is obtained as $\vec{Q}^{\mathrm{slow}}=\frac{\Delta \vec
{k}^{\mathrm{slow}}}{k_{0}}.$

To obtain motion charge $\vec{Q}^{\mathrm{slow}}$ (the corresponding wave
vector $\Delta \vec{p}^{\mathrm{slow}}$), we carefully analyze its shaking of
$\Gamma^{d}$.

The normal direction of boundary of system is $\Gamma^{d}.$ Under the matrix
representation, the boundary fluctuations are characterized by the shaking of
$\Gamma^{d},$ i.e.,
\begin{align*}
\Gamma^{d}  &  \rightarrow(\Gamma^{d})^{\prime}(x,t)=S(x,t)\Gamma
^{d}(S(x,t))^{-1}\\
&  =\alpha_{d}(x,t)\Gamma^{d}+%
{\displaystyle \sum \nolimits_{\mu \neq d}}
\alpha_{\mu}(x,t)\Gamma^{\mu}%
\end{align*}
where these coefficients $\alpha_{d}(x,t)$ and $\alpha_{\mu}(x,t)$ satisfy
$\alpha_{d}^{2}(x,t)+%
{\displaystyle \sum \nolimits_{\mu \neq d}}
\alpha_{\mu}^{2}(x,t)=1,$ and $\alpha_{d}(x,t)\gg%
{\displaystyle \sum \nolimits_{\mu \neq d}}
\alpha_{\mu}(x,t)$. Now, the system is still at EPs. However, the polarization
direction becomes fluctuating.

Then, we derive the motion charge from boundary fluctuations on the changing rates.

For example, we consider the case of $\alpha_{\mu}(x,t)=\alpha_{x}$ and
$(\Gamma^{d})^{\prime}(x,t)=S\Gamma^{d}S^{-1}=\alpha_{d}\Gamma^{d}+\alpha
_{x}\Gamma^{x}$. Here, $\alpha_{d}$ and $\alpha_{x}$ are constant. The
changing of $\Gamma^{d}$ slightly causes the changing of $\Gamma^{x}$
synchronously, i.e.,%
\begin{align*}
\Gamma^{x}  &  \rightarrow(\Gamma^{x})^{\prime}(x,t)=S\Gamma^{x}S^{-1}\\
&  =\alpha_{d}\Gamma^{x}-\alpha_{x}\Gamma^{d}.
\end{align*}

We return to kinetic representation. Now, the Gamma matrices cannot be
changes. The changings of Gamma matrices are replaced by the changings of
changing rates. The changing rate along x-th direction turns into
\begin{equation}
\alpha_{d}k_{0}\simeq(1-\alpha_{x}^{2}/2)k_{0}.
\end{equation}
As a result, we have
\begin{equation}
\mathcal{T}(\delta x)\leftrightarrow \hat{U}^{\mathrm{T}}(\delta \phi
^{x})=e^{i\cdot \delta \phi^{x}\Gamma^{x}},
\end{equation}
where $\delta \phi^{x}=k_{0}^{x}\delta x$ and $k_{0}^{x}=k_{0}-\alpha_{x}%
^{2}/2$. The motion charge is obtained as
\[
Q_{x}^{\mathrm{slow}}=\frac{\alpha_{x}^{2}}{2k_{0}}.
\]

In general, under boundary fluctuations, the changing rates $k_{0}^{\mu}$
along different directions change and the motion charges are obtained as
\[
\vec{Q}^{\mathrm{slow}}=\frac{\left \vert \vec{\alpha}\right \vert ^{2}}{2k_{0}}%
\]

Finally, we derive the effective Hamiltonian for slow motion.

Under kinetic representation for real zero lattice, the system is set to be
flat. Due to gapless nature of boundary fluctuations of gravitational waves,
the excited slow modes are also gapless. The effective Hamiltonian eventually
becomes%
\[
\hat{H}_{(d-1)+1}^{\mathrm{slow}}=c\vec{\Gamma}\cdot \vec{p}^{\mathrm{slow}},
\]
where $\vec{p}^{\mathrm{slow}}=\hbar \Delta \vec{k}^{\mathrm{slow}}=\hbar
k_{0}\vec{Q}^{\mathrm{slow}}=\frac{\left \vert \vec{\alpha}\right \vert
^{2}\hbar}{2}$.

\subsubsection{Summary}

In this section, we developed a CFT for real zero lattice. In particular, for
slow motion from gravitational wave along boundary, the $\mathrm{\tilde
{S}\tilde{O}(d+1)}$ non-unitary physical variant is regarded as a many-body
system rather than a quantum spacetime. Now, the fluctuation of Gamma matrices
lead to finite motion charge. The low energy physics is described by the
effective Hamiltonian of \textrm{\~{S}\~{O}((d-1)+1)} quantum rotor $\hat
{H}_{(d-1)+1}^{\mathrm{slow}}=c\vec{\Gamma}\cdot \vec{p}^{\mathrm{slow}}.$
According to the effective Hamiltonian $\hat{H}_{(d-1)+1}^{\mathrm{slow}}%
,$\ the excitations becomes gapless that can be regarded as residue processes
of gravitation waves on boundary of the system.\

\subsection{AdS/CFT correspondence}

In above sections, we have developed two theories (AdS or CFT) to characterize
the same $\mathrm{\tilde{S}\tilde{O}(d+1)}$ non-unitary physical variant. The
first theory about AdS comes from the geometry representation for the d+1
dimensional complex zero lattice. The information unit (or elementary
particle) is just the zero of the complex zero lattice. Under the geometry
representation, the theory is similar to that for unitary physical variant.
The second theory about CFT comes from (d-1)+1 dimensional real zero lattice.
Now, there doesn't exist the zero solution along d-th direction with amplitude
changing. The information unit (or the elementary particle) becomes the zero
of real zero lattice. Under kinetic representation, to characterize the slow
motion, we have a CFT on (d-1)+1 dimensional spacetime.

\emph{What's relationship between them? }

The (d-1)+1 dimensional spacetime can be regarded as a dimensional reduction
on d+1 dimensional complex zero lattice by projecting the d-th direction under
a global non-Hermitian polarization effect. The equivalence relation between
the first theory (AdS) on d+1 dimensional complex zero lattice and the second
theory (CFT) on (d-1)+1 dimensional real zero lattice is just AdS/CFT
correspondence. In this section, we explore the underlying mechanism for
AdS/CFT correspondence\cite{ma}. A fundamental principle of AdS/CFT
correspondence\cite{witten} is obtained by the equivalence of both theories:

\textit{AdS/CFT correspondence -- In thermodynamic limit of the }%
(d+1)\textit{-dimensional} \textrm{\~{S}\~{O}(d+1)} \textit{non-unitary
physical variant }$V_{\mathrm{\tilde{S}\tilde{O}(d+1)},d+1}(\Delta \phi^{\mu
},\Delta x^{\mu},k_{0},\omega_{0})$\textit{, the CFT representation is
equivalence to the AdS representation for the boundary of the system.}

In particular, the fast motion in CFT corresponds to the quantum motion of
elementary particles on the boundary of AdS; the slow motion in CFT
corresponds to the quantum motion of gravitation waves on the boundary of AdS.
In the following parts, we provide the AdS/CFT correspondence in detail.

\subsubsection{Correspondence between the spacetime}

Firstly, we consider the correspondence between the zero lattice of CFT and
that of the boundary of AdS.

Because each zero of complex zero lattice on the boundary in AdS corresponds
to each zero of real zero lattice, the number of zero lattice of boundary in
AdS is equal to the number of real zeroes in CFT. As a result, the number of
elementary particles on the boundary of AdS is equal to the number of
elementary particles in CFT.

Next, we consider the correspondence between the variability of CFT and that
of the boundary of AdS.

Now, under complex knot projection, the (d+1)-dimensional \textrm{\~{S}%
\~{O}(d+1)} non-unitary physical variant\textit{ }$V_{\mathrm{\tilde{S}%
\tilde{O}(d+1)},d+1}(\Delta \phi^{\mu},\Delta x^{\mu},k_{0},\omega_{0})$ is
reduced to a complex zero lattice. The boundary of system is regarded as a
sub-system that is outermost side of the (d+1) dimensional complex zero
lattice with a finite width $\Delta \tilde{x}^{d}=l_{0}$ along d-th direction.

For the boundary of uniform non-unitary physical variant, we have 1-th order
variability. Along the spatial direction except for the d-th direction, we
have
\begin{equation}
\mathcal{T}(\delta x^{i})\leftrightarrow \hat{U}^{\mathrm{T}}(\delta \phi
^{i})=e^{i\cdot \delta \phi^{i}\Gamma^{i}},\text{ }i=x_{1},x_{2},\text{...}%
,x_{d-1},
\end{equation}
where $\delta \phi^{i}=k_{0}\delta x^{i}$ and $\Gamma^{i}$ are the Gamma
matrices obeying Clifford algebra $\{ \Gamma^{i},\Gamma^{i}\}=2\delta^{ij}$;
Along tempo direction, the 1-th order variability along time direction is
described by
\begin{equation}
\mathcal{T}(\delta t)\leftrightarrow \hat{U}^{\mathrm{T}}(\delta \phi
^{t})=e^{i\cdot \delta \phi^{t}\Gamma^{t}},
\end{equation}
where $\delta \phi^{t}=(\omega_{0}+\Delta \omega)\delta t$ and $\Gamma^{t}$ is
another Gamma matrix anticommuting with $\Gamma^{i},$ $\{ \Gamma^{i}%
,\Gamma^{t}\}=2\delta^{it}$.

On the other hand, under real knot projection, the (d+1)-dimensional
\textrm{\~{S}\~{O}(d+1)} non-unitary physical variant\textit{ }%
$V_{\mathrm{\tilde{S}\tilde{O}(d+1)},d+1}(\Delta \phi^{\mu},\Delta x^{\mu
},k_{0},\omega_{0})$ is reduced to a real zero lattice. For the uniform case,
we have the same 1-th order variability.

Finally, we point out that the equivalence of variabilities indicates the
equivalence of physical laws of two theories (AdS and CFT).

\subsubsection{Correspondence between the matters}

In this part, we consider the correspondence between the matter in AdS and
that in CFT.

\paragraph{Correspondence between the sizes of elementary particles of AdS and
those of CFT}

Firstly, we consider the sizes of elementary particles of AdS and those of CFT.

On the one hand, for the theory of AdS, the elementary particle is a complex
zero. For a complex zero, the size is $l_{p}$ along an arbitrary direction on
the boundary and $il_{p}$ along $x^{d}$-th direction. Along tempo direction,
the size of the elementary particle is $l_{p}/c$. In addition, along tempo
direction, there exists finite motion charge $Q_{t}$ proportional to mass $m$.

On the other hand, for the theory of CFT, the elementary particle is a real
zero. For a real zero, size is $l_{p}$ along the directions of the boundary
and $iL_{d}=i\frac{L_{d}}{l_{p}}l_{p}$ along $x^{d}$-th direction. Along tempo
direction, the size of the elementary particle is $\frac{2\pi}{\omega_{0}^{R}%
}$.

Therefore, the complex zeroes on the boundary of AdS and the real zeroes of
CFT are almost same each other except for the size along tempo direction.

\paragraph{Correspondence between non-Hermitian effect of AdS and that of CFT}

Secondly, we consider the non-Hermitian effect of AdS and that of CFT.

On the one hand, for the theory of AdS, the non-unitary variability along d-th
direction $\hat{U}(\delta \phi^{d})=e^{k_{0}x^{d}\Gamma^{d}}$ can be considered
as a non-unitary operation on AdS. The relative weight of elementary particles
between the different eigenstates of $\Gamma^{d}$ is changed. For the
elementary particles on the boundary of system, the non-unitary operation
becomes maximum, i.e., $\hat{U}\sim e^{k_{0}L_{d}\Gamma^{d}}$. In the limit of
$L_{d}\rightarrow \infty,$ the amplitude of eigenstates with positive
elgenvalues of $\Gamma^{d}$ diverge. As a result, the degrees of freedom for
the elementary particles becomes fully polarized on the boundary. The quantum
states of elementary particles on the boundary of the system are at EPs. This
non-Hermitian effect of AdS is named spacetime skin effect.

On the other hand, for the theory of CFT, the non-unitary variability along
d-th direction $\hat{U}(\delta \phi^{d})=e^{k_{0}x^{d}\Gamma^{d}}$ can be also
considered as a non-unitary operation and also changes the relative weight of
elementary particles between the different eigenstates of $\Gamma^{d}$. For
the elementary particles of real zero lattice, the global non-unitary
operation is $\hat{U}\sim e^{\frac{L_{d}^{2}}{2l_{0}^{2}}\Gamma^{d}}$. In the
limit of $L_{d}\rightarrow \infty,$ the amplitude of eigenstates with positive
elgenvalues of $\Gamma^{d}$ diverge. As a result, the degrees of freedom for
the real zero becomes fully polarized. The quantum states of elementary
particles of real zero lattice are also at EPs. This non-Hermitian effect of
AdS is named non-Hermitian polarization effect. However, due to the
integrating non-Hermitian effect along $x^{d}$-direction, the non-Hermitian
polarization effect in CFT from $\hat{U}\sim e^{\frac{L_{d}^{2}}{2l_{0}^{2}%
}\Gamma^{d}}$ is more robust than the spacetime skin effect in AdS from
$\hat{U}\sim e^{k_{0}L_{d}\Gamma^{d}}$.

\paragraph{Geometry quantization for the elementary particles}

In this part, we study the geometric property for elementary particles on
boundary of AdS and those in CFT and show their geometry quantization.

In CFT, because we use kinetic representation, the spacetime is always flat.
The elementary particles have trivial geometric property, i.e., the volume of
each elementary particle in CFT is proportional to $l_{0}^{d-1}$. So, we focus
on the case of elementary particles on boundary of AdS.

According to above discussion, there exists spacetime skin effect in AdS. The
quantum states on the boundary of AdS are at EPs under singular non-unitary
similar transformation $\hat{U}\sim e^{k_{0}L_{d}\Gamma^{d}}$ ($L_{d}%
\rightarrow \infty$)$.$ This fact indicates that the boundary of fact AdS
becomes a surface $\mathcal{S}$ with a normal direction denoted by constant
$\Gamma^{d}$. With constant normal direction (or constant $\Gamma^{d}$),
surface $\mathcal{S}$ can be regarded as Geodesic sub-manifold. As a result,
it has minimum area.\ For the 2D case, the surface is denoted by the lines
that is orthogonal to $\Gamma^{d}$, i.e., $\Gamma^{\perp}$. The line along the
$\Gamma^{\perp}$ is Geodesic line has minimum length. For other cases in
higher dimensions, we have similar situation.

\paragraph{Holographic Entanglement entropy}

In this part, we derive the holographic entanglement entropy that was firstly
derived by S. Ryu and T. Takayanagi (RT)\cite{rt}.

To calculate the entanglement entropy in the CFT, we divide the boundary
$\mathcal{S}$ (including time) into two sub-regions, $\mathcal{S}_{A}$ and
$\mathcal{S}_{B}$. The boundary of $\mathcal{S}_{A}$ is $\partial A$. Notice
that $\mathcal{S}_{A}$ is a surface with minimum area.

We then consider quantized geometry of $\mathcal{S}_{A}$ as a sub-system with
$N_{U}$ unit cell. Now, we apply the theory of quantized geometry for quantum
flat spacetime.

On the other hand, the entanglement entropy $S_{A}$ is defined by smearing out
the region $\mathcal{S}_{B}$. The smearing process produces the information
loss for the observer and that should be measured by $\mathcal{S}_{A}$. The
information loss indicates a random distribution of the $N_{U}$ unit cell on
the surface $\mathcal{S}_{A}$. The physical variant becomes stochastic. With
considering the fixed number of unit cell on the surface $\mathcal{S}_{A}$,
the statistics of probability distribution of unit cells is given by
\[
\Omega=\frac{(N_{U})^{N_{U}}}{(N_{U})!}.
\]
In thermodynamic limit $N_{U}\rightarrow \infty$, we have the holographic
entanglement entropy $S_{A}$ to be%
\begin{align*}
S_{A}  &  =\ln \Omega=\ln(\frac{(N_{U})^{N_{U}}}{(N_{U})!})\\
&  \simeq N_{U}+\frac{1}{2}\ln(2\pi N_{U})+...\\
&  \simeq N_{U}.
\end{align*}

Finally, in continuum limit, we derive the RT formula of holographic
entanglement entropy $S_{A}$ in CFT \cite{rt}
\begin{equation}
S_{A}\simeq N_{U}={\frac{\mathrm{Area}(\mathcal{S}_{A})}{l_{0}^{2}}=}%
\frac{\mathrm{Area}(\mathcal{S}_{A})}{4l_{p}^{2}}%
\end{equation}
where the sub-manifold $\mathcal{S}_{A}$ is the $d$-dimensional minimal area
surface in AdS. Its area is denoted by $\mathrm{Area}(\mathcal{S}_{A})$.

\subsubsection{Correspondence between the motion}

In this section, we consider the correspondence of motion in AdS and that in CFT.

According to above discussion, the elementary particles are fully polarized
and their quantum states are at EP. Therefore, the fast motion for elementary
particles are frozen. We focus on the correspondence of slow motion in AdS and
that in CFT. In AdS, slow motion comes from fluctuating of gravitational waves
along boundary of the system; in CFT, the slow motion comes from shaking of
the normal direction $\Gamma^{d}$.

\paragraph{Correspondence between effective Hamiltonians}

Firstly, we consider the correspondence between effective Hamiltonians from
both sides.

In AdS, there are two types of motions: one is about the motion of elementary
particles, the other is about motion of gravitational waves.

The total action in bulk is given by
\begin{align}
S  &  =\mathcal{S}_{\mathrm{4D}}+S_{\mathrm{EH}}\\
&  =\int \sqrt{-g(\tilde{x})}\bar{\Psi}(e_{a}^{\mu}\gamma^{a}\hat{D}_{\mu
}-m)\Psi \text{ }d^{4}\tilde{x}\nonumber \\
&  +\frac{1}{16\pi G}\int \sqrt{-g}\tilde{R}\text{ }d^{4}\tilde{x}.\nonumber
\end{align}
This EH action is reduced to a non-Abelian Chern-Simon action on (d-1)+1
dimensional surface
\begin{equation}
-\frac{1}{16\pi G}(l_{0})^{2}%
{\displaystyle \int}
\epsilon_{abcd}\, \mathrm{Tr}(\Gamma^{z}\omega^{ab}\wedge F^{cd}).
\end{equation}
On the boundary of the system, the effective Hamiltonian for elementary
particles is reduced in to a (d-1)+1 dimensional massive Dirac model,
\[
\hat{H}=\vec{\Gamma}\cdot \Delta \vec{p}+m\Gamma^{t}.
\]

On the other hand, for the CFT, there also are two types of motions: one is
about the fast motion of elementary particles, the other is about slow motion
of gravitational waves.

The fast motion is described by the following effective Hamiltonian%
\[
\hat{H}_{(d-1)+1}^{\mathrm{fast}}=\vec{\Gamma}\cdot \Delta \vec{p}+m_{R}%
\Gamma^{t}%
\]
where $m_{R}=L_{d}/l_{0}m$. According to above Hamiltonian, for the case of
fast motion of an elementary particle, the energy is $\pm \sqrt{\left \vert
\Delta \vec{p}\right \vert ^{2}+m_{R}^{2}}.$ In the thermodynamic limit
$L_{d}\rightarrow \infty$, the mass turns to infinite, i.e., $m_{R}=L_{d}%
/l_{0}m\rightarrow \infty.$ The quantum processes for fast motion of elementary
particles are irrelevant to low energy physics.

The slow motion is described by the following effective Hamiltonian%
\begin{equation}
\hat{H}_{(d-1)+1}^{\mathrm{slow}}=%
{\displaystyle \sum \nolimits_{\mu \neq d,t}}
c\Gamma^{\mu}p^{\mu}.
\end{equation}
For excited elementary particle, the energy is $\pm c\left \vert \Delta \vec
{p}\right \vert .$ The slow motion is the residue effect of the gravitational
waves on the boundary of system. Due to gapless nature of boundary
fluctuations of gravitational waves, the excitation is gapless.

Therefore, we have the following correspondences, i.e.,%
\begin{align*}
&  \text{Motion of elementary particles described }\\
&  \text{by }\mathcal{H}\text{ on boundary of AdS }\\
&  \Leftrightarrow \text{\ Fast motion described by }\mathcal{H}_{(d-1)+1}%
^{\mathrm{fast}}\text{ of CFT}%
\end{align*}

and%
\begin{align*}
&  \text{Residue effect of gravitational waves }\\
&  \text{on boundary of AdS }\\
&  \Leftrightarrow \text{\ Slow motion described by }\mathcal{H}_{(d-1)+1}%
^{\mathrm{slow}}\text{ of CFT.}%
\end{align*}

\paragraph{Correspondence between boundary metric $g_{\mu \nu}$ in AdS and
motion tensor in CFT}

In this part, we consider the correspondence between boundary metric
$g_{\mu \nu}$ in AdS and the motion tensor $M_{\mu \nu}$ in CFT. This
correspondence is really an intrinsic relationship between shape changing of
the boundary in AdS and expansion/contraction of the matter in CFT.

Firstly, we use matrix representation to characterize the boundary
fluctuations of AdS.

From the above discussion, it was known that a quantum spacetime is uniquely
characterized by the spatial/tempo translation operators $\mathcal{T}(\Delta
x^{\mu})\leftrightarrow \hat{U}=e^{i\Gamma^{\mu}k_{0}\Delta x^{\mu}}.$ Under
matrix representation, the shape changings of AdS is characterized by the
changings of matrix network,
\begin{equation}
\mathcal{T}(\Delta x^{\mu})\rightarrow \mathcal{T}((\Delta x^{\mu})^{\prime
})\leftrightarrow \hat{U}=e^{i(\Gamma^{\mu})^{\prime}k_{0}\Delta x^{\mu}}%
\end{equation}
where $k_{0}$ and $\Delta x^{\mu}$\ are constant, $(\Gamma^{\mu})^{\prime}$
become vector field of matrices. We then record its information of curving
spacetime\ by local spatiotemporal operations, $\hat{S}(x)$ that are all
$4\times4$ matrices under matrix representation, i.e.,
\begin{align}
\mathcal{T}((\Delta x^{\mu})^{\prime})  &  \leftrightarrow \hat{U}%
=e^{i\Gamma^{\mu}k_{0}(\Delta x^{\mu})^{\prime}}=e^{i(\Gamma^{\mu})^{\prime
}k_{0}\Delta x^{\mu}}\nonumber \\
&  =\hat{S}(x)\mathcal{T}(\Delta x^{\mu})(\hat{S}(x))^{-1},
\end{align}
\ where the operation $\hat{S}(x)=e^{i\phi_{\mu}(x)\Gamma^{\mu}}$
characterizes the local changes of spatial/tempo translation operators.

When the shape of boundary of AdS at $x^{d}\rightarrow \infty$ changes, the
external normal direction $\Gamma^{d}$ of the surface $\mathcal{S}$\ is no
more fixed and becomes fluctuating. Now, we have
\begin{equation}
\Gamma^{d}\rightarrow(\Gamma^{d})^{\prime}=\hat{S}(x)\Gamma^{d}(\hat
{S}(x))^{-1}.
\end{equation}
We then focus on the case of $d=3$. Within the representation of $\Gamma
^{d}=\Gamma^{z}=\gamma^{0}$, we have
\begin{equation}
(\gamma^{0}(x))^{\prime}=\hat{S}(x)\gamma^{0}(\hat{S}(x))^{-1}=%
{\displaystyle \sum \nolimits_{a}}
\gamma^{a}n^{a}(x),
\end{equation}
where $n^{a}(x)=(n^{1},n^{2},n^{3},n^{0})$ is a unit \textrm{SO(4)}
vector-field in $\gamma$-matrix representation.

Secondly, we show the relationship between matrix representation and geometry representation.

The vierbein fields $e^{a}(x)$ is defined as%
\begin{equation}
e^{a}(x)=dx^{a}(x)\text{ and }e_{\mu}^{a}(x)=\frac{\partial x^{a}(x)}%
{\partial \xi_{\mu}},
\end{equation}
where $\xi_{\mu}$ denotes the coordinate variable of the flat topological
lattice. For the smoothly perturbed vector-fields $n^{a}(x)\ll1$, we have
\begin{align}
\frac{dx^{a}(x)}{l_{0}}  &  =\frac{d\phi^{a}(x)}{2\pi}=\mathrm{tr}[\gamma
^{0}d\gamma^{a}(x)]\nonumber \\
&  =A^{a0}(x),\text{ }a=1,2,3.
\end{align}
Thus, the relationship between $e^{a}(x)$ and $A^{a0}(x)$ is obtained as
\[
e^{a}(x)\equiv l_{0}A^{a0}(x),\text{ }a=1,2,3.
\]
Then, according to the definition of induced metric $g_{\mu \nu}\,=%
{\displaystyle \sum \limits_{a}}
(e_{\mu}^{a}e_{\nu}^{a})$, we have
\begin{align*}
\delta g_{\mu \nu}  &  =l_{0}^{2}%
{\displaystyle \sum \limits_{a}}
(\delta A_{\mu}^{a0}\delta A_{\nu}^{a0})\\
&  =l_{0}^{2}[%
{\displaystyle \sum \limits_{a}}
(\partial_{\mu}n^{a}(x))(\partial_{\nu}n^{a}(x))]\\
&  =l_{0}^{2}((%
{\displaystyle \sum \limits_{a}}
(\partial_{\mu}n^{a}(x)))(%
{\displaystyle \sum \limits_{b}}
(\partial_{\nu}n^{b}(x)))).
\end{align*}

Thirdly, we use kinetic representation to characterize the boundary
fluctuations of AdS that corresponds to the slow motion in CFT.

According to above discussion, the slow motion in CFT is characterized by the
changing of wave vector $\Delta k^{\mu}$, i.e., $k_{0}^{\mu}\rightarrow
k_{0}^{\mu}+\Delta k_{\mathrm{slow}}^{\mu}.$ The motion charge along given
spatial direction $\vec{Q}_{\mathrm{slow}}^{R}=\frac{\Delta \vec{k}%
_{\mathrm{slow}}}{k_{0}}$ characterizes the slow motion. Now, the locally
change of spatial/tempo translation operators comes from the changing of
changing rate $(k^{\mu})^{^{\prime}}$
\begin{align}
\mathcal{T}(\Delta x^{\mu})  &  \rightarrow \mathcal{T}((\Delta x^{\mu
})^{\prime})\leftrightarrow \hat{U}=e^{i(\Gamma^{d})^{\prime}k_{0}\Delta x^{d}%
}\nonumber \\
&  =e^{i\Gamma^{\mu}(k^{\mu})^{^{\prime}}\Delta x^{\mu}}.
\end{align}
Now, on a fixed, flat spacetime, the changing rate $k^{\mu}$ becomes a vector
field that can fluctuate.

To characterize the slow motion, we introduce a new physical quantity, i.e.,
\emph{motion tensor} that is defined by%
\begin{equation}
M_{\mu \nu}=\mathrm{Tr}[(\hat{U}^{-1}\partial_{\mu}\hat{U})\cdot(\hat{U}%
^{-1}\partial_{\nu}\hat{U})]
\end{equation}
where $\hat{U}$ is considered to be an operation of usual many-body system.
Therefore, $M_{\mu \nu}$ characterizes the slow motion of real zero lattice.
The energy-momentum tensor $T_{\mu \nu}$ for fluctuating vector field $k^{\mu}$
of CFT is defined as the changing of motion tensor, i.e.,
\[
T_{\mu \nu}=\delta M_{\mu \nu}=M_{\mu \nu}-M_{0,\mu \nu}%
\]
where $M_{0,\mu \nu}$ is the motion tensor for ground state. So, we have
\begin{align*}
T_{\mu \nu}  &  =(k_{\mu}^{\prime}k_{\nu}^{\prime})-k_{\mu}k_{\nu}\\
&  =(k_{\mu}+\delta k_{\mu}^{\mathrm{slow}})(k_{\nu}+\delta k_{\nu
}^{\mathrm{slow}})-k_{\mu}k_{\nu}%
\end{align*}
where $\delta k_{\mu}^{\mathrm{slow}}$ and $\delta k_{\nu}^{\mathrm{slow}}$
are assumed to be very tiny. As a result, we have%
\begin{align*}
T_{\mu \nu}  &  \simeq k_{\nu}\delta k_{\mu}^{\mathrm{slow}}+k_{\mu}\delta
k_{\nu}^{\mathrm{slow}}\\
&  =k_{0}(e_{\nu}\delta k_{\mu}^{\mathrm{slow}}+e_{\mu}\delta k_{\nu
}^{\mathrm{slow}}).
\end{align*}
For finite wave vector along $\mu$-th direction, we have a finite momentum
$\delta T_{\mu0}=k_{0}\delta k_{\mu}^{\mathrm{slow}}.$

Finally, we derive the correspondence between metric of boundary of AdS
$g_{\mu \nu}$ and the motion tensor $M_{\mu \nu}$ in CFT (not energy-momentum
tensor $T_{\mu \nu}$). From the equation
\begin{align*}
&  \mathrm{Tr}[(\hat{U}^{-1}\partial_{\mu}\hat{U})\cdot(\hat{U}^{-1}%
\partial_{\nu}\hat{U})]\\
&  =\mathrm{Tr}(\partial_{\mu}(\Gamma^{d})^{\prime}\partial_{\nu}(\Gamma
^{d})^{\prime}).\\
&  =\mathrm{Tr}(\partial_{\mu}\gamma^{0}(x)\partial_{\nu}\gamma^{0}(x)).\\
&  =%
{\displaystyle \sum \limits_{a}}
(\partial_{\mu}n^{a}(x)))(%
{\displaystyle \sum \limits_{b}}
(\partial_{\nu}n^{b}(x)).
\end{align*}
we have%
\[
g_{\mu \nu}=l_{0}^{2}%
{\displaystyle \sum \limits_{a}}
(\delta A_{\mu}^{a0}\delta A_{\nu}^{a0})=l_{0}^{2}M_{\mu \nu}.
\]

In the end of this part, we give a brief explanation on above equation.

On the one hand, under geometry representation, the shape changings of
boundary of AdS is characterized by changings of coordinates $(\Delta x^{\mu
})^{\prime}$. As a result, we have
\[
g_{\mu \nu}=l_{0}^{2}\mathrm{Tr}[(\hat{U}^{-1}\partial_{\mu}\hat{U})\cdot
(\hat{U}^{-1}\partial_{\nu}\hat{U})]
\]
where $\hat{U}=e^{i\Gamma^{\mu}k_{0}(\Delta x^{\mu})^{\prime}}$ with fixed
$\Gamma^{\mu}$ and $k_{0}$. On the other hand, under kinetic representation,
the shape changings of boundary of AdS is characterized by mapping changings
between group-changing space and Cartesian space,
\begin{equation}
\mathcal{T}(\Delta x^{\mu})\rightarrow \mathcal{T}((\Delta x^{\mu})^{\prime
})\leftrightarrow \hat{U}=e^{i\Gamma^{\mu}(k_{0}^{\mu})^{\prime}\Delta x^{\mu}}%
\end{equation}
where $\Gamma^{\mu}$ and $\Delta x^{\mu}$\ are constant, $(k_{0}^{\mu
})^{\prime}$ become a vector field that can change. As a result, we have the
\[
M_{\mu \nu}=\mathrm{Tr}[(\hat{U}^{-1}\partial_{\mu}\hat{U})\cdot(\hat{U}%
^{-1}\partial_{\nu}\hat{U})]
\]
where $\hat{U}=e^{i\Gamma^{\mu}(k_{0}^{\mu})^{\prime}\Delta x^{\mu}}$ with
fixed $\Gamma^{\mu}$ and $\Delta x^{\mu}$. Combining the two together, we have
the correspondence between fluctuation of boundary metric $g_{\mu \nu}$ of AdS
and motion tensor of slow motion in CFT ($g_{\mu \nu}=l_{0}^{2}M_{\mu \nu}$).

\paragraph{Correspondence between particle's mass in AdS and anomalous
dimension of correlation functions in CFT}

In traditional quantum field theory, the correlation functions are important
functions that describe how microscopic variables, such as spin and density,
at different positions. It was known that according the result of AdS/CFT
correspondence, the two-point correlation function $G_{E}(x-y)=\,
\langle \mathcal{O}(x)\, \mathcal{O}(y)\rangle$ can be derived by using the
formula of classical gravity in bulk. As a result, people can easily obtain
the correlation functions on the boundary of AdS. For the correlation
functions of massive Dirac particles, a dimension/mass relation is obtained
as
\begin{equation}
\Delta={\frac{d}{2}}+\nu
\end{equation}
where the anomalous dimension $\nu=|mL_{d}|$ is determined by the particle's
mass $m$ in AdS.

\emph{How to understand the dimension/mass relation?} Let give an explanation
on it.

In this part, we consider the CFT as a quantum many-body system with finite
density of elementary particles. Based on the quantum many-body system,
fluctuations from Gamma matrix $\Gamma^{d}(d)$ lead to fluctuations of
energy-momentum tensor.

Because the energy for fast motion diverges in thermodynamic limit, it is
irrelevant to dimension/mass relation. We focus on the slow motion that is the
residue effect of the gravitational waves on the boundary of AdS and becomes
relevant to dimension/mass relation.

It was known that each real zero in CFT corresponds to $N_{d}=L_{d}/l_{0}$
complex zero in AdS. The phase changing rate $\tilde{\omega}_{0}$ along tempo
direction of elementary particles is different from that $\omega_{0}$ in AdS,
\[
\tilde{\omega}_{0}=\lambda \omega_{0}=(1+N_{d}\frac{mc^{2}}{\omega_{0}\hbar
})\omega_{0}.
\]
The changing of the changing rate along tempo direction lead to a changing of
definition of the elementary particles in CFT. The situation is same the
quasi-particles in 1D Luttinger liquid. In particular, an elementary particle
in CFT obtains an additional phase changing along tempo direction
\[
\Delta \varphi_{CFT}=(1+N_{d}\frac{mc^{2}}{\omega_{0}\hbar})\Delta \varphi_{AdS}%
\]
where $\Delta \varphi_{CFT}\ $and $\Delta \varphi_{AdS}$ are the phase changing
in CFT and that in AdS, respectively. The ratio between the total phase
changing in CFT and that in AdS is $(1+N_{d}\frac{mc^{2}}{\omega_{0}\hbar})$.

By using the approach in 1D Luttinger liquid, from redefining elementary
particle $\psi \rightarrow \psi^{(1+N_{d}\frac{mc^{2}}{\omega_{0}\hbar})},$ we
can calculate the correlation function in CFT side. As a result, there exists
an anomalous dimension
\begin{align*}
\nu &  =N_{d}\frac{mc^{2}}{\omega_{0}\hbar}\\
&  =N_{d}\frac{l_{0}}{2\pi}m=\frac{L_{d}m}{2\pi}.
\end{align*}
Because the size of $L_{d}$ is the perimeter of the whole AdS $L_{d}=2\pi L$,
we have%
\[
\nu=\frac{L_{d}m}{2\pi}=mL.
\]
The result is consistent to that from conjecture of AdS/CFT correspondence.

In summary, we have a correspondence between particle's mass of AdS and
anomalous dimension of correlation functions in CFT. In particular, the
underlying mechanism of this correspondence is the re-definition the
elementary particles in both sides. The anomalous dimension plays the role of
the ratio of the tempo changing rate $\omega_{0}$ of AdS and that of CFT
$k_{0}^{t}=\tilde{\omega}_{0}$, i.e., $1+\nu=\frac{\tilde{\omega}_{0}}%
{\omega_{0}}$. In addition, $\frac{\tilde{\omega}_{0}}{\omega_{0}}$ is also
the ratio of the particle's volume of AdS and that of CFT.

\paragraph{Correspondence between gauge field in AdS and current in CFT}

In this part, we give a brief discussion on the correspondence between gauge
field of AdS and current in CFT. In general, Abelian/non-Abelian gauge fields
characterize the dynamics of global/relative loop currents on spacetime. We
take \textrm{U}$^{\mathrm{em}}$\textrm{(1)} gauge field as an example to show
the correspondence.

On dS, the \textrm{U}$^{\mathrm{em}}$\textrm{(1)} gauge field $A_{\mu}$
characterizes the phase changings on spacetime and the strength of gauge field
$F_{\mu \nu}$\ characterizes the changing of loop current along $\mathcal{C}$
on spacetime. We then consider \textrm{U}$^{\mathrm{em}}$\textrm{(1)} gauge
field on AdS. The situation is quite different from that on dS.

For a loop from boundary to bulk, the loop current is always reduced to
current on boundary. See the illustration in Fig.12. As shown in Fig.12, for
the closed loop, there are four line segment, AB, BC, CD, and DA,
respectively. AB is on the boundary, CD is on the opposite side, BC and CD are
all along $x^{d}$-th direction.

On the line segments of BC and CD, because this is the direction with only
amplitude changing, the phase of gauge fields cannot be changed. Therefore, On
the loops of BC and DA, there doesn't exist finite gauge fields that
characterizes phase changings. On the line segment of CD, its weight of
quantum states becomes infinite small. Therefore, the contribution for all
physical processes can be negligible. On the line segment of AB, we have usual
phase changings that is current. As a result, the loop current around ABCD for
the gauge fields is reduced to the current on line segment AB that is on the
boundary of the AdS.

\begin{figure}[ptb]
\includegraphics[clip,width=0.75\textwidth]{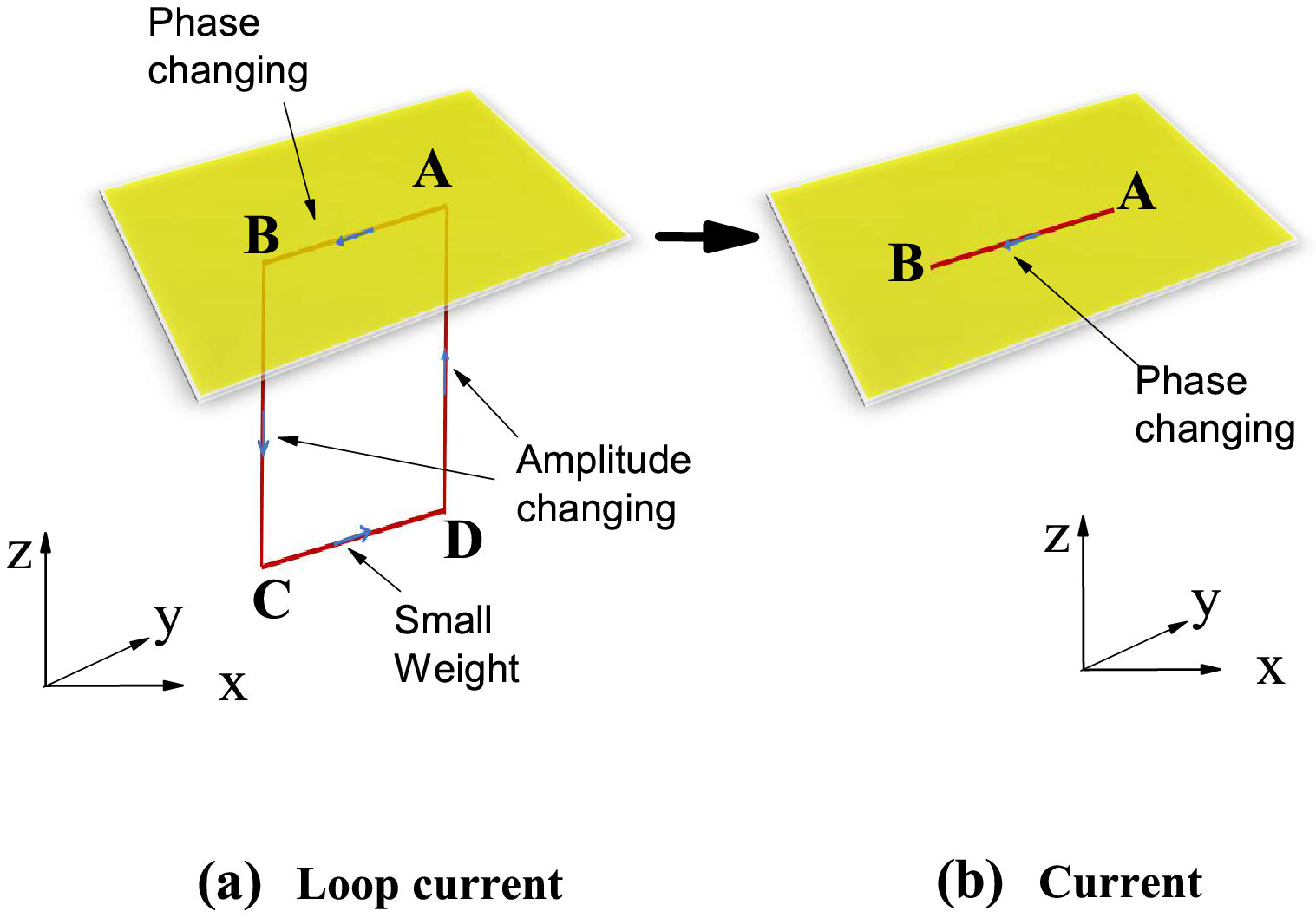}\caption{(Color online)
An illustration of the reduction of loop current around ABCD in AdS to current
from A to B in CFT }%
\end{figure}

Finally, we have
\[
\text{Loop currents in AdS }\leftrightarrow \text{ Currents in CFT. }%
\]

\subsection{Gravity/N-gauge equivalence}

In above sections, we had discussed the AdS/CFT correspondence. We found that
the fast motion in CFT for elementary particles corresponds to the quantum
motion of elementary particles on the boundary of AdS, and the slow motion in
CFT for elementary particles corresponds to the quantum motion of gravitation
waves along boundary of AdS. We may ask a question: \emph{does there exist an
equivalent relationship between AdS and CFT including the bulk effect of AdS
rather than only considering boundary effect}? In this section, we update the
AdS/CFT correspondence to gravity/N-gauge equivalence. Here, N-gauge indicates
"non-Hermitian gauge theory".

\textit{Gravity/N-gauge equivalence -- For the }(d+1)\textit{-dimensional}
\textrm{\~{S}\~{O}(d+1)} \textit{non-unitary physical variant }%
$V_{\mathrm{\tilde{S}\tilde{O}(d+1)},d+1}(\Delta \phi^{\mu},\Delta x^{\mu
},k_{0},\omega_{0})$\textit{, the representation of ((d-1)+1)-dimensional
non-Hermitian gauge theory (NGT) on flat spacetime is equivalence to the
representation of (d+1)-dimensional AdS.}

Here, the NGT representation is a non-Hermitian gauge theory that corresponds
to the bulk dynamics of AdS. When we reduce the NGT to the unitary physical
processes of the system, AdS/NGT equivalence is reduced to usual AdS/CFT
correspondence between the theory for boundary of AdS and CFT. Let us give
detailed discussion on this issue in the following parts.

\subsubsection{Non-Hermitian gauge theory for AdS}

A key point of Gravity/N-gauge equivalence is non-Hermitian gauge field.

It was known that the elementary particle of real zero corresponds to
$L_{d}/l_{0}$ zeroes of complex zero lattice, that is the lattice number along
d-th direction with imaginary lattice number. According to this fact, to
derive a complete theory based on real zero lattice, we consider the zero of
real zero lattice to be a composite zero with $L_{d}/l_{0}$ internal, level-2
zeroes. Therefore, an effective gauge fields emerge under Kaluza-Klein
compactification \cite{kk}.

In additional, along $x^{d}$-th direction, we have non-Hermitian polarization
effect. The non-unitary variability along d-th direction $\hat{U}(\delta
\phi^{d})=e^{k_{0}x^{d}\Gamma^{d}}$ can be considered as a global non-unitary
operation $U\sim e^{\frac{L_{d}^{2}}{2l_{0}^{2}}\Gamma^{d}}$ for real zeroes
that changes the relative weight of elementary particles. In the limit of
$L_{d}\rightarrow \infty,$ the amplitude of eigenstates with positive
elgenvalues of $\Gamma^{d}$ diverge while the amplitude of eigenstates with
negative elgenvalues of $\Gamma^{d}$ turns to zero. Therefore, the degrees of
freedom for the real zero becomes fully polarized on the boundary and for each
real zeroes, quantum states are at EPs. From point view of level-2 zeroes, the
non-Hermitian polarization effect becomes non-Hermitian skin effect on a 1D
chain under open boundary condition\cite{skin}. Therefore, an effective gauge
fields become non-Hermitian. The situation can be regarded as a non-Hermitian
generalization of Kaluza-Klein compactification.

It is known that under dimensional reduction in usual Kaluza-Klein theory, the
changings of fifth dimensional space with periodic boundary condition turns
into the \textrm{U}$^{\mathrm{em}}$\textrm{(1)} gauge fields. In this section,
we point out that under dimensional reduction in non-Hermitian Kaluza-Klein
theory, along the fifth dimension with open boundary condition, the changing
of fifth space turns into a non-unitary \textrm{U(0,1)}$\times$\textrm{SU(0,}%
$\left \vert \lambda^{\lbrack12]}\right \vert $\textrm{)} ($\left \vert
\lambda^{\lbrack12]}\right \vert =L_{d}/l_{0}$) gauge fields. See below discussion.

\paragraph{Non-unitary 2-th order Physical variant}

A usual (d+1)-dimensional 2-th order\textit{ }\textrm{\~{S}\~{O}(d+1)}
physical variant is a higher-order mapping between\textit{ }$\mathrm{C}%
_{\mathrm{\tilde{U}}^{[2]}\mathrm{(1)}}^{[2]},$\textit{ }\textrm{\~{S}%
\~{O}(d+1)} Clifford group-changing space\textit{ }$\mathrm{C}_{\mathrm{\tilde
{S}\tilde{O}(d+1)},d+1}^{[1]}$\textit{\ }and a rigid spacetime\textit{
}$\mathrm{C}_{d+1}$, of which the ratio between the changing rates of two
levels is $\lambda^{\lbrack12]}=\frac{\delta \phi^{\lbrack2]}}{\delta
\phi_{\mathrm{global}}^{[1]}}$\cite{kou1}. Under K-projection, each of lattice
site of level-1 zero lattice corresponds to a level-2 zero lattice with
$\lambda^{\lbrack12]}$ level-2 zero.

In this part, we generalize the concept of 2-th order\textit{ }\textrm{\~{S}%
\~{O}(d+1)} physical variant to a non-unitary one, of which the level-2
group-changing space is non-unitary and level-1 group-changing space is
unitary. Therefore, the original 1-th order \textrm{\~{S}\~{O}(d+1)}
non-unitary physical variant turns into a ((d-1)+1)-dimensional 2-th order
\textrm{\~{S}\~{O}((d-1)+1)} non-unitary physical variants. Now, we have a
higher-order mapping between $\mathrm{C}_{\mathrm{\tilde{U}}_{\mathrm{open}%
}^{[2]}\mathrm{(0,1)}}^{[2]}$, \textrm{(d-1)+1} Clifford group-changing space
$C_{\mathrm{(d-1)+1},(d-1)+1}^{[1]}$\ and a rigid spacetime $\mathrm{C}%
_{(d-1)+1},$\textit{ }i.e.,%
\begin{align}
V_{\mathrm{\tilde{U}}_{\mathrm{open}}^{[2]}\mathrm{(0,1)},\mathrm{\tilde
{S}\tilde{O}}^{[1]}\mathrm{((d-1)+1)},(d-1)+1}^{[2]}  &  :\mathrm{C}%
_{\mathrm{\tilde{U}}_{\mathrm{open}}^{[2]}\mathrm{(0,1)}}^{[2]}\nonumber \\
&  \Longleftrightarrow \mathrm{C}_{\mathrm{\tilde{S}\tilde{O}}^{[1]}%
\mathrm{((d-1)+1)},(d-1)+1}^{[1]}\nonumber \\
&  \Longleftrightarrow \mathrm{C}_{(d-1)+1}%
\end{align}
\textit{ }where $\Leftrightarrow$\ between $\mathrm{C}_{\mathrm{\tilde{U}%
}_{\mathrm{open}}^{[2]}\mathrm{(0,1)}}^{[2]}$ and\textit{ }$\mathrm{C}%
_{\mathrm{\tilde{S}\tilde{O}}^{[1]}\mathrm{((d-1)+1)},(d-1)+1}^{[1]}$\textit{
}denotes an ordered mapping under fixed ratio between the changing rates
$\lambda^{\lbrack12]}=\frac{\delta \phi^{\lbrack2]}}{\delta \phi
_{\mathrm{global}}^{[1]}}=iL_{d}/l_{0}$, $\Leftrightarrow$\ between\textit{
}$\mathrm{C}_{\mathrm{\tilde{S}\tilde{O}}^{[1]}\mathrm{(d+1)},d+1}^{[1]}%
$\textit{ }and $\mathrm{C}_{d+1}$\textit{ }denotes an ordered mapping under
fixed changing rate of integer multiple $k_{0}$ or $\omega_{0}^{[1]},$\ and
$\mu$ labels the spatial direction. In particular, $\mathrm{C}_{\mathrm{\tilde
{U}}_{\mathrm{open}}^{[2]}\mathrm{(0,1)}}^{[2]}$ is non-unitary\textit{
}$\mathrm{\tilde{U}}_{\mathrm{open}}^{[2]}\mathrm{(0,1)}$\textit{
}group-changing space under open boundary condition. We have set light speed
$c=1$.

\paragraph{2-th order variability}

For this non-unitary 2-th order\textit{ }\textrm{\~{S}\~{O}}((d-1)+1) physical
variant $V_{\mathrm{\tilde{U}}_{\mathrm{open}}^{[2]}\mathrm{(0,1)}%
,\mathrm{\tilde{S}\tilde{O}}^{[1]}\mathrm{((d-1)+1)},(d-1)+1}^{[2]}$, there
exists the 2-th order variability, i.e.,
\begin{align}
\mathcal{T}(\delta x^{\mu})  &  \leftrightarrow \hat{U}^{[1]}((\delta
\phi^{\lbrack1]\mu}))=\exp(i(T^{\mu}\delta \phi^{\lbrack1]\mu}))\\
&  =\exp(i(T^{\mu}k_{0}^{\mu}\delta x^{\mu})),\nonumber
\end{align}
and%
\begin{equation}
\hat{U}^{[1]}(\delta \phi_{\mathrm{global}}^{[1]})\leftrightarrow \hat{U}%
^{[2]}(\delta \phi^{\lbrack2]})=\exp(i\lambda^{\lbrack12]}\delta \phi
_{\mathrm{global}}^{[1]}\Gamma^{d})
\end{equation}
where the ratio $\lambda^{\lbrack12]}=\frac{\delta \phi^{\lbrack2]}}{\delta
\phi_{\mathrm{global}}^{[1]}}=iL_{d}/l_{0}$ becomes an imaginary number. In
particular, the non-unitary Abelian group $\mathrm{\tilde{U}}_{\mathrm{open}%
}^{[2]}\mathrm{(0,1)}$ describes internal non-unitary operations of
$\Gamma^{d}$ that doesn't commutate with the unitary operations along spatial directions.

Under K-projection, each of lattice site of level-1 zero lattice corresponds
to a level-2 zero lattice with $\left \vert \lambda^{\lbrack12]}\right \vert
=L_{d}/l_{0}$ level-2 imaginary zero.

\paragraph{Matter -- classification with complex topological invariant}

Next, we discuss the matter for 2-th order non-unitary physical variants
$V_{\mathrm{\tilde{U}}_{\mathrm{open}}^{[2]}\mathrm{(0,1)},\mathrm{\tilde
{S}\tilde{O}}^{[1]}\mathrm{((d-1)+1)},(d-1)+1}^{[2]}$ with imaginary changing
rate $\lambda^{\lbrack12]}=iL_{d}/l_{0}.$

Matter corresponds to globally expand or contract the group-changing space
$\mathrm{C}_{\mathrm{\tilde{S}\tilde{O}}^{[1]}\mathrm{(d+1)},d+1}^{[1]}$ or
the group-changing space $\mathrm{\tilde{U}}_{\mathrm{open}}^{[2]}%
\mathrm{(0,1)}$ with changing their corresponding sizes. Therefore, an object
is classified by real two integer numbers $n^{[1]}$ and $n^{[2]}$: the number
of level-1 real zeroes $n^{[1]}$ (a real number) and that of level-2 imaginary
zeroes $n^{[2]}$ (a real number), respectively. We point out that $n^{[1]}$
denotes the number of elementary particles of real zeroes. We then classify
the types of elementary particles by $n^{[2]}$ that denotes color charge. For
a level-2 zero, the electric charge is $\frac{1}{\lambda^{\lbrack12]}}$ that
is an imaginary number. We label different types of elementary particles
different level-2 imaginary zeroes. After for $n^{[2]}$ level-2 zeroes, the
color charge is $n^{[2]}$ and the electric charge is $\mathrm{e}_{0}%
=in^{[2]}/\left \vert \lambda^{\lbrack12]}\right \vert $.

So, there are $\left \vert \lambda^{\lbrack12]}\right \vert $ types of
elementary particles: one is electron with one level-1 zero, $\left \vert
\lambda^{\lbrack12]}\right \vert $ level-2 zeroes and unit electric charge,
quark-1 with one level-1 zero, one level-2 zeroes and $1/\lambda^{\lbrack12]}$
electric charge, quark-2 with one level-1 zero, two level-2 zeroes and
$2/\lambda^{\lbrack12]}$ electric charge, quark-3 with one level-1 zero, three
level-2 zeroes and $3/\lambda^{\lbrack12]}$ electric charge, ...
quark-($\left \vert \lambda^{\lbrack12]}\right \vert -1$) with one level-1 zero,
$(\left \vert \lambda^{\lbrack12]}\right \vert -1)$ level-2 zeroes and
$(\lambda^{\lbrack12]}-1)/\lambda^{\lbrack12]}$ electric charge.

\paragraph{Quantum states and symmetry for motion of level-2 zeroes}

Firstly, we discuss the quantum states for the level-2 zero of a level-1 zero.
We call these quantum states to be internal quantum states of the elementary particle.

We "generate" an extra ($i$-th) level-2 non-unitary group-changing element
$\delta \varphi_{I^{[2]},I^{[1]},i}^{[2]}$ on the position $\varphi
_{I^{[2]},I^{[1]},i}^{[2]}$ of $I^{[2]}$-th level-2 zero and on the position
$\varphi_{I^{[1]},i}^{[1]}$ of $I^{[1]}$-th level-1 zero, i.e., $\hat
{U}(\delta \varphi_{I^{[2]},I^{[1]},i}^{[2]}(\varphi_{I^{[1]},i}^{[1]}%
))=e^{i((\delta \varphi_{I^{[2]},I^{[1]},i}^{[2]})\cdot \hat{K})}$ and $\hat
{K}=-i\frac{d}{d\varphi^{\lbrack2]}}.$\ Here, the i-th infinitesimal
non-unitary group-changing operation $\hat{U}(\delta \varphi_{i}^{[2]})$
generates a level-2 non-unitary group-changing element on $I$-th level-1 zero
with real phase $\varphi_{I^{[1]},i}^{[1]}$ and imaginary phase $\varphi
_{I^{[2]},I^{[1]},i}^{[2]}$.\ Therefore, the "wave function" for a system with
$n^{[1]}$ level-1 zeroes and $n^{[2]}$ level-2 zeroes is described by the
information of level-2 imaginary phase $\delta \varphi_{I^{[2]},I^{[1]}}^{[2]}$
and level-1 real phase $\varphi_{I^{[1]}}^{[1]}$. Here, $I^{[2]}$ and
$I^{[1]}$ label the level-2 zero and level-1 zero, respectively.

The motion of level-2 non-unitary group-changing space comes from its local
expansion and contraction on different level-1 zeroes.

If there exist $N^{[2]}$ level-2 zeroes, the total size of all level-2
group-changing elements is $\pm iN^{[2]}\pi$, i.e.,
\begin{equation}%
{\displaystyle \sum}
\delta \varphi_{I^{[2]},I^{[1]}}^{[2]}=\pm iN^{[2]}\pi.
\end{equation}
The local expansion and contraction of level-2 non-unitary group-changing
space changes level-2 imaginary phase $\delta \varphi_{I^{[2]},I^{[1]}}^{[2]}$
and real phase $\varphi_{I^{[1]}}^{[1]}$ on I-th level-1 zero, or changing the
position of lattice sites of level-2 group-changing space on I-th level-1
zero. Therefore, the motion for level-2 group-changing space is defined by the
changings of the configuration of level-2 phase $\delta \varphi_{I^{[2]}%
,I^{[1]}}^{[2]}$ and real level-2 phase $\varphi_{I^{[1]}}^{[1]}$ for
different level-1 zeroes. Because there are total $\lambda^{\lbrack12]}$
lattice sites for level-2 zeroes of a level-1 zero, we have $\left \vert
\lambda^{\lbrack12]}\right \vert $\ level-2 phases $\delta \varphi
_{I^{[2]},I^{[1]}}^{[2]}$ for a level-2 group-changing element of a level-1
zero. In particular, we point out that the $\left \vert \lambda^{\lbrack
12]}\right \vert $\ level-2 phases $\delta \varphi_{I^{[2]},I^{[1]}}^{[2]}$ are
all imaginary. So, we split them into two groups -- a global level-2 imaginary
phase $\delta \varphi_{\mathrm{global},I^{[1]}}^{[2]}$ and $\left \vert
\lambda^{\lbrack12]}\right \vert -1$ relative imaginary phases. In sometime, we
may use the Abbreviation $\varphi^{\lbrack2]}$ to denote $\varphi
_{\mathrm{global}}^{[2]}$ without "$\mathrm{global}$"$\mathrm{.}$

To characterize these $\left \vert \lambda^{\lbrack12]}\right \vert $\ level-2
imaginary phases $\delta \varphi_{I^{[2]},I^{[1]}}^{[2]}$, we must define
$\left \vert \lambda^{\lbrack12]}\right \vert $ references, $\delta
\varphi_{0,I^{[2]},I^{[1]}}^{[2]}.$ For the global imaginary phase
$\delta \varphi_{\mathrm{global},I^{[1]}}^{[2]}$, the reference is
$\delta \varphi_{0,\mathrm{global},I^{[1]}}^{[2]}$. It was known that according
to level-2 variability $\hat{U}_{I}^{[1]}(\delta \phi_{I,\mathrm{global}}%
^{[1]})\leftrightarrow \hat{U}_{I}^{[2]}(\delta \phi_{I}^{[2]}),$ the changing
of reference $\delta \varphi_{0,\mathrm{global},I^{[1]}}^{[2]}$ for global
imaginary phase is same to the changing of the reference of level-1 global
phase $\varphi_{0,I}^{[1]}$%
\[
\delta \varphi_{0,I}^{[1]}=\left \vert \lambda^{\lbrack12]}\right \vert
\delta \varphi_{0,\mathrm{global},I^{[1]}}^{[2]}%
\]
where $\delta \varphi_{0,I}^{[1]}=((\varphi_{0,I}^{[1]})^{\prime}-\varphi
_{0,I}^{[1]})$ and $\delta \varphi_{0,\mathrm{global},I^{[1]}}^{[2]}%
=((\varphi_{0,\mathrm{global},I^{[1]}}^{[2]})^{\prime}-\varphi
_{0,\mathrm{global},I^{[1]}}^{[2]}).$ This becomes the local \textrm{U(0,1)}
non-unitary gauge transformation.

On the other hand, there are $\lambda^{\lbrack12]}-1$ references for relative
imaginary phases. To set these references for $\lambda^{\lbrack12]}-1$
relative imaginary phases, we define the reference state based on the
representation of compact\textrm{ SU(0,N)} group.

We consider a level-2 zero to be an internal level-2 elementary particle, and
label the sites of the level-2 zero lattice by $1$, $2$, ..., $\left \vert
\lambda^{\lbrack12]}\right \vert .$ Now, an extra level-2 zero on $I^{[1]}$-th
level-1 zero is characterized a $\left \vert \lambda^{\lbrack12]}\right \vert
$-component "field", i.e.,
\begin{equation}
\left(
\begin{array}
[c]{c}%
\left \vert \psi_{1^{[2]},I^{[1]}}^{[2]}\right \rangle \\
\left \vert \psi_{2^{[2]},I^{[1]}}^{[2]}\right \rangle \\
...\\
\left \vert \psi_{(\left \vert \lambda^{\lbrack12]}\right \vert )^{[2]},I^{[1]}%
}^{[2]}\right \rangle
\end{array}
\right)  .
\end{equation}
Here, $\left \vert \psi_{I^{[2]},I^{[1]}}^{[2]}\right \rangle $ denotes the
quantum state of the level-2 zero on the $I^{[2]}$-th lattice site of level-2
zero lattice of $I^{[2]}$-th level-1 zero. Because quantum states of the
internal imaginary zero on different sites of the level-2 imaginary
zero-lattice are orthogonal, i.e.,%
\begin{equation}
\langle \psi_{J^{[2]},I^{[1]}}^{[2]}\left \vert \psi_{I^{[2]},I^{[1]}}%
^{[2]}\right \rangle =\delta_{J^{[2]}I^{[2]}},
\end{equation}
$\left \vert \psi_{1^{[2]},I^{[1]}}^{[2]}\right \rangle ,$ $\left \vert
\psi_{2^{[2]},I^{[1]}}^{[2]}\right \rangle ,$ ..., $\left \vert \psi_{\left \vert
\lambda^{\lbrack12]}\right \vert ,I^{[1]}}^{[2]}\right \rangle $ make up a
complete basis.

Therefore, in general, we can re-label the corresponding states of level-2
elementary particles by a new one $\left \vert \psi_{1^{[2]},I^{[1]}}%
^{[2]}\right \rangle ^{\prime},$ $\left \vert \psi_{2^{[2]},I^{[1]}}%
^{[2]}\right \rangle ^{\prime},$ ..., $\left \vert \psi_{\left \vert
\lambda^{\lbrack12]}\right \vert ,I^{[1]}}^{[2]}\right \rangle ^{\prime}%
$.\textbf{\ }The relationship between the two basis is
\begin{equation}
\left(
\begin{array}
[c]{c}%
\left \vert \psi_{1^{[2]},I^{[1]}}^{[2]}\right \rangle ^{\prime}\\
\left \vert \psi_{2^{[2]},I^{[1]}}^{[2]}\right \rangle ^{\prime}\\
...\\
\left \vert \psi_{(\left \vert \lambda^{\lbrack12]}\right \vert )^{[2]},I^{[1]}%
}^{[2]}\right \rangle ^{\prime}%
\end{array}
\right)  =\hat{U}_{\mathrm{SU(0,N)}}^{[2]}\left(  x,t\right)  \left(
\begin{array}
[c]{c}%
\left \vert \psi_{1^{[2]},I^{[1]}}^{[2]}\right \rangle \\
\left \vert \psi_{2^{[2]},I^{[1]}}^{[2]}\right \rangle \\
...\\
\left \vert \psi_{(\left \vert \lambda^{\lbrack12]}\right \vert )^{[2]},I^{[1]}%
}^{[2]}\right \rangle
\end{array}
\right)
\end{equation}
where $\hat{U}_{\mathrm{SU(0,N)}}^{[2]}\left(  x,t\right)  =e^{i\Theta \left(
\vec{x},t\right)  }$ is the matrix of the representation of $\mathrm{SU(0,N)}$
group. The imaginary number $\Theta \left(  x,t\right)  =i\sum_{a=1}%
^{\mathrm{N}^{2}-1}\theta^{a}\left(  x,t\right)  \tau^{a}$ and real numbers
$\theta^{a}$ are a set of $\mathrm{N}^{2}-1$ constant parameters, and
$\tau^{a}$ are $\mathrm{N}^{2}-1$ $\mathrm{N}\times \mathrm{N}$ matrices
representing the $\mathrm{N}^{2}-1$ generators of the Lie algebra of
$\mathrm{SU(N)}$. The global imaginary phase of $\left(
\begin{array}
[c]{c}%
\left \vert \psi_{1^{[2]},I^{[1]}}^{[2]}\right \rangle \\
\left \vert \psi_{2^{[2]},I^{[1]}}^{[2]}\right \rangle \\
...\\
\left \vert \psi_{(\lambda^{\lbrack12]})^{[2]},I^{[1]}}^{[2]}\right \rangle
\end{array}
\right)  $ is $\delta \varphi_{I^{[1]}}^{[2]},\ $of which the reference is
$\delta \varphi_{\mathrm{global},I^{[1]}}^{[2]}.$\ The reference of relative
imaginary phase can be defined by a fixed group element of \textrm{SU(0,N)}
group, i.e., $U_{\mathrm{SU(0,N)}}^{[2]}\left(  x,t\right)  =e^{i\Theta
_{0}\left(  \vec{x},t\right)  }$ where $\Theta_{0}\left(  \vec{x},t\right)  $
is an imaginary number.

Finally, we write down the effective Hamiltonian for a level-2 zero.

We define generation operator $(c_{I^{[2]}}^{[2]})^{\dagger}$ of a level-2
zero at the site $I^{[2]}$ by $(c_{i^{I[2]}}^{[2]})^{\dagger}\left \vert
0\right \rangle =\left \vert I^{[2]}\right \rangle $. Here, $I^{[2]}$ is an
imaginary integer number. The hopping term between two nearest neighbor sites
$I^{[2]}$ and $J^{[2]}$ on topological lattice becomes
\begin{equation}
\mathcal{H}_{\left \{  i,j\right \}  }^{[2]}=\mathcal{J}^{[2]}(c_{I^{[2]}}%
^{[2]})^{\dagger}\mathbf{T}_{\left \{  I^{[2]},J^{[2]}\right \}  }c_{J^{[2]}%
}^{[2]}(t)
\end{equation}
where $\mathbf{T}_{\left \{  I^{[2]},J^{[2]}\right \}  }$ is the transfer matrix
between two nearest neighbor sites $I^{[2]}$ and $J^{[2]}$ and $c_{I^{[2]}%
}^{[2]}(t)$ is the annihilation operator of elementary particle at the
imaginary site $I^{[2]}$. On the other hand, we consider the terms from
on-site potential%
\begin{equation}
\mathcal{H}^{[2]}=%
{\displaystyle \sum \limits_{I^{[2]}}}
\mathcal{H}_{I^{[2]}}^{[2]}=V%
{\displaystyle \sum \limits_{I^{[2]}}}
(c_{I^{[2]}}^{[2]})^{\dagger}c_{I^{[2]}}^{[2]}+h.c..
\end{equation}
As a result, the total Hamiltonian is
\[
\mathcal{H}^{[2]}=\mathcal{J}^{[2]}%
{\displaystyle \sum \limits_{\{I^{[2]},J^{[2]}\}}}
(c_{I^{[2]}}^{[2]})^{\dagger}\mathbf{T}_{\left \{  I^{[2]},J^{[2]}\right \}
}c_{J^{[2]}}^{[2]}(t)+%
{\displaystyle \sum \limits_{I^{[2]}}}
\mathcal{H}_{I^{[2]}}^{[2]}.
\]
In general, we have $V\equiv0$.

Then, we discuss the case from non-Hermitian Kaluza-Klein compactification.\

When there exist an excited level-1 elementary particle, an extra level-2 zero
(or a parton) must appear synchronously. The quantum states for the level-2
zero is described by imaginary momenta along the fifth direction under open
boundary condition,
\begin{equation}
p_{5}=i\frac{2\pi \hbar}{L}n,\text{ }n\in \mathbb{Z}.
\end{equation}
In particular, due to the imaginary momentum $p_{5}=i\frac{2\pi \hbar}{L}n,$
under open boundary condition, we have \emph{non-Hermitian skin effect}, i.e.,
the level-2 zeroes localized on the boundary of the level-1 zero. The result
is consistent to that from spacetime skin effect. In addition, for the quantum
states with imaginary momentum $p_{5}$ under open boundary condition, the
energy doesn't change due to the effect of non-Hermitian similar
transformation,
\[
E\equiv0!
\]

\paragraph{Quantum states and symmetry for motion of level-1 zeroes}

In this part, we discuss the quantum states and symmetry for motion of
a\ level-1 zero (or an elementary particle) with $n^{[2]}$ level-1 zeroes (or
$n^{[2]}$ partons).

Firstly, we consider the case of $n^{[2]}=\left \vert \lambda^{\lbrack
12]}\right \vert .$ We use one component field $\left \vert \psi_{I^{[1]}}%
^{[1]}\right \rangle $ to characterize it. This is just the case of real zero
of CFT that has been discussed in above sections. Now, the elementary
particles only couple the \textrm{U(0,1)} non-unitary gauge field. The
electric charge is $\mathrm{e}=il_{p}\hbar c$ with $n^{[2]}=\left \vert
\lambda^{\lbrack12]}\right \vert .$ In the following parts, we will show that
the fluctuations of \textrm{U(0,1)} non-unitary gauge field plays the role of
the residue gravitational waves along boundary of the system in CFT. Or,
\textrm{U(0,1)} non-unitary gauge field characterizes the slow motion.

Secondly, we consider the case of $n^{[2]}=1$. We use another $\left \vert
\lambda^{\lbrack12]}\right \vert $-component field
\begin{equation}
\left(
\begin{array}
[c]{c}%
\left \vert \psi_{1^{[2]},I^{[1]}}^{[1]}\right \rangle \\
\left \vert \psi_{2^{[2]},I^{[1]}}^{[1]}\right \rangle \\
...\\
\left \vert \psi_{(\left \vert \lambda^{\lbrack12]}\right \vert )^{[2]},I^{[1]}%
}^{[1]}\right \rangle
\end{array}
\right)  .
\end{equation}
to describe the quantum states of level-1 zero (or an elementary particle).
The global phase of it is $\varphi_{I}^{[1]}$ that is changed synchronously
with the global imaginary phase of $\varphi_{\mathrm{global,}I}^{[2]}.$
$\left \vert \psi_{I^{[2]},I^{[1]}}^{[1]}\right \rangle $ denotes the quantum
state of its level-2 imaginary zero on the $I^{[2]}$-th lattice site of
level-2 imaginary zero lattice inside $I^{[1]}$-th level-1 zero. Therefore,
changing of relative imaginary phase of level-2 elementary particle leads to
corresponding changing of relative imaginary phase of level-1 zero, i.e,
\begin{equation}
\left(
\begin{array}
[c]{c}%
\left \vert \psi_{1^{[2]},I^{[1]}}^{[2]}\right \rangle ^{\prime}\\
\left \vert \psi_{2^{[2]},I^{[1]}}^{[2]}\right \rangle ^{\prime}\\
...\\
\left \vert \psi_{(\left \vert \lambda^{\lbrack12]}\right \vert )^{[2]},I^{[1]}%
}^{[2]}\right \rangle ^{\prime}%
\end{array}
\right)  =\hat{U}_{\mathrm{SU(0,N)}}^{[2]}\left(  x,t\right)  \left(
\begin{array}
[c]{c}%
\left \vert \psi_{1^{[2]},I^{[1]}}^{[2]}\right \rangle \\
\left \vert \psi_{2^{[2]},I^{[1]}}^{[2]}\right \rangle \\
...\\
\left \vert \psi_{(\left \vert \lambda^{\lbrack12]}\right \vert )^{[2]},I^{[1]}%
}^{[2]}\right \rangle
\end{array}
\right)
\end{equation}
and
\begin{equation}
\left(
\begin{array}
[c]{c}%
\left \vert \psi_{1^{[2]},I^{[1]}}^{[1]}\right \rangle ^{\prime}\\
\left \vert \psi_{2^{[2]},I^{[1]}}^{[1]}\right \rangle ^{\prime}\\
...\\
\left \vert \psi_{(\left \vert \lambda^{\lbrack12]}\right \vert )^{[2]},I^{[1]}%
}^{[1]}\right \rangle ^{\prime}%
\end{array}
\right)  =\hat{U}_{\mathrm{SU(0,N)}}^{[1]}\left(  x,t\right)  \left(
\begin{array}
[c]{c}%
\left \vert \psi_{1^{[2]},I^{[1]}}^{[1]}\right \rangle \\
\left \vert \psi_{2^{[2]},I^{[1]}}^{[1]}\right \rangle \\
...\\
\left \vert \psi_{(\left \vert \lambda^{\lbrack12]}\right \vert )^{[2]},I^{[1]}%
}^{[1]}\right \rangle
\end{array}
\right)  .
\end{equation}
This provides a non-Abelian variability constraint, i.e.,%
\[
\hat{U}_{\mathrm{SU(0,N)}}^{[1]}\left(  x,t\right)  \equiv \hat{U}%
_{\mathrm{SU(0,N)}}^{[2]}\left(  x,t\right)  .
\]
This non-Abelian variability constraint plays important role in non-Abelian
non-unitary gauge symmetry for Yang-Mills field. In addition, the elementary
particles couple \textrm{U(0,1)} non-unitary gauge field. The electric charge
is $\mathrm{e}=\mathrm{e}_{0}=i\frac{l_{p}}{{L}}\hbar c.$

Thirdly, we consider the case of $n^{[2]}>1$. There are $C_{\lambda
^{\lbrack12]}}^{n^{[2]}}$ internal quantum states. As a result, we use a
$C_{\lambda^{\lbrack12]}}^{n^{[2]}}$-component field to characterize it. The
elementary particles couple both \textrm{U(0,1) }non-unitary gauge field and
non-unitary $\mathrm{SU(0,N)}$ Yang-Mills gauge field.

\paragraph{Variability constraints}

There are two types of variability constraints - one is global variability
constraint, the other is relative variability constraint.

On the one hand, we discuss the global variability constraint. According to
above discussion, due to level-2 variability, the changings of references
$\varphi_{0,\mathrm{global,}I}^{[1]}$ and $\varphi_{0,\mathrm{global,}I}%
^{[2]}$ for the two group-changing spaces must be synchronously,
$\delta \varphi_{0,\mathrm{global,}I}^{[1]}=i\delta \varphi_{0,\mathrm{global,}%
I}^{[2]}$.

On the other hand, we discuss the relative variability constraint. By trapping
level-2 zeroes, there exist different types of elementary particles. Due to
effect of state nesting, the internal states of the level-1 zero (or
elementary particle) are defined by the quantum states of the internal level-2
zeroes. The wave functions of quantum states of the level-1 zero are functions
of the wave functions of quantum states of the level-2 zeroes, i.e.,%
\[
\psi_{1^{[2]},I^{[1]}}^{[1]}(\psi_{1^{[2]},I^{[1]}}^{[2]}).
\]
According to the condition of state nesting, we have relative variability
constraint. When the quantum states of level-2 zeroes change, there exists
corresponding operation of \textrm{SU(0,N)} group on $\psi_{1^{[2]},I^{[1]}%
}^{[2]},$%
\[
\psi_{1^{[2]},I^{[1]}}^{[2]}\rightarrow(\psi_{1^{[2]},I^{[1]}}^{[2]})^{\prime
}=U_{\mathrm{SU(0,N)}}^{[2]}\left(  x,t\right)  \psi_{1^{[2]},I^{[1]}}^{[2]}.
\]
Because the internal states of level-1 zero is determined by the quantum
states of level-2 zero, the changings of quantum states of level-2 zero lead
to the changings of internal states of level-1 zero, i.e.,
\begin{align*}
\psi_{1^{[2]},I^{[1]}}^{[1]}(\psi_{1^{[2]},I^{[1]}}^{[2]})  &  \rightarrow
(\psi_{1^{[2]},I^{[1]}}^{[1]}(\psi_{1^{[2]},I^{[1]}}^{[2]}))^{\prime}\\
&  =\psi_{1^{[2]},I^{[1]}}^{[1]}((\psi_{1^{[2]},I^{[1]}}^{[2]})^{\prime})\\
&  =\psi_{1^{[2]},I^{[1]}}^{[1]}(U_{\mathrm{SU(0,N)}}^{[2]}\left(  x,t\right)
\psi_{1^{[2]},I^{[1]}}^{[2]})\\
&  =U_{\mathrm{SU(0,N)}}^{[2]}\left(  x,t\right)  \psi_{1^{[2]},I^{[1]}}%
^{[1]}(\psi_{1^{[2]},I^{[1]}}^{[2]})\\
&  =U_{\mathrm{SU(0,N)}}^{[1]}\left(  x,t\right)  \psi_{1^{[2]},I^{[1]}}%
^{[1]}(\psi_{1^{[2]},I^{[1]}}^{[2]}).
\end{align*}
Therefore, we have the relative variability constraint,
\[
U_{\mathrm{SU(0,N)}}^{[2]}\left(  x,t\right)  \equiv U_{\mathrm{SU(0,N)}%
}^{[1]}\left(  x,t\right)  .
\]

\paragraph{Local non-unitary gauge symmetries}

In this part, we discuss the local non-unitary gauge symmetry in detail.

There are two types of local non-unitary gauge symmetries, one is Abelian,
non-unitary gauge symmetry for global motion of level-2 zeroes of a level-1
zero, the other is non-Abelian, non-unitary gauge symmetry for relative motion
of level-2 zeroes of a level-1 zero.

On the one hand, the level-2 invariance is reduced to non-unitary gauge
invariant under the operations of non-unitary \textrm{U}$_{\mathrm{global}%
}^{[2]}$\textrm{(0,1)} group and translation invariant $\mathcal{T}^{[2]}$ on
the level-2 zero lattice with $\left \vert \lambda^{\lbrack12]}\right \vert $
lattice sites, i.e.,
\[
\tilde{U}^{[2]}\rightarrow \hat{U}_{\mathrm{U}_{\mathrm{global}}\mathrm{(0,1)}%
}^{[2]}\otimes \mathcal{T}^{[2]}.
\]
$\hat{U}_{\mathrm{U}_{\mathrm{global}}\mathrm{(0,1)}}^{[2]}$ indicates a
non-Hermitian\ similar transformation. So, it doesn't change the energy of
given states.

On the other hand, we discuss the non-Abelian non-unitary gauge symmetry for
relative motion. The symmetry for relative motion is correspondingly reduced
\[
\mathcal{T}^{[2]}\rightarrow \hat{U}_{\mathrm{SU(0,}\left \vert \mathrm{\lambda
^{\lbrack12]}}\right \vert \mathrm{)}}^{[2]}.
\]
Due to non-Hermitian skin effect, such an invariant under the operation of
\textrm{SU}$^{[2]}$\textrm{(0,}$\left \vert \mathrm{\lambda^{\lbrack12]}%
}\right \vert $\textrm{)} group means that the system with different quantum
states have same energy. Due to the relative variability constraint from state
nesting effect $\hat{U}_{\mathrm{SU(0,\left \vert \mathrm{\lambda^{\lbrack12]}%
}\right \vert )}}^{[2]}\left(  x,t\right)  \equiv \hat{U}%
_{\mathrm{SU(0,\left \vert \mathrm{\lambda^{\lbrack12]}}\right \vert )}}%
^{[1]}\left(  x,t\right)  ,$ we have a local $\mathrm{SU(0,\left \vert
\mathrm{\lambda^{\lbrack12]}}\right \vert )}$ symmetry that denotes the
indistinguishable internal quantum states of the elementary particle,
\begin{align*}
&  \hat{U}_{\mathrm{SU(0,\left \vert \mathrm{\lambda^{\lbrack12]}}\right \vert
)}}(x,t)\\
&  =\hat{U}_{\mathrm{SU(0,\left \vert \mathrm{\lambda^{\lbrack12]}}\right \vert
)}}^{[2]}\left(  x,t\right)  \equiv \hat{U}_{\mathrm{SU(0,\left \vert
\mathrm{\lambda^{\lbrack12]}}\right \vert )}}^{[1]}\left(  x,t\right)  .
\end{align*}
For simplicity, we have
\begin{align}
\psi_{1^{[2]},I^{[1]}}^{[1]}(\psi_{1^{[2]},I^{[1]}}^{[2]})  &  \rightarrow
(\psi_{1^{[2]},I^{[1]}}^{[1]}(\psi_{1^{[2]},I^{[1]}}^{[2]}))^{\prime
}\nonumber \\
&  =\hat{U}_{\mathrm{SU(0,\left \vert \mathrm{\lambda^{\lbrack12]}}\right \vert
)}}(x,t)\psi_{1^{[2]},I^{[1]}}^{[1]}(\psi_{1^{[2]},I^{[1]}}^{[2]}).
\end{align}
The symmetry of the different internal zeroes leads to the symmetry of the
internal quantum states of elementary particles.

In summary, we have%
\begin{align*}
&  \text{Level-2 variability with }\left \vert \lambda^{\lbrack12]}\right \vert
>1\\
&  \rightarrow \text{ \textrm{U(0,1)} local non-unitary gauge symmetry }\\
&  \text{+ \textrm{SU(0,}}\left \vert \mathrm{\lambda^{\lbrack12]}}\right \vert
\text{\textrm{)} local non-unitary gauge symmetry, }%
\end{align*}%
\begin{align*}
&  \text{\textrm{U(0,1)} local non-unitary gauge symmetry }\\
&  \text{=}\text{ Level-2 variability under global variability constraint. }%
\end{align*}
and
\begin{align*}
&  \text{Local \textrm{SU(0,}}\left \vert \mathrm{\lambda^{\lbrack12]}%
}\right \vert \text{\textrm{)} gauge symmetry }\\
&  \text{=}\text{ Two global SU(0,}\left \vert \mathrm{\lambda^{\lbrack12]}%
}\right \vert \text{) group with relative variability }\\
&  \text{constraint due to state nesting effect. }%
\end{align*}

\paragraph{\textrm{U(0,1)}$\times$\textrm{SU(0,}$\left \vert \mathrm{\lambda
^{\lbrack12]}}\right \vert $\textrm{)\ }non-unitary gauge fields and their
effective Hamiltonians}

According to above discussion, for a level-2 zero, the momenta are imaginary.
Therefore, under open boundary condition, due to the existence of
non-Hermitian skin effect inside a level-1 zero, the energies for different
states are always zero and become degenerate! This leads to non-Hermitian
gauge symmetry! On the other hand, different quantum states with different
imaginary momenta have different complex electric charges $\mathrm{e}%
=n\mathrm{e}_{0}=inc\hbar^{\lbrack2]}=i\frac{l_{p}}{{L}}n\hbar c$. The result
can be straightforwardly obtained by the approach similar to that in 2-th
order unitary physical variant. Then, we have \textrm{U(0,1)}$\times
$\textrm{SU(0,}$\left \vert \lambda^{\lbrack12]}\right \vert $\textrm{)}
($\lambda^{\lbrack12]}=iL_{d}/l_{0}$) non-unitary gauge fields that
characterize the dynamics of $n^{[2]}$ level-2 zeroes inside an elementary
particle of real zero. In particular, the gauge charges of \textrm{U(0,1)} and
\textrm{SU(0,}$\left \vert \lambda^{\lbrack12]}\right \vert $\textrm{)}
non-unitary gauge fields are all imaginary.

In this section, we provide a detailed discussion on \textrm{U(0,1)}$\times
$\textrm{SU(0,}$\left \vert \mathrm{\lambda^{\lbrack12]}}\right \vert
$\textrm{)} gauge field that characterizes the quantum fluctuations of level-2
non-unitary group-changing space.

The \textrm{U(0,1)} gauge field comes from the non-uniform distribution of
level-2 non-unitary group-changing elements on level-1 zero lattice.

As a result, the vector field $A_{I,I^{\prime}}$ that characterizes the local
position perturbation of effective level-2 zero-lattice plays the role of
\textrm{U(0,1)} gauge field. To illustrate the local \textrm{U(0,1)} gauge
symmetry, we do a local \textrm{U(0,1)} gauge transformation $\hat
{U}_{I,\mathrm{U(0,1)}}=e^{i\Delta \varphi_{0,I}}$ where $\Delta \varphi_{0,I}$
is imaginary. Under the local \textrm{U(0,1)} non-unitary gauge
transformation, we have
\begin{equation}
\psi_{I}\rightarrow \psi_{I}^{\prime}=\hat{U}_{I,\mathrm{U(0,1)}}\psi
_{I}=e^{i\mathrm{e}_{0}\Delta \varphi_{0,I}}\psi_{I},
\end{equation}
and
\begin{align}
\mathrm{e}_{0}A_{I,I^{\prime}}  &  \rightarrow \mathrm{e}_{0}A_{I,I^{\prime}%
}^{\prime}\nonumber \\
&  =\mathrm{e}_{0}A_{I,I^{\prime}}-(\Delta \varphi_{0,I}-\Delta \varphi
_{0,I^{\prime}}).
\end{align}

We then introduce non-unitary loop current $\Delta \Phi_{\left \langle
IJKL\right \rangle }^{[2]}$ on the plaquette of $IJKL$ lattice sites, i.e.,%
\begin{align}
\Delta \Phi_{\left \langle IJKL\right \rangle }^{[2]}  &  =A_{Ij}-A_{KL}%
\nonumber \\
&  =\frac{1}{\mathrm{e}_{0}}(-\Delta \varphi_{I}^{[2]}+\Delta \varphi_{J}%
^{[2]}+\Delta \varphi_{K}^{[2]}-\Delta \varphi_{L}^{[2]}).
\end{align}
$\Delta \Phi_{\left \langle ijkl\right \rangle }^{[2]}$\ is an imaginary number.
The quantum state of dynamic fluctuations (locally expanding and contracting)
for level-2 non-unitary group-changing space are described by
\[
\{ \Delta \Phi_{\left \langle IJKL\right \rangle }^{[2]},\text{ }\left \langle
IJKL\right \rangle \in \mathrm{All}\}.
\]
For the imaginary loop current, there doesn't exist usual action term,
$\mathcal{S}\propto%
{\displaystyle \sum \limits_{\left \langle IJKL\right \rangle }}
\cos(\Delta \mathcal{\Phi}_{\left \langle IJKL\right \rangle }^{[2]}).$

In continuum limit, we have $\hat{U}_{\mathrm{U(0,1)}}(I)\rightarrow \hat
{U}_{\mathrm{U(0,1)}}(x,t)$, $A_{I,I^{\prime}}\rightarrow A_{\mu}(x).$ The
Abelian gauge symmetry is represented by%
\begin{equation}
\psi^{\prime}\rightarrow \hat{U}(x,t)\psi
\end{equation}
and
\begin{align}
A_{\mu}(x,t)  &  \rightarrow A_{\mu}(x,t)+i\left(  \partial_{\mu}\hat
{U}_{\mathrm{U(0,1)}}(x,t)\right)  \left(  \hat{U}_{\mathrm{U(0,1)}%
}(x,t)\right)  ^{-1}\nonumber \\
&  =A_{\mu}(x,t)+\frac{1}{\mathrm{e}_{0}}\partial_{\mu}\varphi(x,t).
\end{align}
In continuum limit, $\Delta \Phi_{\left \langle IJKL\right \rangle }^{[2]}$ is
reduced to the strength of non-unitary gauge field $F_{\mu \nu}.$

The \textrm{SU(0,}$\left \vert \lambda^{\lbrack12]}\right \vert $\textrm{)}
non-unitary gauge field also comes from the non-uniform distribution of
level-2 non-unitary group-changing elements. Now, we use the vector field of
$\left \vert \lambda^{\lbrack12]}\right \vert \times \left \vert \lambda
^{\lbrack12]}\right \vert $ matrix $\mathcal{A}_{I,I^{\prime}}=%
{\displaystyle \sum \limits_{a}}
A_{I,I^{\prime}}T^{a}$ to characterize the local position perturbation of
level-2 zero-lattice Here, $T^{a}$ is generate of \textrm{SU(}$\left \vert
\lambda^{\lbrack12]}\right \vert $\textrm{)} group along a-th direction. The
vector field $\mathcal{A}_{I,I^{\prime}}$ plays the role of \textrm{SU(0,}%
$\left \vert \lambda^{\lbrack12]}\right \vert $\textrm{)} gauge field.

To illustrate the local \textrm{SU(0,}$\left \vert \lambda^{\lbrack
12]}\right \vert $\textrm{)} gauge symmetry, we do a local \textrm{SU(0,}%
$\left \vert \lambda^{\lbrack12]}\right \vert $\textrm{)} gauge transformation
$\hat{U}_{I,\mathrm{SU(N)}}=\exp(i%
{\displaystyle \sum \limits_{a}}
\Delta \varphi_{0,I}^{a}T^{a})$ via changing the initial imaginary phases along
$a$-th direction%
\begin{equation}
\varphi_{0,I}^{a}\rightarrow(\Delta \varphi_{0,I}^{a})^{\prime}=\varphi
_{0,I}^{a}+\Delta \varphi_{0,I}^{a}.
\end{equation}
Here, $\varphi_{0,I}^{a}$\ is an imaginary number. Under the local
\textrm{SU(0,}$\left \vert \lambda^{\lbrack12]}\right \vert $\textrm{)} gauge
transformation $\hat{U}_{I,\mathrm{SU(N)}}$, we have
\begin{align}
\mathcal{A}_{I,I^{\prime}}  &  \rightarrow \mathcal{A}_{I,I^{\prime}}^{\prime
}\nonumber \\
&  =\mathcal{A}_{I,I^{\prime}}-\frac{1}{g}%
{\displaystyle \sum \limits_{a}}
(\Delta \varphi_{0,I}^{a}-\Delta \varphi_{0,I^{\prime}}^{a})T^{a}%
\end{align}
where $g$ denotes the coupling constant.

We denote them by colored imaginary loop current $\Delta \Phi_{\left \langle
IJKL\right \rangle }^{[2]}$ on the plaquette of $\left \langle IJKL\right \rangle
$ lattice site, i.e.,%
\begin{align}
\Delta \mathcal{\Phi}_{\left \langle IJKL\right \rangle }^{[2]}  &  =%
{\displaystyle \sum \limits_{a}}
(\Delta \mathcal{\Phi}_{\left \langle IJKL\right \rangle }^{a,[2]}T^{a}%
)=\mathcal{A}_{IJ}-\mathcal{A}_{KL}\nonumber \\
&  =\frac{1}{g}%
{\displaystyle \sum \limits_{a}}
(-\Delta \varphi_{I}^{a,[2]}+\Delta \varphi_{J}^{a,[2]}\nonumber \\
&  +\Delta \varphi_{K}^{a,[2]}-\Delta \varphi_{L}^{a,[2]})T^{a}.
\end{align}
The quantum state for dynamic fluctuations (locally expanding and contracting)
for level-2 non-unitary group-changing space are described by colored loop
current on the plaquette $\left \langle IJKL\right \rangle $,
\[
\{ \Delta \mathcal{\Phi}_{\left \langle IJKL\right \rangle }^{[2]}(I),\text{
}\left \langle IJKL\right \rangle \in \mathrm{All}\}.
\]
For the imaginary loop current, there doesn't exist usual action term
$\mathcal{S}\propto$\textrm{Tr}$(%
{\displaystyle \sum \limits_{a}}
(%
{\displaystyle \sum \limits_{\left \langle IJKL\right \rangle }}
T^{a}\cos(\Delta \mathcal{\Phi}_{\left \langle IJKL\right \rangle }^{a,[2]}))).$

Because there doesn't exist usual action term $\mathcal{S},$ we derive the new formula.

For the case of $n^{[2]}=\left \vert \lambda^{\lbrack12]}\right \vert $, in
continuum limit, we have the effective Hamiltonian as
\[
\hat{H}_{(d-1)+1}=\vec{\Gamma}\cdot(\mathrm{e}\vec{A}_{U(0,1)})+\Gamma
^{t}(\mathrm{e}A_{t,U(0,1)})
\]
The boundary fluctuations of gravitational waves turn into that of non-unitary
$\mathrm{U(0,1)}$ gauge fields. Now, the Gamma matrices become fixed. The slow
motion is characterized by quantum fluctuations of non-unitary gauge fields
$A_{\mu,U(0,1)}$. The finite non-unitary gauge fields $\vec{A}_{U(0,1)}$\ and
$A_{t,U(0,1)}$ give contribution to motion charge. The situation is quite
different from the case of usual unitary gauge fields.

For the case of $n^{[2]}=1,$ we must consider the fluctuations of non-unitary
Yang-Mills gauge field. Now, elementary particles have $\left \vert
\lambda^{\lbrack12]}\right \vert $ components and couples non-unitary
Yang-Mills gauge fields $\mathcal{A}_{\mu}^{a}$. The slow motion comes from
the bulk fluctuations of gravitational waves. In continuum limit, we have the
effective Hamiltonian as
\begin{align*}
\hat{H}_{(d-1)+1}  &  =\vec{\Gamma}\cdot(\mathrm{e}\vec{A}_{U(0,1)}%
+g\mathcal{\vec{A}})\\
&  +\Gamma^{t}(\mathrm{e}A_{t,U(0,1)}+g\mathcal{A}_{t}).
\end{align*}

\paragraph{Absent of quark confinement}

In the end, we give a comment on the effect of quark confinement in the
non-unitary Yang-Mills gauge fields.

The generation/annihilation of an elementary particle (real zero) leads to
contraction/expansion $\pi$-phase changing of Clifford group-changing space
along an arbitrary direction. This leads to the longitudinal changings of
quantum spacetime, the total volume will increase or decrease. For usual
Yang-Mills fields, there exists induced particle number $\frac{n^{[2]}%
}{\lambda^{\lbrack12]}}$ of level-2 zeroes for quarks. The induced particle
number $\frac{n^{[2]}}{\lambda^{\lbrack12]}}$ provides extra 3-volume on
quantum spacetime that disturbs the spacetime. This leads to the well known
effect of quark confinement. An question is whether there exists confinement
for non-Hermitian Yang-Mills field. Our answer is "no".

For non-Hermitian Yang-Mills field, there also exists induced particle number
$\frac{n^{[2]}}{\lambda^{\lbrack12]}}.$ However, this value $\frac{n^{[2]}%
}{\lambda^{\lbrack12]}}$ is imaginary rather than a real one. The imaginary
particle number changes the size along $x^{d}$-th direction rather than the
size of the boundary of the system. As a result, the dynamical processes from
\textrm{SU(0,}$\left \vert \lambda^{\lbrack12]}\right \vert $\textrm{)}
non-unitary gauge fields slightly change the amplitude of eigenstates of
$\Gamma^{d}$ for the given real zero,
\begin{align*}
\hat{U}  &  =%
{\displaystyle \prod \limits_{x^{d}}}
\hat{U}(\delta \phi^{d})\\
&  \rightarrow \hat{U}^{\prime}=%
{\displaystyle \prod \limits_{x^{d}}}
\hat{U}^{\prime}(\delta \phi^{d}).
\end{align*}
In thermodynamic limit, the real zero is almost fully polarized. The slightly
changing of the amplitude of different eigenstates of $\Gamma^{d}$ can be
always neglected, i.e.,
\[
\hat{U}\simeq \hat{U}^{\prime}.
\]

In summary, there doesn't exist the effect of quark confinement.

\subsubsection{Equivalence between the spacetime}

In this part, we consider the equivalence between the two representations (AdS
and NGT).

Firstly, we consider the equivalence between the zero lattice of NGT and that
of the AdS.

Under complex K-projection, the (d+1)-dimensional \textrm{\~{S}\~{O}(d+1)}
non-unitary physical variant\textit{ }$V_{\mathrm{\tilde{S}\tilde{O}%
(d+1)},d+1}(\Delta \phi^{\mu},\Delta x^{\mu},k_{0},\omega_{0})$ is reduced to a
complex zero lattice. Under real K-projection, the (d+1)-dimensional
\textrm{\~{S}\~{O}(d+1)} non-unitary physical variant\textit{ }%
$V_{\mathrm{\tilde{S}\tilde{O}(d+1)},d+1}(\Delta \phi^{\mu},\Delta x^{\mu
},k_{0},\omega_{0})$ is reduced to a (d-1)+1 dimensional real zero lattice.
After considering the composite nature, each real zero has $\left \vert
\lambda^{\lbrack12]}\right \vert $ level-2 internal zeroes. Therefore, for the
uniform system, under the two representations, the number of complex zeroes
$N$ is \emph{equal} to the product of the number of real zeroes $N^{F}$ and
the number of level-2 zeroes for same real zero $\left \vert \lambda
^{\lbrack12]}\right \vert $, i.e.,
\[
N=N^{F}\cdot \left \vert \lambda^{\lbrack12]}\right \vert .
\]

Next, we consider the correspondence between the variability of NGT and that
of the boundary of AdS.

Along the spatial direction except for the d-th direction, for both
representations, we have same 1-th order variability
\begin{equation}
\mathcal{T}(\delta x^{i})\leftrightarrow \hat{U}^{\mathrm{T}}(\delta \phi
^{i})=e^{i\cdot \delta \phi^{i}\Gamma^{i}},\text{ }i=x_{1},x_{2},\text{...}%
,\tilde{x}_{d},
\end{equation}
where $\delta \phi^{i}=k_{0}\delta x^{i}$ and $\Gamma^{i}$ are the Gamma
matrices obeying Clifford algebra $\{ \Gamma^{i},\Gamma^{i}\}=2\delta^{ij}$;
Along tempo direction, the 1-th order variability along time direction is
described by
\begin{equation}
\mathcal{T}(\delta t)\leftrightarrow \hat{U}^{\mathrm{T}}(\delta \phi
^{t})=e^{i\cdot \delta \phi^{t}\Gamma^{t}},
\end{equation}
where $\delta \phi^{t}=(\omega_{0}+\Delta \omega)\delta t$ and $\Gamma^{t}$ is
another Gamma matrix anticommuting with $\Gamma^{i},$ $\{ \Gamma^{i}%
,\Gamma^{t}\}=2\delta^{it}$. In particular, we have $\tilde{x}^{d}=ix^{d}$.

Along the d-th spatial direction, for both representations, we also have same
1-th order variability.

On the side of AdS, along the d-th direction, we have a 1-th order non-unitary
spatial variability
\[
\mathcal{T}(\delta x^{d})\leftrightarrow \hat{U}(\delta \phi^{\mu}%
)=e^{i\cdot \delta \phi^{d}\Gamma^{d}}=e^{k_{0}x^{d}\Gamma^{d}}.
\]

On the side of NGT, we have a non-unitary 2-th order\textit{ }\textrm{\~{S}%
\~{O}}((d-1)+1) physical variant $V_{\mathrm{\tilde{U}}_{\mathrm{open}}%
^{[2]}\mathrm{(0,1)},\mathrm{\tilde{S}\tilde{O}}^{[1]}\mathrm{((d-1)+1)}%
,(d-1)+1}^{[2]}$. For this non-unitary 2-th order\textit{ }\textrm{\~{S}%
\~{O}}((d-1)+1) physical variant $V_{\mathrm{\tilde{U}}_{\mathrm{open}}%
^{[2]}\mathrm{(0,1)},\mathrm{\tilde{S}\tilde{O}}^{[1]}\mathrm{((d-1)+1)}%
,(d-1)+1}^{[2]}$, there exists the 2-th order variability, i.e.,
\begin{align}
\mathcal{T}(\delta x^{\mu})  &  \leftrightarrow \hat{U}^{[1]}((\delta
\phi^{\lbrack1]\mu}))\\
&  =\exp(i(T^{\mu}\delta \phi^{\lbrack1]\mu}))\\
&  =\exp(i(T^{\mu}k_{0}^{\mu}\delta x^{\mu})),\nonumber
\end{align}
and%
\begin{align}
\tilde{U}^{[1]}(\delta \phi_{\mathrm{global}}^{[1]})  &  \leftrightarrow \hat
{U}^{[2]}(\delta \phi^{\lbrack2]})\nonumber \\
&  =\exp(i\lambda^{\lbrack12]}\delta \phi_{\mathrm{global}}^{[1]}\Gamma^{d}).
\end{align}
In particular, we emphasize that the elements of level-2 group-changing space
don't commutate those of level-1 group-changing space.

With the help of non-Hermitian generalization of Kaluza-Klein
compactification, the 2-th order variability on the side of NGT is equal to
the 1-th order variability on the side of AdS. Now, the fifth dimension on the
side of AdS becomes the internal space on the side of NGT.

\subsubsection{Equivalence between the matter}

Secondly, we consider the equivalence of matter under the two representations
(AdS and NGT).

A complex zero of AdS obviously is a level-1 real zero with an extra level-2
imaginary zero of NGT. Now, we have $n^{[2]}=\left \vert \lambda^{\lbrack
12]}\right \vert -1$. Or, the elementary particle in AdS corresponds to the
elementary particle with a level-2 imaginary zero that has $(\left \vert
\lambda^{\lbrack12]}\right \vert -1)/\left \vert \lambda^{\lbrack12]}\right \vert
$ imaginary electric charge and unit color charge.

In above section, we had studied a special type of elementary particle in NGT
without extra level-2 imaginary zero. Now, we have $n^{[2]}=\left \vert
\lambda^{\lbrack12]}\right \vert $. Or, the elementary particle in AdS becomes
the elementary particle with unit imaginary electric charge and zero color
charge. This is an elementary particle without coupling non-Hermitian
\textrm{SU(0,N)} Yang-Mills gauge fields. In the limit of $\left \vert
\lambda^{\lbrack12]}\right \vert \rightarrow \infty$, $\left \vert \lambda
^{\lbrack12]}\right \vert -1\simeq \left \vert \lambda^{\lbrack12]}\right \vert .$
The degrees of freedom for two types of elementary particles (one with
$n^{[2]}=\left \vert \lambda^{\lbrack12]}\right \vert $, the other with
$n^{[2]}=\left \vert \lambda^{\lbrack12]}\right \vert -1$) can be regarded as
same. Without considering bulk fluctuations from non-Hermitian
\textrm{SU(0,N)} Yang-Mills gauge fields, two types of elementary particles
(one with $n^{[2]}=\left \vert \lambda^{\lbrack12]}\right \vert $, the other
with $n^{[2]}=\left \vert \lambda^{\lbrack12]}\right \vert -1$) is exactly equal
each other.

In summary, an elementary particle on the side of AdS is equivalence to that
on the side of NGT.

\subsubsection{Equivalence between the motion}

According to Gravity/N-gauge equivalence, for a (d+1)-dimensional
\textrm{\~{S}\~{O}(d+1)} non-unitary physical variant $V_{\mathrm{\tilde
{S}\tilde{O}(d+1)},d+1}(\Delta \phi^{\mu},\Delta x^{\mu},k_{0},\omega_{0})$,
the representation of ((d-1)+1)-dimensional non-Hermitian gauge theory (NGT)
on flat spacetime is equivalence to the representation of (d+1)-dimensional
AdS. In AdS, slow motion comes from the fluctuations of gravitational waves;
in NGT, slow motion comes from fluctuations of non-Hermitian \textrm{U(0,1)}%
$\times$\textrm{SU(0,N)} gauge fields.

For AdS, the slow motion about gravitational waves is described by the
Einstein-Hilbert action,
\begin{equation}
S_{\mathrm{EH}}=\frac{1}{16\pi G}\int \sqrt{-g}\tilde{R}\text{ }d^{4}\tilde
{x}.\nonumber
\end{equation}
On the other hand, for the NGT, the slow motion about non-Hermitian
\textrm{U(0,1)}$\times$\textrm{SU(0,N)} gauge fields is described by the
following Hamiltonian,
\[
\hat{H}_{(d-1)+1}^{\mathrm{slow}}=\vec{\Gamma}\cdot(\mathrm{e}\vec{A}%
_{U(0,1)}+g\mathcal{\vec{A}})+\Gamma^{t}(\mathrm{e}A_{t,U(0,1)}+g\mathcal{A}%
_{t}).
\]
In particular, the fluctuations of non-Hermitian \textrm{U(0,1)} Abelian gauge
field describe the shape changings of boundary of AdS. Without considering
non-Hermitian \textrm{SU(0,N)} non-Abelian gauge fields, NGT is reduced to a
non-Hermitian \textrm{U(0,1)} gauge fields that is just the theory of CFT.

\subsection{Summary}

\begin{figure}[ptb]
\includegraphics[clip,width=0.9\textwidth]{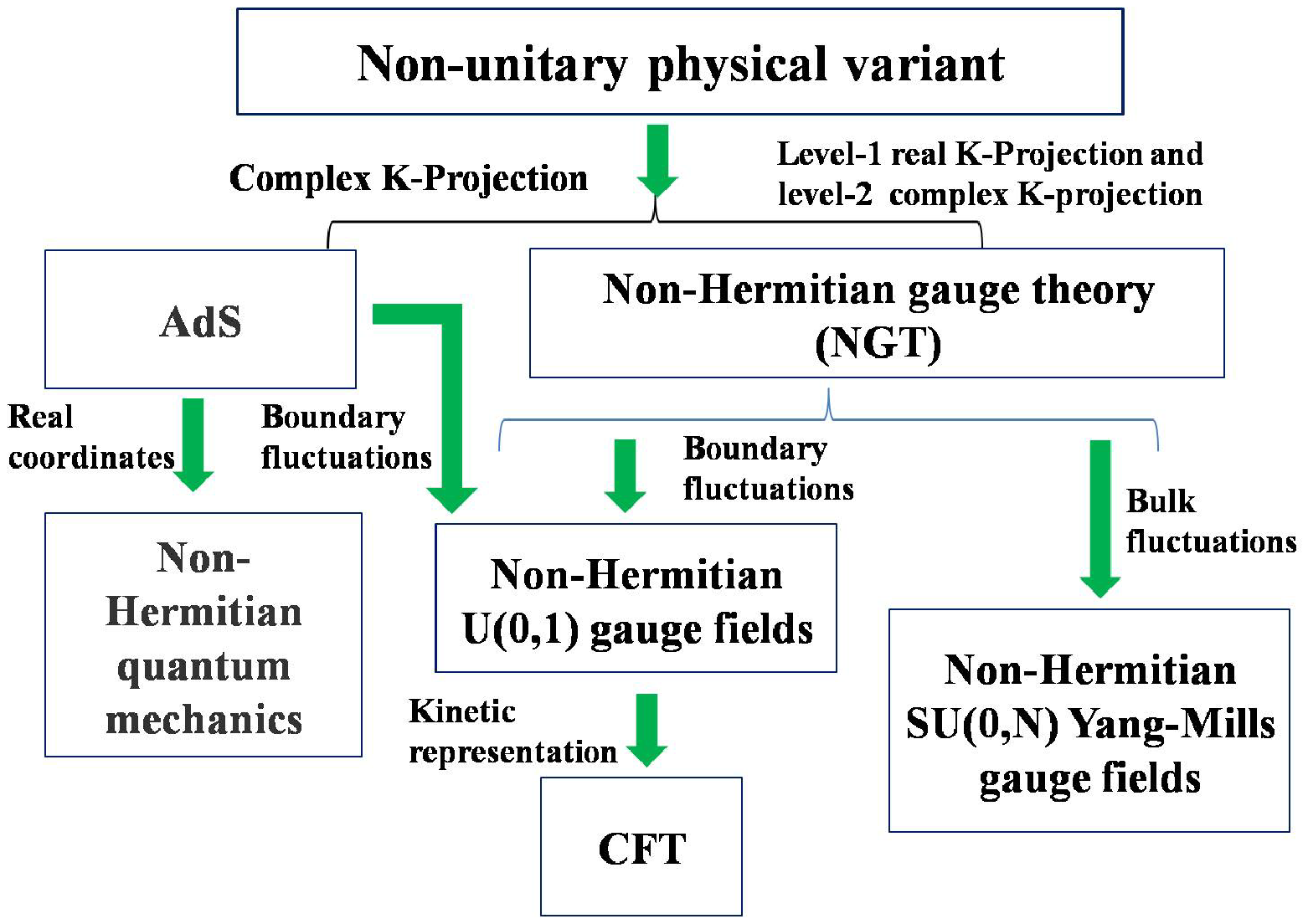}\caption{The logical
structure of the theory of non-unitary physical variant for AdS/CFT
correspondence}%
\end{figure}

In the end, we draw the conclusion.

The\emph{ starting point} of this theory is very simple -- (d+1)-dimensional
\textrm{\~{S}\~{O}(d+1)} non-unitary physical variant\textit{ }%
$V_{\mathrm{\tilde{S}\tilde{O}(d+1)},d+1}(\Delta \phi^{\mu},\Delta x^{\mu
},k_{0},\omega_{0})$. The non-unitary physical variant is characterized by
1-th order non-unitary spatial variability along the d-th direction
\[
\mathcal{T}(\delta x^{d})\leftrightarrow \hat{U}(\delta \phi^{\mu}%
)=e^{k_{0}x^{d}\Gamma^{d}}.
\]
Based on the simple starting point, we develop a microscopic theory for
AdS/CFT correspondence and its updated version -- AdS/NGT equivalence. When we
only consider unitary physical processes on the boundary of AdS, the AdS/NGT
equivalence is reduced to usual AdS/CFT correspondence. See the logical
structure of the part in Fig.13.

In our world, quantum mechanics is Hermitian theory characterizing unitary
time evolution processes. However, we point out that to characterize AdS, the
equivalent theory is non-Hermitian quantum physics including non-Hermitian
quantum mechanics and Non-Hermitian gauge theory.\emph{\emph{ }}

In the end of this part, we answer all six questions at beginning and show how
the troubles disappear:

1. What's the \emph{exact} rule of AdS/CFT correspondence within the framework
of quantum gravity rather than just a conjecture?

\textbf{The answer:}

We found that AdS/CFT correspondence characterizes the equivalence for the
slow motion in CFT and that on the boundary of AdS. In brief, the key point of
AdS/CFT correspondence is spacetime skin effect due to non-unitary variability
along d-th spatial direction. According to the spacetime skin effect, the
dynamics for (d-1)+1 dimensional real zero lattice is almost equal to the that
for the outermost side of the d+1 dimensional complex zero lattice.

2. Why the perturbative metric fluctuations $g_{\mu \nu}$ of AdS correspond to
a boundary stress tensor $T_{\mu \nu}$ in CFT within the framework of quantum gravity?

\textbf{The answer:}

This is really a correspondence between shape changing of boundary in AdS and
expansion/contraction in CFT. The exact correspondence between metric
fluctuations in AdS and the motion tensor $M_{\mu \nu}$ are given by $g_{\mu
\nu}=l_{0}^{2}%
{\displaystyle \sum \limits_{a}}
(\delta A_{\mu}^{a0}\delta A_{\nu}^{a0})=l_{0}^{2}M_{\mu \nu}.$ It is the
changing of motion tensor $M_{\mu \nu}$ is equal to energy-momentum tensor
$T_{\mu \nu}$ rather than $M_{\mu \nu}$ itself.

3. According to the dictionary from AdS/CFT correspondence, the particle's
mass $m$ in AdS plays the role of anomalous dimension $\nu$ in correlation
functions. Is it correct within the framework of quantum gravity? Why?

\textbf{The answer:}

We indeed have a correspondence between particle's mass $m$ of AdS and
anomalous dimension $\nu$ of correlation functions in CFT. So, it is correct.
The underlying mechanism of this correspondence is the re-definition the
elementary particles in both sides. The anomalous dimension plays the role of
the ratio of the size of an elementary particle in AdS and that in CFT.

4. According to AdS/CFT correspondence, the gauge fields $A_{\mu}$ in AdS
correspond to usual current in CFT $J^{\mu}$. What does it mean within the
framework of quantum gravity?

\textbf{The answer:}

Abelian/non-Abelian gauge fields characterize the dynamics of global/relative
loop currents on spacetime. In AdS, the loop currents for the gauge fields is
reduced to the current of CFT on the boundary of the AdS, i.e.,
\[
\text{Loop currents in AdS }\leftrightarrow \text{ Currents in CFT. }%
\]

5. According to AdS/CFT correspondence, there exists Ryu-Takayanagi's formula
of the holographic entangled entropy. Is it correct within the framework of
quantum gravity? What's underlying mechanism of Ryu-Takayanagi's formula?

\textbf{The answer:}

The underlying mechanism of holographic entangled entropy in AdS/CFT
correspondence really comes from the geometry quantized for quantum flat
spacetime. As a result, each unit cell of quantum flat spacetime in CFT carry
area $l_{0}^{2}$. When one smears out the information of the unit cells, the
entropy is just the RT formula of the holographic entangled entropy.

6. How to characterize quantum fluctuations from gravitational waves in the
bulk of AdS by CFT beyond the boundary formula?

\textbf{The answer:}

In this part, we update the AdS/CFT correspondence to gravity/N-gauge
equivalence. Based on gravity/N-gauge equivalence, the quantum fluctuations
from gravitational waves both in bulk and on boundary of AdS can all be
characterized by non-Hermitian \textrm{U(0,1)}$\times$\textrm{SU(0,N)} gauge
fields. When we reduce the NGT to its unitary physical processes on boundary
of the system, AdS/NGT equivalence is reduced to usual AdS/CFT correspondence
between the theory for boundary of AdS and CFT.

\newpage

\section{Black Hole -- A Physical Variant with Topological Defects}

\subsection{Introduction}

In classical physics, as the collapse of a spherical star, a black hole
becomes a region of spacetime in which the gravitational potential $\frac
{GM}{r}$, exceeds the square of the speed of light, $c^{2}$. In modern
physics, the boundary of the black hole is called event horizon, beyond which
the stellar matter continues to collapse into a singularity of zero volume and
infinite density at $r=0$. Once a black hole has formed, and after all the
matter disappeared into the singularity, the geometry of spacetime itself
continues to collapse towards the singularity. With the help of general
relativity, people make much deeper insight into black holes and fundamental
relationship between gravitation, thermodynamics, and quantum theory is
explored. Hawking's discovery of the thermal radiation from black holes
provides a deep connection between gravity and quantum mechanics\cite{haw}.
The relation between geometrical properties of the event horizon and
thermodynamic quantities provides a clear indication that there is a relation
between properties of the spacetime geometry and some kind of quantum physics.
Another progress is about Sachdev--Ye--Kitaev (SYK) model that is exactly
solvable in the large $N$ and IR limit\cite{Kitaev-talks,ye}. The SYK model is
believed to describe the behavior of correlation functions near horizon of
extremal black hole\cite{mac}.

However, black hole still a big beast to be recognized, of which there are a
lot of unsolved mysteries:

\begin{enumerate}
\item What's the exact \emph{microstructure} of spacetime around black hole
near Planck length? What's the exact \emph{microstructure} of spacetime inside
black hole? And, how to characterize it?

\item The object in $r=0$ is the source of the gravitational field and is
called the \emph{singularity}. Everything that crosses the event horizon will
end at the singularity. Since the singularity does not belong to the
spacetime, it simply cannot be described or represented in the framework of
general relativity. What is the exact solution for the singularity
problem\cite{sin}?

\item In quantum theory, \emph{black holes} emit Hawking radiation with a
perfect thermal spectrum. This allows a consistent interpretation of the laws
of black hole mechanics as physically corresponding to the ordinary laws of
thermodynamcs\cite{haw}. The classical laws of black hole mechanics together
with the formula for the temperature of Hawking radiation allow one to
identify a quantity associated with black holes --- namely $\frac{A}{4}$ in
general relativity--- as playing the mathematical role of entropy. A major
goal of research in quantum gravity is to provide a derivation of the formula
for the entropy of a black hole. What is the exact approach to derive the
entropy of black hole? Why black hole has finite temperature?

\item Another issue related to black hole is the \textquotedblleft \emph{black
hole information paradox}\textquotedblright. According to Hawking radiation,
during the evaporation process, an initial pure state may evolve to a mixed
state, i.e., \textquotedblleft information\textquotedblright \ will lost.
However, it is known that in quantum mechanics, an isolated pure state will
never evolve a final mixed state. Therefore, the issue of whether a pure state
can evolve to a mixed state in the process of black hole formation and
evaporation is usually referred to as the \textquotedblleft black hole
information paradox\textquotedblright \cite{haw}. How to solve this paradox? Is
quantum mechanics wrong, or is general relativity wrong? Or both wrong? Is
Page curve for Hawking radiation correct?

\item SYK model is relevant to physics of black hole\cite{Kitaev-talks,ye}.
What does this model really mean? How to provide a derivation of the formula
for SYK model?
\end{enumerate}

All above puzzles are all relevant to a complete theory of quantum gravity for
black hole. In this part, we develop a complete theory to characterize black
hole. Within the new theory, we answer above five questions and interpret the
black hole by using the concepts of the microscopic properties of a new
physical framework, i.e.,%
\begin{align*}
&  \text{Black hole (a phenomenological theory)}\\
&  \Longrightarrow \text{A physical variant with }\\
&  \text{topological defects (a microscopic theory).}%
\end{align*}
In other words, the physical reality of black hole is really a physical
variant with topological defects. All physical processes of our world be
intrinsically described by the processes of the changings of physical variants.

\subsection{Topological defects of variant}

\subsubsection{Review on topological defects of usual fields (non-changing
structures)}

In the first part, before discussing the topological defects of variant, we
review theory for topological defects of usual fields.

Topological defects like domain walls, vortices and monopoles arise in a
variety of different areas of physics, such as condensed matter physics,
particle physics, astrophysics and cosmology\cite{min}. With the help of
homotopy theory, from topological properties of the vacuum manifold of the
underlying field theory, the topological defects for usual fields can be
classified\cite{kib} .

For a system with spontaneous symmetry breaking, there exists order parameter
that characterizes the existence of the (traditional) long range order. The
order parameter is defined by the expectation value in the ground state
$\left \vert 0\right \rangle $, i.e., $\left \langle 0|\hat{A}|0\right \rangle
\>=A_{0},$ where $\hat{A}$ is an operator with a non-vanishing ground-state
expectation value which transforms non-trivially under group $G$. As a result,
from spontaneous symmetry breaking $\hat{U}(g)\left \vert 0\right \rangle
\neq \left \vert 0\right \rangle $, for some $g\in G$, the order parameter
changes with the changing of the ground state
\begin{equation}
\left \langle 0|\hat{U}^{-1}(g)\hat{A}\hat{U}(g)|0\right \rangle \>=D(g)A_{0}%
\neq A_{0}.
\end{equation}
In general, not all elements of $G$ lead to distinct ground states. There may
be some subgroup $H$ of elements such that $D(h)A_{0}=A_{0}\quad$for$\;h\in
H.$ The distinct degenerate ground states correspond to the distinct values of
$A=D(g)A_{0}$. Hence they are in one-to-one correspondence with the left
cosets of $H$ in $G$ (sets of elements of the form $gH$). These cosets are the
elements of the quotient space $M=G/H.$ This space may be regarded as the
vacuum manifold or manifold of degenerate ground states.

Homotopy theory is an approach to classify topological defects\cite{Vilenkin}.
Let us consider the structure with given base point $x\in \mathcal{M}$ in a
given topological space $\mathcal{M}$. The homotopy group of the topological
space $\mathcal{M}$ with base point $x$ is denoted by
\begin{equation}
\pi_{n}\left(  \mathcal{M}\,,x\right)
\end{equation}
that characterizes the equivalent classes of maps from $n$-spheres into
$\mathcal{M}$.

For the system with spontaneous symmetry breaking, there exist topological
defects due to topologically mapping between group space and geometric space.
The general conditions for the existence of defects can be expressed in terms
of the topology of the vacuum manifold $M$, specifically its \emph{homotopy
groups}. The homotopy classes constitute the elements of a group, the
\emph{fundamental group} of $M$, denoted by
\begin{equation}
\pi_{d}(M)=\pi_{d}(G/H).
\end{equation}
As a result, domain walls occur if the vacuum manifold has disconnected
components, that is $\pi_{0}\left(  G/H\right)  \neq I$. Vortex lines occur if
the vacuum manifold contains unshrinkable loops that is $\pi_{1}\left(
G/H\right)  \neq I$. The monopoles are characterized by $\pi_{2}\left(
G/H\right)  \neq I$ that is unshrinkable $2$-spheres.

We take domain wall as an example. For a real scalar field described
$\left \langle \hat{\phi}(\mathbf{r})\right \rangle =\phi,$ there are two
degenerate ground states, i.e., $\phi=\pm \phi_{0}.$ The topological defect is
kink that is domain wall separating the regions with different degenerate
ground states, for example, $\phi=\phi_{0}$ and $\phi=-\phi_{0}$. For one
dimensional (1D) $\phi^{4}$-field, such a topological domain wall can be
described by a soliton solution, $\phi(x)=\phi_{0}\tanh(x).$

\subsubsection{Kinetic representation for unitary/non-unitary variants}

Unitary/non-unitary variant describes a structure of phase/amplitude changings
that is denoted by\ a mapping between a d-dimensional unitary/non-unitary
group-changing space $\mathrm{C}_{\mathrm{\tilde{G},}d}$ with total size
$\Delta \phi^{\mu}$\ and Cartesian space $\mathrm{C}_{d}$\ with total size
$\Delta x^{\mu}$\cite{kou1}. For the case of unitary variant, we have real
$\delta \phi^{\mu}$; while for non-unitary one, we have complex $\delta
\phi^{\mu}=e^{i\varphi^{\mu}}\left \vert \delta \phi^{\mu}\right \vert $ with
$\varphi^{\mu}\neq0,\pi$. Here, $\delta \phi^{\mu}$ denotes group-changing
element along $\mu$-direction (or element of non-unitary Clifford
group-changing space along $\mu$-direction). In this part, we focus on the
non-unitary variant with a pure imaginary $\varphi^{\mu=d}=\pm \frac{\pi}{2}$.

In this part, we firstly provide an alternative representation for
unitary/non-unitary variants -- kinetic representation.

Now, the corresponding group-changing space of the non-unitary variant has an
imaginary phase $\delta \phi^{d}=e^{\pm i\frac{\pi}{2}}\left \vert \delta
\phi^{d}\right \vert =\pm i\left \vert \delta \phi^{d}\right \vert $ along d-th
direction. A unitary/non-unitary variant $V_{\mathrm{\tilde{G},}d}[\Delta
\phi^{\mu},\Delta x^{\mu},k_{0}^{\mu}]$ is denoted by\ a unitary/non-unitary
mapping between a d-dimensional unitary group-changing space $\mathrm{C}%
_{\mathrm{\tilde{G},}d}$ with total size $\Delta \phi^{\mu}$\ and Cartesian
space $\mathrm{C}_{d}$\ with total size $\Delta x^{\mu}$, i.e.,%
\begin{equation}
V_{\mathrm{\tilde{G},}d}[\Delta \phi^{\mu},\Delta x^{\mu},k_{0}^{\mu
}]:\mathrm{C}_{\mathrm{\tilde{G},}d}=\{ \delta \phi^{\mu}\} \Longleftrightarrow
\mathrm{C}_{d}=\{ \delta x^{\mu}\}
\end{equation}
where $\Longleftrightarrow$\ denotes an ordered mapping under fixed changing
rate of integer multiple $k_{0}$.\ In particular, $\delta \phi^{\mu}$ denotes
group-changing element along $\mu$-direction (or element of group-changing
space along $\mu$-direction) rather than group element (or element of group).
For the cases of the unitary variant, the changing rate $k_{0}$ is real; for
the cases of non-unitary variants, it becomes complex.

Based on kinetic representation, we define unitary/non-unitary variants.

Now, we take a 1D unitary/non-unitary variant $V_{\mathrm{\tilde{U}(1),}%
1}[\Delta \phi,\Delta x,k_{0}]$ as an example to show the concept.

$V_{\mathrm{\tilde{U}(1),}1}[\Delta \phi,\Delta x,k_{0}]$ describes the
unitary/non-unitary mapping between 1D unitary group-changing space
$\mathrm{C}_{\mathrm{\tilde{U}(1)},1}(\Delta \phi)$\textit{ }and Cartesian
space $\mathrm{C}_{1}$, i.e.,
\[
V_{\mathrm{\tilde{U}(1),}1}[\Delta \phi,\Delta x,k_{0}]:\mathrm{C}%
_{\mathrm{\tilde{U}(1)},1}(\Delta \phi)=\{ \delta \phi \} \Longleftrightarrow
\mathrm{C}_{1}=\{ \delta x\}.
\]
According to above definition,\ for a 1D unitary/non-unitary variant
$V_{\mathrm{\tilde{U}(1),}1}[\Delta \phi,\Delta x,k_{0}],$ we have $\delta
\phi_{i}=k_{0}n_{i}\delta x_{i}$ where $k_{0}$ is a constant real/complex
number. For a higher-dimensional case $V_{\mathrm{\tilde{G},}d}[\Delta
\phi^{\mu},\Delta x^{\mu},k_{0}^{\mu}]$, along different directions (for
example, $\mu$-direction), the situation is similar to the 1D case by
considering real or imaginary changing rate along d-th direction.{}

\subsubsection{Topological defects of variant}

In this section, we discuss the topological defects of variants based on
kinetic representation. With topological defects, the variant cannot be
uniform. To consider simple situations, we focus on the variants with 1D
topological defects that are domain walls between unitary/non-unitary
variants. See the illustration in Fig.14.

\begin{figure}[ptb]
\includegraphics[clip,width=0.75\textwidth]{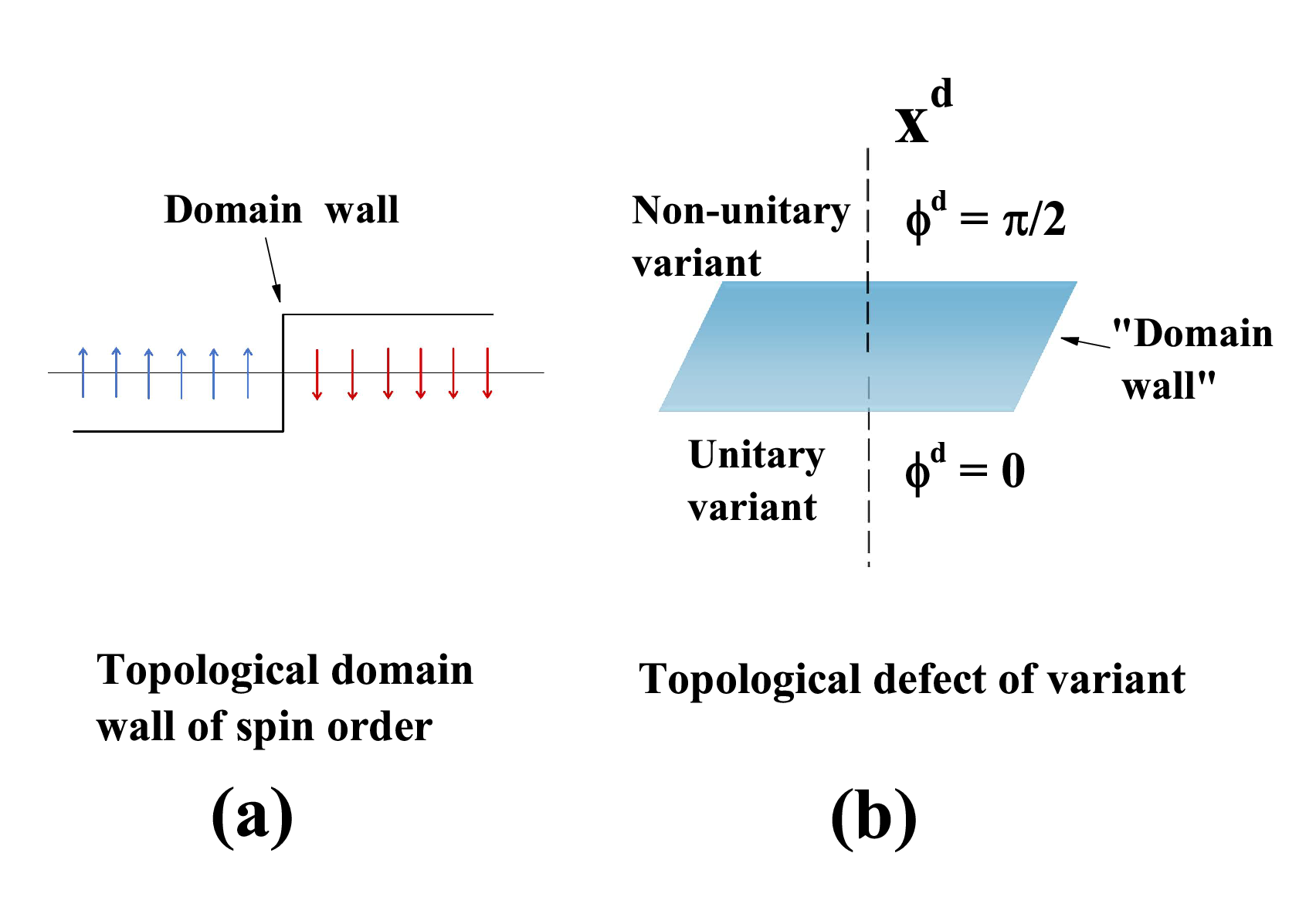}\caption{(a) An
illustration of topological defect of spin order between two degenerate ground
states; (b) An illustration of topological defect between unitary variant\ and
non-unitary variant. Now, the phase change of the changing rate $k_{0}^{\mu
=d}$ along d-th direction is $\pm \frac{\pi}{2}.$}%
\end{figure}

We give the definition of topological defects in variants.

\textit{Definition -- A topological defect is a domain wall between two
unitary/non-unitary variants }$V_{\mathrm{\tilde{G},}d}^{A}(k_{A,0}^{\mu
=d},k_{0}^{\mu \neq d})$\textit{ and }$V_{\mathrm{\tilde{G},}d}^{B}%
(k_{B,0}^{\mu=d},k_{0}^{\mu \neq d})$\textit{ along d-th direction. There
exists a sudden phase change of the (complex) changing rate }$k_{0}^{\mu=d}$
\textit{along d-th direction} \textit{from }$k_{A,0}^{\mu=d}$\textit{ to
}$k_{B,0}^{\mu=d}$\textit{. Along other directions }$k_{0}^{\mu \neq d}%
$\textit{ are constant. }

So, the topological defects of variants become singularities on complex plane
of changing rate $k_{0}^{\mu}$, i.e.,%
\begin{align*}
&  \text{Topological defects of variants }\\
&  =\text{Singularities on complex plane of the changing rates }k_{0}^{\mu
}\text{.}%
\end{align*}

Then, we classify the classes of topological defects in variants.

In general, due to the two types of variants (unitary and non-unitary ones),
there are three different classes topological defects of variants: U-U class
denotes the domain wall between two unitary variants, N-N class denotes the
domain wall between two non-unitary variants, U-N (or N-U) class denotes the
domain wall between a unitary variant and a non-unitary one. For U-U class and
N-N class, the phase change of the changing rate $k_{0}^{\mu=d}$ along d-th
direction from $k_{A,0}^{\mu=d}$ to $k_{B,0}^{\mu=d}$ is $\pi.$

We take a special U-N class of topological defects as an example.

Now, the phase change of the (complex) changing rate $k_{0}^{\mu=d}$ along
d-th direction from $k_{A,0}^{\mu=d}$ to $k_{B,0}^{\mu=d}$ is $\pm \frac{\pi
}{2}.$ We may assume a kink-like structure for the function of $(k_{0}^{\mu
=d})^{2},$ i.e.,
\[
(k_{0}^{\mu=d})^{2}=(k_{0}^{\mu \neq d})^{2}\tanh(x^{d}).
\]
In the limit of $x^{d}\rightarrow \infty$, we have a unitary variant
$(k_{0}^{\mu=d})^{2}=(k_{0}^{\mu \neq d})^{2}$ or $k_{0}^{\mu=d}=k_{0}^{\mu \neq
d};$ In the limit of $x^{d}\rightarrow-\infty$, we have a non-unitary variant
$(k_{0}^{\mu=d})^{2}=-(k_{0}^{\mu \neq d})^{2}$ or $k_{0}^{\mu=d}=ik_{0}%
^{\mu \neq d}.$ At the domain wall $x^{d}=0$, the changing rate $k_{0}^{\mu=d}$
is zero.

In the following parts, we will show that the event horizon of black holes
belongs to U-N class topological defect between a unitary variant and a
non-unitary one. It is really "domain wall" with a $\frac{\pi}{2}$ branch cut
on complex plane of the changing rates $k_{0}^{\mu}$.

\subsubsection{Higher-order variability for variant with topological defect}

For variants with topological defects, the original variability is always
reduced to its sub-variability. We take a U-N class of topological defects in
\textrm{\~{S}\~{O}(d)} unitary/non-unitary variant\textit{ }$V_{\mathrm{\tilde
{S}\tilde{O}(d)},d}(\Delta \phi^{\mu},\Delta x^{\mu},k_{0},\omega_{0})$ as an
example. There are two regions of the system -- one is \textrm{\~{S}\~{O}(d)}
unitary variant, the other is \textrm{\~{S}\~{O}(d)} non-unitary variant.

In the region of \textrm{\~{S}\~{O}(d)} unitary variant, the spatial-tempo
variability is determined by the following equation,
\begin{equation}
\mathcal{T}(\delta x^{\mu})\leftrightarrow \hat{U}(\delta \phi^{\mu}),
\end{equation}
where $\hat{U}(\delta \phi^{\mu})=e^{i\cdot \delta \phi^{\mu}\Gamma^{\mu}}$ and
$\delta \phi^{\mu}=k_{0}x^{\mu}$ is the corresponding phase angle\textit{. }

In the region of \textrm{\~{S}\~{O}(d)} non-unitary variant, the spatial-tempo
variability is determined by the following equation,
\begin{equation}
\mathcal{T}(\delta x^{\mu})\leftrightarrow \hat{U}(\delta \phi^{\mu}),
\end{equation}
where $\hat{U}(\delta \phi^{\mu})=e^{i\cdot \delta \phi^{\mu}\Gamma^{\mu}}$ and
$\delta \phi^{\mu \neq d}=\pm \left \vert \Delta \phi^{d}\right \vert =\pm \left \vert
k_{0}x^{\mu}\right \vert $ and $\delta \phi^{\mu=d}=\pm i\left \vert k_{0}%
x^{d}\right \vert $ is the corresponding phase angle\textit{.}

In addition, we show the higher-order variability on the topological defect.

The topological defect between unitary variant and non-unitary variant is the
interface between them that is described by a ($d-1$)-dimensional
\textrm{\~{S}\~{O}(d-1)} non-unitary variant\textit{ }$V_{\mathrm{\tilde
{S}\tilde{O}(d-1)},d-1}(\Delta \phi^{\mu},\Delta x^{\mu},k_{0},\omega_{0})$
with $\mu \neq d$. Therefore, spatial-tempo variability is determined by the
following equation,
\begin{equation}
\mathcal{T}(\delta x^{\mu})\leftrightarrow \hat{U}(\delta \phi^{\mu}),\text{
}\mu \neq d
\end{equation}
where $\hat{U}(\delta \phi^{\mu})=e^{i\cdot \delta \phi^{\mu}\Gamma^{\mu}}$ and
$\delta \phi^{\mu}=k_{0}x^{\mu}$ is the corresponding phase angle ($\mu \neq
d$)\textit{.} We call the higher-order variability of topological defects to
be \emph{residue higher-order variability}\textit{.}

In particular, along d-th direction on topological defect, we have
\begin{equation}
\mathcal{T}(\delta x^{d})\leftrightarrow \hat{U}(\delta \phi^{d})
\end{equation}
where $\tilde{U}(\delta \phi^{d})=e^{i\cdot \delta \phi^{d}\Gamma^{d}}$ and
$\delta \phi^{d}=k_{0}^{d}x^{d}$ with $k_{0}^{d}=0$. That means along d-th
direction, the order of variability is reduce to 0-th order! In other words,
along d-th direction, it is "non-changing" structure that cannot be described
by usual variant.\textit{ }

\subsubsection{Representations}

In this section, we discuss the representations for a variant with topological
defects. We focus on U-N class topological defects of \textrm{\~{S}\~{O}(d)}
unitary/non-unitary variant $V_{\mathrm{\tilde{S}\tilde{O}(d)},d}[\Delta
\phi^{\mu},\Delta x^{\mu},k_{0}^{\mu}]$. This is a domain wall, of which the
phase change of the complex changing rate $k_{0}^{\mu=d}$ along d-th direction
from $k_{A,0}^{\mu=d}$ to $k_{B,0}^{\mu=d}$ is $\pm \frac{\pi}{2}.$

Firstly, we consider the representation under complex knot projection, by
which both phase changings and amplitude changings are characterized.

To derive complex knot projection (K-projection) away from the topological
defect, we replace the real coordinates $x$ by complex ones $\tilde
{x}=e^{i\varphi(x)}x$. Under complex K-projection, according to the zero
equation $\hat{P}_{\theta}[\mathrm{z}(\tilde{x}^{i})]\equiv \xi_{\theta}%
(\tilde{x}^{i})=\cos(k_{0}^{i}\cdot \tilde{x}^{i})=0$, we have a complex zero
lattice, $\tilde{x}^{i}=[l_{0}\cdot N^{i}/2+\frac{l_{0}^{i}}{2\pi}%
(\theta+\frac{\pi}{2})]$. Along $i$-th spatial direction of the zero lattice,
the lattice site is labeled by $N^{i}.$

On the topological defect, under complex K-projection, we have
(d-1)-dimensional zero lattice $\tilde{x}^{i\neq d}=[l_{0}\cdot N^{i\neq
d}/2+\frac{l_{0}^{i\neq d}}{2\pi}(\theta+\frac{\pi}{2})].$ Without changing
rate on the topological defect along d-th direction, there doesn't exist zero
along $\tilde{x}^{d}$-th direction on the topological defect.

In addition, one can use matrix network to characterize a variant with
topological defect. In the region of unitary variant, we have a Hermitian
matrix network; in the region of non-unitary variant, we have a non-Hermitian
matrix network. In particular, for topological defect, we have reduced matrix
network, of which there doesn't exist the component of $\Gamma^{d}$.

Secondly, we consider geometry representation under real K-projection, by
which only phase changings are characterized.

For the representation under real K-projection, according to the zero equation
$\hat{P}_{\theta}[\mathrm{z}(\tilde{x}^{i})]\equiv \xi_{\theta}(\tilde{x}%
^{i})=\cos(k_{0}^{i}\cdot \tilde{x}^{i})=0$, we have
\begin{align*}
\cos(k_{0}^{i}e^{i\varphi^{i}}\cdot x^{i})  &  =\cos(\cos(\varphi^{i}%
)k_{0}^{i}x^{i}+i\sin(\varphi^{i})k_{0}^{i}x^{i})\\
&  =\cos(\cos(\varphi^{i})k_{0}^{i}x^{i})\cosh(\sin(\varphi^{i})k_{0}^{i}%
x^{i})\\
&  -\sin(\cos(\varphi^{i})k_{0}^{i}x^{i})\sinh(\sin(\varphi^{i})k_{0}^{i}%
x^{i})\\
&  =0.
\end{align*}
In the region of unitary variant, along $x^{d}$-th direction, due to
$\varphi^{i}=0,$ we have
\[
\cos(k_{0}^{i}e^{i\varphi^{i}}\cdot x^{i})=\cos(k_{0}^{i}x^{i})=0,
\]
of which the zero lattice is usual; In the region of non-unitary variant,
along $x^{d}$-th direction, due to for the case of $\varphi^{i}=\pm \frac{\pi
}{2},$ we have
\[
\cos(k_{0}^{i}e^{i\varphi^{i}}\cdot x^{i})=\cosh(k_{0}^{i}x^{i})=0.
\]
Now, there doesn't exist real zero solutions at all. Therefore, along $x^{d}%
$-th direction we only have real zero lattice in the region of unitary
variant; along other directions, we have real lattices in both regions.

On the topological defect, using similar approach, we have (d-1)-dimensional
real zero lattice $\tilde{x}^{i\neq d}=[l_{0}\cdot N^{i\neq d}+\frac
{l_{0}^{i\neq d}}{2\pi}(\theta+\frac{\pi}{2})].$ Without changing rate on the
topological defect along d-th direction, there also doesn't exist zero along
$\tilde{x}^{d}$-th direction on the topological defect.

Thirdly, we consider the representation under imaginary K-projection, by which
only amplitude changings are characterized.

For the representation under imaginary K-projection, $\hat{P}_{\theta
}[\mathrm{z}(\tilde{x}^{i})]\equiv \xi_{\theta}(\tilde{x}^{i})=\cos(k_{0}%
^{i}\cdot \tilde{x}^{i})=0,$ we consider its imaginary solutions where
$\tilde{x}^{i}=ix^{i}$. Now, we have
\begin{align*}
\cos(k_{0}^{i}e^{i(\varphi^{i}-\frac{\pi}{2})}\cdot ix^{i})  &  =\cos
(k_{0}^{i}e^{i(\varphi^{i}-\frac{\pi}{2})}\cdot \tilde{x}^{i})\\
&  =\cos(\cos(\varphi^{i}-\frac{\pi}{2})k_{0}^{i}\tilde{x}^{i}\\
&  +i\sin(\varphi^{i}-\frac{\pi}{2})k_{0}^{i}\tilde{x}^{i})\\
&  =\cos(-\sin \varphi^{i}k_{0}^{i}\tilde{x}^{i}+i\cos \varphi^{i}k_{0}%
^{i}\tilde{x}^{i}).
\end{align*}

With help of imaginary K-projection, in the region of unitary variant, along
$x^{d}$-th direction, due to $\varphi^{i}=0,$ we have
\[
\cos(ik_{0}^{i}\tilde{x}^{i})=\cosh(k_{0}^{i}x^{i})=0.
\]
Now, there doesn't exist imaginary zero solutions at all; In the region of
non-unitary variant, along $x^{d}$-th direction, due to for the case of
$\varphi^{i}=\pm \frac{\pi}{2},$ we have
\[
\cos(k_{0}^{i}\cdot \tilde{x}^{i})=0.
\]
This is an imaginary zero lattice. Along other directions for the whole
system, we don't have imaginary zero lattice.

\subsection{Black hole as topological defect between unitary physical variant
and non-unitary physical variant}

In this section, we discuss the theory about black hole based on a
\textrm{\~{S}\~{O}(3+1)} physical variant with topological defects. In brief,
we found that black hole is a special physical variant with U-N class of
topological defects between unitary physical variant and non-unitary physical
variant. To correctly derive a topological defect of \textrm{\~{S}\~{O}(3+1)}
physical variant, we must solve Einstein equation. In this part, we take the
Schwarzschild solution as an example to learn the nature of black hole.

\subsubsection{Schwarzschild solution and event horizon}

The Schwarzschild solution for a black hole with mass $M$ in spherical
coordinates $(t,r,\theta,\phi)$ is given by
\begin{align}
ds^{2}  &  =(1-\frac{2GM}{rc^{2}})c^{2}dt^{2}\\
&  -(1-\frac{2GM}{rc^{2}})^{-1}dr^{2}-r^{2}(d\theta^{2}+\sin^{2}d\phi^{2}).
\end{align}
According to above metric, there seems to be two singularities at which the
metric diverges: one at $r=0$ and the other at $r_{\mathrm{Schw}}=\frac
{2GM}{c^{2}}.$ $r_{\mathrm{Schw}}$ is know as the Schwarzschild radius.

It is easy to see that strange things occur close to $r_{\mathrm{Schw}}$. For
the proper time we get:
\begin{equation}
d\tau=\left(  1-\frac{2GM}{rc^{2}}\right)  ^{1/2}\;dt. \label{time}%
\end{equation}
When $r\longrightarrow \infty$ both times ($t$ and $\tau$) agree, so $t$ is
interpreted as the proper time measure from an infinite distance. As the
system with proper time $\tau$ approaches to $r_{\mathrm{Schw}}$, $dt$ tends
to infinity according to Eq. (\ref{time}). As a result, an object will never
reach the Schwarszchild surface when seen by an infinitely distant observer.
The closer the object is to the Schwarzschild radius, the slower it moves for
the external observer. Therefore, on Schwarzschild radius, one may guess that
there doesn't exist clock and all matter are static and cannot move any more.

A direct physical consequence of the difference introduced by gravity in the
local time respect to the time of an observer at infinity is that the
radiation that escapes from a given $r>r_{\mathrm{Schw}}$ will be redshifted
when received by a distant and static observer. Events that occur at
$r<r_{\mathrm{Schw}}$ are disconnected from the rest of the universe. Hence,
the surface determined by $r=r_{\mathrm{Schw}}$ is called an \emph{event
horizon}. Whatever crosses the event horizon will never return. This is the
origin of the expression \textquotedblleft black hole\textquotedblright,
introduced by John A. Wheeler in the mid 1960s. The black hole is the region
of spacetime inside the event horizon. It was known that the metric is
non-singular at $r=2GM/c^{2}$. The only real singularity is at $r=0$, where
the Riemann tensor diverges. It looks like that General Relativity is
incomplete and cannot provide a full description of the gravitational behavior
of singularity at $r=0$.

\subsubsection{Black hole as a physical variant with topological defect}

Without black hole, the spacetime is an \textrm{\~{S}\~{O}(3+1)} unitary
physical variant\textit{ }$V_{\mathrm{\tilde{S}\tilde{O}(3+1)},3+1}(\Delta
\phi^{\mu},\Delta x^{\mu},k_{0},\omega_{0})$, that is a mapping
between\textrm{ \~{S}\~{O}(3+1)} unitary Clifford group-changing space\textit{
}$\mathrm{C}_{\mathrm{\tilde{S}\tilde{O}(3+1)},3+1}$\textit{\ }and a rigid
spacetime $\mathrm{C}_{3+1},$\textit{ }i.e.,\textit{ }%
\begin{equation}
V_{\mathrm{\tilde{S}\tilde{O}(3+1)},3+1}[\Delta \phi^{\mu},\Delta x^{\mu}%
,k_{0}^{\mu}]:\{ \delta \phi^{\mu}\} \Leftrightarrow \{ \delta x^{\mu}\}
\end{equation}
where $\Leftrightarrow$\ denotes an ordered mapping with fixed changing rate
of integer multiple $k_{0}$ or $\omega_{0},$\ and $\mu$ labels the spatial direction.

\begin{figure}[ptb]
\includegraphics[clip,width=0.6\textwidth]{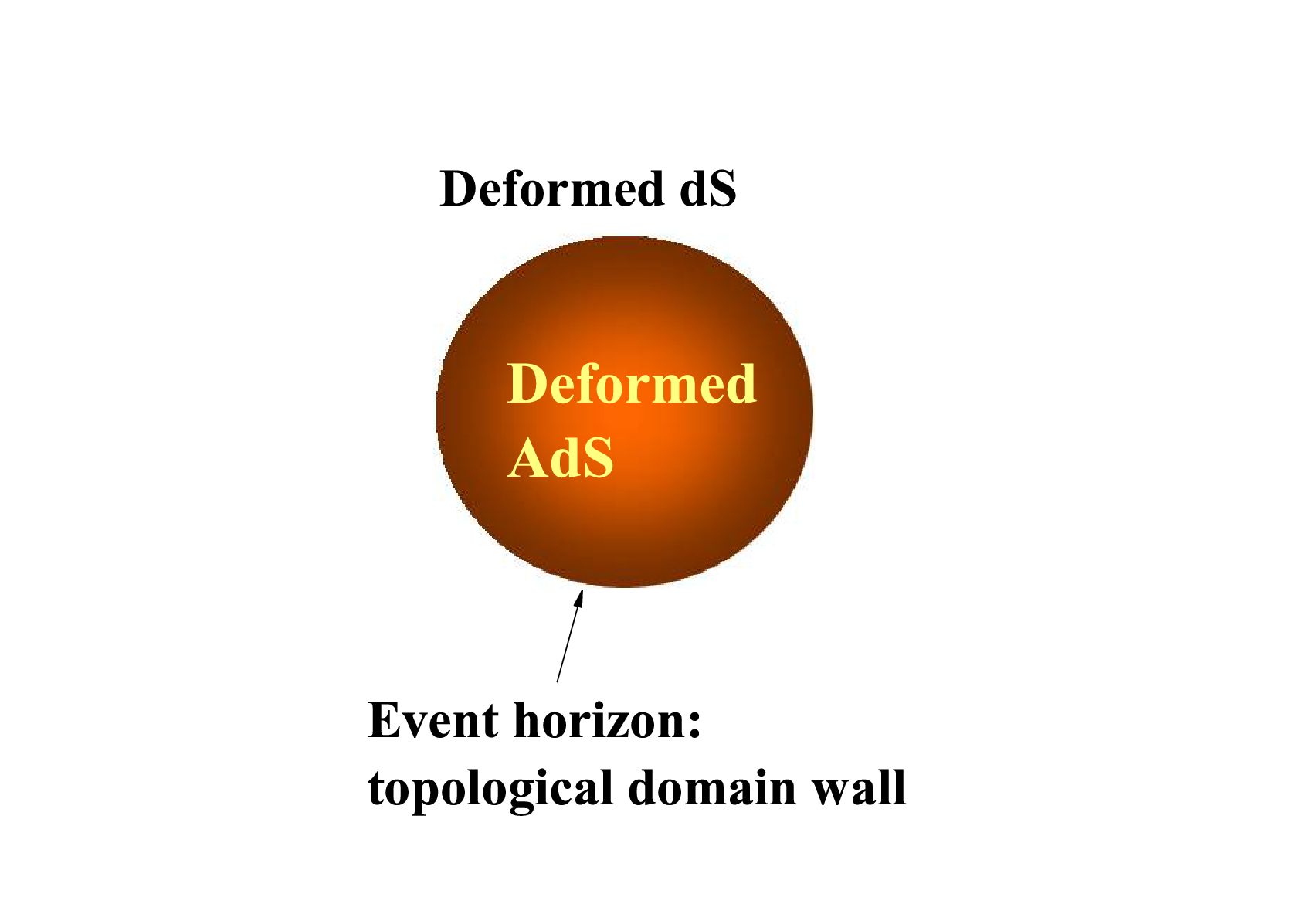}\caption{Black hole
becomes a physical variant with topological defect between unitary physical
variant\ (dS) and non-unitary physical variant (AdS)}%
\end{figure}

When there exists black hole, the situation changes. Black hole becomes a
physical variant with topological defect between unitary physical variant and
non-unitary one. See the illustration in Fig.15.

Out of black hole $1-\frac{2GM}{rc^{2}}>0$, we have
\begin{align}
ds^{2}  &  =\left \vert 1-\frac{2GM}{rc^{2}}\right \vert c^{2}dt^{2}\\
&  -\left \vert 1-\frac{2GM}{rc^{2}}\right \vert ^{-1}dr^{2}-r^{2}(d\theta
^{2}+\sin^{2}d\phi^{2}).
\end{align}
This is the region of a deformed \textrm{\~{S}\~{O}(3+1)} unitary physical
variant. However, inside black hole, we have
\begin{align}
ds^{2}  &  =-\left \vert 1-\frac{2GM}{rc^{2}}\right \vert c^{2}dt^{2}\\
&  +\left \vert 1-\frac{2GM}{rc^{2}}\right \vert ^{-1}dr^{2}-r^{2}(d\theta
^{2}+\sin^{2}d\phi^{2}).
\end{align}
This is the region of a deformed \textrm{\~{S}\~{O}(3+1)} non-unitary physical
variant, of which along radial direction and tempo direction, the charge rates
turn into imaginary. Therefore, we use a deformed AdS to characterize the
physical processes inside black hole. On the event horizon, the changing rate
along tempo direction is zero, i.e,
\[
(1-\frac{2GM}{rc^{2}})c^{2}dt^{2}\rightarrow0\text{ at }r=r_{\mathrm{Schw}}.
\]

To characterize black hole more clear, we use Eddington-Finkelstein
coordinates,%
\[
r_{\ast}=r+\frac{2GM}{c^{2}}\log \left \vert \frac{r-2GM/c^{2}}{2GM/c^{2}%
}\right \vert .
\]
Then, we have a new metric for Schwarzschild solution, i.e.,
\[
ds^{2}=\left(  1-\frac{2GM}{rc^{2}}\right)  (c^{2}dt^{2}-dr_{\ast}^{2}%
)-r^{2}d\Omega^{2}%
\]
where
\[
d\Omega^{2}=d\theta^{2}+\sin^{2}\theta d\phi^{2}.
\]
With help of the new radial coordinate $r_{\ast},$ the radial null rays
satisfy $d(ct\pm r_{\ast})=0$. The singularity at $r=r_{\mathrm{Schw}}$ is removed.

According to above description of black hole, we have deformed physical
variant with a topological defect at event horizon $r=r_{\mathrm{Schw}}$, of
which the phase change of the changing rate $k_{0}^{\mu=d}$ along tempo
direction and radial direction are all $\pm \frac{\pi}{2}.$ Out of the event
horizon, due to $1-\frac{2GM}{rc^{2}}>0,$ we have a deformed unitary physical
variant; inside the event horizon, due to $1-\frac{2GM}{rc^{2}}<0,$ we have a
deformed non-unitary physical variant. In particular, on the event horizon,
due to $1-\frac{2GM}{rc^{2}}=0,$ the changing rate along tempo direction turn
to zero. Now, the metric is reduced into a 2D one.\

\subsubsection{Higher-order variability for black hole}

According to above discussion, black hole becomes a physical variant with
topological defects.\emph{ What's higher-order variability?}

To characterize the higher-order variability of black hole, we use
Eddington-Finkelstein coordinates for Schwarzschild solution, $ds^{2}=\left(
1-\frac{2GM}{rc^{2}}\right)  (c^{2}dt^{2}-dr_{\ast}^{2})-r^{2}d\Omega^{2}$
with $d\Omega^{2}=d\theta^{2}+\sin^{2}\theta d\phi^{2}.$

On one hand, out of black hole, we have \textrm{\~{S}\~{O}(3+1)} unitary
variant\textit{ }$V_{\mathrm{\tilde{S}\tilde{O}(d+1)},d+1}(\Delta \phi^{\mu
},\Delta x^{\mu},k_{0},\omega_{0})$. The spatial-tempo variability is
determined by the following equation,
\begin{equation}
\mathcal{T}(\delta x^{\mu})\leftrightarrow \hat{U}(\delta \phi^{\mu}),
\end{equation}
where $\hat{U}(\delta \phi^{\mu})=e^{i\cdot \delta \phi^{\mu}\Gamma^{\mu}}$ and
$\delta \phi^{\mu}=k_{0}\delta x^{\mu}$ is the corresponding phase
angle\textit{.} The coordinates becomes variables, $\delta x^{\mu}%
\rightarrow \delta x^{\mu}(x^{\mu}).$

On the other hand, inside the black hole, we have \textrm{\~{S}\~{O}(3+1)}
non-unitary variant\textit{ }$V_{\mathrm{\tilde{S}\tilde{O}(3+1))},3+1}%
(\Delta \phi^{\mu},\Delta x^{\mu},k_{0},\omega_{0})$. The spatial-tempo
variability is determined by the following equation,
\begin{equation}
\mathcal{T}(\delta x^{\mu})\leftrightarrow \hat{U}(\delta \phi^{\mu}),
\end{equation}
where $\hat{U}(\delta \phi^{\mu})=e^{i\cdot \delta \phi^{\mu}\Gamma^{\mu}}$ and
$\delta \phi^{\mu \neq r,t}=\pm \left \vert \Delta \phi^{\mu}\right \vert
=\pm \left \vert k_{0}\delta x^{\mu}\right \vert $ and $\delta \phi^{\mu=r,t}=\pm
i\left \vert k_{0}\delta x^{\mu}\right \vert $ is the corresponding phase angle.

The event horizon plays role of topological defect between unitary variant and
non-unitary variant. We then use a two dimensional (2D) \textrm{\~{S}%
\~{O}((3-1)+1)} non-unitary variant\textit{ }$V_{\mathrm{\tilde{S}\tilde
{O}((3-1)+1)},\mathrm{(3-1)+1}}(\Delta \phi^{\mu},\Delta x^{\mu},k_{0}%
,\omega_{0})$ to characterize its spatial-tempo variability. The spatial-tempo
variability is determined by the following equation,
\begin{equation}
\mathcal{T}(\delta x^{\mu})\leftrightarrow \hat{U}(\delta \phi^{\mu}),\text{
}\mu \neq r,t
\end{equation}
where $\hat{U}(\delta \phi^{\mu})=e^{i\cdot \delta \phi^{\mu}\Gamma^{\mu}}$ and
$\delta \phi^{\mu}=k_{0}x^{\mu}$ is the corresponding phase angle ($\mu \neq
r,t$)\textit{.} In particular, along radial direction and tempo direction, it
is "non-changing" structure that cannot be described by usual variant, and the
variability is reduce to 0-th order.

The residue higher-order variability is determined by the following equation,
\begin{equation}
\mathcal{T}(\delta x^{\mu})\leftrightarrow \hat{U}(\delta \phi^{\mu}),\text{
}\mu \neq r,t
\end{equation}
where $\hat{U}(\delta \phi^{\mu})=e^{i\cdot \delta \phi^{\mu}\Gamma^{\mu}}$ and
$\delta \phi^{\mu}=k_{0}x^{\mu}$ is the corresponding phase angle ($\mu \neq
r,t$)\textit{.} The residue higher-order variability becomes key connection
between the different regions of the spacetime separated by the event horizon
of the black hole.

\subsubsection{A summary for representation of black hole}

A black hole becomes a physical variant with an U-N class topological defect,
of which the phase change of the changing rate $k_{0}^{\mu=r/t}$ along
radial/tempo direction is $\pm \frac{\pi}{2}.$ In other words, the event
horizon of a black hole is really a topological domain wall between a unitary
physical variant (or a dS) and a non-unitary physical variant (or an AdS).

\subsection{Theory for spacetime out of black hole}

In this section, we develop the theory for spacetime out of black hole.

Out of the event horizon, due to $1-\frac{2GM}{rc^{2}}>0,$ we have a deformed
unitary physical variant. By using Eddington-Finkelstein coordinates for
Schwarzschild solution, $ds^{2}=\left(  1-\frac{2GM}{rc^{2}}\right)
(c^{2}dt^{2}-dr_{\ast}^{2})-r^{2}d\Omega^{2}$ with $r_{\ast}=r+\frac
{2GM}{c^{2}}\log \left \vert \frac{r-2GM/c^{2}}{2GM/c^{2}}\right \vert $ and
$d\Omega^{2}=d\theta^{2}+\sin^{2}\theta d\phi^{2},$ the spatial-tempo
variability is determined by the following equation, $\mathcal{T}(\delta
x^{\mu})\leftrightarrow e^{i\cdot k_{0}\delta x^{\mu}(x^{\mu})\Gamma^{\mu}}%
$\textit{.}

For the case far from black hole, we return to usual flat quantum spacetime
$ds^{2}\rightarrow(c^{2}dt^{2}-dr_{\ast}^{2})-r^{2}d\Omega^{2}$. When we
approach the event horizon, without considering quantum nature of spacetime,
the traditional theory (general relativity) becomes incomplete. So,

Firstly, we consider the representation under complex K-projection and get a
deformed zero lattice. Near the event horizon, along the radial direction, we
have $(r_{\ast}\sqrt{1-\frac{2GM}{rc^{2}}})=[l_{0}\cdot n^{r_{\ast}}%
+\frac{l_{0}}{\pi}(\theta+\frac{\pi}{2})]$; along tempo direction, we have
$(t\sqrt{1-\frac{2GM}{rc^{2}}})=[l_{0}\cdot n^{t}+\frac{l_{0}}{\pi}%
(\theta+\frac{\pi}{2})].$ Along tangential directions, we have a uniform zero lattice.

Secondly, we discuss the properties of elementary particles out of black hole.
It was known that that a zero is an elementary particle.

According to general relativity, for the proper time we have $d\tau=\left(
1-\frac{2GM}{rc^{2}}\right)  ^{1/2}\;dt.$ Near event horizon $1-\frac
{2GM}{rc^{2}}\rightarrow0$, the size of an elementary particle along tempo
direction turns to infinite, i.e., $\Delta t=\frac{t_{p}}{\sqrt{1-\frac
{2GM}{rc^{2}}}}.$ However, according to quantum mechanics, the situation
becomes complex. Because the size of an elementary particle turns to infinite
near event horizon, the internal structure of an elementary particle becomes
\emph{extremely amplified}. Due to this extremely amplification effect, the
quantum fluctuations become exposed. Then, what's the physical consequence?
The answer is "\emph{randomness}".

We then review the emergence of probability in quantum mechanics.

In quantum mechanics, a pure state is denoted by a group of group-changing
elements with ordered distribution and a mixed state is denoted by a group of
group-changing elements with random distribution\cite{kou1}. To characterize
the order/disorder property of group-changing elements for an elementary
particle, we had introduced a concept of "quantum ensemble" that is an
ensemble of a lot of same elementary particle, of which all space-changing
elements (for example, the number is $N$) are identical and cannot be
distinguishable. Therefore, without additional internal information, due to
indistinguishability each space-changing elements has the same probability
(that is $\frac{1}{N}$) to find an elementary particle.

For a mixed state, we have a group of group-changing elements with random
distribution, each of which is $\frac{1}{N}$ particle. We consider a lot of
sample of the given mixed state (for example, $N_{F}$ particle, $N_{F}%
\rightarrow \infty$). This is a system with $N_{F}\times N$ identical
group-changing elements. Such a quantum ensemble is characterized by a group
of group-changing elements for $N_{F}$ elementary particles. Among
$N_{F}\times N$ group-changing elements, arbitrary $N$ group-changing elements
correspond to a particle. If the density of group-changing elements is
$\rho_{\mathrm{piece}},$ the density of group-changing elements $\frac{1}%
{N}\rho_{\mathrm{particle}}\ $becomes the probability to find a particle in a
given region $\psi^{\ast}(x,t)\psi(x,t)\Delta V$. In addition, the probability
in quantum mechanics also appears during K-projection with random projection
angle $\theta$. Now, the density of group-changing elements $\frac{1}{N}%
\rho_{\mathrm{particle}}\ $is just the probability to find a zero in a given
region $\psi^{\ast}(x,t)\psi(x,t)\Delta V.$

Finally, we discuss the emergent probability near event horizon in quantum mechanics.

It was known that the size of an elementary particle turns to infinite. That
means the local detection measures single group-changing element rather than
the whole elementary particle (or a group of group-changing elements). Because
each space-changing elements has the same probability (that is $\frac{1}{N}$)
to find an elementary particle, due to the extremely amplification effect,
quantum fluctuations become "\emph{randomness}"!

\subsection{Theory for event horizon of black hole}

In this section, we develop the theory for the event horizon of a black hole.

\subsubsection{Non-variability and randomness of event horizon}

On the event horizon, the changing rate along tempo direction is exact zero.
So, when we do a local operation $\hat{U}(\delta \phi^{t}(x,t))=e^{i\delta
\phi^{t}(x,t)\Gamma^{t}},$ the group-changing space becomes globally shifting
without changing its size. As a result, the event horizon doesn't change any
more, i.e.,
\[
\hat{U}(\delta \phi^{t})\rightarrow1.
\]
In other words, event horizon is a very special \textquotedblleft%
\emph{non-changing}" structure. Therefore, non-variability of event horizon
indicates the phase angles of all group-changing elements of elementary
particles become random numbers, i.e., $\phi^{\mu}(x)\in \operatorname{rand}%
(0,k_{0}L\cdot2\pi).$ This is a characteristic of classical object. So, we say
that \emph{the event horizon is a classical object}.

The randomness from non-variability of event horizon is consistent to that for
particle's motion out of the black hole. The size of an elementary particle
turns to infinite near the event horizon. That means the local detection
measures single group-changing element rather than whole elementary particle
(or a group of group-changing elements). Because each space-changing elements
has the same probability (that is $\frac{1}{N}$) to find an elementary
particle, due to this extremely amplification effect, the effect of quantum
fluctuations become "randomness". As a result, \emph{the event horizon is an
classical object}.

\subsubsection{Stochastic variant}

To complete characterize the "randomness" of a black hole, we introduce the
concept of stochastic variant, i.e.,

\textit{Definition: A stochastic variant }$V_{\mathrm{\tilde{G},}d}[\Delta
\phi^{\mu},\Delta x^{\mu},k_{0}^{\mu}]$\textit{ is denoted by\ a stochastic
mapping between a d-dimensional group-changing space }$\mathrm{C}%
_{\mathrm{\tilde{G},}d}$\textit{ with total size }$\Delta \phi^{\mu}%
$\textit{\ and \textit{Cartesian }space }$\mathrm{C}_{d}$\textit{\ with total
size }$\Delta x^{\mu}$\textit{, i.e.,}%
\begin{align}
&  V_{\mathrm{\tilde{G},}d}[\Delta \phi^{\mu},\Delta x^{\mu},k_{0}^{\mu
}]\nonumber \\
&  :\mathrm{C}_{\mathrm{\tilde{G},}d}=\{ \delta \phi^{\mu}\}
\Longleftrightarrow \mathrm{C}_{d}=\{ \delta x^{\mu}\}
\end{align}
\textit{where }$\Longleftrightarrow$\textit{\ denotes disordered mapping under
randomized changing rate of integer multiple }$k_{0}^{\mu}$\textit{.\ }%
$\delta \phi^{\mu}$\textit{ denotes group-changing element along }$\mu
$\textit{-direction (or element of group-changing space along }$\mu
$-direction\textit{) rather than group element (or element of group). The
total sizes of variant }$\Delta \phi^{\mu}$\textit{ is fixed as topological
invariables. In particular, the changing rates }$k_{0}^{\mu}$ are random
values. \textit{ }

We take 1D example $V_{\mathrm{\tilde{U}(1),}1}[\Delta \phi,\Delta x,k_{0}]$ to
show stochastic variant.

According to above definition,\ for a 1D stochastic variant $V_{\mathrm{\tilde
{U}(1),}1}[\Delta \phi,\Delta x,k_{0}],$ we have%
\begin{equation}
\delta \phi_{i}=k_{0}n_{i}\delta x_{i}%
\end{equation}
where $k_{0}$ is a constant real number and $n_{i}$ is a random integer
number. $k_{0}n_{i}$ is changing rate for $i$-th space element, i.e.,
$k_{0}n_{i}=\delta \phi_{i}/\delta x_{i}$. Under the mapping, each of the
infinitesimal element of $\mathrm{C}_{\mathrm{\tilde{U}(1)},1}(\Delta \phi)$ is
marked by a given position $x_{i}$ in 1D Cartesian space $\mathrm{C}_{1},$
i.e., $\delta \phi_{i}\rightarrow \delta \phi_{i}(x_{i})$ or $n_{i}\rightarrow
n_{i}(x_{i})$. As a result, in some sense, a stochastic variant can be
described by random distribution of $n_{i}.$

For higher dimensional stochastic variants, an infinitesimal element of
group-changing space has $d$ components. Because the randomly changings of
changing rate, i.e., $\frac{\delta \phi^{\mu}}{\delta x^{\mu}}=nk_{0}^{\mu}$
where $n$ is a random integer number, we have $d$ series of random numbers of
infinitesimal elements, i.e.,
\begin{align}
V_{\mathrm{\tilde{G},}d}[\Delta \phi^{\mu},\Delta x^{\mu},k_{0}^{\mu}]  &
:\{n_{i}^{\mu}\} \nonumber \\
&  =(...n_{1}^{\mu},n_{2}^{\mu},n_{3}^{\mu},n_{4}^{\mu},n_{5}^{\mu},n_{6}%
^{\mu},...).
\end{align}

In summary, event horizon of black hole becomes an example of 2D stochastic
variant in (3+1)D spacetime.

\subsubsection{Ensemble and statistics of quantum spacetime}

To characterize the physical property of a stochastic variant for event
horizon, we introduce the \emph{statistical ensemble} of a black hole. For
microcanonical ensemble of black hole, the key point is \emph{microcanonical
partition function} (MPF).

In general, we can consider a microcanonical ensemble of a lot of black holes
described by the same Schwarzschild solution. In thermodynamic limit (the
total energy $E$ and the area $S$ turn to infinite with fixed $E/S$), we have
the rule of a new quantum statistical theory for event horizon.

For the microcanonical ensemble, one has to calculate the MPF which is usually
defined as the number of states with a definite value $E$ of total energy:
\begin{equation}
\Omega \equiv \sum_{\mathrm{states}}\delta(E-E_{\mathrm{state}}).
\end{equation}
For a quantum system, the MPF is the trace of the operator $\delta(E-\hat{H}%
)$:
\begin{equation}
\Omega=\mathrm{tr}\delta(E-\hat{H}) \label{mpf0}%
\end{equation}
with proper normalization of the basis states.

For instance, for one non-relativistic free particle, one has to calculate the
trace summing over plane waves normalized:
\begin{equation}
\Omega=\mathrm{tr}(E-\hat{H})=\sum_{\mathbf{p}}\delta \left(  E-\frac
{\mathrm{p}^{2}}{2m}\right)  \langle \mathbf{p}|\mathbf{p}\rangle.
\end{equation}
Thereby, one recovers the well known classical expression implying that the
MPF is the number of phase space cells with size $h^{3}$ and given energy $E$.
In the thermodynamic limit $E\rightarrow \infty$ and $V\rightarrow \infty$, by
replacing the sum over discrete levels with a phase space integration
$\sum_{\mathrm{cells}}\underset{V\rightarrow \infty}{\longrightarrow}\frac
{V}{(2\pi)^{3}}\int \mathrm{d}^{3}\mathrm{p},$ we have $\Omega=\frac{1}%
{(2\pi)^{3}}\int \mathrm{d}^{3}\mathrm{x}\int \mathrm{d}^{3}\mathrm{p}\,
\delta \left(  E-\frac{\mathrm{p}^{2}}{2m}\right)  .$ The phase space cells
with size $h^{3}$ becomes hidden.

Let us use similar assumption by considering the cells of space with size
$h^{2}$ for event horizon of a black hole. That is just the cell of (3-1)
dimensional real zero lattice of event horizon of a uniform physical variant
without considering randomness.

Now, on each cell unit of space, we have an area $l_{0}^{2}$. Fig.16(a) show
an event horizon with a uniform distribution of quantized fluxes. Because the
total size of the event horizon is topological invariable, the total 2-volume
(or area) is also topological invariable. However, the statistics of space
doesn't obey usual fermionic statistics due to violating the condition of
perturbative uniform variant and becomes a new one. Let us explore the new formula.

Now, we have $N_{U}$ unit cells. According to an assumption of the stochastic
variant, the $N_{U}$ unit cells have a randomized distribution on these
original $N_{U}$ unit cells with fixed number. As a result, the statistics of
spacetime for event horizon is given by the following MPF, i.e.,
\[
\Omega=\frac{(N_{U})^{N_{U}}}{(N_{U})!}.
\]
We\quad call this quantum statistics to be \emph{spacetime statistics} to
distinguish usual Fermi-Dirac statistics, Bose-Einsten statistics and
Boltzmann statistics. Different quantum states that correspond to different
geometric structures of the event horizon have the same probability. This is
\emph{Principle of equal probability} for spacetime! Now, the coordinates and
wave vectors (momentums) on event horizon become fluctuating. The definition
of usual spacetime on event horizon becomes invalid.

\paragraph{Entropy and temperature}

Firstly, we try to obtain the \emph{entropy} of a black hole.

In thermodynamic limit $N_{U}\rightarrow \infty$, according to spacetime
statistics, we have the entropy $S_{A}$ to be%
\begin{align*}
S_{A}  &  =k_{B}\ln \Omega=k_{B}\ln(\frac{(N_{U})^{N_{U}}}{(N_{U})!})\\
&  \simeq k_{B}N_{U}+\frac{1}{2}k_{B}\ln(2\pi N_{U})\\
&  \simeq k_{B}N_{U}.
\end{align*}
In thermodynamic limit, the formula of entropy $S_{A}$ of a black hole is
obtained as%
\begin{equation}
S_{A}\simeq k_{B}N_{U}=k_{B}{\frac{S}{l_{0}^{2}},}\text{ }l_{0}=2l_{p}{.}%
\end{equation}
This is just the Bekenstein-Hawking formula of black hole entropy\cite{haw}.

\begin{figure}[ptb]
\includegraphics[clip,width=0.8\textwidth]{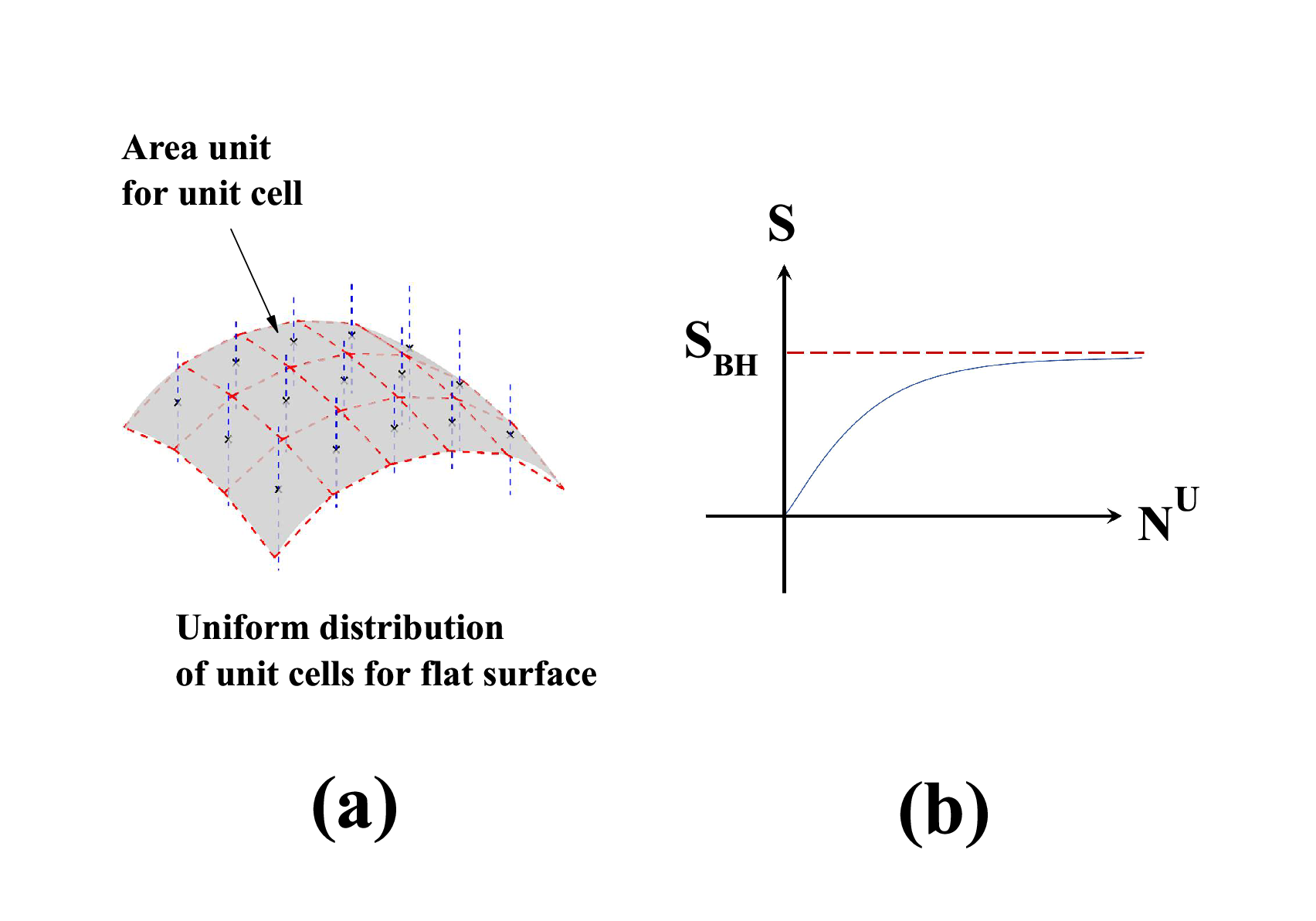}\caption{(a) Event
horizon with a uniform distribution of quantized fluxes; (b) The entropy
$S_{A}$ of a black hole via $N^{U}.$ In thermodynamic limit $N^{U}%
\rightarrow \infty$, the result becomes the Bekenstein-Hawking formula of black
hole entropy.}%
\end{figure}

An important physical quantity is \emph{temperature}.

To derive the value of temperature, we variate the total energy $E$ of the
black hole by its entropy $S_{A}\simeq k_{B}N_{U}=k_{B}{\frac{S}{l_{0}^{2}}}$,
and have
\begin{align*}
T  &  =\frac{\delta E}{\delta S}=c^{2}\frac{\delta M}{\delta S}\\
&  =c^{2}(\frac{\delta S}{\delta M})^{-1}=c^{2}(\frac{8\pi k_{B}GM}{hc}%
)^{-1}\\
&  =\frac{hc^{3}}{8\pi k_{B}GM}.
\end{align*}
Here, we have used $S=4\pi r_{s}^{2}=\frac{16\pi G^{2}M^{2}}{c^{4}}$. This
result is consistent to Hawking temperature without surprising.

Therefore, we have usual Boltzmann distribution for the black holes,
\begin{equation}
f_{m}=\frac{e^{-\beta E}}{Z} \label{eq: star}%
\end{equation}
where the partition function is $Z=\sum_{E}e^{-\beta E}.$

In the end of this part, we point out that temperature and thermalization
effect of a black hole are emergent phenomena in the limit of $N^{U}%
\rightarrow \infty.$ In other words, \emph{more is difference.} According to
the statistics of spacetime $S_{A}=k_{B}\ln \Omega=k_{B}\ln(\frac
{(N_{U})^{N_{U}}}{(N_{U})!}),$ for the case of $N_{U}=1,$ the concepts of
"temperature" and "thermalization" are misleading. As a result, for a 1+1
dimensional black hole with $N_{U}=1,$ there doesn't exist the concept of
Hawking temperature or Hawking radiation at all. This issue will be addressed
again in following parts.

\paragraph{Thermal fluctuations for black hole}

Due to finite temperature, black hole becomes thermally fluctuating. In this
part, we discuss the property of thermal fluctuations for black hole.

In statistical theory, the thermal fluctuation for physical quantity $A$ is
characterized by mean squared deviation,
\[
\left \langle (\Delta A)^{2}\right \rangle =\left \langle (A)^{2}\right \rangle
-(\left \langle A\right \rangle )^{2}.
\]
For example, we consider the thermal fluctuation for total energy $E$. Then,
we have
\[
\left \langle (\Delta E)^{2}\right \rangle \sim \frac{1}{N_{U}}.
\]
Therefore, the thermal fluctuation for the Schwarzschild radius
$r_{\mathrm{Schw}}$ is obtained as
\[
\left \langle (\Delta r_{\mathrm{Schw}})^{2}\right \rangle \sim \frac{1}{N_{U}}.
\]
By using similar approach, one can calculate other physical quantities. Here,
$\langle \cdots \rangle_{\beta}$ denotes the averaging over the thermal
distribution together with the quantum averaging:
\begin{equation}
\langle \cdots \rangle_{\beta}\equiv \frac{\text{tr}\left[  e^{-\beta E}%
\cdots \right]  }{\text{tr}\left[  e^{-\beta E}\right]  }.
\end{equation}

Next, we write down the probability distribution of a given physical quantity
$\rho(A)$.

In general, we consider the case of the thermodynamic limit $N_{U}%
\rightarrow \infty.$ Now, under the assumption of the maximum entropy
principle, the probability distribution of a given physical quantity $\rho(A)$
is always described by usual Gaussian distribution, i.e.,
\[
\rho(A)=\frac{1}{\sqrt{2\pi A^{2}}}\exp(-\frac{\left \langle (\Delta
A)^{2}\right \rangle }{2\left \langle A\right \rangle ^{2}}).
\]
For example, for the number of unit cells on event horizon $N_{U}$, we have
\[
\rho(N_{U})=\frac{1}{\sqrt{2\pi \left \langle N_{U}\right \rangle ^{2}}}%
\exp(-\frac{(\Delta N_{U})^{2}}{2\left \langle N_{U}\right \rangle ^{2}}).
\]
Because the number of unit cell is proportional to the area of the event
horizon, the area $S$ has similar probability distribution.

\paragraph{Hawking radiation effect and the possible Hartle-Hawking state}

In this part, we discuss Hawking radiation effect and the possible
Hartle-Hawking state.

If we complexify this time coordinate by $t\rightarrow i\tau$, we obtain the
Euclidean metric
\begin{align}
ds^{2}  &  =\left(  1-\frac{2M}{r}\right)  d\tau^{2}+\left(  1-\frac{2M}%
{r}\right)  ^{-1}dr^{2}\nonumber \\
&  +r^{2}(d\theta^{2}+\sin^{2}\theta d\phi^{2}).
\end{align}
In this metric, $r=2M$ is an origin in the $r$, $\tau$ plane. The spacetime is
smooth there if $\tau$ is an angular coordinate with period $\beta=2\pi
/\kappa$ where $\kappa=1/4M$ is the black hole's surface gravity. That becomes
the Hartle-Hawking state, a thermal state at temperature $T=\kappa/2\pi=1/8\pi
M$\cite{har}. Therefore, on event horizon, without tempo changing rate, we
have an \emph{imaginary coordinate of time} with periodic boundary condition.

Another fact about a black hole is \emph{nonequilibrium state}. For the
Schwarzschild black hole, its specific heat is negative, i.e., $C_{V}=\partial
M/\partial T<0$. A black hole will emit thermal radiation at late times ---
the true \emph{Hawking radiation effect}. Thus, if the mass fluctuates
downwards, the temperature rises, and the black hole will radiate more than it
absorbs from the thermal bath, further lowering its mass. So this equilibrium
state for Schwarzschild is un-physical; real black holes will never reach this equilibrium.

\subsubsection{Information properties of black hole}

In this part, we discuss the information properties of black hole and solve
black hole information paradox.

The randomness from non-variability of event horizon leads to thermalization
and decoherence of the quantum states near event horizon. The event horizon
can be regarded as a classical object with finite temperature. When a quantum
object reaches the classical object, quantum measurement occurs. Therefore,
there exists "wave-function collapse" during measurement process that
corresponds to \textrm{R}-process. The original quantum object melts and
becomes part of the black hole. Therefore, the quantum information disappear
and a pure quantum state evolves to a mixed state. Hence, the
\textquotedblleft \textit{black hole information paradox}\textquotedblright \ is
completely solved. This indicates usual quantum mechanics becomes invalid near
event horizon!

Finally, we give a comment on the result about \emph{Page curve} for Hawking
evaporation process.

It was known that an isolated black hole will \textquotedblleft
evaporate\textquotedblright \ completely via the Hawking process within a
finite (but very long) time. If black hole evaporation is a unitary process,
the entanglement entropy between the outgoing radiation and the quantum state
associated to the remaining black hole is characterized by Page
curve\cite{page}. At the beginning, the entanglement entropy monotonically
increases via time which comes from the coarse grained thermal entropy of the
radiation that has been emitted up to that point. When the coarse grained
entropy of the radiation exceeds the coarse grained entropy of the remaining
black hole, the black hole's entropy becomes a decreasing function of time.
The time when the entanglement entropy transitions from increasing to
decreasing is called to be Page time. If one can reproduce the Page curve
without explicitly assuming unitary, Hawking's black hole information paradox
is then solved and the information doesn't loss.

Recently, Page curve was indeed obtained by using semi-classical methods for
black holes in asymptotically AdS spacetime coupled to a CFT reservoir. The
result is related to the Ryu-Takayanagi formula \cite{ryu} and the possible
extremal hypersurfaces terminating on so-called islands behind the event
horizon\cite{alm}. For eternal AdS black holes, with the islands extended
outside the horizon, one may derive the curve as predicted by Page. We
emphasize that these results are always obtained based on 1+1 dimensional
dilaton gravity\cite{alm}.

Our result shows that the final state of a black hole is always a mixed state,
i.e., \textquotedblleft information\textquotedblright \ will lost. In general,
the entropy of final state is maximum. What's wrong about above theoretical
results? The key point is \emph{theoretical reliability of quantum mechanics
inside black hole}. Our answer is that for a black hole, traditional quantum
mechanics \emph{fails}. The results for derive Page curve based on usual
quantum mechanics are all not reliable. To correctly answer this question, we
must seek help from theory of physical variant.

According to above discussion, inside black hole, the usual Hermitian quantum
mechanics is invalid. Instead, to characterize the dynamical processes inside
a black hole, one must use non-Hermitian quantum mechanics. On the other hand,
on the event horizon of a black hole, quantum mechanics is also invalid.
Hence, page curve cannot characterize the information process for Hawking
evaporation of black hole. In addition, in above part, we had show that for a
1+1 dimensional black hole with $N_{U}=1,$ there doesn't exist finite Hawking
temperature or the phenomenon of Hawking radiation. Therefore, the
calculations based on 1+1 dimensional dilaton gravity cannot be applied to
explain the information process of higher dimensional black holes. And, there
doesn't "islands" behind event horizon at all.

\subsection{Theory for spacetime inside black hole}

In this section, we develop the theory to characterize the spacetime inside a
black hole.

\subsubsection{Dynamical theory}

\paragraph{Theory for AdS}

Inside the event horizon, due to $1-\frac{2GM}{rc^{2}}>0,$ we have a
(deformed) non-unitary physical variant. By using Eddington-Finkelstein
coordinates for Schwarzschild solution, $ds^{2}=-\left \vert 1-\frac
{2GM}{rc^{2}}\right \vert (c^{2}dt^{2}-dr_{\ast}^{2})-r^{2}d\Omega^{2}$ with
$r_{\ast}=r+\frac{2GM}{c^{2}}\log \left \vert \frac{r-2GM/c^{2}}{2GM/c^{2}%
}\right \vert $ and $d\Omega^{2}=d\theta^{2}+\sin^{2}\theta d\phi^{2},$ the
spatial-tempo variability is determined by the following equation,
$\mathcal{T}(\delta x^{\mu})\leftrightarrow e^{i\cdot k_{0}\delta x^{\mu
}(x^{\mu})\Gamma^{\mu}}$\textit{.} In particular, along radial direction and
tempo direction, we have $k_{0}=\pm i\left \vert k_{0}\right \vert $ and
$\omega_{0}=\pm i\left \vert \omega_{0}\right \vert $ that indicates a deformed
non-unitary transformation.

With help of complex coordinates $x^{\mu}\rightarrow \tilde{x}^{\mu}=\pm
ix^{\mu}$, we derive the geometry representation for curved AdS that are same
to those out of the black hole. Based on geometry representation under
D-projection and K-projection, a deformed non-unitary physical variant is
reduced into a deformed complex zero lattice.

We also assume that each zero corresponds to an elementary particle and
becomes the information unit for the system of "changings". Each elementary
particle corresponds to an zero with $\pi$-phase changing along an arbitrary
direction on the complex zero lattice. The effective action is
\begin{align}
S  &  =\int \sqrt{-g(\tilde{x})}\bar{\Psi}(e_{a}^{\mu}\gamma^{a}\hat{D}_{\mu
}-m)\Psi \text{ }d^{4}\tilde{x}\nonumber \\
&  +\frac{1}{16\pi G}\int \sqrt{-g}\tilde{R}\text{ }d^{4}\tilde{x}.\nonumber
\end{align}

Under kinetic representation, we replace the complex coordinates $\tilde
{x}^{\mu}=e^{i\varphi^{\mu}}\cdot x^{\mu}$ by the real coordinates $x\ $and
replace the real changing rate by the complex one,
\[
k_{0}\rightarrow \tilde{k}_{0}^{\mu}=e^{i\varphi^{\mu}}\cdot k_{0}.
\]
Now, Gamma matrices $\Gamma^{\mu}$ are Hermitian.

Along $\mu$-th ($\mu \neq r,t$) directions, the matter comes from the phase
changings; while Along $\mu$-th ($\mu=r,t$) directions, the matter comes from
amplitude changings. As a result, along the direction with real changing rate,
the elementary particle becomes a unitary zero changing phase $e^{i\pi}$ and
obey usual fermionic statistics; along the direction with imaginary changing
rate, the elementary particle becomes a non-unitary zero changing amplitude
$e^{i\pi \cdot e^{i\varphi}}$ ($\varphi=\frac{\pi}{2}$) and obey non-Hermitian
fermionic statistics.

The effective non-Hermitian Hamiltonian for elementary particles on spacetime
with fully real coordinates is written as
\[
\mathcal{H}=\int(\bar{\Psi}^{\dagger}(\mathbf{x})\hat{H}\Psi(\mathbf{x}%
))d^{3}x
\]
where $\hat{H}=\Gamma \cdot \Delta \tilde{p}+m\Gamma^{t}$ with $\Delta \tilde
{p}^{\mu}=\hbar \Delta \tilde{k}^{\mu}=\hbar(k^{x},k^{y},ik^{z})$.
$\Psi^{\dagger}(\mathbf{x})$ denotes the generalized creation operation for
non-Hermitian elementary particle, of which the amplitude changes $e^{\pi}$
along radial and tempo directions and phase changes $e^{i\pi}$ along other directions.

Near the center of the black hole (or singularity), the metric becomes
defective, i.e.,
\begin{align*}
ds^{2}  &  =-\left \vert 1-\frac{2GM}{rc^{2}}\right \vert (c^{2}dt^{2}-dr_{\ast
}^{2})-r^{2}d\Omega^{2}\\
&  \rightarrow \frac{2GM}{rc^{2}}(c^{2}dt^{2}-dr_{\ast}^{2}).
\end{align*}
By solving the zero equation, we find that the lattice constants of the
complex zero lattice along radial and tempo directions turns to zero, i.e.,
\[
\Delta r_{\ast}=\frac{2GM}{rc^{2}}l_{0}\rightarrow0,\text{ }c\Delta
t=\frac{2GM}{rc^{2}}l_{0}\rightarrow0
\]
This leads to divergence of curvature and called singularity puzzle of spacetime.

To solve this puzzle, the key point is to be aware of the \emph{imaginary}
nature of the coordinates along radial direction.

When we transform the imaginary coordinate to real one, we get a non-Hermitian
matrix network and the curvature becomes imaginary. In particular, the
non-uniform non-unitary variability along radial direction is described by
\begin{align*}
\hat{U}(\delta r)  &  =e^{k_{0}r_{\ast}\Gamma^{r}}\\
&  =\exp(k_{0}(r+\frac{2GM}{c^{2}}\log \left \vert \frac{r-2GM/c^{2}}{2GM/c^{2}%
}\right \vert )\Gamma^{r})\\
&  =\Gamma^{r}\left \vert \frac{r-2GM/c^{2}}{2GM/c^{2}}\right \vert
^{\frac{2GMk_{0}}{c^{2}}}\exp(k_{0}r\Gamma^{r}).
\end{align*}
$\hat{U}(\delta r)$ can be considered as a non-unitary operation on the
elementary particles and changes the weight of elementary particles. The
weight is $0$ at $r=r_{\mathrm{Schw}}$ and becomes maximum at $r=0.$ In
particular, near the center of the black hole $r=0,$ instead of the existence
of singularity, we have a usual non-unitary transformation $\hat{U}(\delta
\phi^{r})\sim \exp(k_{0}r\Gamma^{r}),$ $r\rightarrow0$.\

In addition, the coordinates along tempo direction are also imaginary. Under
time evolution, there appears additional non-Hermitian polarization effect
under matrix $\Gamma^{t}$.

Another relevant issue is \emph{cosmic censorship hypothesis}\cite{cos}. The
cosmic censorship hypothesis guarantees that any spacetime singularity will be
surrounded by the event horizon. If this cosmic censorship hypothesis is
correct, all singularity occurs in a spacetime with imaginary coordinates (or
in AdS). For an observer in a spacetime with real coordinates, there must
exist an event horizon (a topological defect of physical variants) around the
singularity. However, if the singularity occurs in a spacetime with real
coordinates, cosmic censorship hypothesis is incorrect. For this case, the
change rates along certain directions turns to infinite. The theory based on
physical variant is invalidity.

\paragraph{Theory for CFT}

In this part, we develop the theory for the inside structure of a black hole
under real K-projection.

Inside the event horizon $1-\frac{2GM}{rc^{2}}>0,$ we do real K-projection and
have real zero lattice. Along $i$-th spatial direction of the real zero
lattice, the lattice site is labeled by $N^{i}.$ Along radial direction or
tempo direction, there doesn't exist zero lattice at all. Therefore, we get a
2D zero lattice with real lattice number and Hermitian $\Gamma^{\mu}$. This 2D
zero lattice plays the role of quantum spacetime of approaching event horizon,
of which the external normal lines are denoted by $\Gamma^{r_{\ast}}.$ If we
consider $\Gamma^{r_{\ast}}$ to be a fixed, constant Gamma matrix, the
corresponding spacetime must be flat and cannot be curved. This results the
theory of CFT.

Under geometry representation on real zero lattice, we also assume that each
zero corresponds to an elementary particle and becomes the information unit
for the system of "changings". Each elementary particle corresponds to an zero
with $\pi$-phase changing along different directions on the boundary of the
black hole. Therefore, these elementary particles obey fermionic statistics.

However, along radial direction, the total size of non-unitary group-changing
space about the elementary particle is same to the radius $L_{r}$ of the black
hole. Now, each zero of real zero lattice corresponds to $L_{r}/l_{0}$ zeroes
of complex zero lattice, that is the lattice number along radial direction
with imaginary lattice number. On the other hand, if the total mass of the
black hole is $M$ and the number of real zeroes is $N$, the elementary
particle corresponding to each real zero has a large mass to be $m_{R}=M/N$.

Along radial direction, the non-unitary variability $\hat{U}(\delta \phi
^{r})=e^{k_{0}r_{\ast}\Gamma^{r}}=\left \vert \frac{r-2GM/c^{2}}{2GM/c^{2}%
}\right \vert ^{\frac{2GMk_{0}}{c^{2}}}\exp(k_{0}r\Gamma^{r})$ can be also
considered as a global non-unitary operation on the real zero and changes the
weight of elementary particles. Then, we derive the global non-unitary
operation,
\[
\hat{U}_{\mathrm{global}}=\exp(\frac{1}{l_{0}}%
{\displaystyle \int}
i(k_{0}r_{\ast}\Gamma^{r})dr_{\ast})=\exp(\frac{L_{r^{\ast}}^{2}}{2l_{0}^{2}%
}\Gamma^{r})).
\]
In the limit of $r_{\ast}/l_{0}\rightarrow \infty,$ due to $L_{r^{\ast}%
}\rightarrow \infty,$ the amplitude of eigenstates with positive elgenvalues of
$\Gamma^{r}$ diverge while the amplitude of eigenstates with negative
elgenvalues of $\Gamma^{r}$ turns to zero. The degrees of freedom for the real
zero becomes fully polarized on the boundary and for each real, its quantum
states are at EPs. By introducing global non-unitary operation on a real zero,
the role of singularity becomes less important.

Next issue is about the \emph{geometric} property for elementary particles (or
real zeroes).

The quantum spacetime for real zero lattices of black hole is always flat. The
elementary particles have trivial geometric property, i.e., the area of each
elementary particle in CFT is proportional to $l_{0}^{2}/4$. So, we can study
the geometric property of elementary particles on boundary of AdS by AdS/CFT
correspondence\cite{ads}. The surface $\mathcal{S}$ is defined as the boundary
of the black hole, of which the external normal direction is $\Gamma^{r}.$

Finally, we discuss the \emph{motion} inside the black hole.

There are two types of motions, one is about the fast motion of elementary
particles, the other is about slow motion as the residue effect of
gravitational waves on the boundary of the black hole.

The fast motion is described by the following effective Hamiltonian
\[
\mathcal{H}_{(3-1)}^{\mathrm{fast}}=\int(\Psi^{\dagger}(\mathbf{x})\hat
{H}_{(3-1)+1}\Psi(\mathbf{x}))d^{2}x
\]
where $\hat{H}_{(3-1)+1}=\vec{\Gamma}\cdot \Delta \vec{p}+m_{R}\Gamma^{t}$
($m_{R}=mL_{r}/l_{0}$). According to above Hamiltonian, for the case of fast
motion of an elementary particle, the energy is $\pm \sqrt{\left \vert
\Delta \vec{p}\right \vert ^{2}+m_{R}^{2}}.$ In the thermodynamic limit
$L_{r}\rightarrow \infty$, the mass turns to infinite, i.e., $m_{R}=L_{r}%
/l_{0}m\rightarrow \infty.$ The quantum processes for fast motion of elementary
particles are irrelevant to low energy physics.

The slow motion is described by the following effective Hamiltonian%
\[
H_{(3-1)+1}^{\mathrm{slow}}=%
{\displaystyle \sum \nolimits_{\mu \neq d}}
ck^{\mu}\Gamma^{\mu}.
\]
Now, the Gamma matrices become fluctuating. The energy is given by $\left \vert
ck^{\mu}\right \vert .$ However, in next section, we point out that this is
incorrect! Due to randomness of the event horizon, the true CFT comes from the
boundary of (1+1)D Euclidean AdS rather than the usual boundary of (3+1)D AdS
of black hole. In the following parts, we will give detailed discussion.

\paragraph{Theory for non-Hermitian gauge theory}

In this part, we use non-Hermitian gauge theory to characterize the inner
spacetime of black hole.

Because the spacetime inside black hole is AdS. For the $\mathrm{\tilde
{S}\tilde{O}(3+1)}$ non-unitary physical variant $V_{\mathrm{\tilde{S}%
\tilde{O}(3+1)},3+1}(\Delta \phi^{\mu},\Delta x^{\mu},k_{0},\omega_{0})$, the
representation of ((3-1)+1)-dimensional non-Hermitian gauge theory (NGT) on
flat spacetime is equivalence to the representation of (3+1)-dimensional AdS.
We then use Gravity/N-gauge equivalence to characterize its structure. In AdS,
slow motion is described by quantum fluctuations of gravitational waves; in
NGT, the slow motion is described by quantum fluctuations of non-Hermitian
\textrm{U(0,1)}$\times$\textrm{SU(0,N)} gauge fields.

Now, for the slow motion, the effective Hamiltonian is%
\begin{align*}
H_{(d-1)+1}^{\mathrm{slow}}  &  =\vec{\Gamma}\cdot(\mathrm{e}\vec{A}%
_{U(0,1)}+g\mathcal{\vec{A}})\\
&  +\Gamma^{t}(\mathrm{e}A_{t,U(0,1)}+g\mathcal{A}_{t})
\end{align*}
where $A_{\mu,U(0,1)}$ and $\mathcal{A}_{\mu}$ are the non-Hermitian
\textrm{U(0,1)} gauge fields\textrm{ }and non-Hermitian \textrm{SU(0,N)} gauge
fields, respectively. Due to gapless nature of fluctuations of non-Hermitian
\textrm{U(0,1)}$\times$\textrm{SU(0,N)} gauge fields, the excitation is gapless.

In particular, when we reduce the NGT to the unitary physical processes of the
system, AdS/NGT equivalence is reduced to usual AdS/CFT correspondence between
the theory for boundary of AdS and CFT. Because the low energy degrees of
freedom is dominated by gapless gravitational waves on the boundary of the AdS
(that is approaching the event horizon infinitely), it is described by
fluctuations from non-unitary \textrm{U(0,1)} Abelian gauge field
$A_{\mu,U(0,1)}.$

\subsubsection{Thermodynamics theory}

In this section, we develop the thermodynamics theory for the black hole
inside event horizon. We focus on the slow motion induced by gravitational
waves on the event horizon.

\paragraph{Euclidean physical variant}

In this part, we show that there exists an \emph{Euclidean physical variant}
with emergent variability on imaginary time, $t\rightarrow it=\tau$. Let give
a detailed discussion on this issue.

It was known that near event horizon, the changing rate along tempo direction
becomes disappear. This fact leads to randomness of the event horizon and the
temperature becomes finite, $T\neq0$ or $\hbar \beta \neq0$ ($\beta=\frac
{1}{k_{B}T}$). In particular, we assume that the\ temperature of the black
hole inside even horizon is also $T.$ For a usual quantum system with finite
temperature, we have weight changings for different quantum states,
\[
\left \vert \Psi \right \rangle _{n}\rightarrow e^{-\beta E_{n}}\left \vert
\Psi \right \rangle _{n}=e^{-i\Delta \tau E_{n}}\left \vert \Psi \right \rangle
_{n}.
\]
This leads to uniform phase changing along imaginary tempo direction. As a
result, Euclidean physical variant emerge. Let us show it.

The total metric of the black hole inside event horizon can be regarded as the
sum of 1+1 dimensional Euclidean AdS $ds_{\mathrm{slow}}^{2}$ for slow
variables and the others (or $S_{2}$) for fast variables $d\Omega
_{\mathrm{fast}}^{2}$, i.e.,
\begin{align}
ds^{2}  &  \approx ds_{\mathrm{slow}}^{2}+d\Omega_{\mathrm{fast}}^{2}\\
&  =h_{ij}(x^{0},x^{1})dx^{i}dx^{j}+\Phi^{2}(x^{0},x^{1})d\Omega^{2}%
\end{align}
where $i,j=0,1,$ $x^{0}=\tau,$ $x^{1}=r.$ When we reduce it to a 1+1
dimensional AdS, Jackiw-Teitelboim gravity emerges. In particular, we derive
this metric by splitting the fast/slow variables rather than introducing
un-physical fine-tuned "\emph{magnetic charges}" in extremal black hole,
$E=M-\frac{Q}{l_{p}}=0$.

A (1+1) dimensional Euclidean physical variant has higher order variability.

Along radial direction, the local spatial variability is non-unitary
\begin{equation}
\mathcal{T}(\delta x^{r})\leftrightarrow \hat{U}(\delta \phi^{r}),
\end{equation}
where $\hat{U}(\delta \phi^{r})=e^{i\cdot \delta \phi^{r}\Gamma^{r}}$ and
$\delta \phi^{r}=\pm i\left \vert k_{0}\delta x^{r}\right \vert $. Along
imaginary tempo direction unitary, we have unitary variability,
\begin{equation}
\mathcal{T}(\delta x^{\tau})\leftrightarrow \hat{U}(\delta \phi^{\tau}),
\end{equation}
where $\hat{U}(\delta \phi^{\tau})=e^{i\cdot \delta \phi^{\tau}\Gamma^{\tau}}$
and $\delta \phi^{\tau}=E\delta \tau$. Here, the energy $E$ is the total energy
of the black hole and the size along the imaginary time is $\beta \hbar.$ In
particular, there exists $\beta E/2\pi$ zeroes along imaginary time direction.

For the Euclidean physical variant, there are two types of motion, one for the
fast motion for the real zeroes that characterizes the expansion and
contraction of the event horizon, the other for slow motion from boundary
gravitational waves that characterizes the shape changing of the event
horizon. Due to very large mass, the degrees of freedom of particles with fast
motion can be regarded as fast variables. The slow motion from random boundary
gravitational waves that characterize the fluctuations of the shape of event
horizon are slow variables. If we focus on the dynamics of shape changings of
event horizon, we integrate fast variables and get effective model. The
effective model has three equivalent forms: one is effective Jackiw-Teitelboim
gravity\cite{jt} under geometric representation, second is effective SYK
model\cite{Kitaev-talks,ye} under matrix representation, third is effective 1D
gauge theory under kinetic representation.

\paragraph{Geometric representation and emergent Jackiw-Teitelboim gravity}

In this part, we discuss the dynamics of a black hole inside event horizon
based on effective Jackiw-Teitelboim gravity under geometric
representation\cite{jt}.

From the Euclidean physical variant, we may assume that the dynamics of the
complex zeroes of the boundary (the outermost side) of AdS is same to that on
event horizon. Then, the key point is to integrate the fast variables from the
fast motion for real zeroes that characterize the expansion and contraction of
the event horizon.

Then, we do complex K-projection.

Under complex K-projection, the (1+1) dimensional Euclidean physical variant
is reduced into a complex zero lattice, $x^{i}=[l_{0}\cdot N^{i}/2+\frac
{l_{0}^{i}}{2\pi}(\theta+\frac{\pi}{2})]e^{-i\varphi^{i}}$. Along $i$-th
spatial direction of the zero lattice, the lattice site is labeled by $N^{i}.$
In addition, we have a zero lattice along imaginary time direction. Along
$\tau$-th direction, due to $\varphi^{\tau}=0,$ we have
\[
\cos(E\cdot \tau/\hbar)=0,
\]
of which the lattice constant $\epsilon$ is $\frac{2\pi \hbar}{E}.$ On the
other hand, to characterize this (1+1) dimensional ((1+1)D) Euclidean physical
variant, we can also use real knot projection and get kinetic representation
(or CFT representation).

Under complex K-projection, in continuum limit, the (1+1)D Euclidean AdS is
described by curved spacetime, i.e.,%
\[
ds_{\mathrm{slow}}^{2}=h_{ij}(x^{0},x^{1})dx^{i}dx^{j}%
\]
($i,j=0,1$, $x^{0}=\tau$, $x^{1}=r$) that characterizes the (1+1)D Euclidean
physical variant. This is a spacetime with boundary. One can use Poincare
coordinates to characterize the (1+1)D Euclidean AdS by introducing the
variable $z$,
\begin{equation}
z=\frac{L^{2}}{r-r_{\mathrm{Schw}}},
\end{equation}
where $L$ is the radius of the (1+1)D Euclidean AdS and is proportional to the
number of imaginary zeroes inside a level-1 zero. The metric turns into
\begin{equation}
ds^{2}=\frac{L^{2}}{z^{2}}\left(  -dt^{2}+dz^{2}\right)  .
\end{equation}
\begin{figure}[ptb]
\includegraphics[clip,width=0.8\textwidth]{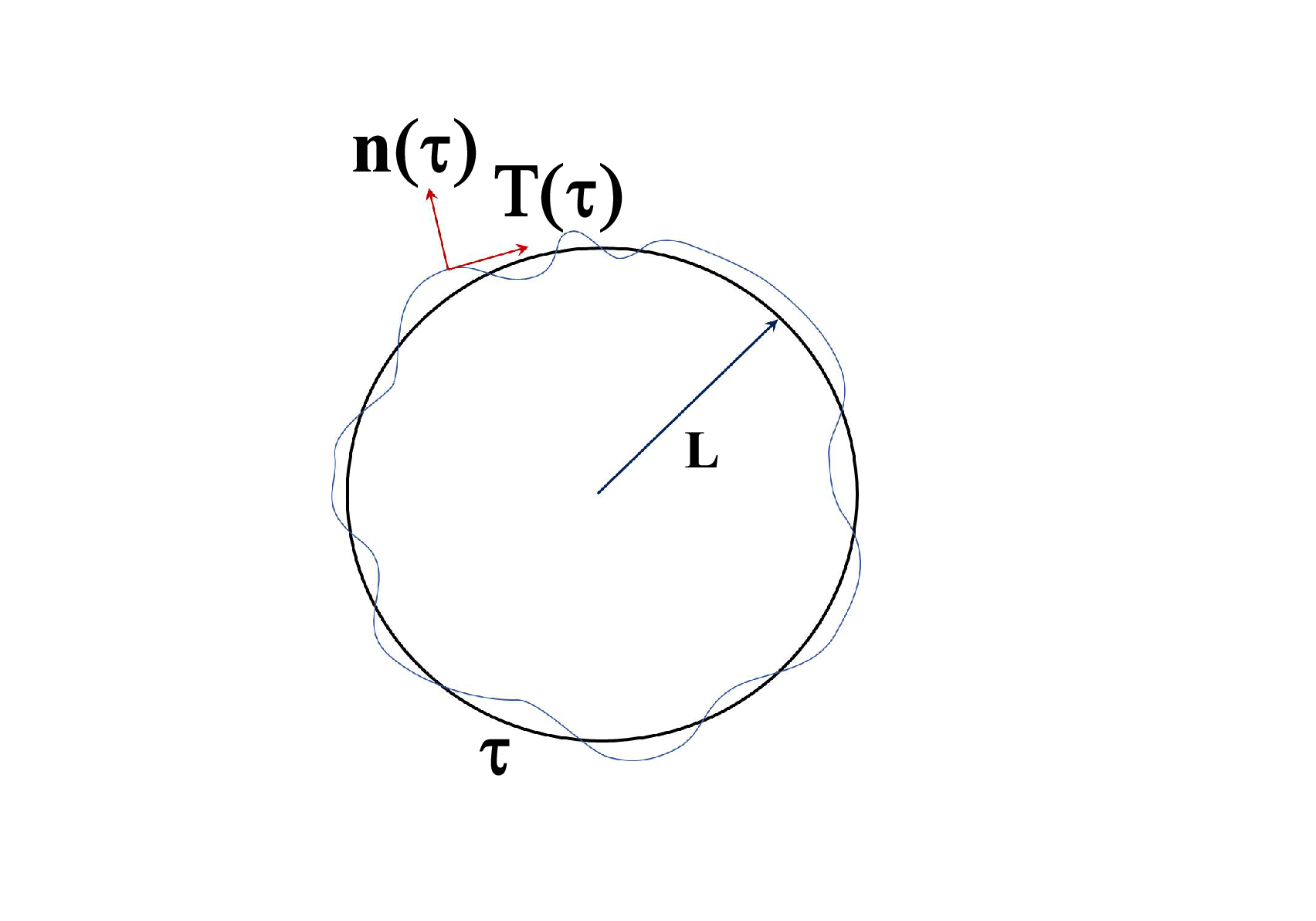}\caption{An illustration
of (1+1)D Euclidean AdS: The function $t(\tau)$ determines the boundary curve
along imaginary time $\tau$ and the shape of event horizon of the (1+1)D
Euclidean AdS. $\mathcal{T}$ and $n$ are tangent and unit normal vectors to
the boundary curve of 1+1 dimensional Euclidean AdS}%
\end{figure}

The fluctuations of total energy (or particle number) leads to the dynamics
for the changings of the imaginary time $\tau \rightarrow t(\tau)$. The
function $t(\tau)$ determines both the boundary curve along imaginary time
$\tau$ and the shape of event horizon of the (1+1)D Euclidean AdS. See the
illustration in Fig.17. To deal with the boundary, the value of the metric is
fixed to be
\begin{align}
ds  &  \mid_{\mathrm{bdy}}=\sqrt{\frac{ds^{2}}{d\tau^{2}}}d\tau \\
&  =\sqrt{\frac{(t^{\prime})^{2}+(z^{\prime})^{2}}{z^{2}}}d\tau,
\end{align}
which the cutoff $\epsilon$ is just the lattice constant $\frac{\pi}{E}$.
According to $\frac{(t^{\prime})^{2}+(z^{\prime})^{2}}{z^{2}}=\frac
{1}{\epsilon^{2}},\quad$we have$\quad$%
\[
z(\tau)=\epsilon t^{\prime}(\tau)+\mathcal{O}(\epsilon^{3})
\]
and
\begin{equation}
ds\mid_{\mathrm{bdy}}=\frac{d\tau}{\epsilon},
\end{equation}
The boundary metric is $g_{\tau \tau}=\frac{1}{\epsilon^{2}}$.\ The number of
zero lattice along imaginary time direction is
\begin{align}
N_{\tau}  &  =\int ds=\int_{0}^{\beta \hbar}\frac{d\tau}{\epsilon}\\
&  =\frac{\beta \hbar}{\epsilon}=\beta E/2\pi \equiv N_{F}.
\end{align}
This implies that $N_{\tau}$ is exactly equal to the number of real zeroes
$N_{F}$ on event horizon. This result is remarkable! We have $\epsilon
=\frac{\beta \hbar}{N_{F}}$ or $\epsilon=\frac{2\pi \hbar}{E}.$

In the limit $\epsilon \rightarrow0$, the AdS is invariant of isometry group
$\mathrm{SO(2,1)}\simeq \mathrm{SL(2,R)}/\mathrm{Z}_{2}$. Therefore, the
functions $t(\tau)$ and $\tilde{t}(\tau)$ describe the same geometry under a
transformation:
\[
t(\tau)\rightarrow \tilde{t}(\tau)=\frac{at(\tau)+b}{ct(\tau)+d},\quad
\]
where$\quad ad-bc=1\quad$and$\quad a,b,c,d\in \mathbb{R}$.

In addition, we must take the effect of fast variables from the fast motion
for real zeroes into consideration.

The gravity for the (1+1)D Euclidean AdS comes from its shape changings. Due
to thermal fluctuations, the fast motion along transverse directions provided
a contribution to its shape changings. After considering spherical symmetry,
the only approach to characterize the fast variables is to introduce the
dilaton field $\Phi(x^{0},x^{1})$ that locally changes the size of the event
horizon on the (1+1)D Euclidean AdS. Now, we have
\begin{equation}
d\Omega_{\mathrm{fast}}^{2}=\Phi^{2}(x^{0},x^{1})d\Omega^{2}.
\end{equation}
The effect of the fast variables is fully characterized by $\Phi(x^{0}%
,x^{1})=\Phi$ that has thermal fluctuations of different wave vector on event
horizon. We just focus on the fluctuations of $\Phi$ and have{\normalsize
\[
\Phi^{2}=\Phi_{0}^{2}+\delta \phi,\quad \delta \phi \ll \Phi_{0}^{2}.
\]
}A finite changing of the dilaton field $\delta \phi$ indicates a finite
changing of total energy. Therefore, with finite changing of the dilaton field
$\delta \phi,$ the processes for slow variables become physical.

The situation is similar to the effective SYK model in matrix representation.
In SYK\ model, the dilaton field $\delta \phi$ in geometric representation
plays the role of $\left \langle \delta N_{k}\right \rangle $ in matrix representation.

Then, after expanding the total action up to the second order in $\frac
{\delta \phi}{\Phi_{0}^{2}},$ the effective action for the (1+1) dimensional
Euclidean AdS is obtained as
\[
S_{\mathrm{JT}}=S_{bulk}-\frac{1}{8\pi G}\int_{bdy}\delta \phi_{b}\mathcal{K},
\]
where
\begin{equation}
S_{bulk}=-\frac{1}{16\pi G}\int d^{2}x\sqrt{h}\, \delta \phi \left(
R_{h}+2\right)  .
\end{equation}
$\mathcal{K}$\ is the extrinsic curvature,
\[
\mathcal{K}=-\frac{h_{ab}T^{a}T^{c}\nabla_{c}n^{b}}{h_{ab}T^{a}T^{b}},
\]
where $T^{a}$ and $n^{a}$ are tangent and unit normal vectors to the boundary
curve of 1+1 dimensional Euclidean AdS. $\delta \phi_{b}$ is the boundary value
of $\delta \phi,$ i.e., $\delta \phi \mid_{\mathrm{bdy}}=\delta \phi_{b}$. This is
just action for Jackiw--Teitelboim gravity\cite{jt}.

The equation of motion for the dilaton in bulk leads to $R_{h}+2=0$ that
describes the metric of (1+1)D AdS. The equations of motion for the metric are
given by{\normalsize
\begin{equation}
T_{ij}^{\delta \phi}\equiv \frac{1}{8\pi G}\left(  \nabla_{i}\nabla_{j}%
(\delta \phi)-h_{ij}\nabla^{2}(\delta \phi)+h_{ij}(\delta \phi)\right)  =0,
\end{equation}
}which determines the dilaton field $\delta \phi$. Near boundary, we define a
\textquotedblleft renormalized\textquotedblright \ boundary dilaton field
$\delta \phi_{r}(\tau)$, $\delta${\normalsize $\phi_{b}\approx \frac{\delta
\phi_{r}(\tau)}{\epsilon}$}.

Then, we evaluate the boundary term on the clipped Poincar\'{e} disk and
obtain the 1D theory with Schwarzian action.

The tangent and normal vectors to the curve $\left(  t(\tau),z(\tau)\right)  $
in the Poincar\'{e} metric are $\mathcal{T}=%
\begin{pmatrix}
t^{\prime}\\
z^{\prime}%
\end{pmatrix}
$ and $n^{a}=\frac{z}{\sqrt{(t^{\prime})^{2}+(z^{\prime})^{2}}}%
\begin{pmatrix}
-z^{\prime}\\
t^{\prime}%
\end{pmatrix}
,$ respectively. Therefore, the extrinsic curvature is obtained
as{\normalsize
\begin{align*}
\mathcal{K}  &  =\frac{d\mathcal{T}}{ds}=\frac{t^{\prime}\left(  t^{\prime
2}+z^{\prime2}+zz^{\prime \prime}\right)  -zz^{\prime}t^{\prime \prime}}{\left(
t^{\prime2}+z^{\prime2}\right)  ^{3/2}}\\
&  =1+\epsilon^{2}\mathrm{Sch}\left[  t(\tau),\tau \right]  +\mathcal{O}%
(\epsilon^{4})
\end{align*}
}where the Schwarzian derivative is defined as
\[
\mathrm{Sch}(t(u),u)=\frac{2t^{\prime}t^{\prime \prime \prime}-3{t^{\prime
\prime}}^{2}}{2{t^{\prime}}^{2}}.
\]
Integrating over the time on the boundary, we obtain the following
action:{\normalsize
\begin{align*}
S_{\mathrm{JT}}^{min}  &  =-\frac{1}{8\pi G}\int_{\mathrm{bdy}}ds\frac
{\delta \phi_{r}(\tau)}{\epsilon}\mathcal{K}\\
&  \simeq-\frac{1}{8\pi G}\int_{0}^{\beta}\frac{d\tau}{\epsilon}\frac
{\delta \phi_{r}(\tau)}{\epsilon}\\
&  \times \{1+\epsilon^{2}\mathrm{Sch}\left[  t(\tau),\tau \right]  \}.
\end{align*}
}The divergent term of "1" corresponds to the linear term in matrix
representation $\mathcal{\hat{P}}_{r}[%
{\displaystyle \sum \limits_{k}}
\left \langle \delta N_{k}\right \rangle \Gamma_{k}^{r}]=%
{\displaystyle \sum \limits_{k}}
\left \langle \delta N_{k}\right \rangle \mathcal{\hat{P}}_{r}(\Gamma_{k}^{r})$
and can be removed. Thus, in the leading order in $\epsilon$ we obtain the
following action:{\normalsize
\begin{equation}
S_{\mathrm{JT}}^{min}\approx-\frac{1}{8\pi G}\int_{0}^{\beta \hbar}d\tau
(\delta \phi_{r}(\tau)\text{\textrm{Sch}}\left[  t(\tau),\tau \right]  ).
\end{equation}
}

In addition, the time dependence of the $\delta \phi_{r}(\tau)$ can be removed
by the rescaling the time on the boundary theory with a new coordinate
$\tilde{\tau}$, i.e., $d\tilde{\tau}=\frac{\delta \bar{\phi}_{r}d\tau}%
{\delta \phi_{r}(\tau)}$, where $\delta \bar{\phi}_{r}$ is some positive
dimensionless constant. Or, we directly assume the boundary value of the
dilaton to be a constant $\delta \phi_{r}(\tau)=\delta \bar{\phi}_{r}.$ The
action of the Schwarzian is obtained as\cite{mac}
\[
S_{\mathrm{bdy}}\approx-\frac{\delta \bar{\phi}_{r}}{8\pi G}\int_{0}%
^{\tilde{\beta}\hbar}d\tilde{\tau}(\mathrm{Sch}\left[  t(\tilde{\tau}%
),\tilde{\tau}\right]  ).
\]
where
\begin{equation}
\mathrm{Sch}\left[  f\left(  g(\tau)\right)  ,\tau \right]  =(g^{\prime
2}\mathrm{Sch}\left[  f(g),g\right]  +\mathrm{Sch}[g,\tau]).
\end{equation}
The integral of the second term, $\delta \phi_{r}\mathrm{Sch}\left[
\tilde{\tau},\tau \right]  =-2\delta \phi_{r}^{\prime \prime}$, is zero due to
the periodicity $\delta \phi_{r}^{\prime}(\tau+\beta)=\delta \phi_{r}^{\prime
}(\tau)$ (the boundary curve is smooth and closed). So, we may consider
$\delta \bar{\phi}_{r}$ to be constant boundary values of the dilaton and get
action for the deformation of boundary of the (1+1) dimensional Euclidean AdS
$t(\tilde{\tau})$.

It is also convenient to change to the Rindler coordinates\cite{rin} using the
map $t(\tau)=\tan \frac{\varphi(\tau)}{2}$, which follows from the
near-boundary limit of the identities:
\begin{equation}
\mathrm{Sch}\left[  t,\tau \right]  =\mathrm{Sch}\left[  \varphi,\tau \right]
+\frac{(\varphi^{\prime})^{2}}{2}.
\end{equation}
Varying the corresponding action by $\varphi,$ we obtain the following
equation of motion:
\[
\frac{\mathrm{Sch}\left[  \varphi,\tau \right]  ^{\prime}}{\varphi^{\prime}%
}-\varphi^{\prime \prime}=0,
\]
which has a linear in time solution:{\normalsize
\[
\varphi(\tau)=\frac{2\pi \tau}{\beta \hbar}.
\]
}We choose the coefficient of the linear dependence in such a way that the
Rindler time is periodic with the period $2\pi$, $\varphi \sim \varphi+2\pi$.
This solution can be associated to the boundary theory at the temperature
$\beta$. This leads to the growth saturating the \textquotedblleft bound on
chaos\textquotedblright \ for the regularized out-of-time-ordered correlation
function (OTOC)\cite{otoc}.

\paragraph{Matrix representation and emergent SYK model}

In this part, we discuss the matrix representation of the (3+1)D physical
variant for black hole inside event horizon. Under matrix representation, we
have a fixed flat spacetime but fluctuating Gamma matrices $\Gamma^{\mu}$ that
describe the fluctuations of the shape of event horizon. To exactly
characterize fluctuating Gamma matrices $\Gamma^{\mu},$ the key point is to
integrate the fast variables.

The slow motion for boundary fluctuations of gravitational waves is described
by the following effective Hamiltonian%
\[
\mathcal{\hat{H}}_{(3-1)+1}^{\mathrm{slow}}=\int(\Psi_{R}^{\dagger}%
(\mathbf{x})\hat{H}_{(3-1)+1}^{\mathrm{slow}}\Psi_{R}(\mathbf{x}))d^{2}x
\]
where
\[
\hat{H}_{(3-1)+1}^{\mathrm{slow}}=%
{\displaystyle \sum \nolimits_{\mu \neq r}}
\Gamma^{\mu}\delta p^{\mu}.
\]
Now, the Gamma matrices become fluctuating.

On the other hand, according to above discussion, there exists thermal
fluctuation for particle number of real zero on event horizon $N_{F},$ i.e.,%
\[
\rho(N_{F})=\frac{1}{\sqrt{2\pi \left \langle N_{F}\right \rangle ^{2}}}%
\exp(-\frac{(\Delta N_{F})^{2}}{2\left \langle N_{F}\right \rangle ^{2}}).
\]

Next, under matrix representation, we integrate the massive particles and
consider their renormalization on the effective Hamiltonian of Gamma matrices.

Because the normal direction of boundary of flat AdS is $\Gamma^{r},$ under
the matrix representation the boundary fluctuations are characterized by the
changings of $\Gamma^{r},$ i.e.,
\[
\Gamma^{r}\rightarrow(\Gamma^{r})^{\prime}(x,t)=S\Gamma^{r}S^{-1}=\alpha
_{r}\Gamma_{0}^{r}+%
{\displaystyle \sum \nolimits_{\mu \neq d}}
\alpha_{\mu}\Gamma^{\mu}%
\]
where these coefficients $\alpha_{r}$ and $\alpha_{\mu}$ satisfy $\alpha
_{r}^{2}+%
{\displaystyle \sum \nolimits_{\mu \neq d}}
\alpha_{\mu}^{2}=1,$ and $\alpha_{r}\gg \alpha_{\mu}$. Now, the system is still
at EPs. However, the direction of the polarization becomes fluctuating.

Then, we re-write the effective Hamiltonian of slow motion from $\mathcal{\hat
{H}}_{(3-1)+1}^{\mathrm{slow}}=\int(\Psi_{R}^{\dagger}(\mathbf{x})(%
{\displaystyle \sum \nolimits_{\mu \neq r}}
\Gamma^{\mu}\delta p^{\mu})\Psi_{R}(\mathbf{x}))d^{2}x$ to
\begin{align*}
\mathcal{\hat{H}}_{(3-1)+1}^{\mathrm{slow}}  &  =\mathcal{\hat{P}}_{r}%
[\int(\Psi_{R}^{\dagger}(\mathbf{x})\Gamma^{r}(x,t)\Psi_{R}(\mathbf{x}%
))d^{2}x]\\
&  \simeq \mathcal{\hat{P}}_{r}[\int \Psi_{R}^{\dagger}(\mathbf{x})\Gamma
^{r}(x,t)\Psi_{R}(\mathbf{x}))d^{2}x]
\end{align*}
where $%
{\displaystyle \sum \nolimits_{\mu \neq r}}
\alpha_{\mu}\Gamma^{\mu}=\mathcal{\hat{P}}_{r}[\Gamma^{r}(x,t)]$ and
$\mathcal{\hat{P}}_{r}$\ is projected operator that gets rid of the component
of $\Gamma_{0}^{r}$ from the rotor $\Gamma^{r}(x,t)$. In general, the
projected operator $\mathcal{\hat{P}}_{r}$\ is defined as
\[
\mathcal{\hat{P}}_{r}(\hat{A})=\hat{A}-\Gamma_{0}^{r}\mathrm{Tr}(\Gamma
_{0}^{r}\hat{A}).
\]
We have%
\[%
{\displaystyle \int}
[d\Psi_{R}^{\dagger}][d\Psi_{R}][d\Gamma^{r}]\exp(-\beta \mathcal{\hat{H}%
}_{(3-1)+1}^{\mathrm{slow}})
\]
where $\beta=\frac{1}{k_{B}T}.$

Because there exists residue unitary variability along transverse directions
on event horizon, the wave vector $k$ (or transverse momentum $p$) is good
quantum number. Hence, we perform Fourier decomposition and study the slow
motion in momentum space.

In momentum space, we integrate massive fermions for different wave vectors
and get
\begin{align}
\mathcal{H}_{(d-1)+1}^{\mathrm{slow}}  &  =\mathcal{\hat{P}}_{r}[%
{\displaystyle \sum \limits_{k}}
\left \langle \delta N_{F}^{k}\right \rangle \Gamma_{k}^{r}]\\
&  -\frac{1}{2}\mathcal{\hat{P}}_{r}[%
{\displaystyle \sum \limits_{k,k^{\prime}}}
\left \langle \delta N_{F}^{k}\delta N_{F}^{k^{\prime}}\right \rangle
(\Gamma_{k}^{r})(\Gamma_{k^{\prime}}^{r})]+...\nonumber
\end{align}

Under the projection operator $\mathcal{\hat{P}}_{r},$ the leading term about
$\Gamma_{0}^{r}$\ disappears and
\[
\mathcal{\hat{P}}_{r}(\Gamma_{k}^{r})\simeq \mathcal{\hat{P}}_{r}(\Gamma
_{k}^{r})=0.
\]
Then, we have
\[
\mathcal{\hat{P}}_{r}[%
{\displaystyle \sum \limits_{k}}
\left \langle \delta N_{F}^{k}\right \rangle \Gamma_{k}^{r}]=%
{\displaystyle \sum \limits_{k}}
\left \langle \delta N_{F}^{k}\right \rangle \mathcal{\hat{P}}_{r}(\Gamma
_{k}^{r})\simeq0
\]
for perturbative random wave vectors on event horizon. Under the projected
operation $\mathcal{\hat{P}}_{r}$, the second term with the coupling between
two $\Gamma_{k}^{r}$ can be finite. The projected operator $\mathcal{\hat{P}%
}_{r}$ for the coupling between different Gamma matrices $\Gamma_{k}^{r}$ with
same wave vectors plays the role of "trace", i.e.,%
\begin{align*}
&  \mathcal{\hat{P}}_{r}[%
{\displaystyle \sum \limits_{k,k^{\prime}}}
\left \langle \delta N_{F}^{k}\delta N_{F}^{k^{\prime}}\right \rangle
(\Gamma_{k}^{r})(\Gamma_{k}^{r})]\\
&  =\mathcal{\hat{P}}_{r}\{[%
{\displaystyle \sum \limits_{k,k^{\prime},\mu,\nu}}
\left \langle \delta N_{F}^{k}\delta N_{F}^{k^{\prime}}\right \rangle
(\Gamma_{k}^{r})^{\mu}(\Gamma_{k^{\prime}}^{r})^{\nu}]\} \\
&  =%
{\displaystyle \sum \limits_{k,k^{\prime},\mu}}
\left \langle \delta N_{F}^{k}\delta N_{F}^{k^{\prime}}\right \rangle
(\Gamma_{k}^{r})^{\mu}(\Gamma_{k^{\prime}}^{r})^{\mu}\\
&  =\mathrm{Tr}[%
{\displaystyle \sum \limits_{k,k^{\prime}}}
\left \langle \delta N_{F}^{k}\delta N_{F}^{k^{\prime}}\right \rangle
(\Gamma_{k}^{r})(\Gamma_{k^{\prime}}^{r})].
\end{align*}
Ignoring higher-order terms, we have
\[
\mathcal{H}_{(d-1)+1}^{\mathrm{slow}}\simeq-\frac{1}{2}\mathrm{Tr}[%
{\displaystyle \sum \limits_{k,k^{\prime}}}
\left \langle \delta N_{F}^{k}\delta N_{F}^{k^{\prime}}\right \rangle
(\Gamma_{k}^{r})(\Gamma_{k^{\prime}}^{r})].
\]

In the infinite mass limit $m_{R}\gg ck$, we assumed that for different wave
vectors, the fluctuations are all same and equal to $\left \langle \delta
N_{F}\right \rangle /N,$ of which $\left \langle \delta N_{F}\right \rangle $
denotes the fluctuations of total particle number. Under this ansatz, for
$k\neq k^{\prime},$ we have$\  \left \langle \delta N_{F}^{k}\delta
N_{F}^{k^{\prime}}\right \rangle =\left \langle \delta N_{F}^{k}\right \rangle
\left \langle \delta N_{F}^{k^{\prime}}\right \rangle =\left \langle \delta
N_{F}^{k}\right \rangle ^{2}\simeq \left \langle \delta N_{F}\right \rangle ^{2}$
and get
\begin{align*}
\mathcal{H}_{(3-1)+1}^{\mathrm{slow}}  &  \simeq-\frac{1}{2}\mathrm{Tr}[%
{\displaystyle \sum \limits_{k,k^{\prime}}}
\left \langle \delta N_{F}^{k}\delta N_{F}^{k^{\prime}}\right \rangle
(\Gamma_{k}^{r})(\Gamma_{k^{\prime}}^{r})]\\
&  \simeq-\frac{1}{2}\mathrm{Tr}[%
{\displaystyle \sum \limits_{k,k^{\prime}}}
\left \langle \delta N_{F}\right \rangle ^{2}(\Gamma_{k}^{r})(\Gamma_{k^{\prime
}}^{r})]\\
&  =-\frac{1}{2}%
{\displaystyle \sum \limits_{k,k^{\prime},\mu}}
J_{kk^{\prime}}(\Gamma_{k}^{r})^{\mu}(\Gamma_{k^{\prime}}^{r})^{\mu}.
\end{align*}
If $k=k^{\prime},$ the corresponding terms become constant and can be emitted.
Here, the couplings $J_{kk^{\prime}}=\left \langle \delta N_{F}\right \rangle
^{2}$ are distributed randomly and independently, i.e. accordingly to the
Gaussian distribution with the following probability density function:
\begin{equation}
P(J_{kk^{\prime}})=\exp \left(  -\frac{J_{kk^{\prime}}^{2}}{\left \langle
N_{k}\right \rangle ^{2}}\right)  \quad \text{for every}\quad J_{kk^{\prime}}.
\end{equation}
Here, $\left \langle \delta N_{F}\right \rangle ^{2}=J_{kk^{\prime}}$ plays the
role of $\delta \bar{\phi}_{r}$ in emergent Jackiw-Teitelboim gravity under the
geometric representation. This is just action of an effective complex SYK
model with $q=2.$

In 3+1 dimensional spacetime, $\Gamma_{k}^{r}$ is reduced to usual Pauli
matrices. $\mathcal{H}_{(3-1)+1}^{\mathrm{slow}}$ becomes a Heisenberg model
with random interaction. By using slave particle approach $\Gamma_{\alpha
\beta}^{r}=\psi_{\alpha}^{\dagger}\psi_{\beta},$ $%
{\displaystyle \sum \limits_{\alpha=1}^{2}}
\psi_{\alpha}^{\dagger}\psi_{\alpha}=1$, we have a complex SYK model with
random Gaussian four-fermion coupling,
\[
\mathcal{H}_{2}^{\mathrm{slow}}\simeq-\frac{1}{2}%
{\displaystyle \sum \limits_{\alpha,\beta=1}^{2}}
{\displaystyle \sum \limits_{k,k^{\prime}}}
J_{kk^{\prime}}(\psi_{k\alpha}^{\dagger}\psi_{k\beta}\psi_{k^{\prime}\alpha
}^{\dagger}\psi_{k^{\prime}\beta}).
\]

Let us discuss the emergence of the effective (complex) SKY model on event horizon.

For above effective SYK model, the matrix $\Gamma_{k}^{r}$ denotes the
external normal direction. Due to the thermalization condition, it becomes
fluctuating. The index $k$ labeling different $\Gamma_{k}^{r}$ is wave vector
rather than spatial position on event horizon. Because $\left \langle \delta
N_{k}\right \rangle =\frac{\left \langle \delta N_{F}\right \rangle }{N_{F}}$ is
the uniform fluctuated particle number for whole black hole, there exists
random couplings between the matrices $\Gamma_{k}^{r}$ for arbitrary two
modes. The equivalence for the random couplings between the matrices
$\Gamma_{k}^{r}$ for two modes with different wave vectors comes from the
infinite mass for the fermionic particles of real zeroes. The slave particle
denoted by $\psi_{k\alpha}^{\dagger}$ is not real one. Instead, it is an
auxiliary one.

The SYK model and its various generalizations have received much attention in
the recent years. In the large $N$ limit, the SYK model is dominated by
melonic graphs \cite{Kitaev-talks,ye}. This allows us to find correlations
using functional methods. The two-point function obeys the Schwinger-Dyson
equation, reflecting the fact that the leading correction to the propagator
comes from inserting a \textquotedblleft melon\textquotedblright. This makes
the system amenable to mean-field approaches. It turns out that at the
mean-field level the infinite dimensional conformal symmetry gets broken by
the interaction self-energy down to the conformal group $\mathrm{SL}(2,R)$ of
rational transformations,
\begin{align*}
t  &  \rightarrow t^{\prime}=\tfrac{at+b}{ct+d},\\
ad-bc  &  =1.
\end{align*}
This leads to a classic symmetry breaking scenario and the emergence of
Goldstone modes whose fluctuations become unhampered in the long time limit
where the \emph{explicit} symmetry breaking (represented by the time
derivative $\partial_{t}$ present in the system's action) becomes negligible.
The situation bears similarity to that in a magnet, with the important
difference that the dimension of the Goldstone mode manifold is infinite,
while the spatial dimension is zero. The dynamics of the pseudo-Goldstone
boson which is associated to this broken symmetry (so-called \textquotedblleft
soft mode\textquotedblright) is approximately described by the Schwarzian
action \cite{mac},
\[
I_{\mathrm{bdy}}\approx-\frac{\bar{\phi}_{r}}{8\pi G}\int_{0}^{\tilde{\beta
}\hbar}d\tilde{\tau}\mathrm{Sch}\left[  t(\tilde{\tau}),\tilde{\tau}\right]
.
\]
This action is same to that from Jackiw-Teitelboim gravity in geometric representation.

In addition, we discuss the issue of additional mode associated with $U(1)$
charge for the "complex" SYK model. According to above discussion, the
effective model is a complex SYK with an additional global $U(1)$ symmetry.
However, the situation is complex. It looks like that there exists a global
Abelian symmetry by rotating along the direction of $\Gamma_{k}^{r}.$
Remember, along the direction of $\Gamma_{k}^{r},$ the group is non-unitary.
Or it is about the changing amplitude rather than phase. The corresponding
$U(1)$ charge is imaginary and isn't conserved. As a result, there doesn't
exist such a global $U(1)$ symmetry. The pseudo-Goldstone mode of the complex
SYK is same that of the real one, i.e., a $h=2$ mode of \cite{mac}.

In addition, we give a comment on the correspondence between complex SYK model
in geometric representation and Jackiw-Teitelboim gravity in geometric representation.

On the one hand, in matrix representation, the physical process comes from the
random coupling between Gamma matrices $\Gamma_{k}^{r}$ for different modes
with wave vectors $k.$ Now, the spacetime is flat. We integrate the fast
variables from the fast motion for real zeroes that characterize the expansion
and contraction of the event horizon. Then, the coupling between Gamma
matrices $\Gamma_{k}^{r}$ for different modes become renormalized. The low
energy effective model is described by Schwarzian action.

On the other hand, in geometric representation, the physical process comes
from the shape changings of (1+1) dimensional Euclidean AdS. This is described
by Jackiw-Teitelboim gravity, of which the fast variables is characterized by
a dilaton field. Now, the spacetime is curved. Instead, the Gamma matrices are
all constant. In geometric representation, the fluctuations of total size
along $r$-th direction (or the direction with imaginary coordinates
$\tau \rightarrow \tau^{\prime}=t(\tau)$) is relevant to the dilaton field. The
low energy effective model is also described by Schwarzian action.

\paragraph{CFT representation and 1D non-Hermitian gauge theory}

In this part, we can use non-Hermitian gauge theory to characterize the
boundary of (1+1)D Euclidean AdS under Gravity/N-gauge equivalence.

For the (1+1)D \textrm{\~{S}\~{O}(1+1)} non-unitary physical variant
$V_{\mathrm{\tilde{S}\tilde{O}(1+1)},1+1}(\Delta \phi^{\mu},\Delta x^{\mu
},k_{0},\omega_{0})$, the representation of 1D non-Hermitian gauge theory
(NGT) on flat spacetime is equivalence to the representation of (1+1)D
Euclidean AdS. When we reduce the NGT to the unitary physical processes of the
system, AdS/NGT equivalence is reduced to usual AdS/CFT correspondence between
the theory for boundary of (1+1)D Euclidean AdS and 1D CFT. The key point is
the existence of internal imaginary zeroes inside a real zero and each
internal imaginary zero plays the role of a level-2 imaginary zero.

Now, the slow motion from the fluctuations of gravitational waves is described
by the non-Hermitian \textrm{U(0,1)}$\times$\textrm{SU(0,N)} gauge fields. The
effective Hamiltonian becomes 0D, i.e.,
\[
\mathcal{H}_{0}=\Psi^{\dagger}\hat{H}_{1}\Psi
\]
where $\hat{H}_{0}=(\mathrm{e}A_{\tau,U(0,1)}+g\mathcal{A}_{\tau})\Gamma
^{\tau}.$ The excitation is gapless. We have results of CFT.

Because the low energy degrees of freedom is dominated by gapless
gravitational waves on the boundary of the AdS (that is approaching the event
horizon infinitely), we ignore the \textrm{SU(0,N)} non-Abelian gauge fields
and focus on non-Hermitian \textrm{U(0,1)} Abelian gauge field $A_{\tau
,U(0,1)}$. Now, the effective model is reduced into
\[
\mathcal{H}_{0}=\mathrm{e}\rho_{\mathrm{bdy}}A_{\tau,U(0,1)}\Gamma^{\tau}%
\]
where $\rho_{\mathrm{bdy}}=\Psi^{\dagger}\Psi \mid_{\mathrm{bdy}}$ is the
density of elementary particles at the boundary of (1+1)D Euclidean AdS. Under
\textrm{U(0,1)} gauge transformation, $\rho_{\mathrm{bdy}}$ and $A_{\tau
,U(0,1)}$ change simultaneously. The Hamiltonian $\mathcal{H}_{0}$ is invariant.

Next, we consider $A_{\tau,U(0,1)}$\ from the fluctuations of $\Gamma
^{r}(r,\tau)$ and have
\begin{align*}
\mathcal{\hat{H}}_{(3-1)+1}^{\mathrm{slow}}  &  =\mathcal{\hat{P}}%
_{r}(\mathrm{e}\rho_{\mathrm{bdy}}\Gamma^{r}(r,\tau))\\
&  =\mathcal{\hat{P}}_{r}(\mathrm{e}\rho_{\mathrm{bdy}}A_{\tau,U(0,1)}%
\gamma^{0}\gamma^{r0}),\text{ }0=\tau.
\end{align*}
With same $\gamma^{r0}$, $A_{\tau,U(0,1)}$ really becomes $\omega^{r0}$ that
is the connection between two orthogonal frames. If we consider $\omega^{r0}$
to be non-Abelian gauge field, $S=e^{i\gamma^{r0}\delta \vartheta}$ becomes
gauge transformation along z-th direction (that is orthogonal to other two
frames). $\delta \vartheta(r,\tau)$ is the phase angle of gauge field
$A_{\tau,U(0,1)}$ on the perfect circle that describes the fluctuations of the
boundary of (1+1)D Euclidean AdS. Without strength of gauge fields,
non-Hermitian \textrm{U(0,1)} Abelian gauge field $A_{\tau,U(0,1)}$ becomes
pure gauge and is determined by $\delta \vartheta(r,\tau)$ along $\tau$.

On the other hand, according to above discussion, the extrinsic curvature
$\mathcal{K}$ is obtained as $\mathcal{K}=\frac{d\mathcal{T}}{ds}$ where
$\mathcal{T}$ is the tangent vector to the curve $\left(  t(\tau
),z(\tau)\right)  $. Under matrix representation, the tangent vector and
normal vector to the curve $\left(  t(\tau),z(\tau)\right)  $ become matrix
$\Gamma^{\tau}$ and matrix $\Gamma^{r},$ respectively. Except for an initial
value $\theta_{0}$, the tangential angle $\theta$ of the curve is equal to the
angle of the direction for tangent matrix $\vartheta$. So, we have
\[
\int_{\mathrm{bdy}}A_{U(0,1)}=\int_{\mathrm{bdy}}\delta \vartheta
=\int_{\mathrm{bdy}}\mathcal{K}%
\]
where $\mathcal{K}$, $\delta \theta$, and $A_{U(0,1)}$\ are all 1-form. In
addition, we point out that the dilaton field $\frac{\delta \bar{\phi}_{r}%
}{8\pi G}$ corresponds to the density of elementary particles on event
horizon,
\[
\mathrm{e}\rho_{\mathrm{bdy}}\sim-\frac{\delta \bar{\phi}_{r}}{8\pi G}.
\]

As a result, on the boundary of (1+1)D Euclidean AdS along the coordinates
$\tau$, we map the theory for non-Hermitian \textrm{U(0,1)} Abelian gauge
field $A_{\tau,U(0,1)}$ to another Hermitian \textrm{U(1)} Abelian gauge field.

Finally, on the boundary of (1+1)D Euclidean AdS, the Schwarzian action is
obtained as
\begin{align*}
S_{\mathrm{bdy}}  &  =\int_{\mathrm{bdy}}\mathrm{e}\rho_{\mathrm{bdy}%
}A_{U(0,1)}\\
&  =-\frac{\delta \bar{\phi}_{r}}{8\pi G}\int_{\mathrm{bdy}}\frac{\delta \theta
}{ds}d\tau \\
&  =-\frac{\delta \bar{\phi}_{r}}{8\pi G}\int_{\mathrm{bdy}}\mathcal{K}\\
&  \approx-\frac{\delta \bar{\phi}_{r}}{8\pi G}\int_{0}^{\tilde{\beta}\hbar
}d\tilde{\tau}(\mathrm{Sch}\left[  t(\tilde{\tau}),\tilde{\tau}\right]  ).
\end{align*}

\subsubsection{Summary}

In the end of this section, we give a summary.

Due to the "non-changing" structure along tempo direction, there exists random
distribution of geometry structure. Under an assumption of Principle of equal
probability and the constraint of energy (or particle number), we have a new
statistics of spacetime. In continuum limit, from it, the Hawking entropy,
Hawking temperature are exactly derived. The SYK model (rather than
Schrodinger's equation) or Jackiw-Teitelboim gravity effectively characterizes
the dynamics of quantum geometry for black hole inside event horizons.

\subsection{Other relevant issues}

\subsubsection{Unruh effect and quantum thermodynamics for accelerated
systems}

In this section, we study the Unruh effect associated to quantum
thermodynamics for accelerated systems. This result was originally derived by
Unruh~\cite{Unruh}, and is therefore called the Unruh effect.

Thermal phenomena appear with respect to the Rindler time\cite{rin}. By
transforming the usual Cartesian coordinates $(T,X)$ on flat space to the
Rindler coordinates $(x,t)$,
\begin{equation}
X=x\cosh \kappa t,\quad T=x\sinh \kappa t,
\end{equation}
we have the metric
\begin{align}
ds^{2}  &  =-dT^{2}+dX^{2}+ds_{\mathbb{R}^{(d-2)}}^{2}\nonumber \\
&  =-\kappa^{2}x^{2}dt^{2}+dx^{2}+ds_{\mathbb{R}^{(d-2)}}^{2}.
\end{align}
Another useful coordinate system can be defined by setting $x=e^{\kappa \rho}$.
Now, the metric turns into
\begin{equation}
ds^{2}=\kappa^{2}e^{2\kappa \rho}(-dt^{2}+d\rho^{2})+ds_{\mathbb{R}^{(d-2)}%
}^{2}.
\end{equation}
This spacetime also describes a special physical variant with topological
defect at its Killing horizon.

Near horizon, because traditional quantum mechanics fails, the results from
the path integral approach on Euclidean spacetime are all not reliable. To
answer this question, one must seek help from theory of physical variant.

According to above metric, there exists event horizon at $\rho=\frac{1}{k}\ln
x\rightarrow-\infty.$ Now, the metric is reduced to a two dimensional one. At
the event horizon, the changing rates of the corresponding physical variant
along motion direction and tempo direction turn to zero. This leads to
randomness on the horizon. So, the spacetime along transverse directions
become a stochastic variant. We have a statistics for spacetime, i.e,
\begin{align*}
S_{A}  &  =k_{B}\ln \Omega=k_{B}\ln(\frac{(N_{U})^{N_{U}}}{(N_{U})!})\\
&  \simeq k_{B}N_{U}+\frac{1}{2}k_{B}\ln(2\pi N_{U})\\
&  \simeq k_{B}N_{U}.
\end{align*}
In continuum limit, we derive the formula of entropy $S_{A}$ that is same to
that of black hole, i.e.,%
\begin{equation}
S_{A}\simeq k_{B}N_{U}=k_{B}{\frac{S}{l_{0}^{2}}.}%
\end{equation}

For accelerated quantum particle, the true spacetime is flat and has no
topological defect. \emph{Does a pure accelerated\ quantum state evolute into
a mixed state by simply making a change of coordinates?}

Firstly of all, this issue is relevant to the case outside the horizon. In
particular, the answer depends on observations. For Rindler observers, near
horizon, the size of an quantum particle turns to infinite and the internal
structure of an elementary particle becomes extremely amplified. Due to this
extremely amplification effect, the effect of quantum fluctuations become
exposed and the quantum measurement leads to randomness. The experiment in
Ref.\cite{chin} is within a framework for the simulation of quantum physics in
a non-inertial frame, based on Bose--Einstein condensates under time-evolution
by the frame transformation. Because this is Rindler observer, effective Unruh
effect is observed. However, for Minkowski observers, the size of an quantum
particle is always very small. Without the extremely amplification effect,
there doesn't exist Unruh effect.

\subsubsection{$\mathrm{ER=EPR?}$}

"It from Qubit" is a new idea about understanding the origins of spacetime. To
follow the idea of "It from Qubit", there are two different methodologies: One
is Reductionism from top to down, the other is Emergence from down to up.
Following the methodology of Reductionism, people try to understand the nature
of spacetime by studying the quantum entanglement of spacetime. An example is
about the conjecture of $\mathrm{ER=EPR}$\cite{er}. Following the methodology
of Emergence, people try to understand the nature of spacetime by constructing
certain many-body models and studying its ground states and excitations. In
this section, we study the quantum entanglement of spacetime by Reductionism.
The key point to answer the question of $\mathrm{ER=EPR.}$

Firstly, I review the issue about $\mathrm{ER=EPR.}$

The starting point is the AdS/CFT correspondence that is an equivalence
between CFT and asymptotically AdS spacetime. According to AdS/CFT
correspondence, people may guess that the entanglement of quantum states of
CFT side correspond to the connection of spacetime of AdS side. Then, the
entanglement between the microstates of these black holes plays similar role
to an Einstein-Rosen (ER) bridge (or wormhole) connecting two black holes.
This idea is just \textquotedblleft$\mathrm{ER=EPR}$\textquotedblright%
\cite{er}. It was suggested that an AdS wormhole is dual to two uncorrelated
but entangled CFTs in a \textquotedblleft thermofield double\textquotedblright%
\ state $|\text{TFD}\rangle$ \cite{raa}.

Now, one considers a spacetime with two equivalent asymptotically AdS regions,
suggesting that the dual description should involve two copies of the CFT. An
observer in either asymptotic region sees the Schwarzschild AdS black hole
spacetime, which corresponds to the thermal state of CFT. On the other hand,
tracing over the degrees of freedom of one of the CFTs, one finds that the
density matrix for the remaining CFT is exactly the thermal density matrix:
\[
\rho_{T}=\mathrm{Tr}_{2}(|\psi \rangle \langle \psi|)=\sum_{i}e^{-\beta E_{i}%
}|E_{i}\rangle \langle E_{i}|.
\]

The presence of horizons in the black hole spacetime which forbid
communication between the two asymptotic regions may be naturally associated
with the absence of interactions between the two CFTs. If we consider two
correlated black hole by a wormhole. The situation changes. Let us consider a
CFT on a spatially infinite line. The \textquotedblleft thermofield
double\textquotedblright \ state $|\text{TFD}\rangle$ is defined by an
entangled pure state of two copies of thermal CFT:
\begin{equation}
|\text{TFD}\rangle=\sum_{n}e^{-\frac{\beta}{2}E_{n}}|n_{L}\rangle \otimes
|n_{R}\rangle.
\end{equation}
Here, $\beta^{-1}$ is the temperature, and $|n_{L,R}\rangle$ are the $n$-th
energy eigenstates of individual systems. Note that each copy of CFTs is in
the mixed thermal state
\begin{equation}
e^{-\beta H}=\mathrm{Tr}_{L}|\text{TFD}\rangle \langle \text{TFD}|=\mathrm{Tr}%
_{R}|\text{TFD}\rangle \langle \text{TFD}|.
\end{equation}

It is believed that the spacetime subregion associated with the entanglement
between $C$ and $D$ is the entanglement wedge \cite{Czech:2012bh}, the
geodesic, referred as the entanglement wedge cross-section (EWCS) \cite{ewc}.
And, EWCS is equal to the horizon area of the wormhole.

Let us check this statement in variant theory by using two black holes in dS
rather than AdS as example to discuss.

In variant theory, the black hole is physical variant with a 2D U-N class
topological defect, of which the phase change of the changing rate $k_{0}%
^{\mu=r/t}$ along radial/tempo direction is $\pm \frac{\pi}{2}$. In other
words, the event horizon of black hole is a domain wall between a unitary
physical variant (or a dS) and a non-unitary physical variant (or an AdS). Due
to the "non-changing" structure along tempo direction, the event horizon of
the black hole becomes a stochastic variant, of which the information unit is
unit cell. According to assumption of the stochastic variant, the $N_{U}$ unit
cells have a randomized distribution on these original $N_{U}$ unit cells with
fixed $N_{U}$. The statistics of spacetime for event horizon is given by the
following MPF, i.e., $\Omega=\frac{(N_{U})^{N_{U}}}{(N_{U})!}.$ The event
horizon becomes a classical object with finite temperature $T$.

Now, we consider two black holes. See the illustration in Fig.18.

When the black holes are disconnected, they may have different Hawking
temperatures. However, when the black holes are connected by a wormhole, the
situation changes. We point out that the ER bridge (or wormhole) connecting
the two black holes has a dumbbell handle shaped event horizon. See Fig.18.
So, as shown in Fig.18, we have a dumbbell-shaped event horizon for the whole
system with two connected black holes. Inside the dumbbell-shaped event
horizon, the spacetime becomes AdS. Or we have a non-unitary physical variant.
For the dumbbell-shaped event horizon, the information unit is also unit cell
with unit area $l_{0}^{2}$.

Then, the information units of both black holes are unit cells that could move
from its horizon to the other. Without considering the area of dumbbell handle
from ER bridge, we approximatively have the total entropy $S$ to be
\begin{align}
S  &  =k_{B}\ln \Omega \nonumber \\
&  =k_{B}\ln(\frac{(N_{L}^{U}+N_{R}^{U})^{(N_{L}^{U}+N_{R}^{U})}}{(N_{L}%
^{U}+N_{R}^{U})!})\nonumber \\
&  \simeq k_{B}(N_{L}^{U}+N_{R}^{U}).
\end{align}
We may also assume the validity of the Principle of equal probability and
unique Hawking temperatures
\begin{equation}
T=T_{R}=T_{L}.
\end{equation}
\begin{figure}[ptb]
\includegraphics[clip,width=0.8\textwidth]{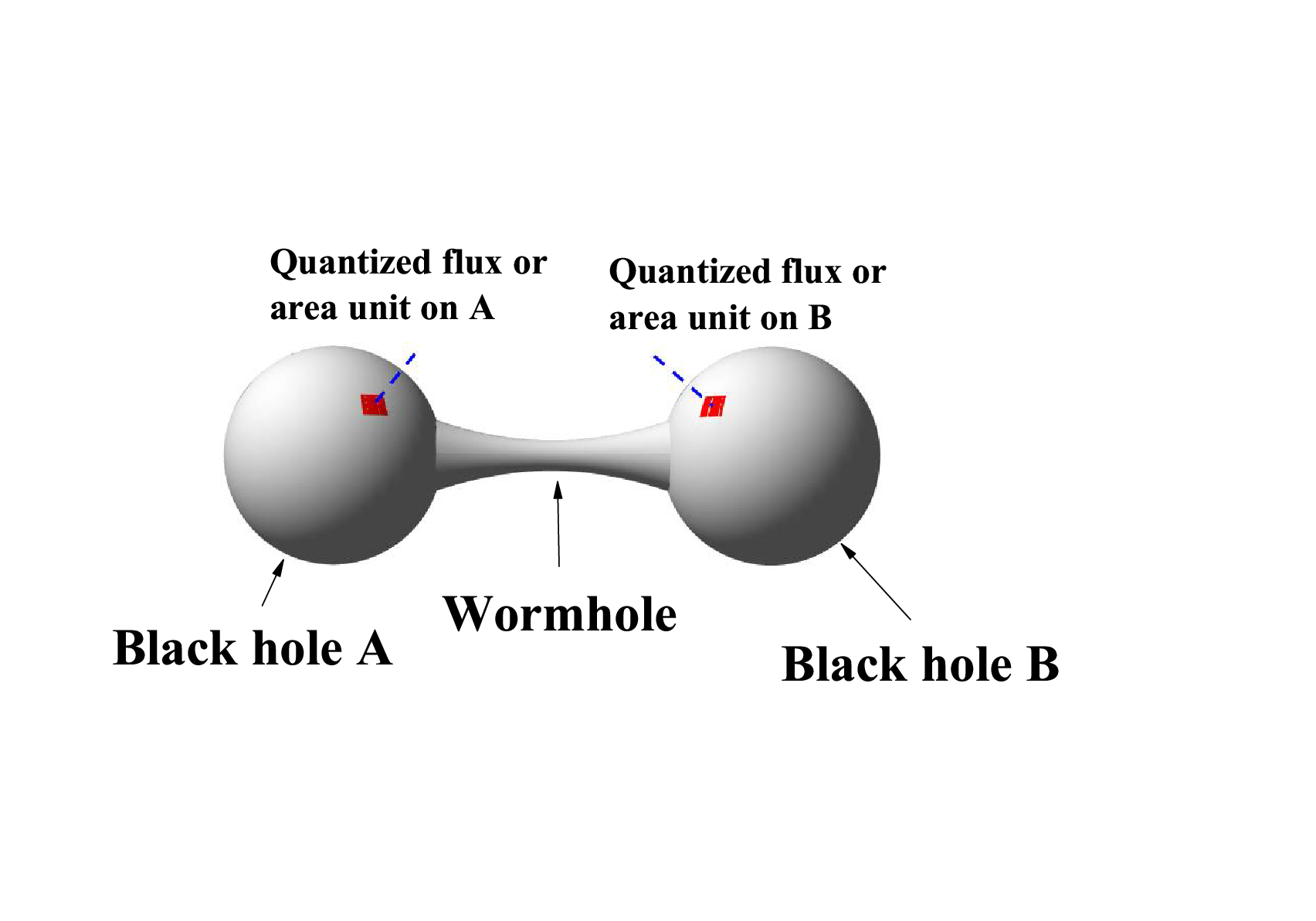}\caption{An illustration
of two correlated black hole by a wormhole. The surface of the whole system is
the event horizon.}%
\end{figure}

This is a description for two connected black holes on dS space. For them on
AdS, the situation doesn't change.

Finally, we draw conclusion. Although for same sub-spacetime (or two black
holes), we have same information unit, the information units of different
black hole cannot be regarded as "entangled states". Instead, in both picture
(AdS or CFT) they are thermalized states with single temperature
$T=T_{R}=T_{L}$. This is underlying physics of $\mathrm{ER=EPR.}$ I don't
think $\mathrm{ER}$ of two black holes provides valuable clues about the
essence of quantum entanglement for EPR.

\subsection{Discussion and conclusion}

In the final section, we draw the conclusion.

We developed a complete theory for black hole based on physical variant with
topological defects. The key point is%
\begin{align*}
&  \text{Black hole (a phenomenological theory)}\\
&  \Longrightarrow \text{Physical variant with topological defect}\\
&  \text{ (a microscopic theory).}%
\end{align*}
In particular, the event horizon of black hole is a 2D U-N class of
topological defect, of which the phase change of the changing rate $k_{0}%
^{\mu=r/t}$ along radial/tempo direction is $\pm \frac{\pi}{2}$. Now, the event
horizon of a black hole becomes a topological domain wall between a unitary
physical variant (or a dS) and a non-unitary physical variant (or an AdS). See
the logical structure of this part in Fig.19.

\begin{figure}[ptb]
\includegraphics[clip,width=0.9\textwidth]{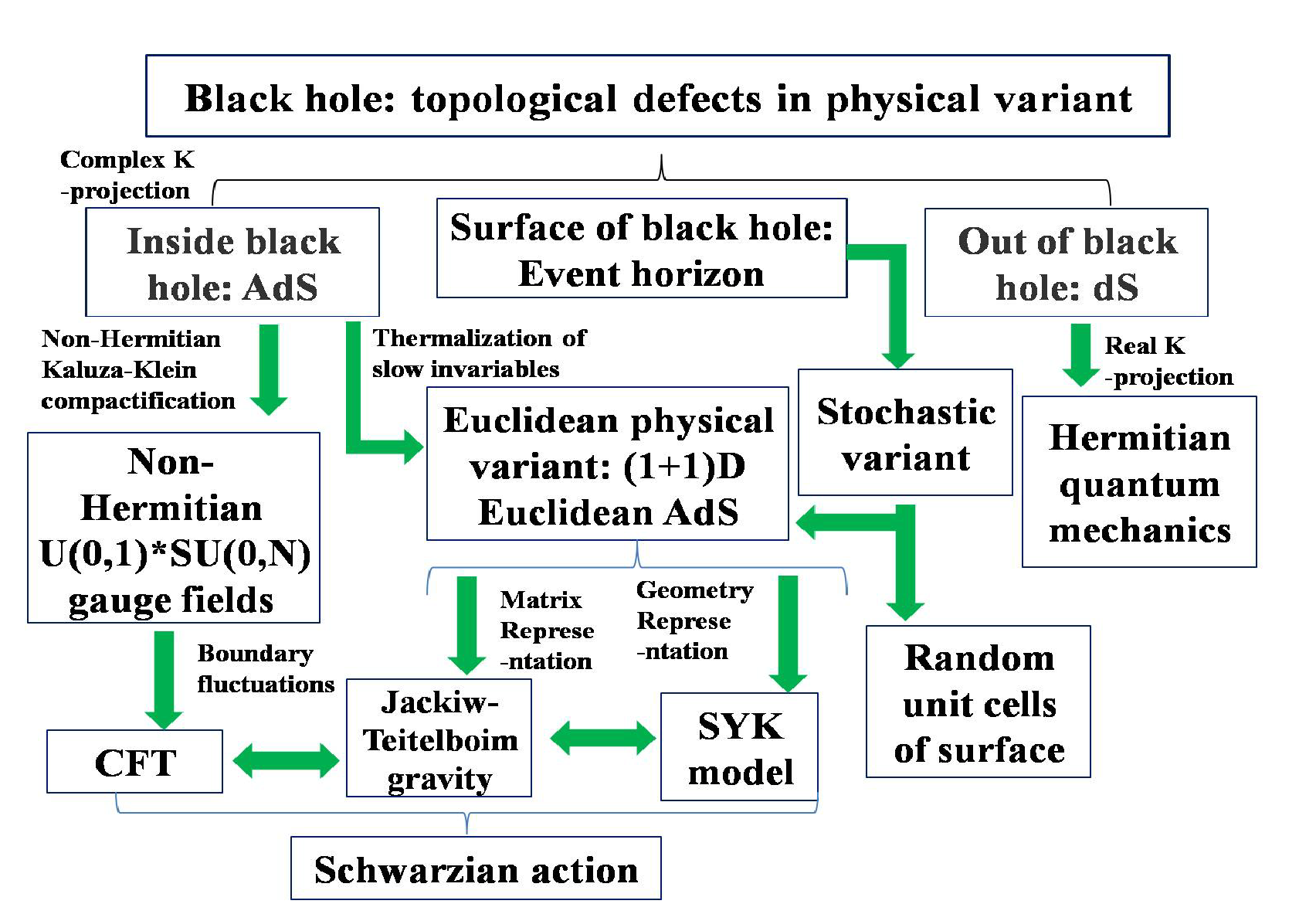}\caption{The logical
structure of theory for black hole}%
\end{figure}

In the end of this part, we answer all five questions at beginning and show
how the troubles about black hole disappear:

1. What's the exact \emph{microstructure} of spacetime around black hole near
Planck length? What's the exact \emph{microstructure} of spacetime inside
black hole? And, how characterize it?

\textbf{Answer:}

Now, the event horizon of a black hole becomes a topological domain wall
between a unitary physical variant (or a dS) and a non-unitary physical
variant (or an AdS). Because the spacetime inside black hole is AdS, we use
Gravity/N-gauge equivalence to characterize its dynamics. Now, the physical
processes for slow motion come from non-Hermitian \textrm{U(0,1)}$\times
$\textrm{SU(0,N)} gauge fields. By integrating fast variables, we get
effective model for slow variables. The effective model has three equivalent
forms: one is effective Jackiw-Teitelboim gravity under geometric
representation, second is effective SYK model under matrix representation,
third is effective 1D gauge theory under kinetic representation.

2. What is the exact solution for the singularity problem?

\textbf{Answer:}

The key point is the imaginary nature of the coordinates along radial
direction inside a black hole. Near the singularity, the curvature of
spacetime becomes imaginary. So, by using non-Hermitian quantum mechanics, the
trouble about singularity doesn't exist at all.

3. A major goal of research in quantum gravity is to provide a derivation of
the formula for the entropy of a black hole. What is the exact approach to
derive the entropy of black hole? Why black hole has finite temperature?

\textbf{Answer:}

In variant theory, the black hole is a U-N class d-2 dimensional topological
defect. Due to the "non-changing" structure along tempo direction, the event
horizon of the black hole becomes a stochastic variant with a random
distribution of unit cells. Under an assumption of Principle of equal
probability and the constraint of energy (or particle number), we have a new
statistics of spacetime $\Omega=\frac{(N_{U})^{N_{U}}}{(N_{U})!}$ where
$N_{U}$ is the number of unit cells. As a result, in thermodynamic limit, a
black hole becomes a classical object with finite temperature. From the
statistics of spacetime, the Hawking entropy, Hawking temperature are exactly derived.

4. How to solve the black hole information paradox? Is quantum mechanics
wrong, or is general relativity wrong? Or both wrong? Is Page curve for
Hawking radiation correct?

\textbf{Answer:}

According to above discussion, the randomness from non-variability of event
horizon leads to thermalization and decoherence of the quantum states near
event horizon. The event horizon can be regarded as a classical object with
finite temperature. When a quantum object reaches the classical object,
quantum measurement occurs. Therefore, the quantum information disappear and a
pure quantum state evolves to a mixed state. This indicates usual quantum
mechanics becomes invalid near event horizon! Hence, the \textquotedblleft%
\textit{black hole information paradox}\textquotedblright \ is solved. Our
results indicate that Page curve cannot characterize the information process
for Hawking evaporation of black hole.

5. SYK model is relevant to physics of black hole. What does this model really
mean? How to provide a derivation of the formula for SYK model?

\textbf{Answer:}

In matrix representation, the shape fluctuations of the event horizon become
the fluctuations of the external normal directions (or Gamma matrices
$\Gamma_{k}^{r}$). By integrating fast invariable with different wave vectors,
we obtain an effectively coupling between Gamma matrices $\Gamma_{k}^{r}$.
Then, the low energy effective model becomes SYK model.\ So, the SYK model
characterizes the random coupling between Gamma matrices $\Gamma_{k}^{r}$ on
event horizons. The formula can be applied to all kinds of black hole rather
than only extremal one with its fine-tuned magnetic charge.

\newpage

\section{Theory for Scattering Amplitudes -- from Dynamical Physics to Event
Physics}

\subsection{Introduction}

Scattering amplitudes are the central predictions in theories of fundamental
interactions. By detecting scattering amplitudes in experiments, people can
obtain the information of the input particles. A standard approach about
scattering amplitudes in perturbation theory is to use Feynman diagrams.
However, it is very difficult to obtain the exact results of scattering
amplitudes by directly calculating Feynman diagrams. Fortunately, in certain
systems, there may exist a shortcut obtaining the exact results of scattering
amplitudes without using the diagrammatic expansion.

In 2003, Witten developed the theory \cite{Witten:2003nn} that provides a
strikingly compact formula~\cite{rsv} for tree--level scattering amplitudes in
four-dimensional (4D) Yang-Mills theory in terms of an integral over the
moduli space of maps from the $n$-punctured sphere in momentum space
\cite{Witten:2003nn,rsv,ca,Cachazo:2012da,Cachazo:2012kg,Huang:2012vt} An
important progree is about gravitational amplitudes that become the square of
Yang-Mills amplitudes (or the so-called double copy)\cite{Kawai:1985xq}. Then,
the duality between colour and kinematics was explored\cite{Bern:2010ue}. In
Ref.\cite{chy1,chy2,chy3,chy4}, Cachazo, He and Yuan (CHY) equation was
proposed, by which the scattering amplitudes of massless particles of spins 0,
1 or 2 in arbitrary dimension are obtained.

On the other hand, the Britto-Cachazo-Feng-Witten (BCFW) recursion relations
were obtained\cite{BCFW1, BCFW2}. By the BCFW recursion relations, people can
represent the amplitude as a sum over basic building blocks. The existence of
building block for scattering amplitudes indicates a new structure in
algebraic geometry, that was known as the positive Grassmannian\cite{ar1,ar2}.
The recursion relations can be solved in many different ways, and the final
amplitude can be expressed as a sum of on-shell processes. The on-shell
diagrams satisfy identities from their association with cells of the positive
Grassmannian. A new geometric representation for the amplitude was then
discovered called \textquotedblleft Amplituhedron", of which \textquotedblleft
dual volume" of \textquotedblleft certain canonical region" with different
\textquotedblleft triangulations" of "certain space"\cite{ar3}.

Furthermore, it was known that these representations are supported on
solutions of the scattering equations by using cohomology classes on
ambitwistor space\cite{am1}. Then, the amplitudes for particles of different
spins (the scalar, Yang-Mills and gravitational amplitudes) arise from the
bosonic, `heterotic' and `type II' ambitwistor strings, respectively.

Despite significant progresses, the whole picture about the scattering
amplitudes are still not complete and there are a lot of unsolved mysteries:

\begin{enumerate}
\item What's the exact \emph{microstructure} of the scattering amplitudes for
different particles?

\item Why \emph{ambitwistor strings}? The bosonic and heterotic models of
strings are problematic because the gravitational amplitudes they contain do
not seem to correspond to Einstein gravity.

\item Why \emph{double copy}?

\item Why \emph{amplituhedron}? The connection between the amplituhedron and
scattering amplitudes is still a conjecture.

\item How to calculate \emph{loop} amplitudes?
\end{enumerate}

All above puzzles are relevant to the theory of quantum gravity. In this part,
we develop a new theory beyond "quantum field theory" to calculate the
scattering amplitudes. All physical processes of scattering amplitudes are
intrinsically described by the processes of the changings of \emph{angular
variant.} The angular variant is defined by a mapping between angular
group-changing space and angular space, i.e.,
\begin{align*}
&  \text{Scattering amplitudes }\\
&  \Longrightarrow \text{Event processes on angular space.}%
\end{align*}

So, another important concept is "\emph{event physics}". During the scattering
processes, the information of outcome (or the final states) is determined by
the initial state. This introduces the physics of event process. In this part,
we will point out that the event processes and corresponding theory are quite
different from those for dynamical processes. Within the new theory, we answer
above five questions.

\subsection{Event processes in physics}

\subsubsection{Events processes: concept and classification}

In physics, measurement is a very important issue. People obtain the
information of certain systems through experiments and test the rationality of
physical laws. During measurement, there occur \emph{event processes}. For
event processes, people only concern about the information of final states
from given initial states that are respectively the state at infinite future
and past. People don't know the detailed structure of the intermediate
processes under time evolution. Therefore, event processes can be regarded as
dynamical processes under projection, i.e.,%
\begin{align*}
&  \text{Event process = Projected dynamic processes }\\
&  \text{without knowing the detailed structure under time evolution.}%
\end{align*}
It looks like, without knowing the detailed structure under time evolution,
the theory about event processes is simpler than dynamical ones. However,
without the detailed structure under time evolution, the theories for event
processes always look strange and become \emph{counter-intuitive}.

Next, we classify event processes.

In our world, there exist two types of different objects, \emph{classical}
objects or \emph{quantum} objects. Classical object is a \textquotedblleft
non-changing" object with disordered group-changing elements and classical
motion describes certain globally motion of a quantum/classical object with
ordered/disordered group-changing elements; quantum object is a
\textquotedblleft changing" object with ordered group-changing elements and
quantum motion describes the ordered relative motion between group-changing
elements of the elementary particles\cite{kou1}. Therefore, there are totally
three types of event processes (or measurement) in our world, classical to
classical event (\textrm{CC}-event), quantum to classical event (\textrm{QC}%
-event), quantum to quantum event (\textrm{QQ}-event).

In the following parts, we simplify "event processes" by "event".

\subsubsection{CC event}

\textrm{CC}-event denotes a process from classical initial states to classical
final states without knowing the detailed structure of the intermediate
processes under time evolution.

During the processes of CC-event, we may assume that there at least exist
three physical objects -- object to be measured (classical object \textrm{A}
with velocity $\vec{v}_{A}$), the surveyors or instruments (classical object
\textrm{B} with velocity $\vec{v}_{B}$), and rigid spacetime as reference with
zero velocity. We consider two objects \textrm{A} and \textrm{B} doing
classical motion. We assume that for the observers \textrm{A}, the rulers and
clocks are independent of the physical properties of the measured object
\textrm{B}. We may denote the \textrm{CC}-event by a mapping between the two
classical objects on rigid spacetime, i.e.,
\begin{equation}
\mathrm{CC}\text{-event: }\tilde{V}_{A}\longrightarrow \tilde{V}_{A^{\prime}}.
\end{equation}

In particular, the theory for CC-event depends on the dispersion of elementary
particles. For example, the case of the linear dispersion is quite different
from that of quadratic one. For the case of the linear dispersion, we have the
$\mathrm{SO(1,3)}$ Lorentz group. Now, the correct theory that characterizes
the CC-event is just the special relativity. In this part, we focus on this case.

During CC-event, without knowing the detail dynamical processes, the global
information of classical object A with velocity $\vec{v}_{A}$ can be obtained
by the surveyors or instruments (classical object B with velocity $\vec{v}%
_{B}$). According to special relativity, clocks at different points can only
be synchronized in the given frame. If we want to know the relation between
the times between these ticks as measured in both objects, we have $\Delta
t^{\prime}=\gamma \Delta t$ (for events in which $\Delta x=0)$ that is larger
than the time $\Delta t$ between these ticks as measured in the rest frame of
the clock. This phenomenon is called time dilation. The length $\Delta
x^{\prime}$ in the 'moving' frame $S^{\prime}$ is shorter than the length
$\Delta x$ in its own rest frame. This phenomenon is called length contraction
or Lorentz contraction.

As typical CC-event, these effects are not merely appearances. However, the
detailed structure of \textrm{CC}-event is characterized by a classical motion
of time evolution in general relativity. By using the framework of general
relativity, CC-event returns to a dynamical process, of which the
corresponding effects (time dilation or length contraction) are no more counter-intuitive.

\subsubsection{QC events}

\textrm{QC}-event denotes the event process from quantum initial states to
classical final states. \textrm{QC}-event is defined by a mapping between a
quantum state and a classical one, i.e,%
\begin{align*}
&  \text{QC type of event physical process }\\
&  \text{=}\text{ A mapping between quantum state }\\
&  \text{and classical state.}%
\end{align*}

Quantum measurement is a typical \textrm{QC}-event from an unknown quantum
state to classical states of instruments \textrm{B}, i.e.,%
\begin{equation}
\text{Quantum measurement: }\tilde{V}_{A}\Longrightarrow \tilde{V}_{B}.
\end{equation}
During quantum measurement there must exist a \textrm{R}-process that denotes
a process from a quantum object to a classical one. This is called decoherence
in traditional quantum physics. As a result, a regular distribution of the
group-changing elements for a quantum object suddenly changes into a
disordered distribution of the group-changing elements for a classical object.

During QC-event, without the detail dynamical processes, the information of
quantum object \textrm{A} is obtained by the surveyors or instruments
(classical object \textrm{B}). The results is consistent to those predicted by
quantum mechanics without considering the master equation. From point view of
quantum mechanics, the probability in quantum mechanics occurs.

In principle, one can derive the detailed results of the \textrm{QC}-event by
solving the master equation.

\subsubsection{QQ events}

\paragraph{Review on scattering processes and scattering matrix}

Before discussing QQ events, we firstly review the scattering processes and
scattering matrix.

In quantum field theory, for a scattering process in flat spacetime, we define
$n_{in}$ original states and $n_{out}\,=\,n-n_{in}$ final states to be
$|p^{1}\ldots p^{n_{in}}\rangle_{in}{}$and $|p^{1}\ldots p^{n_{out}}%
\rangle_{out}$.\ Then, the elements of scattering matrix (S-matrix) describe
the transition amplitudes from initial states to final states
\[
{}_{out}\langle p^{1}\ldots p^{n_{out}}|p^{1}\ldots p^{n_{in}}\rangle
_{in}=\langle p^{1}\ldots p^{n_{out}}|\hat{S}|p^{1}\ldots p^{n_{in}}\rangle.
\]
The S-matrix operator can be conveniently written as $\hat{S}\,=1+i\hat{T}$,
with the operator $\hat{T}$ defining the scattering amplitude
\begin{align}
\langle p^{1}\ldots p^{n_{out}}|i\hat{T}|p^{1}\ldots p^{n_{in}}\rangle &
=M_{n}\left \{  p^{1}\ldots p^{n_{in}}\right \} \nonumber \\
&  \rightarrow \left \{  p^{1}\ldots p^{n_{out}}\right \}  .
\end{align}
The S-matrix operator $\hat{S}$ is unitary, i.e., $\hat{S}\hat{S}^{\dagger
}\,=1=\hat{S}^{\dagger}\hat{S}$. If we assume that all states are incoming,
the scattering amplitude becomes symmetric,
\[
M_{n}\,=\,M_{n}(p^{1},\, \ldots \,p^{n}).
\]

Because the scattering amplitude $M_{n}$ is invariant under the Poincar{\'{e}}
group, we add a $\delta$-function to guarantee\ momentum conservation and
consider the correct dispersion to ensure the Lorentz invariant.

The physical information for the massless representation of the Poincar{\'{e}}
group are encoded in the light-like momenta $p_{\mu}^{i}$ and in the
polarization tensors $\varepsilon_{\mu_{1}\ldots \mu_{s}}^{i}$. One can map a
Lorentz four-vector to a bi-spinor as
\begin{equation}
p_{\mu}\rightarrow p_{a\dot{a}}=\sigma_{a\dot{a}}^{\mu}p_{\mu}=\lambda
_{a}\tilde{\lambda}_{\dot{a}},
\end{equation}
where $\sigma_{a\dot{a}}^{\mu}\,=(1_{a\dot{a}},\, \overrightarrow{\sigma
}_{a\dot{a}})$ are the Pauli matrices. Now, the bi-spinor is denoted as a
direct product of two spinors $\lambda_{a}$ and $\tilde{\lambda}_{\dot{a}}$
that transform in the $(1/2,\,0)$ and $(0,\,1/2)$ representations of $SL(2,\,
\mathbb{C})$ and carry helicity $-1/2$ and $+1/2$, respectively.

Thus, the physical data about the external states of an amplitude can be
encoded in the pairs of spinors $(\lambda^{i},\, \tilde{\lambda}^{i})$ and the
helicities $h_{i}=\pm s_{i}$:
\begin{equation}
M_{n}=M_{n}\left(  \{ \lambda^{i},\, \tilde{\lambda}^{i};\,h_{i}\} \right)  .
\end{equation}
Helicity amplitudes with $h=4-n$ are called $\overline{\text{MHV}}$
amplitudes. A typical example is $n$-gluon MHV amplitudes at tree level. The
simplest non-vanishing helicity amplitudes with $h=n-4$, are called MHV
amplitudes that are characterized by the well known \emph{Parke-Taylor}
formula\cite{pt,zhang}
\[
M(1^{+},\ldots,i^{-},\ldots,j^{-},\ldots,n^{+})=\frac{\langle ij\rangle^{4}%
}{\langle12\rangle \langle23\rangle \cdots \langle n1\rangle}.
\]

\paragraph{Scattering processes as QQ-events}

\textrm{QQ}-event denotes an event process from some quantum initial states to
other quantum final states, that is defined by a mapping between different
quantum states, i.e.,%
\begin{align*}
\text{ QQ event }  &  \text{=}\text{ A mapping between }\\
&  \text{different quantum states.}%
\end{align*}
The detailed structure of \textrm{QQ}-event is characterized by a quantum
motion under unitary time evolutions, that is characterized by Sch\"{o}rdinger
equation (particularly, path integral approach).

Quantum scattering process is a typical \textrm{QQ}-event from initial quantum
state $\tilde{V}_{A}$ to the final quantum state $\tilde{V}_{B}$, i.e.,%
\begin{equation}
\text{Scattering process: }\tilde{V}_{A}\Longrightarrow \tilde{V}_{B}.
\end{equation}
\ Fig.20 shows a typical event process for quantum scattering process with
initial quantum state $\tilde{V}_{A}$ ($|p^{1}\ldots p^{n_{in}}\rangle_{in}$)
and the final quantum state $\tilde{V}_{B}$ (or $|p^{1}\ldots p^{n_{out}%
}\rangle_{out}$).\ In Fig.20(a), for all waves in and out, there exists a
\emph{common center}. Therefore, the wave vectors point to the common center,
i.e.,
\[
\Delta \vec{k}=\pm \left \vert \Delta \vec{k}\right \vert \vec{e}_{r}%
\]
where $\vec{e}_{r}$ denotes the radial direction out of the common center and
$\pm$ denotes in and out. \begin{figure}[ptb]
\includegraphics[clip,width=0.8\textwidth]{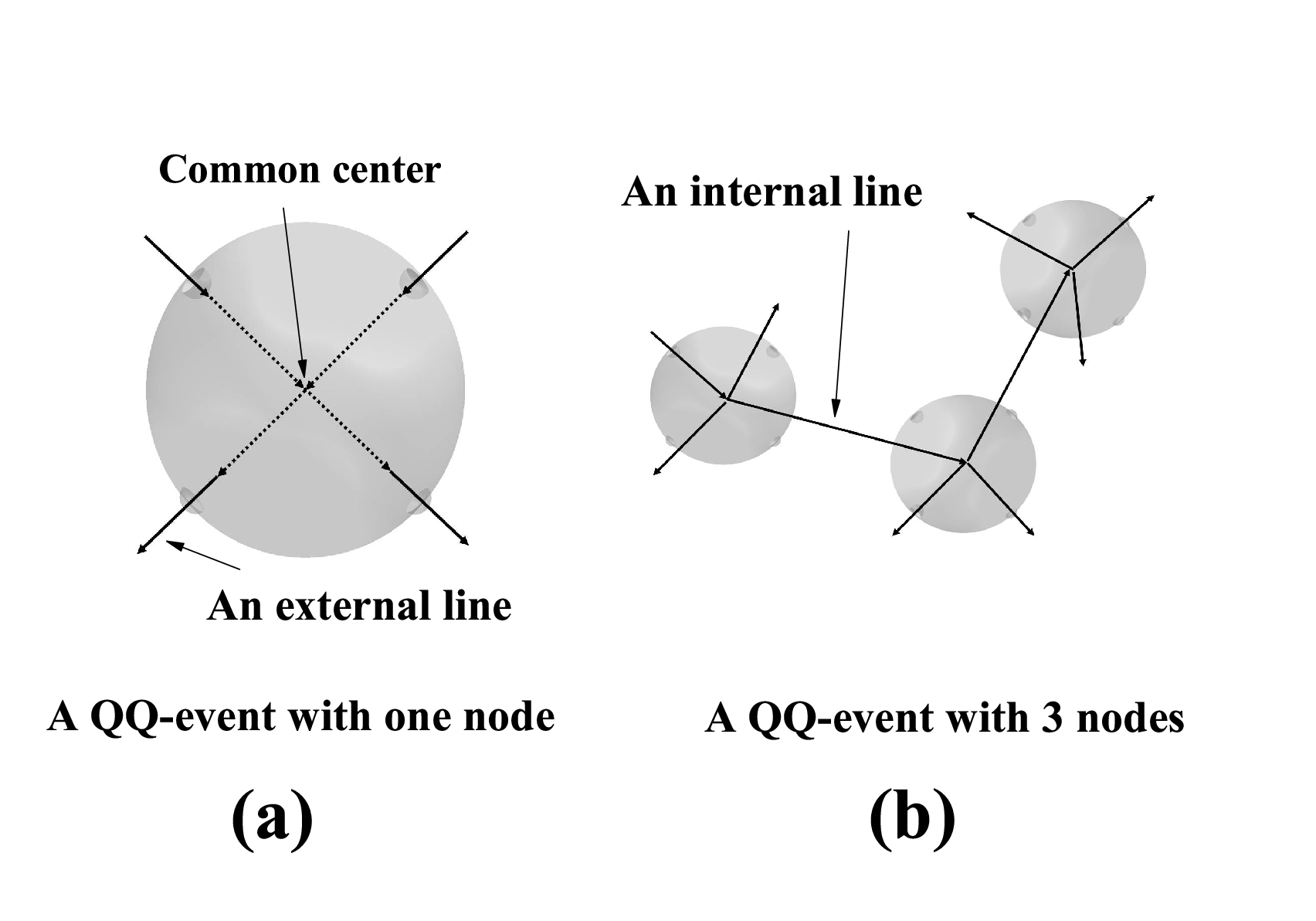}\caption{An illustration
of a typical event process for quantum scattering process with initial quantum
state $\tilde{V}_{A}$ ($|p^{1},p^{2}\rangle_{in}$) and the final quantum state
$\tilde{V}_{B}$ (or $|p^{3},p^{4}\rangle_{out}$).\ For all waves in and out,
there exists a common center. (a) A QQ event with 1 node; (b) A QQ event with
3 nodes.}%
\end{figure}

\paragraph{Classification of scattering processes}

Different QQ-events correspond to different Feynman diagrams. Above figure
shows an example of a QQ-event with a common center. In general, a QQ-event
may have several common centers. See the illustration in Fig.20(b). Then, we
classify the types of different scattering processes.

Firstly, we introduce the following mathematical terms: \emph{nodes},
\emph{external lines}, \emph{internal lines}, \emph{internal loops}. Node is a
common point that connects $n$ ($n\geq3$) external/internal lines; external
lines are the lines that connect only one node; internal lines are the lines
that connect two nodes; Internal loops are closed loop with end-to-end
connecting internal lines.

In particular, each node becomes an element of QQ-events. Or, a node
corresponds to an QQ-event. For a diagram with\ several nodes, we regard it as
a network of several correlated QQ-events. See the examples in Fig.20. For a
given Feynman diagram without internal loops, there only exist single internal
line that connects the two nodes; For a given diagram with internal loops,
there may exist several internal lines that connect the two nodes.

Finally, we classify the scattering processes.

The simplest scattering amplitude for QQ-event is those with single node. We
call them \emph{irreducible tree diagram}. The scattering amplitude for
QQ-event with several nodes and zero internal loop is called \emph{reducible
tree diagram}. The cases of several internal loops is called \emph{loop
diagram}.

\subsection{Fundamental theory for scattering processes -- angular variants}

\subsubsection{Angular variants for single QQ-event}

\paragraph{Definition of angular variant}

We start from a \textrm{QQ}-event with single node with a common center.

To characterize this simple QQ-event, we reduce the original physical
(d+1)-dimensional \textrm{\~{S}\~{O}(d+1)} physical variant\textit{
}$V_{\mathrm{\tilde{S}\tilde{O}(d+1)},d+1}(\Delta \phi^{\mu},\Delta x^{\mu
},k_{0},\omega_{0})$ to a residue (d-1)-dimensional \textrm{\~{S}\~{O}(d-1)}
angular variant\textit{ }$V_{\mathrm{\tilde{S}\tilde{O}(d-1)},d-1}%
^{\text{\textrm{Angular}}}(\Delta \phi^{\mu},\Delta \varphi^{\mu}).$ In the
following parts, we develop the theory for scattering amplitudes based on
angular variant\textit{ }$V_{\mathrm{\tilde{S}\tilde{O}(d-1)},d-1}%
^{\text{\textrm{Angular}}}(\Delta \phi^{\mu},\Delta \theta^{\mu})$.

We define the angular variant.

\textit{Definition: An angular variant }$V_{\mathrm{\tilde{S}\tilde{O}%
(d-1)},d-1}^{\text{\textrm{Angular}}}(\Delta \phi^{\mu},\Delta \theta^{\mu}%
)$\textit{ is a mapping between the }$\mathrm{\tilde{S}\tilde{O}(d-1)}%
$\textit{ group-changing space and the angular space of the original Cartesian
space }$\mathrm{S}_{d-1}^{\text{\textrm{Angular}}}$\textit{, i.e.,}%
\begin{equation}
V_{\mathrm{\tilde{S}\tilde{O}(d-1)},d-1}^{\text{\textrm{Angular}}}[\Delta
\phi^{\mu},\Delta \theta^{\mu},R]:\mathrm{C}_{\mathrm{\tilde{S}\tilde{O}%
(d-1),}d-1}=\{ \delta \phi^{\mu}\} \Longleftrightarrow \mathrm{S}_{d-1}%
^{\text{\textrm{Angular}}}=\{ \delta \theta^{\mu}\}
\end{equation}
\textit{where the d-1 dimensional angular space }$\mathrm{S}_{d-1}%
^{\text{\textrm{Angular}}}$\textit{ is sphere in d dimensional Cartesian space
with a radius }$R$ \textit{(or }$\mathrm{S}_{d-1}^{\text{\textrm{Angular}}}%
$\textit{ manifold). A group-changing space} $\mathrm{C}_{\mathrm{\tilde
{S}\tilde{O}(d-1)},d-1}(\Delta \phi^{a})$\textit{ is a group-changing space of
non-compact }$\mathrm{\tilde{S}\tilde{O}(d-1)}$\textit{ Lie group with fixed
sizes }$\Delta \phi^{a}$\textit{ along different directions.} For simplicity,
we can set the radius to be unit and get the dimensionless space.

For example, for the case of $d=3,$ $V_{\mathrm{\tilde{S}\tilde{O}(2)}%
,2}^{\text{\textrm{Angular}}}[\Delta \phi^{\mu},\Delta \theta^{\mu},R]$ denotes
a two dimensional (2D) group-changing space on a 2D sphere. An infinitesimal
element of group-changing space has $2$ component. To characterize the angular
variant, we have $2$ series of numbers of infinitesimal elements, i.e.,
\begin{equation}
V_{\mathrm{\tilde{G},}d}[\Delta \phi^{\mu},\Delta \theta^{\mu}]\text{: }%
\{n_{i}^{\mu}\},\text{ }(\mu=x,y)\}.
\end{equation}

Angular variant\textit{ }$V_{\mathrm{\tilde{S}\tilde{O}(d-1)},d-1}%
^{\text{\textrm{Angular}}}(\Delta \phi^{\mu},\Delta \theta^{\mu},R)$ is a
sub-variant for the original \textrm{\~{S}\~{O}(d+1)} physical variant\textit{
}$V_{\mathrm{\tilde{S}\tilde{O}(d+1)},d+1}(\Delta \phi^{\mu},\Delta x^{\mu
},k_{0},\omega_{0})$. Without the variability along tempo direction, the
angular variant\textit{ }$V_{\mathrm{\tilde{S}\tilde{O}(d-1)},d-1}%
^{\text{\textrm{Angular}}}(\Delta \phi^{\mu},\Delta \theta^{\mu},R)$ is not a
physical variant. In addition, we point out that the theory for angular
variant will provide a solid physical foundation for \emph{ambitwistor space}
and the \emph{celestial sphere}.

\paragraph{1-th order variability for angular variant}

For uniform angular variant $V_{\mathrm{\tilde{S}\tilde{O}(d-1)}%
,d-1}^{\text{\textrm{Angular}}}(\Delta \phi^{\mu},\Delta \theta^{\mu},R)$, there
exists 1-th order variability of spatial transformation, i.e.,%
\begin{equation}
\mathcal{T}(\delta \theta^{\mu})\leftrightarrow \hat{U}(\delta \phi^{\mu}),\text{
}\mu=x,y
\end{equation}
where $\hat{U}(\delta \phi^{\mu})=e^{i\Gamma^{\mu}\delta \phi^{\mu}}$ with
$\delta \phi^{\mu}=\sqrt{N_{\mathrm{tot}}^{F}}\delta \theta^{\mu}$ are the
translation operations in non-compact \textrm{\~{S}\~{O}(d-1)} Lie group.
$\Gamma^{\mu}$ is Gamma generator $\{ \Gamma^{i},\Gamma^{i}\}=2\delta^{ij}$
and for the 2D case, it is Pauli matrices.\textit{ }$N_{\mathrm{tot}}^{F}$ is
total number of elementary particles inside the angular space $S_{d-1}%
^{\text{\textrm{Angular}}}$ (or the sphere with a radius $R$). Due to the
relationship between particle number and the magnetic charge, there exist
$N_{\mathrm{tot}}^{F}$ inside the angular space $S_{d-1}%
^{\text{\textrm{Angular}}}$.

For a 2D angular variant, there also exists a 1-th order rotation variability
is defined by
\begin{equation}
\hat{U}_{L}^{\mathrm{R}}\leftrightarrow \hat{R}_{\mathrm{space}}%
\end{equation}
where $\hat{U}_{L}^{\mathrm{R}}$ is a rotation operator from one transverse
direction to another on the angular space $S_{d-1}^{\text{\textrm{Angular}}}$.

\paragraph{Representations for angular variant}

Firstly, we characterize an angular variant by geometry representation via
\textquotedblleft \emph{topological lattice}\textquotedblright \ on angular space.

According to the higher order variability $\mathcal{T}(\delta \theta^{\mu
})\leftrightarrow \hat{U}(\delta \phi^{\mu})=e^{i\Gamma^{\mu}\sqrt
{N_{\mathrm{tot}}^{F}}\delta \theta^{\mu}}$, along an arbitrary direction
($\mu=x,y$) after shifting the distance $2\pi/\sqrt{N_{\mathrm{tot}}^{F}}$,
the phase angle of the ground state changes $2\pi.$ The coordinates unit
vectors of angular space $\mathbf{e}^{\mu}$ correspond to Gamma matrices of
non-compact \textrm{\~{S}\~{O}(d-1)} Lie group $\Gamma^{\mu},$ $\mathbf{e}%
^{\mu}\leftharpoondown \Gamma^{\mu}.$

We then do \emph{compactification} on the angular group-changing space
$\mathrm{C}_{\mathrm{\tilde{S}\tilde{O}(d-1)}}$. After compactification, the
coordinate of $\mathrm{C}_{\mathrm{\tilde{S}\tilde{O}(d-1)}}$ along the given
direction is reduced to a compact one, i.e., $\phi^{\mu}(\theta)=2\pi N^{\mu
}(\theta)+\varphi^{\mu}(\theta).$ We relabel a position on angular space by
two numbers ($\varphi(\theta),N(\theta)$): $\varphi^{\mu}(\theta)$ is a small
phase angle $\varphi^{\mu}(\theta)\in \lbrack0,2\pi)$, the other is a very
large integer number $N^{\mu}(\theta)$. Now, we have a theory of
\emph{compact} \textrm{SO(d-1)} group on a crystal labeled by $N^{\mu}%
(\theta)$ and get \emph{\textquotedblleft topological\textquotedblright}
version lattice on angular space.

Next, we characterize the angular variant by matrix representation via a
\textquotedblleft \emph{matrix network}\textquotedblright.

The matrix network is described by $\Gamma^{\{N^{\mu},M^{\mu}\}}$ on the links
between two nearest-neighbor lattice sites $N^{\mu}$ and $M^{\mu}$ of the
topological lattice of spacetime. Or, $\Gamma^{\{N^{\mu},M^{\mu}\}}$ on
different paired links of the topological lattice of spacetime constitute\ a
matrix network.\textit{ }In the continuum limit, the Gamma matrix of matrix
network is reduced to the usual Gamma matrix in the "Dirac equation"
$\Gamma^{\mu}$ for tachyons.\begin{figure}[ptb]
\includegraphics[clip,width=0.75\textwidth]{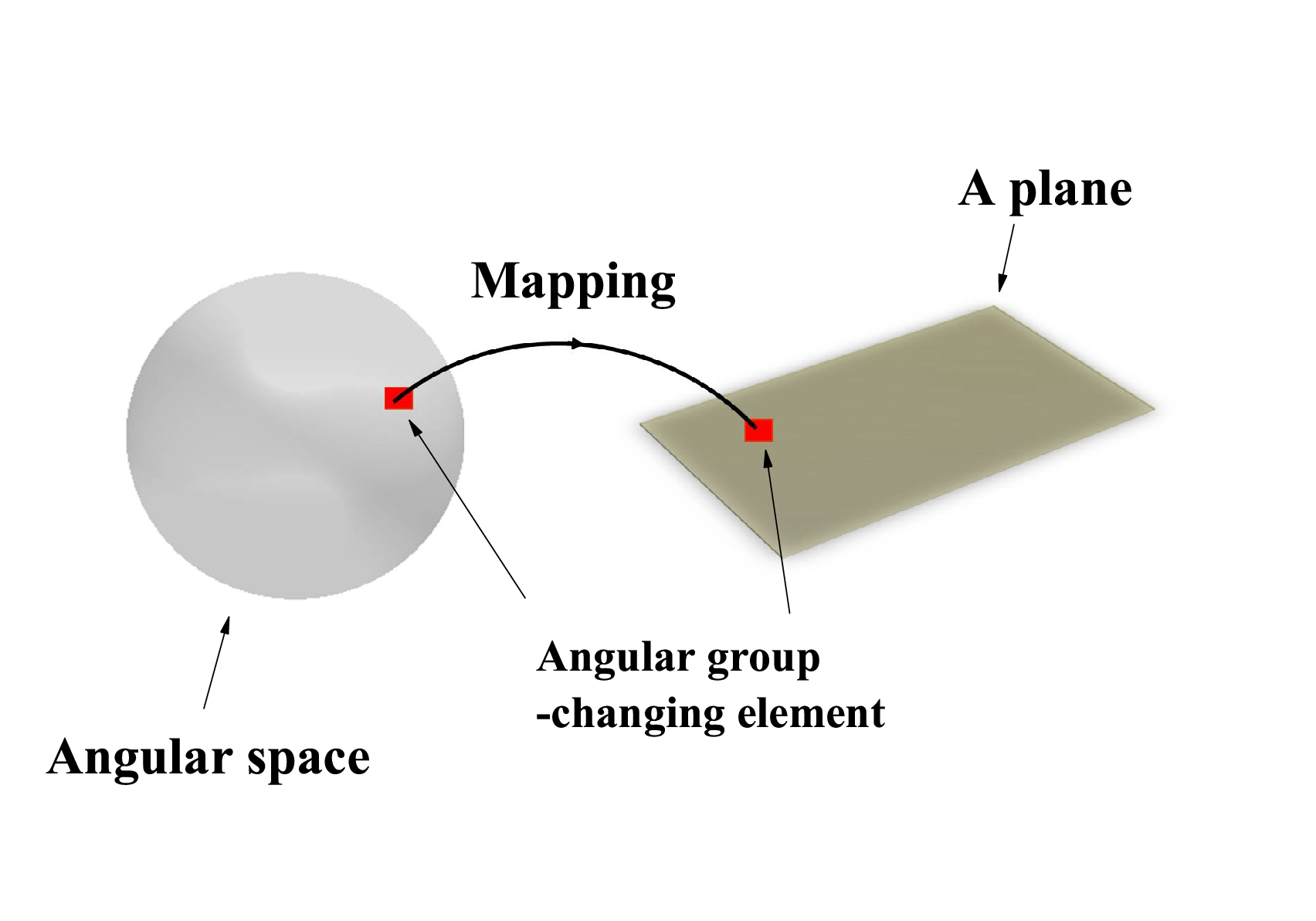}\caption{An
illustration of an angular space that is always mapped onto a plane. The
angular group-changing element is denoted by a red spot.}%
\end{figure}

In this part, we only focus on the case of uniform angular variant that
corresponds to a flat spacetime. See the illustration in Fig.21. Now, the
angular space is locally mapped onto a plane.

\paragraph{Forbidden phase changing from null condition}

On angular space, the physical processes don't have phase changings. This
phenomenon is called\emph{ forbidden phase changing}.

The fact of forbidden phase changing comes from the motion occurs along radial
direction, i.e., the wave vector $\vec{k}=\left \vert k\right \vert \vec{e}_{r}%
$. On angular space, due to the orthogonality relationship, the corresponding
transverse wave vectors are fixed to be zero,
\[
\Delta \vec{k}_{\theta}\equiv0.
\]
Because the transverse wave vectors are really angular momentum of particles,
they are conserved quantities. This is always called \emph{null condition}.

On the other hand, the local phase changings $\delta \varphi(\theta^{\mu})$
must be accompanied by the changing of wave vectors as
\begin{equation}
\delta \varphi(\theta^{\mu})=%
{\displaystyle \sum \limits_{k_{\theta}^{\mu}}}
\delta \varphi_{k_{\theta}^{\mu}}e^{ik_{\theta}^{\mu}\cdot \theta^{\mu}}.
\end{equation}
Without changings from wave vector on angular space (or $\delta k_{\theta
}^{\mu}\equiv0$), except for a global phase factor, the local phase changing
becomes forbidden,
\[
\delta \varphi(\theta^{\mu})\equiv0.
\]

The situation is similar to superconducting systems with Majorana fermions.
Due to phase coherence from order parameter of superconducting pairing, the
local phase changing is also forbidden. As a result, the phyiscal processes of
angular variants are always described by the representation of Majorana
fermions rather than complex ones.

\paragraph{Quantized geometry on angular space}

In this section, we discuss geometric quantities of angular variants by using
matrix (gauge) representation.

In Riemannian geometry, the 2-area for the surface $\mathrm{S}_{d-1}%
^{\text{\textrm{Angular}}}$ is defined by
\begin{equation}
\Delta S=\frac{1}{2}%
{\displaystyle \iint_{\mathrm{S}_{d-1}^{\text{\textrm{Angular}}}}}
\epsilon_{ab}e^{a}\wedge e^{b}.
\end{equation}
For the case of flat quantum spacetime, an area of surface is quantized and
the value of area is topological invariable. Now, the unit of surface is that
with smallest area -- a plaquette with four nearest neighbor lattice sites of
topological lattice. An arbitrary surface can be regarded as a system with a
lot of surface unit. This fact leads to area quantization of a surface.

According to the intrinsic relationship between the gauge representation\ and
the geometric representation, we find that the 2-area $\Delta S$ becomes the
flux number in gauge representation, i.e.,
\begin{align}
\Delta S  &  =\frac{1}{2}%
{\displaystyle \iint_{\mathrm{S}_{d-1}^{\text{\textrm{Angular}}}}}
\epsilon_{ab}e^{a}\wedge e^{b}\\
&  =\frac{1}{2}(l_{0})^{2}%
{\displaystyle \iint_{\mathrm{S}_{d-1}^{\text{\textrm{Angular}}}}}
\epsilon_{ab}A^{a0}\wedge A^{b0}\nonumber \\
&  =-\frac{1}{2}(l_{0})^{2}%
{\displaystyle \iint_{\mathrm{S}_{d-1}^{\text{\textrm{Angular}}}}}
\epsilon_{ab}F^{ab}=-\Delta \Phi(l_{0})^{2}\nonumber
\end{align}
where $\Delta \Phi$ is the flux penetrating the surface $\mathrm{S}%
_{d-1}^{\text{\textrm{Angular}}}.$ Here, we have used the following
equations,
\begin{equation}
e^{a}\wedge e^{b}=(l_{0})^{2}A^{a0}\wedge A^{b0}%
\end{equation}
and
\begin{align}
F^{ab}  &  =dA^{ab}+A^{ac}\wedge A^{cb}\\
&  \equiv-A^{a0}\wedge A^{b0}.\nonumber
\end{align}
Here, $l_{0}$ is the minimum lattice distance along arbitrary direction, i.e.,
$l_{0}=2\pi/\sqrt{N_{\mathrm{tot}}^{F}}$.

As a result, the area (or solid angle) means "flux" of gauge structure on flat
spacetime, i.e.,%
\begin{equation}
\Delta S=-\Delta \Phi(l_{0})^{2}.\nonumber
\end{equation}
The total size of the angular group-changing space is just the total flux
penetrating the surface. The situation is very similar to the FQH states on
Haldane sphere.

\subsubsection{Angular matter}

\paragraph{Definition of information unit}

Matter comes from size changings of group-changing space in a physical
variant. The elementary particle becomes information unit of a physical unit.
For angular variants, situation becomes very different! Matter is no more
usual elementary particles. Instead, they are called \emph{angular matter,}
that comes from size changings of angular group-changing space in an angular
variant. The information unit is quantized flux with unit angular momentum,
for example, photons, or gluons. In the following parts, we focus on the case
of $d=3$.

Firstly, we define the information unit of angular variant:\textit{ }

\textit{Definition: Information unit is the object with quantized angular
momentum (or }$\Delta L=\pm1$\textit{) of a 2D }\textrm{\~{S}\~{O}(2)}
\textit{angular variants }$V_{\mathrm{\tilde{S}\tilde{O}(2)},2}%
^{\text{\textrm{Angular}}}[\Delta \phi^{\mu},\Delta \varphi^{\mu},R]$\textit{.}

To characterize the object with finite angular momentum, we transform the
original XY rectangular coordinates to cylindrical coordinates, i.e.,
\[
(\theta_{x},\theta_{y})\rightarrow(r,\theta)
\]
by $r^{2}=%
{\displaystyle \sum \limits_{\mu}}
\theta_{\mu}^{2},$ $\theta=\arctan \frac{\theta_{x}}{\theta_{y}}.$

Then, based on cylindrical coordinates, we discuss the object with $2\pi$-flux
on angular space.

According to above discussion, its area is $2\pi(l_{0})^{2}.$ Strictly
speaking, the object changes the area $2\pi$.\ The shape of the object can be
arbitrarily changed, as long as the area remains unchanged. If the shape of
the object is circle, its radius $r$ is $\sqrt{2}l_{0}$; if the shape of the
object is semicircle, its radius $r$ is $2l_{0}$. See the illustration in
Fig.23, in which we set $\frac{1}{N_{\mathrm{tot}}^{F}}$ to be unit.

Another fact is that an information with unit flux traps unit of angular
momentum. Let explain it.

Moving around an object with $2\pi$-flux, the changing of phase factor of the
system becomes $2\pi.$ As a result, the angular momentum becomes $2\pi$. As a
result, we have%
\begin{align*}
\text{Information unit }  &  \Longleftrightarrow \text{Unit angular momentum}\\
&  \Longleftrightarrow \text{Changing unit of solid angle. }%
\end{align*}
Therefore, the situation is again similar to superconductors with "p-wave"
Copper pair on angular space. The information unit becomes quantized flux with
unit angular momentum $L=1$. See the illustration of the triangular
equivalence principle for excited modes (for example, gluons, gravitons) on
angular space in Fig.22.\begin{figure}[ptb]
\includegraphics[clip,width=0.67\textwidth]{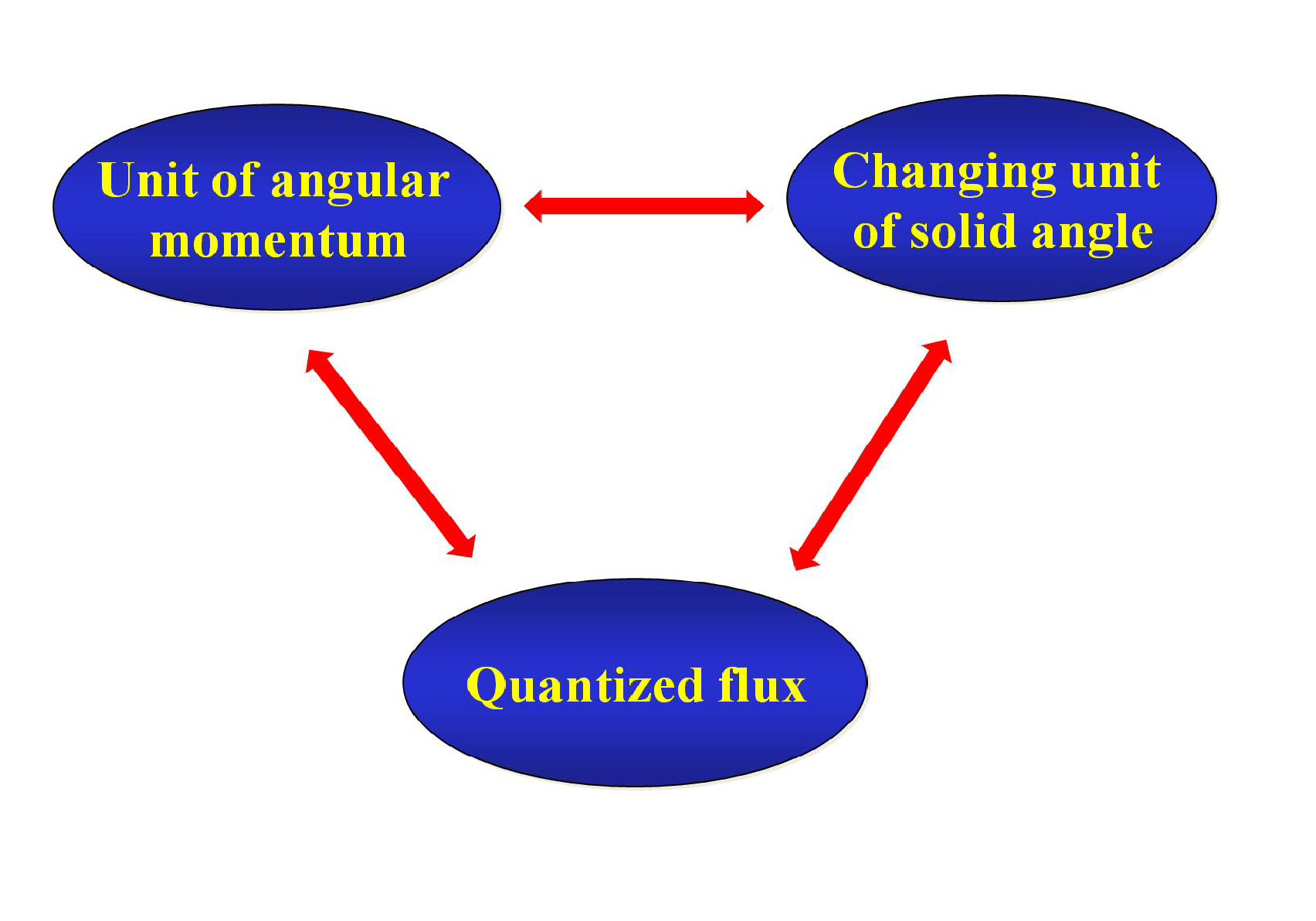}\caption{An
illustration of the triangular equivalence principle for excited modes (for
example, gluons, gravitons) on angular space. This is an intrinsic
relationship between unit of angular momentum, changing unit of solid angle of
angular space and quantized flux. }%
\end{figure}

Finally, we show the physical picture for different excited modes.

Vector fields (photons/gluons) are angular matter with unit angular momentum
$\Delta L=1;$ tensor fields (gravitational waves) are angular matter with
total angular momentum $\Delta L=2$ that can be regarded as a composite object
with two photons of orthogonal polarization directions; Bi-adjoint scalar
field with $\phi^{3}$ self-interaction\cite{chy4} can be regarded as a a
composite object with two photons of opposite angular momenta.

In addition, we point out that for excited mode, the quantum statistics on
angular spacetime is always \emph{different} from the usual quantum statistics
on Cartesian spacetime.

\paragraph{Property of vector fields on angular space}

According to above discussion, vector fields including photons and gluons are
angular matter with angular momentum $\Delta L=1.$ Let us discuss its properties.

On angular space, an excited mode of vector field has fixed area, rather than
a point. The fixed area corresponds to a fixed expansion or contraction of the
angular group-changing space. Due to the conservation of angular momentum, the
area in angular group-changing space cannot be changed. Under fixed changing
rate, the corresponding area in angular space also conserved quantity.

To characterize the shape changings of an excited mode of vector fields, we
introduce an additional degrees of freedom.

Now, to characterize the geometric distribution of group-changing element, we
introduce the $r^{2}$-coordinates ($r^{2},\theta$) on angular space. See the
illustration in Fig.23, in which we set $l_{0}$ to be unit, i.e., $l_{0}=1.$
There are two different configurations for the shapes under $r^{2}%
$-coordinates ($r^{2},\theta$). We denote them by $\left \vert 0\right \rangle $
and $\left \vert 1\right \rangle $, each of which becomes a base. The base
$\left \vert 0\right \rangle $ denotes the circle-like shape in the isotropic
limit; the other $\left \vert 1\right \rangle $ denotes the semicircle-like
shape in the fully anisotropic limit.

\begin{figure}[ptb]
\includegraphics[clip,width=0.8\textwidth]{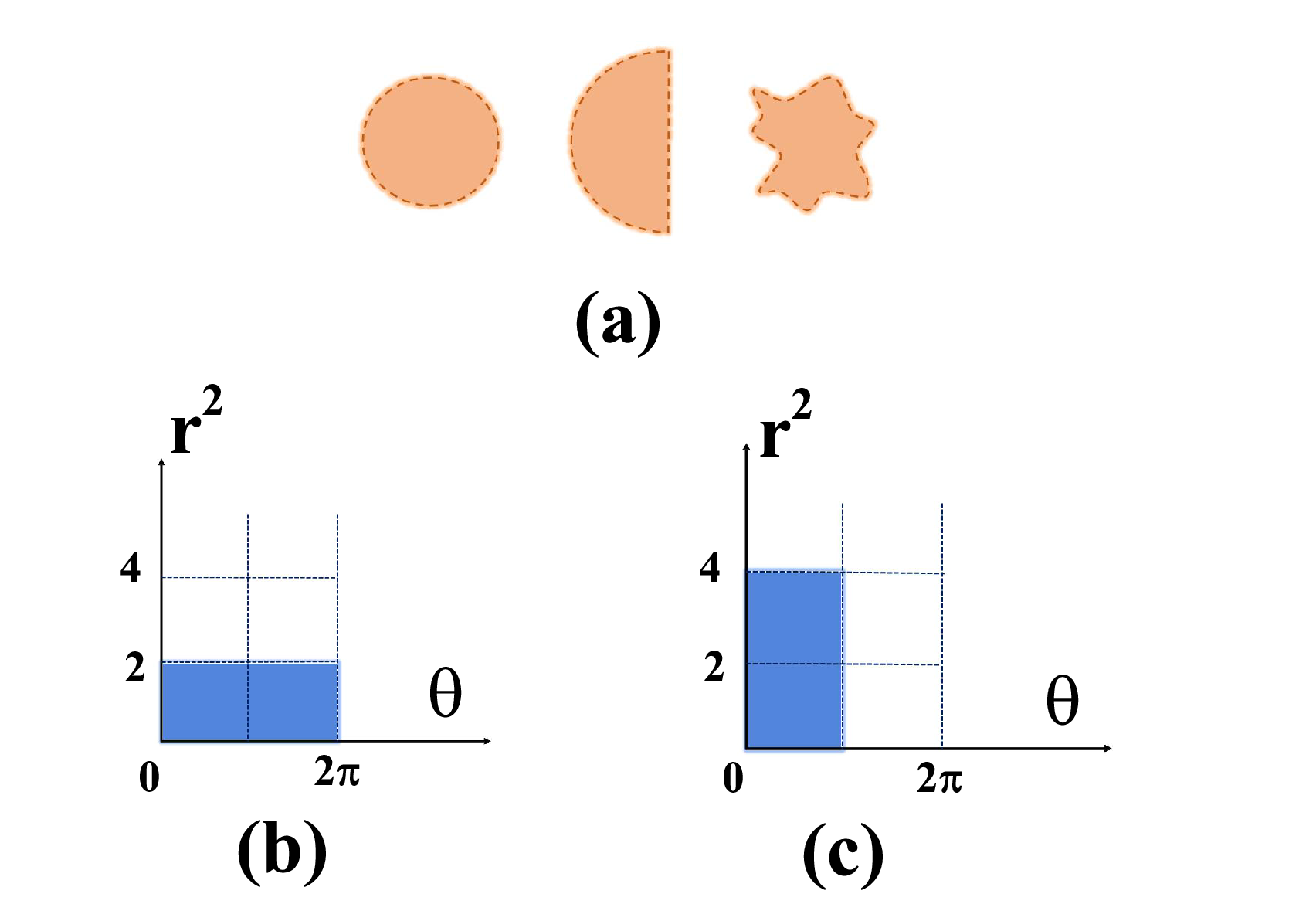}\caption{(a) An
illustration of the shape changings of the excited mode with fixed area and
changing rate; (b) An illustration of an excited mode with spin-1
(photon/gluon) of state $\left \vert 0\right \rangle $ that obeys angular
bosonic statistics; (c) An illustration of an excited mode with spin-1
(photon/gluon) of state $\left \vert 1\right \rangle $ that obeys angular
fermionic statistics. We set $l_{0}$ to be unit, i.e., $l_{0}=1$}%
\end{figure}

Then, we discuss the quantum statistics of vector fields (photons/gluons) on
angular space. To distinguish it from the quantum statistics in usual
spacetime, we call it \emph{angular quantum statistics}.

On the one hand, we study the angular quantum statistics for the state
$\left \vert 0\right \rangle .$ Without $2\pi$-phase change along $\theta
$-direction, the excited modes of $\left \vert 0\right \rangle $ obeys angular
bosonic statistics. When one particle moves around the other, there exists
$2\pi$ extra phase factor. For the state $\left \vert 0\right \rangle $ has
extra $2\pi$ phase factor, the circle-like shape is invariant;

On the other hand, we study the angular quantum statistics for the state
$\left \vert 1\right \rangle .$ Because, the excited mode of $\left \vert
1\right \rangle $ is $\pi$-phase change along $\theta$-direction. Therefore, it
obeys angular fermionic statistics. When one particle moves around the other,
there exists extra $\pi$ phase factor. For the state $\left \vert
1\right \rangle $ has extra $\pi$ phase factor, the left semicircle-like shape
turns into right semicircle-like shape.

In the following parts, we will show that the symmetry between $\left \vert
0\right \rangle $ and $\left \vert 1\right \rangle $ is just \emph{supersymmetry}.

In modern physics, there are two types excited modes of vector fields --
photons for Abelian \textrm{U(1)} gauge fields and gluons for non-Abelian
\textrm{SU(N)} gauge fields. So, we focus on the scattering processes of
self-interacting gluons. For gluons, in addition to the global geometric
degrees of freedom, there exist internal degrees of freedom.

\paragraph{Property of tensor fields on angular space}

Gravitational waves (or gravitons) are typical tensor fields with total
angular momentum $\Delta L=2.$ On angular space, the area of an excited
graviton are twice of that for vector fields (photons or gluons). Therefore,
gravitational waves can be regarded as a composite object with two photons of
orthogonal polarization directions.

\paragraph{Property of Bi-adjoint scalar fields on angular space}

A Bi-adjoint scalar (multi-component scalar mode with $\phi^{3}$
self-interaction) has zero angular momentum $\Delta L=0.$ On angular space,
the area of a excited scalar mode is zero. We can be regarded it as a
composite object with two gluons with opposite angular momenta (or opposite areas).

\subsubsection{Symmetry/invariant on angular space}

Before studying the motion of excited modes on angular space, we discuss the
invariant/symmetry of 2-th level physics structure for motions.

It was known that motion corresponds to locally expand or contract of the
angular group-changing space $\mathrm{C}_{\mathrm{\tilde{S}\tilde{O}%
(d-1)},d-1}$ on angular space. Different states of motions correspond to
different mappings between $\mathrm{C}_{\mathrm{\tilde{S}\tilde{O}(d-1)},d-1}$
and $\mathrm{S}_{d-1}$. If two states (or different mappings between
$\mathrm{C}_{\mathrm{\tilde{S}\tilde{O}(d+1)},d+1}$, and $\mathrm{C}_{d+1}$)
have same energy, we call such an invariance to be \emph{symmetry} of motions.

On the one hand, for uniform physical variant under compactification, the
continuous angular translation operation $\mathcal{T}(\delta \theta^{\mu})$ is
reduced into a discrete spatial angular translation symmetry $T(\delta
\theta^{\mu})$ on the angular zero lattice, i.e., $\mathcal{T}(\delta
\theta^{\mu})\leftrightarrow \hat{T}(\delta N^{\mu}).$ For angular topological
lattice, one lattice site is equivalence to another. As a result, in continuum
limit $l_{p}\rightarrow0$, the 1-th order angular variability is reduced to
continuous spatial translation invariance in rectangular coordinates.

On the other hand, under compactification, the operation $\tilde{U}^{\mu}$ of
non-compact $\mathrm{\tilde{S}\tilde{O}}$\textrm{(d-1)} group is reduced to a
global compact \textrm{U}$_{\mathrm{global}}$\textrm{(1)} group and a residual
compact \textrm{SO(d-1)} group. On each lattice site of zero lattice, we have
an invariant under the global compact \textrm{U}$_{\mathrm{global}}%
$\textrm{(1)} group and the compact \textrm{SO(d-1)} group, i.e.,
\begin{equation}
\tilde{U}^{\mu}\rightarrow \hat{U}_{\mathrm{U}_{\mathrm{global}}\mathrm{(1)}%
}(\delta \varphi)\otimes \hat{U}_{\mathrm{SO(d-1)}}.
\end{equation}
Due to the $U_{\mathrm{U}_{\mathrm{global}}\mathrm{(1)}}$ symmetry, the
particle number (total changing of angular momentum) $N=\Delta L$ becomes a
conserved quantity.

The compact \textrm{SO(d-1)} group is the \emph{Lorentz little group} that is
the subgroup of the Lorentz transformations which leaves the momentum of a
given particle unchanged. The total transformations can be classified by the
two Casimir operators of the Poincar{\'{e}} group, $\hat{P}^{2}$ and $\hat
{W}^{2}$, with $\hat{W}^{\mu}$ being the Pauli-Lubanski pseudo-vector, which
satisfies the commutation relations
\begin{align}
\lbrack \hat{W}^{\mu},\, \hat{P}^{\nu}]\,  &  =\,0,\ [\hat{L}_{\mu \nu},\hat
{W}_{\rho}]\nonumber \\
&  =i\left(  \eta_{\nu \rho}\hat{W}_{\mu}-\eta_{\mu \rho}\hat{W}_{\nu}\right)
,\nonumber \\
\lbrack \hat{W}^{\mu},\, \hat{W}^{\nu}]  &  =i\epsilon^{\mu \nu \rho \sigma}%
\hat{W}_{\rho}\hat{P}_{\sigma}.
\end{align}
Here, the little group transformations are generated by $\hat{W}^{\mu}$.
$\hat{L}_{\mu \nu}$ is the Lorentz generators.

However, the situation becomes complex due to forbidden phase changing. Now,
on angular space, the excited states $\left \vert 1\right \rangle $ obeys
angular fermionic statistics. The local phase changing can only be $0$ or
$\pi.$ Therefore, all phases of particles on angular space are fixed except
for the\emph{ sign} of the states. Hence, on angular space the corresponding
$U_{\mathrm{U}_{\mathrm{global}}\mathrm{(1)}}$\textrm{ }phase rotation
symmetry is broken to \textrm{Z2}$.$

In addition, for excited modes, there exists \emph{supersymmetry} on angular
space that characterizes the invariant/symmetry under the transformation of
particle's shape.

According to above discussion, there exists a (geometric) degrees of freedom
for photons/gluons by different internal states $\left \vert 0\right \rangle $
or $\left \vert 1\right \rangle $. The transformation between the two internal
states $\left \vert 0\right \rangle $ or $\left \vert 1\right \rangle $ changes
the angular quantum statistics for excited modes. Without changing the total
energy of the system, we have an emergent invariant/symmetry. The
corresponding operation changing the shape of the excited modes is denoted by
the super-operator $\hat{Q}$. Now, under the following super-operation,
\[
\delta X^{\mu}=\epsilon \Psi^{\mu},\text{ }\delta \Psi^{\mu}=\epsilon P^{\mu
}\text{, }\delta P_{\mu}=0,
\]
the energy of the system doesn't change. Here, $\delta X^{\mu}$ denotes an
infinitesimal shift, $\Psi^{\mu}$ is an angular fermion for the states
$\left \vert 1\right \rangle $.

This supersymmetry leads to a constraint to the effective action for
gravitons/gluons on angular space. Under this constraint, we will obtain an
effective action that is same to those on ambitwistor space.

\subsubsection{Angular motion}

Finally, we study the motion of excited modes (gravitons/gluons) on angular
space. An interesting result is that our effective models are same to those
about `type II ambitwistor strings.

\paragraph{Motion of photons/gluons on angular space}

In this part, we study the motion of photons/gluons on angular space and
obtain the effective action for them.

For a photon/gluon, there also exists internal geometric degree of freedom
that is characterized by $\left \vert 0\right \rangle $ or $\left \vert
1\right \rangle .$ If the internal state is $\left \vert 0\right \rangle $, it
obey bosonic angular quantum statistics; If the internal state is $\left \vert
1\right \rangle $, it obey fermionic angular quantum statistics.

First, we derive the effective action for a photon/gluon with internal state
$\left \vert 0\right \rangle .$ Now, it obeys\ bosonic angular quantum statistics.

On angular space, due to forbidden phase changing from null condition, the
photons/gluons become massless bosonic particles with zero Hamiltonian $H$ and
zero wave vector $P_{\mu}$. With zero Hamiltonian, to characterize the motion
from one position to another, the action becomes
\[
S_{b}=\frac{1}{2\pi}\int_{\mathrm{A}}P_{\mu}\mathrm{d}X^{\mu}%
\]
where $X^{\mu}$ denotes position and $P_{\mu}$ denotes wave vector for the
photons/gluons on angular space.

Without time dependent evolution on angular space, $\mathrm{d}X^{\mu}$ cannot
be written as $\dot{X}^{\mu}dt.$ $\int_{\mathrm{A}}$\ denotes an integral on
angular space. To enforcing the null constraint $(P^{\mu})^{2}=0,$ we add an
addition term, i.e.,
\[
S_{b}=\frac{1}{2\pi}\int_{\mathrm{A}}P_{\mu}\mathrm{d}X^{\mu}-\frac{e}%
{2}g_{\mu \nu}P^{\mu}P^{\nu}\,
\]
where $e$ is a Lagrange multiplier that plays the role of an effective gauge
field. The gauge transformation is just to change $X$ and $X^{\prime}$ without
changing the result. Now, we have
\[
\delta X^{\mu}=\alpha P^{\mu},\text{ }\delta P_{\mu}=0,\text{ }\delta
e=\mathrm{d}\alpha
\]
that conjugates to the null constraint.

The action is relevant to the \emph{symplectic potential} $\theta=P_{\mu
}\mathrm{d}X^{\mu}$ of the angular space (or the projective ambitwistor space).

Let us give an additional physical explanation on this fact. It was known that
along a given direction of angular space, $P^{\mu}$ and $X^{\mu}$ correspond
to angular momentum $L^{\mu}$ and phase angle $\varphi^{\mu}.$ The physical
meaning of the action is the total phase changing of the whole system induced
by local changings. Here, for an object with angular momentum $L$, the total
phase changing is just $\Delta \varphi=\int_{\mathrm{A}}L^{\mu}\mathrm{d}%
\varphi^{\mu}\sim \int_{\mathrm{A}}P_{\mu}\mathrm{d}X^{\mu}.$ This is the
effective action on angular group-changing space! As a result, it is naturally
\emph{conformal invariant}, i.e., no matter what types of mapping on angular
space, it is invariant.

Next, we consider the other case, of which the internal state is $\left \vert
1\right \rangle .$ Now, the photons/gluons obey angular fermionic statistics.

On the one hand, we consider photons.

For photons, we have the action
\[
S_{f}=g_{\mu \nu}\int_{\mathrm{A}}\Psi^{\mu}\mathrm{d}\Psi^{\nu}%
\]
where $\Psi^{\mu}$ denotes the angular Majorana fermions. In general, on
angular space, we have $g_{\mu \nu}=\delta_{\mu \nu}$. For photons, $\Psi$ is
one component. Quantization of $\Psi^{\mu}$ gives the Dirac matrices and the
quantization of the constraint $\Psi^{\mu}P_{\mu}=0$ is just the massless
Dirac equation. To enforcing the null constraint $P_{\mu}\Psi^{\mu}=0,$ we add
an addition term, i.e.,%
\[
S_{f}=\int_{\mathrm{A}}\Psi^{\mu}\mathrm{d}\Psi^{\mu}+\chi P_{\mu}\Psi^{\mu}%
\]
where $\chi$ is a Lagrange multiplier that also plays the role of an effective
gauge field. In general, one can deal with the gauge freedom by setting $e=0,$
$\chi=0$\ with the introduction of ghosts. This leads to the BRST formula.
However, with self-interaction, there doesn't exist scattering processes for
photons themselves.

The action $\Psi^{\mu}\mathrm{d}\Psi^{\nu}$ is relevant to the super-partner
of \emph{symplectic potential} $\theta=P_{\mu}\mathrm{d}X^{\mu}$ on angular
space (or the projective ambitwistor space). This term gives an additional
contribution on total phase changing of the whole system induced by local
shape changings. Here, for an object with angular momentum $L$, the total
phase changing is just $\Delta \varphi=\int_{\mathrm{A}}P_{\mu}\mathrm{d}%
X^{\mu}+\int_{\mathrm{A}}g_{\mu \nu}\Psi^{\mu}\mathrm{d}\Psi^{\nu}.$

On the other hand, we consider gluons.

For the internal state $\left \vert 1\right \rangle $, we have the following
action
\[
S_{f}=g_{\mu \nu}%
{\displaystyle \sum \limits_{a}}
\int_{\mathrm{A}}\Psi^{a,\mu}\mathrm{d}\Psi^{a,\nu}%
\]
where $\Psi^{a}$ denotes multi-component angular Majorana fermions and
$a=1,2,..N$ labels the internal degrees of freedom. Here, $N$ denotes the
types of gluons. To enforcing the null constraint $P_{\mu}\Psi^{a,\mu}=0,$ we
have%
\[
S_{f}=\int_{\mathrm{A}}g_{\mu \nu}\Psi^{a,\mu}\mathrm{d}\Psi^{a,\nu}+\chi
_{a}P_{\mu}\Psi^{a,\mu}%
\]
where $\chi_{a}$ is a Lagrange multiplier that also plays the role of an
effective gauge field.

For gluons, except for global shape degrees of freedom, there exist internal
degrees of freedom for "fermionic" gluons $\Psi^{a}$ that have a symmetry
under operation of non-Abelian group $G.$ Here, the non-Abelian group $G$ that
can be regarded as "real" version of non-Abelian group $\mathcal{G}$ for
gluons. For example, for non-Abelian gauge fields under \textrm{SU(3)} local
gauge symmetry, $G$\ is global $\mathrm{SO(}3^{2}-1\mathrm{)=SO(8)}$ symmetry.
With the help of the theorem of conformal embedding, a two dimensional model
for non-interacting (complex/Majorana) fermions can be written as WZNW terms.
According to the theorem of conformal embedding, we can define a set of
fractionalization rules for breaking up the free fermion Hamiltonian in terms
of Hamiltonians of different massless models that commute with each other.

By using the standard approach of Bosonization, we have the phenomenon that is
similar to "\emph{spin-charge separation}" for fermionic system on 1+1
dimensional space. Now, we have new group operations of "global" symmetry from
local gauge symmetry for gluons, i.e.,
\begin{align*}
&  \text{Local gauge symmetry of }\mathcal{G}\text{ in usual spacetime}\\
&  \rightarrow \text{ Global symmetry of }G\text{ on angular space.}%
\end{align*}
This leads to an additional current algebra.
\begin{align*}
S_{f}  &  =\int_{\mathrm{A}}\Psi^{a,\mu}\mathrm{d}\Psi^{a,\nu}%
+\text{constraints}\\
&  =S[\text{global}]+S[\text{internal relative}]+\text{constraints}%
\end{align*}
where $S[$global$]=\int_{\mathrm{A}}\Psi^{\mu}\mathrm{d}\Psi^{\nu}$ is about
global current about phase changings from shape changing and $S[$internal
relative$]$ denotes the current $j^{a}$ for internal relative motion of group
$G$. The resulting modes with internal degrees of freedom obey
\[
\lbrack j_{n}^{a},j_{m}^{b}]=if_{c}^{ab}j_{n+m}^{c}+n\delta^{ab}\delta_{n+m}%
\]
where $a,b$ label the different generators of the Lie algebra associated to
$G$ and $f_{c}^{ab}$ are the structure constants of the Lie algebra. $n,m$
label the modes of the current algebra. If we only consider the $n,m=0$
sector, we have the Lie algebra; If we consider all the modes, we have an
infinite dimensional generalization of the Lie algebra -- the \emph{Kac-Moody
algebra}. Therefore, we have a current $J_{a}(z)$ with operator product
expansion (OPE)
\[
j_{a}(z)j_{b}(z^{\prime})=\frac{\delta_{ab}}{(z-z^{\prime})^{2}}+\frac
{f_{ab}^{c}j_{c}}{z-z^{\prime}}+\cdots \ .
\]
Then, we have a current algebra of level-1 \textrm{SO(N)} described by `real'
free fermions $\psi^{a}$, $a=1,\ldots N$, \ and get%
\[
S[\text{internal relative}]=%
{\displaystyle \sum \limits_{a}}
\int_{\mathrm{A}}\psi^{a}\partial_{\mu}\psi^{a}.
\]

Finally, the total action for gluons is obtained as
\[
S=S_{b}+S_{f}%
\]
and
\[
S_{f}=S[\text{global}]+S[\text{internal relative}]
\]
where $S[$internal relative$]$ is the action for the current algebra that
characterizes the internal relative motion and $S[$global$]$ is action for the
global current. This action is same to that about ambitwistor
strings\cite{am1}.

\paragraph{Motion of gravitons on angular space}

In this part, we study the motion of gravitons on angular space.

To characterize motion of gravitons on angular space, we regarded it as a
composite object with two "photons" of orthogonal polarization directions.
Because the two gauge modes of orthogonal polarization directions have same
angular momentum, we can deal with them separately. This will lead to the
phenomenon of \emph{double copy}.

First, we derive the effective action for the gravitons as composite objects
of two photons of\ $\left \vert 0\right \rangle $. The effective action for the
globally shifting of gravitons on angular space is same to that of photons as
\[
S_{b}=\frac{1}{2\pi}\int_{\mathrm{A}}P_{\mu}\mathrm{d}X^{\mu}-\frac{e}%
{2}g_{\mu \nu}P^{\mu}P^{\nu}%
\]
where $X^{\mu}$ denotes position and $P_{\mu}$ denotes extra angular momentum
for the gravitons. $e$ is a Lagrange multiplier that plays the role of an
effective gauge field.

Next, we derive the effective action for the gravitons as composite objects of
two photons of\ $\left \vert 1\right \rangle $. The effective action becomes
\[
S_{f}=\int_{\mathrm{A}}\sum_{r}g_{\mu \nu}\Psi_{r}^{\mu}\mathrm{d}\Psi_{r}%
^{\nu}+\chi_{r}P_{\mu}\Psi_{r}^{\mu}%
\]
where $r$ labels the index of "photons" of the graviton and $\Psi_{r}^{\mu}$
denotes corresponding angular Majorana fermions. $\chi_{r}$ is a Lagrange
multiplier that also plays the role of an effective gauge field. Without
internal degrees of freedom except for the geometric one, there don't exist
the term about current algebras.

Finally, the total action for gravitons is obtained as
\begin{align*}
S  &  =S_{b}+S_{f}\\
&  =\frac{1}{2\pi}\int_{\mathrm{A}}P_{\mu}\mathrm{d}X^{\mu}-\frac{e}{2}%
g_{\mu \nu}P^{\mu}P^{\nu}\\
&  +\int_{\mathrm{A}}\sum_{r}g_{\mu \nu}\Psi_{r}^{\mu}\mathrm{d}\Psi_{r}^{\nu
}+\chi_{r}P_{\mu}\Psi_{r}^{\mu}.
\end{align*}
This supersymmetric effective action is also same to that about ambitwistor
strings\cite{am1}.

\paragraph{Motion of scalar modes on angular space}

In the last part, we study the motion of Bi-adjoint scalar modes on angular space.

On the usual spacetime, the action for the Bi-adjoint scalar modes
(multi-component scalar modes with $\phi^{3}$ self-interaction) is written as%
\begin{align*}
S[\phi^{a\tilde{a}}]  &  =\int_{\mathrm{M}}(\frac{1}{2}\partial_{\mu}%
\phi^{a\tilde{a}}\partial^{\mu}\phi_{a\tilde{a}}\\
&  +\frac{1}{3}f_{abc}\tilde{f}_{\tilde{a}\tilde{b}\tilde{c}}\phi^{a\tilde{a}%
}\phi^{b\tilde{b}}\phi^{c\tilde{c}}).
\end{align*}
The second term denotes the $\phi^{3}$ self-interaction.\thinspace To
characterize motion of scalar modes on angular space, we regard it as a
composite object with two "gluons" of opposite angular momenta.

First, we derive the effective action for the scalar mode of
state\ $\left \vert 0\right \rangle $. The action is obtained as
\[
S_{b}=\frac{1}{2\pi}\int_{\mathrm{A}}P_{\mu}\mathrm{d}X^{\mu}-\frac{e}%
{2}P_{\mu}P^{\mu}.
\]
where $X^{\mu}$ denotes position and $P_{\mu}$ denotes extra angular momentum
for the scalar particles. $e$ is a Lagrange multiplier that plays the role of
an effective gauge field.

Next, we consider the other case $\left \vert 1\right \rangle .$ Due to the
opposite angular momenta, the action of $S[$global$]$ is canceled each other.
With two internal degrees of freedom, there exist the corresponding terms
about current algebras,%

\[
S[\text{internal relative}]=%
{\displaystyle \sum \limits_{r=1}^{2}}
{\displaystyle \sum \limits_{a}}
\int_{\mathrm{A}}\psi_{r}^{a}\mathrm{d}\psi_{r}^{a}%
\]
where $r$ labels the "gluons" with opposite angular momenta and $a$ denotes
the internal degrees of freedom.

\subsection{Scattering Equations}

The scattering amplitude for $n$-particle is defined by the correlation
function $\mathcal{M}(1,\ldots,n)$ for plane waves $e^{i\vec{k}_{i}\cdot
\vec{x}},$ $i=1,2,...n.$ A question is to \emph{determine the positions of all
excited modes on angular space}. In this part, we review the CHY formula about
the scattering equation\cite{chy1,chy2,chy3,chy4} and show how to determine
the positions of all excited modes.

For simplicity, we firstly focus on the scattering amplitudes with one node.

\subsubsection{CHY equation}

On the angular space, the wave vectors are projected to wave scalars as
$\vec{k}_{i}\rightarrow \pm \left \vert \vec{k}_{i}\right \vert =k_{i}.$ Note,
$\vec{k}_{i}$ is no more vector, but a number $k_{i}$, of which the sigh
characterizes the inward or outward. Each external line corresponds to a point
on angular space, of which we project the radial wave vector $\vec{k}_{i}$ to
the corresponding one on angular space.

In addition, for the scattering processes, there exists a constraint from
momentum conservation$,$ i.e., $%
{\displaystyle \sum}
\vec{k}_{i}=0.$ The constraint from momentum conservation on $\vec{k}_{i}$ is
then projected to another constraint on angular space, i.e., $%
{\displaystyle \sum}
k_{i}=0.$ After considering the plane waves in vertex (see detailed discussion
in next section), we can add additional terms $i(\sum_{i=1}^{n}k_{i})\cdot
X_{i}$ in the effective action for excited modes on angular space under the
gauge $e=0,$
\begin{align*}
S  &  =\frac{1}{2\pi}\int_{\mathrm{A}}P_{\mu}\mathrm{d}X^{\mu}+S_{2}%
+i(\sum_{i=1}^{n}k_{i})\cdot X_{i}\\
&  =\frac{1}{2\pi}\int_{\mathrm{A}}P_{\mu}\mathrm{d}X^{\mu}\\
&  +i(\sum_{i=1}^{n}k_{i}\cdot X_{i}\delta(\sigma-X_{i}))+S_{2}.
\end{align*}
Now, the position $X$ is mapped to the position $\sigma$ on the angular space,
i.e., $X\rightarrow \frac{\partial X}{\partial \sigma}\sigma$. The position
$\sigma$ is really the angle on angular space. Correspondingly, the
differential on angular group-changing space $\mathrm{d}$ is mapped to that on
angular space, $\bar{\partial}=\mathrm{d}\bar{\sigma}\, \partial_{\bar{\sigma
}}$.

Then, after integrating out $X$, the zero modes decouple from the kinetic
$P_{\mu}\mathrm{d}X^{\mu}$ and a momentum conserving $\delta$-function
$\delta(\sum P^{\mu})$ appears. However, the non--zero modes are Lagrange
multipliers enforcing the field equation%
\[
\bar{\partial}P_{\mu}=2\pi i\sum_{i}k_{i}\delta(\sigma-\sigma_{i})
\]
where $\bar{\partial}A=\mathrm{d}\bar{\sigma}\, \partial_{\bar{\sigma}}A$ on
angular space coordinate.\ This equation indicates that the finite wave
vectors along radial directions will locally change angular momentum $P_{\mu}$
on angular space. The changing of angular momentum $P_{\mu}$ implies expanding
or contracting of the angular space. This has unique solution
\[
P(X)=d\sigma \sum_{i=1}^{n}\frac{k_{i}}{\sigma-\sigma_{i}}%
\]
\ which may now be substituted into the remaining factors of $P_{\mu}$ in the
vertex operators. Therefore, this term indicates the local flux trapping by them.

In particular, using the on--shell conditions $(k_{i})^{2}=0$, the factors of
$\delta(k_{i}\cdot P(\sigma_{i}))$ impose the scattering equations
\[
\sum_{j\neq i}\frac{k_{i}\cdot k_{j}}{\sigma_{i}-\sigma_{j}}=0
\]
which are sufficient to determine the insertion points $\sigma_{i}$ in terms
of the external momenta. This is consistent to the expressions for massless
amplitudes in~Ref.\cite{chy1,chy2,chy3,chy4}.

\subsubsection{Physical picture for CHY equation}

We then provide a physical explanation on underlying physics of the CHY equation.

For scattering processes, the excited\ modes are described by plane waves
along certain radial directions, $\psi(x,t)=Ce^{-i\Delta \omega \cdot
t+i\Delta \vec{k}\cdot \vec{r}}$. Then, we have finite motion charge along
radial directions (or charge of motion). Now, for the excited modes the
mapping between the \textrm{\~{S}\~{O}(d+1)} Clifford group-changing
space\textit{ }$\mathrm{C}_{\mathrm{\tilde{S}\tilde{O}(d+1)},d+1}$ and
Cartesian spacetime \textrm{C}$_{3+1}$ changes. The changing rate is changed
from $\vec{k}_{0}$ to $\vec{k}_{0}+\Delta \vec{k}$ ($\Delta \vec{k}\ll \vec
{k}_{0})$. The motion charge along radial directions becomes $\vec{Q}%
^{r}=\frac{\Delta \vec{k}}{k_{0}}$.

For excited modes on given position of angular space, the changing of changing
rate along radical direction leads to a changing of the shape of the angular
variant. The locally changings of the radius of the angular space is
proportional to $\Delta \vec{k}$. It is known that the total flux of the
angular space is determined by the total volume of the space inside it.
Therefore, the local changing of the radius of the angular space leads to the
local changing of flux (or angular momentum), i.e.,
\[
\Delta \Phi \sim \Delta \vec{k}.
\]
The situation is similar to the case of extra magnetic flux on an
integer/fractional quantum Hall state.

As a result, there exists 2D classical Coulomb interaction $V(\sigma_{ab})$
between extra fluxes, of which the effective charge is proportional to
$\Delta \vec{k}.$ Finally, we write down the interaction potential
$V(\sigma_{ab})$ for the scattering processes,
\[
V(\sigma_{ab})=\sum_{b\neq a}k_{a}k_{b}\ln \sigma_{ab}%
\]
where $\left \vert \sigma_{ab}\right \vert $ is distance between two excited
modes $\sigma_{ab}$. Using traditional variational method by setting $\delta
V=0$, we can also obtain the same scattering equation, i.e.,
\[
\sum_{j\neq i}\frac{k_{i}\cdot k_{j}}{\sigma_{i}-\sigma_{j}}=0.
\]

Finally, after solving the CHY equation, we can know the exact positions of
all excited modes on angular space.

\subsection{Vertex operators}

Scattering amplitudes are constructed as correlation functions of vertex
operators. So, to calculate scattering amplitudes, we have to write down the
exact formula about the corresponding vertex.

Next, we do projection from usual spacetime $\mathrm{M}$ to angular space
\textrm{A}.

This projection can be also done by the \emph{super geodesic spray, }%
\[
D_{0}=P\cdot \nabla
\]
and
\[
D_{1}=\Psi \cdot \nabla+P\cdot \partial/\partial \Psi.
\]
These projection operators are generates a super null geodesic -- the integral
curves of $D_{0}$ are the horizontal lifts of geodesics with (null) cotangent
vector to the cotangent bundle. Then, under projection on the angular space,
we have $e^{i\vec{k}_{i}\cdot \vec{x}}\rightarrow e^{ik_{i}\cdot X}$ by
reducing the contribution along radial direction.

In particular, to derive the correct formula of vertex operators, we use
\emph{Penrose transform}\cite{pen}\cite{am1}.

The Penrose transform relates deformations of the conformal structure on
spacetime to elements of \emph{Dolbeault cohomology class} on angular space
(or projective ambitwistor space). According to \emph{Theorem of LeBrun
correspondence}\cite{le}, the geometric structure of angular invariant
determines spacetime \textrm{M} and its conformal metric $g_{\mu \nu}$.
Arbitrary small deformations of angular space which preserve super symplectic
potential $\theta$ correspond to small deformations of the conformal structure
on $\mathrm{M}$. This\ is just condition of perturbative angular variant.
Then, to describe a fluctuation in the metric of spacetime we need only
consider a perturbation $\delta \theta$ that is characterized by elements of
the Dolbeault cohomology class.

With the help of Dolbeault representation, we can construct the angular
variant (or super-ambitwistor space) to its symplectic reduction. Now, we have
the super symplectic potential $\theta$ and 2-form $\omega=d\theta$ by%
\begin{align*}
\theta &  =P_{\mu}dx^{\mu}+g_{\mu \nu}\Psi^{\mu}d\Psi^{\nu}/2\,,\\
\omega &  =d\theta=dP_{\mu}\wedge dX^{\mu}+g_{\mu \nu}d\Psi^{\mu}d\Psi^{\nu
}/2\,.
\end{align*}
We then perform the symplectic reduction by both $P^{2}$ and $P\cdot \Psi$.
Thus we set
\[
P^{2}=P\cdot \Psi=0
\]
and quotient by $D_{0}=P\cdot \nabla$ and also $D_{1}=\Psi \cdot \nabla
+P\cdot \partial/\partial \Psi$.

Finally, by using the Penrose transform, we obtain the correct vertex
operators for photons/gluons and gravitons. In the following parts of the this
section, we show the results one by one.

In addition, to derive the correct results, one needs to fixed vertex
operators that correspond to the same type of particles with fixed residual
gauge symmetries. In this paper, we will don't introduce ghosts but borrow the
earlier results about them\cite{am1}.

\subsubsection{Vertex operators for photons/gluons}

In this section, we derive the vertex operators for photons/gluons.

We define $a=\bar{\partial}\alpha$ to be projected gauge field on angular
space under Penrose transformation from photons/gluons $A=A_{\mu}dX^{\mu}$ on
$M$. Here, $\alpha$ to be the corresponding phase changing.

We then consider the Penrose transformation from $D_{1}$ and get
\[
D_{1}\alpha=\Psi^{\mu}A_{\mu}.
\]
According to $D_{0}=D_{1}^{2},$ we have
\[
D_{0}\alpha=D_{1}(\Psi^{\mu}A_{\mu})=P^{\mu}A_{\mu}+\Psi^{\mu}\Psi^{\nu}%
F_{\mu \nu}.
\]
For the excited modes for photons/gluons $A=\mathrm{e}^{ik\cdot X}%
\epsilon_{\mu}dX^{\mu},$ after solving above equation, we get
\[
\alpha=e^{ik\cdot X}\frac{\epsilon \cdot P+\epsilon \cdot \Psi k\cdot \Psi}{k\cdot
P},
\]
and\qquad%
\begin{align*}
a  &  =e^{ik\cdot X}(\epsilon \cdot P+\epsilon \cdot \Psi k\cdot \Psi
)(\bar{\partial}\frac{1}{k\cdot P})\\
&  =e^{ik\cdot X}(\epsilon \cdot P+\epsilon \cdot \Psi k\cdot \Psi)\bar{\delta
}(P\cdot k).
\end{align*}
Here, $\bar{\delta}(k\cdot P)$ is the $\delta$-function on angular space. This
is final result for photons/gluons on angular space by using Penrose
transformation\cite{am1}.

Finally, the integrated vertex operator for gluons becomes
\[
\int_{\mathrm{A}}\mathcal{V}^{a}=\int_{\mathrm{A}}\mathcal{A}_{a}j_{a}%
\]
where
\[
\mathcal{V}^{a}=\bar{\delta}(k\cdot P)\left[  \epsilon \cdot P+\epsilon
\cdot \Psi \,k\cdot \Psi \right]  \, \mathrm{e}^{\mathrm{i}k\cdot X}\,
\ T^{a}j_{a}%
\]
and
\[
\mathcal{A}^{a}=\bar{\delta}(k\cdot P)\, \mathrm{e}^{\mathrm{i}k\cdot
X}(\epsilon \cdot P+\epsilon \cdot \Psi k\cdot \Psi)\,T^{a}.
\]
$a$ denotes the internal degrees of freedom.

One can see that due to the characteristic of transverse wave, the gluons move
along certain direction on angular space that is determined by $P$.

\subsubsection{Vertex operators for gravitons}

Next, we study the vertex operators for gravitons.

The vertex operator for an on-shell linearized graviton corresponds to
variations in the spacetime metric. To describe these momentum eigenstates of
spacetime metric in terms of wave-functions on angular space, we have
\[
\delta g^{\mu \nu}(x)=\epsilon^{\mu \nu}e^{ik\cdot x}.
\]

To characterize gravitons on angular space, we regarded it as a composite
object with two photons of orthogonal polarization directions. The integrated
vertex operator for gravitons becomes
\begin{align*}
\int_{\mathrm{A}}\mathcal{V}  &  =\int_{\mathrm{A}}\bar{\delta}(k\cdot P)\,
\mathrm{e}^{\mathrm{i}k\cdot X}\, \\
&  \times \prod_{r=1}^{2}\left(  \epsilon_{r}\cdot P+\epsilon_{r}\cdot \Psi
_{r}\ k\cdot \Psi_{r}\right)
\end{align*}
where $r$ labels the index of "photons" of the graviton. This result is
consistent to double copy.

One can see that due to the characteristic of composite objects, the two
"photons" of the gravitons move along orthogonal directions on angular space, respectively.

\subsubsection{Vertex operators for Bi-adjoint scalar modes}

To study the vertex operators of Bi-adjoint scalar modes on angular space, we
regarded it as a composite object with two "gluons" of opposite angular momenta.

A deformation to the action on original spacetime, the plane wave of
Bi-adjoint scalar modes is given by $\phi^{a\tilde{a}}=\mathrm{e}^{ik\cdot
X}T^{a}\tilde{T}^{\tilde{a}}$. There are two currents $j_{a}$ and $\tilde
{j}_{a}$ in vertex for the scalar modes due to two gluons. By using the
Penrose transform, the deformation becomes%
\[
\bar{\delta}(k\cdot P)\mathrm{e}^{ik\cdot X}T^{a}\tilde{T}^{\tilde{a}}.
\]
The integrated vertex operator for scalar modes becomes
\[
\int_{\Sigma}\mathcal{V}=\int_{\mathrm{A}}\bar{\delta}(k\cdot P)(T^{a}%
j_{a})\cdot(\tilde{T}^{a}\tilde{j}_{a})\mathrm{e}^{\mathrm{i}k\cdot X},
\]

For the scalar modes, there doesn't exist usual terms in vertex operators for
shape changing due to cancelation effect from two "gluons" of opposite angular
momenta. As a result, we don't have usual terms about polarization $\left(
\epsilon_{r}\cdot P+\epsilon_{r}\cdot \Psi_{r}\ k\cdot \Psi_{r}\right)  .$

\subsection{Scattering amplitudes}

In this section, firstly we will study the scattering amplitude for QQ-event
with single node that is described by the so-called irreducible tree diagram.
Then, we generalize the theory to those with several nodes and zero internal
loop (the so-called reducible tree diagrams). Finally, we consider the cases
of several internal loops (the so-called loop diagrams).

\subsubsection{Irreducible tree-level scattering amplitudes}

In this part, we calculate the tree--level amplitudes with only one node that
characterizes single QQ-event. According to above discussion, we have derived
the effective actions and vertex operators for excited modes (gluons,
gravitons and Bi-adjoint scalar modes). Our results of the tree-level
scattering amplitudes are same to those well known before.

\paragraph{Scattering amplitudes for gravitons}

We firstly calculate tree--level scattering amplitudes for gravitons with only
one node.

The total action for gravitons is
\begin{align*}
S  &  =S_{b}+S_{f}\\
&  =\frac{1}{2\pi}\int_{\mathrm{A}}P_{\mu}\mathrm{d}X^{\mu}-\frac{e}{2}%
g_{\mu \nu}P^{\mu}P^{\nu}\\
&  +\int_{\mathrm{A}}\sum_{r}g_{\mu \nu}\Psi_{r}^{\mu}\mathrm{d}\Psi_{r}^{\nu
}+\chi_{r}P_{\mu}\Psi_{r}^{\mu}\\
&  =\frac{1}{2\pi}\int_{\mathrm{A}}P_{\mu}\bar{\partial}X^{\mu}-\frac{e}%
{2}g_{\mu \nu}P^{\mu}P^{\nu}\\
&  +\int_{\mathrm{A}}\sum_{r}g_{\mu \nu}\Psi_{r}^{a,\mu}\bar{\partial}\Psi
_{r}^{a,\nu}+\chi_{r}P_{\mu}\Psi_{r}^{\mu}.
\end{align*}
This action is same to that from the theory about ambitwistor strings. The
integrated vertex operator for gravitons is
\begin{align*}
\int_{\mathrm{A}}\mathcal{V}\cdot e^{S}  &  =\int_{\mathrm{A}}\bar{\delta
}(k\cdot P)\, \mathrm{e}^{\mathrm{i}k\cdot X}\, \\
&  \times \prod_{r=1}^{2}\left(  \epsilon_{r}\cdot P+\epsilon_{r}\cdot \Psi
_{r}\ k\cdot \Psi_{r}\right) \\
&  \exp(\frac{1}{2\pi}\int_{\mathrm{A}}P_{\mu}\mathrm{d}X^{\mu}-\frac{e}%
{2}g_{\mu \nu}P^{\mu}P^{\nu}\\
&  +\int_{\mathrm{A}}\sum_{r}g_{\mu \nu}\Psi_{r}^{a,\mu}\mathrm{d}\Psi
_{r}^{a,\nu}+\chi_{r}P_{\mu}\Psi_{r}^{\mu})\\
&  =\int_{\mathrm{A}}\bar{\delta}(k\cdot P)\, \mathrm{e}^{\mathrm{i}k\cdot
X}\, \\
&  \times \prod_{r=1}^{2}\left(  \epsilon_{r}\cdot P+\epsilon_{r}\cdot \Psi
_{r}\ k\cdot \Psi_{r}\right) \\
&  \exp(\frac{1}{2\pi}\int_{\mathrm{A}}P_{\mu}\bar{\partial}X^{\mu}-\frac
{e}{2}g_{\mu \nu}P^{\mu}P^{\nu}\\
&  +\int_{\mathrm{A}}\sum_{r}g_{\mu \nu}\Psi_{r}^{a,\mu}\bar{\partial}\Psi
_{r}^{a,\nu}+\chi_{r}P_{\mu}\Psi_{r}^{\mu}).
\end{align*}

Combining the contribution including both sets of Majorana fermions $\Psi^{r}%
$, the scattering amplitude is obtained as
\begin{align*}
\mathcal{M}(1,\ldots,n)  &  =\int_{\mathrm{A}}\mathcal{V}\cdot e^{S}\\
&  =\delta(\sum_{i}k_{i})\int \frac{1}{\mathrm{Vol\,SL(2;C)}}\\
&  \times \mathrm{Pf}^{\prime}(M_{1})\mathrm{Pf}^{\prime}(M_{2}){\prod_{i}%
}\prime \, \bar{\delta}(k_{i}\cdot P(\sigma_{i}))\,,
\end{align*}
where $M_{1}$ is built out of the polarization vectors $\epsilon_{1i}$ and
$M_{2}$ out of the $\epsilon_{2i}$ and where
\[
P(\sigma)=\mathrm{d}\sigma \sum_{i}k_{i}/(\sigma-\sigma_{i}).
\]
This is exactly the expression of CHY formula. The correlations of these
currents lead to the reduced Pfaffians of CHY\cite{chy1,chy2,chy3,chy4}:
\[
\mathrm{Pf}^{\prime}(M)=\frac{1}{\sigma_{1}-\sigma_{2}}\mathrm{Pf}(M_{12}),
\]
where $M$ is the skew $2n\times2n$ matrix with $n\times n$ block
decomposition
\begin{align*}
M  &  =%
\begin{pmatrix}
A & -C^{T}\\
C & B
\end{pmatrix}
\,,\qquad \\
A_{ij}  &  =\frac{k_{i}\cdot k_{j}}{\sigma_{ij}}\,,\\
B_{ij}  &  =\frac{\epsilon_{i}\cdot \epsilon_{j}}{\sigma_{ij}}\,,
\end{align*}
and
\begin{align*}
C_{ij}  &  =\frac{\epsilon \cdot k_{j}}{\sigma_{ij}}\,,\;i\neq j,\\
C_{ii}  &  =-\epsilon_{i}\cdot P(\sigma_{i})\,,
\end{align*}
and $M_{12}$ is $M$ with the first two rows and columns removed.

$\frac{1}{\mathrm{Vol\,SL(2;C)}}$ comes from the usual $c$ ghost path integral
and becomes
\[
\frac{1}{\mathrm{Vol\,SL(2;C)}}=\frac{(\sigma_{12}\sigma_{23}\sigma_{31}%
)}{(d\sigma_{1}d\sigma_{2}d\sigma_{3})}.
\]
This coefficient looks like ultraviolet divergence. However, according to the
definition of the information unit (quantized flux with unit angular
momentum), $d\sigma_{1}$ is really the size of an information unit along given
direction. The finite size is about lattice distance of "topological lattice"
on angular space, not infinitely small.

\paragraph{Scattering amplitudes for gluons}

Secondly, we calculate tree--level scattering amplitudes for gluons with only
one node.

The total action for gluons is
\begin{align*}
S  &  =\int_{\mathrm{A}}\Psi^{a,\mu}\bar{\partial}\Psi^{a,\nu}+\frac{1}{2\pi
}\int_{\Sigma}\,P_{\mu}\bar{\partial}X^{\mu}\\
&  +%
{\displaystyle \sum \limits_{a}}
\int_{\mathrm{A}}\psi^{a}\partial_{\mu}\psi^{a}+\text{constraints}\\
&  =\int_{\mathrm{A}}\Psi^{a,\mu}\bar{\partial}\Psi^{a,\nu}+\frac{1}{2\pi}%
\int_{\Sigma}\,P_{\mu}\bar{\partial}X^{\mu}\\
&  +%
{\displaystyle \sum \limits_{a}}
\int_{\mathrm{A}}\psi^{a}\partial_{\mu}\psi^{a}+\text{constraints.}%
\end{align*}
The integrated vertex operator for gluons is%
\[
\int_{\mathrm{A}}\mathcal{V}^{a}=\int_{\mathrm{A}}\mathcal{A}_{a}j_{a}%
\]
where
\[
\mathcal{V}^{a}=\bar{\delta}(k\cdot P)\left[  \epsilon \cdot P+\epsilon
\cdot \Psi \,k\cdot \Psi \right]  \, \mathrm{e}^{\mathrm{i}k\cdot X}\,
\ T^{a}j_{a}%
\]
and
\[
\mathcal{A}^{a}=\bar{\delta}(k\cdot P)\, \mathrm{e}^{\mathrm{i}k\cdot
X}(\epsilon \cdot P+\epsilon \cdot \Psi k\cdot \Psi)\,T^{a}.
\]
$a$ denotes the internal degrees of freedom. $j_{a}$ denotes the current for
internal relative motion of group $G$.

Finally, the tree--level scattering amplitudes for gluons is obtained
as\cite{pt,zhang}%
\begin{align*}
\mathcal{M}(1,\ldots,n)  &  =\int_{\mathrm{A}}\mathcal{V}\cdot e^{S}\\
&  =\delta(\sum_{i}k_{i})\int \frac{\mathrm{d}^{n}\sigma}{\mathrm{Vol\,SL(2;C)}%
}\\
&  \times{\prod_{i}}\, \bar{\delta}(k_{i}\cdot P(\sigma_{i}))\  \mathrm{Pf}%
^{\prime}(M)\  \\
&  \cdot \left[  \frac{\mathrm{tr}(T_{1}T_{2}\cdots T_{n})}{\sigma_{12}%
\sigma_{23}\cdots \sigma_{n1}}\ +\  \cdots \  \right]  .
\end{align*}

\paragraph{Scattering amplitudes for scalar modes}

Thirdly, we calculate tree--level scattering amplitudes with only one node for
Bi-adjoint scalar modes.

The total action is
\begin{align*}
S  &  =\frac{1}{2\pi}\int_{\mathrm{A}}P_{\mu}\mathrm{d}X^{\mu}-\frac{e}%
{2}P_{\mu}P^{\mu}\\
&  +%
{\displaystyle \sum \limits_{r=1}^{2}}
{\displaystyle \sum \limits_{a}}
\int_{\mathrm{A}}\psi_{r}^{a}\mathrm{d}\psi_{r}^{a}.
\end{align*}
where $r$ labels the opposite angular momenta and $a$ denotes the internal
degrees of freedom. The integrated vertex operator for scalar modes becomes
\[
\int_{\Sigma}\mathcal{V}=\int_{\mathrm{A}}\bar{\delta}(k\cdot P)(T^{a}%
j_{a})\cdot(\tilde{T}^{a}\tilde{j}_{a})\mathrm{e}^{\mathrm{i}k\cdot X},
\]

Finally, the scattering amplitude is\cite{chy4}
\begin{align*}
\mathcal{M}(1,\ldots,n)  &  =\int_{\mathrm{A}}\mathcal{V}\cdot e^{S}\\
&  =\delta(\sum_{i}k_{i})\int \frac{\mathrm{d}^{n}\sigma}{\mathrm{Vol\,SL(2;C)}%
}\\
&  \times{\prod_{i}}\, \bar{\delta}(k_{i}\cdot P(\sigma_{i}))\  \\
&  \times \left[  \frac{\mathrm{tr}(T_{1}T_{2}\cdots T_{n})}{\sigma_{12}%
\sigma_{23}\cdots \sigma_{n1}}\ +\  \cdots \  \right] \\
&  \times \left[  \frac{\mathrm{tr}(\tilde{T}_{1}\tilde{T}_{2}\cdots \tilde
{T}_{n})}{\tilde{\sigma}_{12}\tilde{\sigma}_{23}\cdots \tilde{\sigma}_{n1}%
}\ +\  \cdots \  \right]  .
\end{align*}

\subsubsection{Reducible tree-level scattering amplitudes}

In above section, we have use correlation function on angular space to
characterize the scattering amplitudes of an irreducible tree diagram. In this
section we study reducible tree diagrams.

A reducible tree diagram describes scattering amplitude with $n>1$ nodes but
zero loop. Here, the node is a point of common center, at which, several
(external or internal) lines (more than two) converge. In general, a reducible
tree diagram can be considered as a composite diagram with $n$ irreducible
tree diagram, each of which has its common center. Because an irreducible tree
diagram corresponds to QQ event, a reducible tree diagram describes several
\emph{interconnected }QQ-events. Therefore, we introduce the approach of
coupled \emph{n} angular variants to characterize scattering amplitudes of a
reducible tree diagram with $n$ nodes.

Then, we give an approach to calculate scattering amplitudes $\mathcal{M}$ of
reducible tree diagram with $n$ nodes.

At first step, we map a reducible tree diagram with $n$ nodes to $n$ planes.
Now, each plane denotes a QQ-event and the number of layer of planes is just
$n$; a point on the given layer corresponds to a line of given irreducible
tree diagram; the line connecting two different planes determines the their
relative relationship.

By solving $n$ CHY equations of $n$ nodes, we determine the positions of lines
on $l$-th planes,
\[
P_{l}(X_{l})=d\sigma_{l}\sum_{i=1}^{n}\frac{k_{l,i}}{\sigma_{l}-\sigma_{l,i}}%
\]
\  \ or
\[
\sum_{j\neq i}\frac{k_{l,i}\cdot k_{l,j}}{\sigma_{l,i}-\sigma_{,j}}=0.
\]
Here, $l$ labels the index of plane for given node.

At second step, we calculate the scattering amplitudes $\mathcal{M}_{l}%
(1^{l},\ldots,n^{l})$ for the QQ-event with only one node on different planes.
The results have been obtained in above section.

At third step, the whole scattering amplitude of reducible tree diagrams
$\mathcal{M}$ is finally obtained as%
\[
\mathcal{M=}%
{\displaystyle \prod \limits_{l}}
\frac{1}{(k_{i}^{ll^{\prime}})^{2}}\mathcal{M}_{l}(1^{l},\ldots,n^{l})
\]
where $k_{i}^{ll^{\prime}}$ is the finite momentum between two different nodes
($l$ or $l^{\prime}$), i.e., $k_{i}^{ll^{\prime}}\neq0$.

We point out that both QQ-events for reducible tree diagrams and those for
irreducible tree diagrams are all "\emph{classical}". The word "classical"
means that the positions of all points on different planes are fixed,
predictable. The situation is quite different from those of loop diagrams.

In addition, based on our approach, we give a physical explanation on BCFW
recursive relation\cite{BCFW1, BCFW2}.

BCFW recursive relation provides a notion of constructibility of a theory at
tree level: if one iterates the recursion relations, the $n$-particle
amplitude can be expressed in terms of products of three-particle
amplitudes\cite{BCFW1, BCFW2}. Any intermediate state through which this
factorization can occur is call factorization channel. There exists simple
pole singularity in the amplitude, located in momentum space where the
on-shell condition of the intermediate particle is met,
\begin{equation}
M_{n}\sim \sum_{k}M_{n-k+1}\frac{1}{p_{k}^{2}}M_{k+1},\ p_{k}^{2}\rightarrow0.
\label{factor}%
\end{equation}
Here $M_{n}$ is at a given perturbative order.

Our results are obviously consistent to those from BCFW recursive
relation\cite{BCFW1, BCFW2}.

We take a 1-node tree diagram with four external lines as example. See the
illustration in Fig.24(a). A 1-node tree diagram with four external lines can
be deformed into a tree diagram with two nodes that are connected by a virtual
internal line. Now, the momentum of the virtual internal line between the two
nodes must be zero and corresponds to the "pole", i.e., $k_{i}^{ll^{\prime}%
}\rightarrow0$. Thus, we have the same result as that from BCFW recursive
relation
\begin{equation}
M_{n}\sim \sum_{k}M_{n-k+1}\frac{1}{k_{i}^{ll^{\prime}}}M_{k+1},\ k_{i}%
^{ll^{\prime}}\rightarrow0.
\end{equation}
This argument can be generalized to 1-node tree diagram with $m$ number
external lines ($m>4$).

In our theory, because the angular space has no boundary, we don't worry about
the contribution from boundary terms. As a result, by using BCFW recursive
relation, one can disassemble a complex 1-node tree diagram with a lot of
external lines into several 1-node tree diagram with only three external
lines. Consequently, a plane for single angular space is\ disassembled into
several planes for corresponding angular spaces. In physics, a complex
QQ-event can be considered as a series of simple QQ-event with special constraints.

\subsubsection{Loop scattering amplitudes}

In this section, we focus on the issue about loop scattering amplitudes. In
general, we consider loop scattering amplitudes with $L$ loop diagram, $N$
external lines $M$ internal lines and $n$ nodes.

Firstly, we split the loop scattering amplitudes with $n$ nodes into $n$ tree
scattering amplitudes.

Secondly, we determine the momenta of all internal lines. We immediately
discovered something awkward -- the momenta for internal lines around loops
cannot bee uniquely determined. To calculate the loop scattering amplitudes,
we consider all possible momenta for internal lines and summarize them.

Let us show the details.

Remember, we do calculations on angular space rather than usual spacetime.
Therefore, we must project the usual three dimensional vectorial momenta to
one dimensional scalar momenta that is the size of the original vector. As a
result, the momenta for different lines (either external lines or internal
lines) are \emph{real} number, rather than a three dimensional vector. This
will greatly simplify calculations.

Then, we focus on \emph{fundamental type of loop diagrams.}

For fundamental type of loop diagrams, the number of lines (including external
lines and internal lines) that connect the node around the loop is equal to
$3$. With help of BCFW recursion relation, we decouple arbitrary loop diagrams
to fundamental type of loop diagrams. As a\ result, we may classify the
fundamental type of loop diagrams by the number of nodes (or internal lines).

For a node of a loop diagram, for example, we try to determine the momenta
around a given loop ($l$-th loop) with $n_{L}$ internal lines and $n_{L}$
nodes. There are $n_{L}$ total unknown numbers that correspond to the momenta
of internal lines. For each node, one has a conservation condition for
momenta. Then, under the constraint from conservation condition, one can
firstly determine arbitrary given momentum, for example, $l$-th internal line,
$k_{l}^{l}.$ $k_{l}^{l}$ can be an arbitrary momentum from $-\Lambda$ to
$\Lambda$ where $\Lambda$ is the cutoff of momentum. Then, momenta of others
$k_{1}^{l}$, $k_{2}^{l},$\ ... $k_{l-1}^{l}$ become known.

Thirdly, for each node, we obtain the corresponding scattering amplitude by
using the approach of irreducible tree diagram. The result has been obtained
in earlier parts.

Finally, we summarize the contribution from all nodes and get
\[
\mathcal{M=}%
{\displaystyle \prod \limits_{l}}
{\displaystyle \int \limits_{-\Lambda}^{\Lambda}}
dp_{l}%
{\displaystyle \prod \limits_{a}}
\frac{1}{(k_{i}^{ll^{\prime}})^{2}}\mathcal{M}_{l}(1^{l},\ldots,n^{l}).
\]
The final result is derived by doing the $L$-fold integral. In $\mathcal{M}%
$,\ each integral comes from an uncertain momenta around a loop. There are $N$
factors of $\frac{1}{(k_{i}^{ll^{\prime}})^{2}}.$ Each factor comes from an
internal line.

In the end, we point out that the difficulty to obtain the results comes from
solving a lot of CHY equations by varying discrete $k_{l}^{l}$ rather than
doing integral.

\subsection{The amplituhedron}

Arkani-Hamed et al \cite{ar1,ar2,ar3} discovered the connection between
scattering amplitudes and the Amplituhedron (a generalization of the positive
Grassmanian). The on-shell diagrams constructed by suitably gluing together
the three-particle amplitudes represent physical processes and whole
scattering amplitudes in planar $\mathcal{N}=4$ super-Yang-Mills theory (SYM).
The on-shell diagrams can be associated to a particular configuration among
the boundaries of the positive Grassmannian. Then, the three-particle
amplitudes become building blocks that are glued together. The amplitudes are
identified as the \textquotedblleft volume\textquotedblright \ of the
corresponding object.

In this section, we explore the underlying physics of Amplituhedron based on
angular variant. We take the simplest non-vanishing helicity amplitudes
$A(1^{+},\ldots,i^{-},\ldots,j^{-},\ldots,n^{+})$ with $h=n-4$ as example. It
is always called \emph{MHV amplitudes} and are given by the
\emph{Parke-Taylor} formula~\cite{pt,zhang} $\frac{\langle ij\rangle^{4}%
}{\langle12\rangle \langle23\rangle \cdots \langle n1\rangle}.$

Firstly, we consider the case of irreducible tree diagram with one node for gluons.

To characterize the geometric property of tree-level scattering amplitudes, we
use the bosonic representation for gluons by considering the state of
$\left \vert 0\right \rangle .$ Now, the shape of gluons becomes isotropic.
Because the excited modes make up a perfect circle on different spheres with
different radius, the global structure of the gluons in usual spacetime looks
like a \emph{semi-infinite}, solid tube with fixed radius from infinity to
common center. Therefore, the physical picture of scattering processes with
multi-nodes becomes multi-sphere with interconnected solid tubes. It is very
similar to Riemann surface of world sheet in string theory.

Secondly, we project the external line of gluons with momentum $k_{i}$ onto
angular space with radius $R$. See the illustration in Fig.24(a).

\begin{figure}[ptb]
\includegraphics[clip,width=0.9\textwidth]{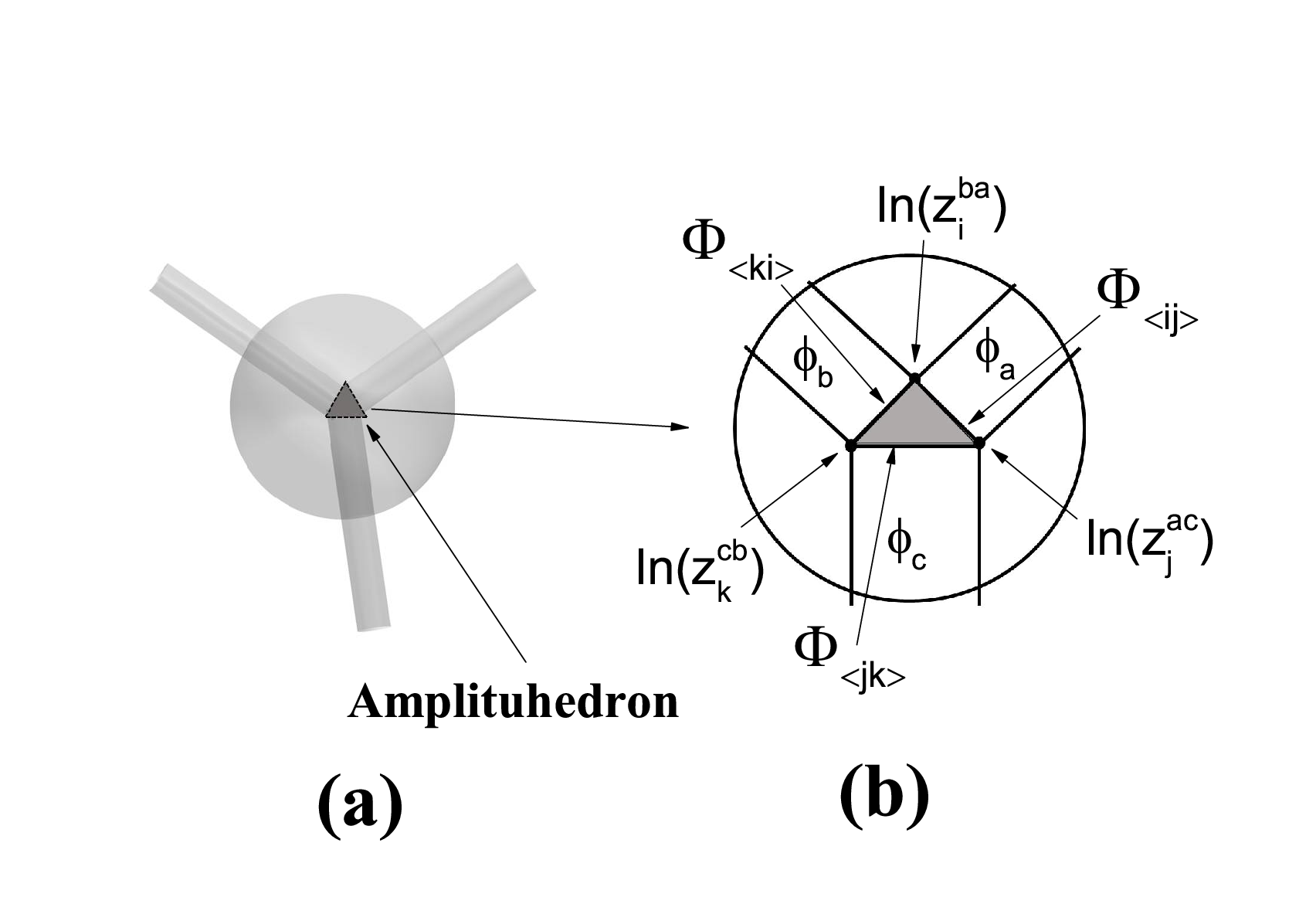}\caption{(a) An
illustration of Amplituhedron; (b) The phases of different regions, different
boundaries, different points between two boundaries around the Amplituhedron.}%
\end{figure}

To project an external line of gluons with momentum $k_{i}$ onto angular
space, we must choose a reference on angular space (for example, $\sigma_{0}$)
that corresponds to a reference angle on celestial sphere. From the angle of
view $\sigma_{0},$ the external lines (or solid tubes with fixed radius) are
projected to \emph{ribbons with fixed width.} The width is just the diameter
of circle for gluons. The starting point $\sigma_{i}$ is just the position of
the gluon on angular space with radius $R$ that is solved by CHY equation.

When we consider three or more excited modes, there exists \emph{common
intersection region} $\mathcal{AM}$\ for the their ribbons with common center.
The shape of common intersection region $\mathcal{AM}$ is certain polygon that
is just the so-called Amplituhedron on angular space!

Thirdly, we determine the phase factors of different regions in the angular
space with radius $R,$ including different areas, different boundaries, and
different points.

It was known that for a ribbon of a projected external line, the global phase
factor is $e^{ik_{i}X(\sigma_{i})}.$ So, the global phase factor in the common
intersection region $\mathcal{AM}$ becomes the product of all ribbons passing
this region, i.e.,%
\begin{equation}%
{\displaystyle \prod \limits_{i}}
e^{ik_{i}X(\sigma_{i})}=e^{%
{\displaystyle \sum \limits_{i}}
ik_{i}X(\sigma_{i})}.
\end{equation}
As a result, the boundaries of the common intersection region turn into branch
cuts. The phase changings on the boundary between $I$-th ribbon with global
phase factor $e^{ik_{I}X(\sigma_{I})}$ and the common intersection region with
global phase factor $e^{%
{\displaystyle \sum \limits_{i}}
ik_{i}X(\sigma_{i})}$ are $e^{%
{\displaystyle \sum \limits_{i\neq I}}
ik_{i}X(\sigma_{i})}=e^{%
{\displaystyle \sum \limits_{i}}
ik_{i}X(\sigma_{i})-k_{I}X(\sigma_{I})}.$ This leads to logarithmic
singularities on all boundaries of the common intersection region. Therefore,
different regions, different boundaries, different points between two
boundaries may have different phase factors. See the illustration in Fig.24(b).

Then, we define the phases of $a$-th external line to be $\phi_{a}^{i}%
=k_{a}X(\sigma_{a}^{i})$. Here, $i$ denotes corner nearby. The phase $\ln
z_{i}$ of $i$-th corner is the difference between phases of two neighbouring
external lines $a$ and $a+1$, $\phi_{a}^{i}$ and $\phi_{a+1}^{i},$ i.e.,
\begin{equation}
\ln z_{i}=\phi_{a}^{i}-\phi_{a+1}^{i}.
\end{equation}
The phase factor of a boundary is defined by
\begin{equation}
\Phi_{\left \langle i,i+1\right \rangle }=\ln \Upsilon_{\left \langle
i,i+1\right \rangle }%
\end{equation}
that denotes phase changing of two neighbouring corners $i$ and $i+1$ on a
boundary of polygon (or the common intersection region). As a result, for each
boundary, we have
\begin{align*}
\ln \Upsilon_{\left \langle i,i+1\right \rangle }  &  =\ln z^{i}-\ln z^{i+1}\\
&  =\phi_{a}^{i}+\phi_{a+2}^{i}-2\phi_{a+1}^{i}.
\end{align*}

Fourthly, we express the amplituhedron differential form $\Omega$ for an
irreducible tree diagram with one node.

There is an associated form with logarithmic singularities on the boundaries
of the polygon
\[
\Omega \sim \prod_{i}d\Phi_{\left \langle i,i+1\right \rangle }=\prod_{i=1}%
^{n-3}d(\ln \Upsilon_{\left \langle i,i+1\right \rangle })
\]
where $\ln \Upsilon_{\left \langle i,i+1\right \rangle }$ denotes phase changing
of two neighbouring points $i$ and $i+1$ on a boundary of polygon. Finally,
the amplituhedron differential form $\Omega=$ $\mathrm{sign}(\Gamma_{n}%
)\prod_{i=1}^{n-3}d(\ln \Upsilon_{\left \langle i,i+1\right \rangle })$ is
obtained as $PT(1,2,...n)$.

This is just the scattering amplitude of Parte-Taylor formula\cite{pt,zhang}.

\begin{figure}[ptb]
\includegraphics[clip,width=0.9\textwidth]{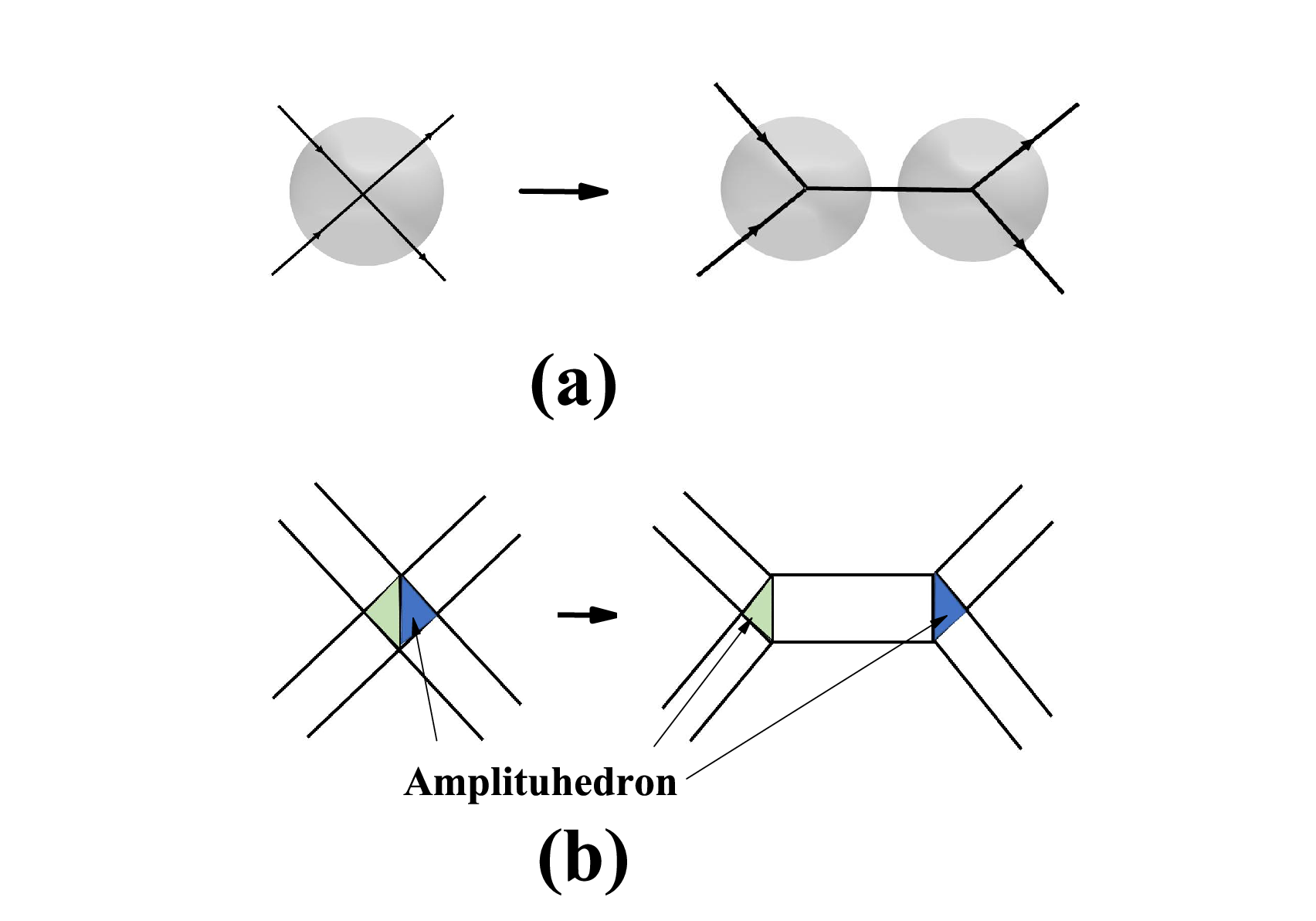}\caption{(a) A 1-node
tree diagram with four external lines is deformed into a tree diagram with two
nodes that are connected by a virtual internal line from BCFW recursive
relation. The momentum of the virtual internal line between the two nodes is
zero. (b) An illustration of Amplituhedron for combining two triangles into
one quadrilateral. The connect line corresponds two boundaries of the
triangles for amplituhedrons.}%
\end{figure}

In addition, we give a brief discussion on the Amplituhedron for reducible
tree diagrams. For the case of reducible tree diagrams with $n$ nodes, we have
$n$ planes. On each plane, we get similar results. We then focus on a connect
line that is shown in Fig.25. On each plane, the corresponding polygon for
amplituhedron is a triangle, of which the boundary is determined by lines for
tree diagram. Each connect line corresponds two boundaries of the polygons for
amplituhedrons on two planes. Now, amplituhedrons become geometric objects
with stereostructure.

\subsection{Conclusion}

In the final section, we draw the conclusion.

We developed a new theory to calculate the scattering amplitudes based on
angular variant that is characterized by 1-th order variability. Now,
scattering process for quantum states is regarded as an event process from
initial quantum states to final quantum states. Based on the framework of
angular variants, the scattering amplitudes are obtained, including tree
diagrams and loop diagrams. In addition, we found that string theory become a
correct framework for event physics on angular space rather than dynamical
physics on usual spacetime. Now, supersymmetry and string structure become
\emph{emergent} phenomena. See the logical structure of the part in Fig.26.

\begin{figure}[ptb]
\includegraphics[clip,width=0.9\textwidth]{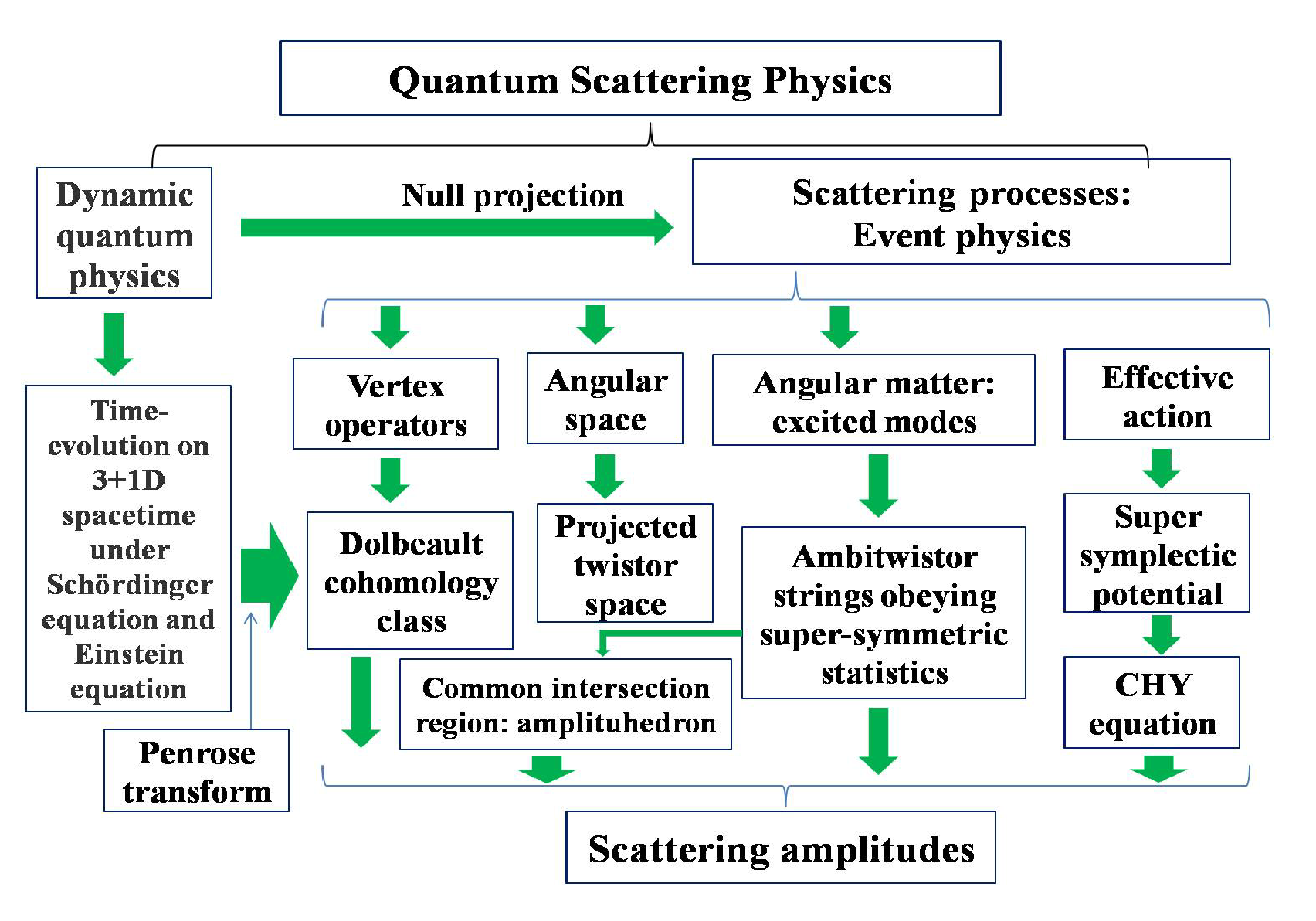}\caption{Logical
structure of the section about scattering amplitudes}%
\end{figure}

In particular, within the new theory, we answer above five questions.

1. What's the exact \emph{microstructure} of the scattering amplitudes for
gravitational waves? And, how characterize it?

\textbf{Answer:}

The\ microscopic structure of scattering amplitudes for \textrm{QQ}-events is
(d-1)-dimensional \textrm{\~{S}\~{O}(d-1)} angular variant\textit{
}$V_{\mathrm{\tilde{S}\tilde{O}(d-1)},d-1}^{\text{\textrm{Angular}}}%
(\Delta \phi^{\mu},\Delta \varphi^{\mu},k_{0})$\textit{ }that is a mapping
between the $\mathrm{\tilde{S}\tilde{O}}$\textrm{(d-1)} angular group-changing
space and the angular space of the original Cartesian space\textit{
}$\mathrm{S}_{d-1}^{\text{\textrm{Angular}}}.$ The angular variant is
characterized by 1-th order variability of spatial transformation, i.e.,%
\begin{equation}
\mathcal{T}(\Delta \theta^{\mu})\leftrightarrow \hat{U}(\delta \phi^{\mu}),
\end{equation}
where $\hat{U}(\delta \phi^{\mu})=e^{i\Gamma^{\mu}\delta \phi^{\mu}}$ with
$\delta \varphi^{\mu}=\sqrt{N_{\mathrm{tot}}^{F}}\delta \theta^{\mu}$.

In addition, we point out that the angular variant provides a solid physical
foundation on ambitwistor space and the celestial sphere. The celestial
conformal symmetry is highly relevant to 1-th order angular variability.

2. Why \emph{ambitwistor strings}?

\textbf{Answer:}

The angular variant provides physical fundation of the ambitwistor space. The
dynamic for shape changings of the excited modes on angular space becomes the
physical mechanism of ambitwistor string. Because excited modes (gravitons or
gluons) have fixed area on angular space, under the constraints from fixed
changing rate and fixed area, the shape of the excited modes can be
characterized by the shape of its boundary, that is a closed string. Different
internal states of excited modes correspond to different closed string. Due to
the energy degneracy on angular space, the invariant of shape changing becomes
an emergent supersymmetry. As a result, the closed string becomes superstring,
more accurately, ambitwistor superstring. Hence, we say that the superstring
exists on angular space for event processes rather on usual spacetime for
dynamical processes.

3. Why \emph{double copy}?

\textbf{Answer:}

Vector fields (photons and gluons) are angular matter with unit angular
momentum $\Delta L=1;$ tensor fields (gravitational waves) are angular matter
with total angular momentum $\Delta L=2.$ As a result, we can be regarded as
tensor field as a composite object with two photons of orthogonal polarization
directions. This leads to the mechanism of double copy.

4. Why \emph{amplituhedron}?

\textbf{Answer:}

To characterize the geometric property of tree-level scattering amplitudes, we
use the bosonic representation by considering the state of $\left \vert
0\right \rangle .$ The geometric structure of external lines \ for the
scattering process becomes ribbons with fixed width. After projected on an
angular space with radius $R$, the common intersection region $\mathcal{AM}$
of several external lines with common center becomes Amplituhedron. After
determining the phase factors of different geometric objects, including areas,
boundaries, and points, amplituhedron differential form $\Omega$ turns into
scattering amplitude.

5. How to calculate \emph{loop} amplitudes?

\textbf{Answer:}

The key point is to split the diagram for loop scattering process with $n$
nodes into $n$ tree scattering amplitudes. The final result is
\[
\mathcal{M=}%
{\displaystyle \prod \limits_{l}}
{\displaystyle \int \limits_{-\Lambda}^{\Lambda}}
dp_{l}%
{\displaystyle \prod \limits_{a}}
\frac{1}{(k_{i}^{ll^{\prime}})^{2}}\mathcal{M}_{l}(1^{l},\ldots,n^{l}).
\]
Here, in $\mathcal{M}$,\ each integral comes from an uncertain momenta around
a loop. There are $N$ factors of $\frac{1}{(k_{i}^{ll^{\prime}})^{2}}.$ Each
factor comes from an internal line. In particular, because on angular space
the momenta for different lines (either external lines or internal lines) are
\emph{real} number, rather than a three dimensional vector, we can easily
determine the momentum of all internal lines and get the loop scattering amplitudes.

In the end, we point out that there are still many open questions for
scattering amplitudes. One is to consider scattering amplitudes including
interacting Dirac fermions with half angular momentum. Another open question
is about the issue of off-shell. To deal with the off-shell processes, we must
use the theory of physical variants on usual 3+1D spacetime. This is very
complex. Can we have a simple approach to deal with these problems about issue
of off-shell? In the future, we will continue research in this area and answer
above questions.

\newpage

\section{Conclusion}

Finally, we draw the conclusion. In this paper, we developed a microscopic
theory of quantum spacetime (or quantum gravity) and unified general
relativity and quantum mechanics into a single theoretical framework. Now, the
relationship between gravity and quantum mechanics becomes clear -- that is
the relationship between transverse changings and longitudinal changings for a
physical variant.

An important point is that the particle is basic block of spacetime and the
spacetime is made of matter. Therefore, according to this idea, the matter is
really certain "changing" of \textquotedblleft spacetime\textquotedblright%
\ itself rather than extra things on it. This is the new idea for the
foundation of quantum gravity and the development of a complete theory. In the
paper, we point out that all physical processes of our world be intrinsically
described by a system "uniform changing" that is an $\mathrm{\tilde{S}%
\tilde{O}}$\textrm{(d+1)} physical variant\textit{ }$V_{\mathrm{\tilde
{S}\tilde{O}(d+1)},d+1}$ with 1-th order variability, $\mathcal{T}(\delta
x^{\mu})\leftrightarrow \hat{U}(\delta \phi^{\mu})=e^{i\cdot k_{0}\delta x^{\mu
}\Gamma^{\mu}}.$ The key point of the new theory is higher-order variability
rather the gauge/global symmetry. Now, the principle of "symmetry induce
interaction" is replaced by the principle of "variability induce interaction".
So, we have a\ "\emph{variability principle of gravity}". According to this
principle, a theory for quantum gravity is developed. Quantum mechanics and
general relativity are unified, i.e.,
\begin{align*}
&  \text{Quantum mechanics + general relativity}\\
&  \Longrightarrow \text{Theory of a physical variant.}%
\end{align*}

The logical structure of the paper in Fig.27. There are two types of physical
variants -- unitary type (dS spacetime) or non-unitary type (AdS spacetime).
The black hole is the domain wall between unitary physical variant and
non-unitary variant. To calculate scattering amplitudes of gravitons, we
introduce angular variant that is projected physical variant by considering
event physics.

\begin{figure}[ptb]
\includegraphics[clip,width=0.9\textwidth]{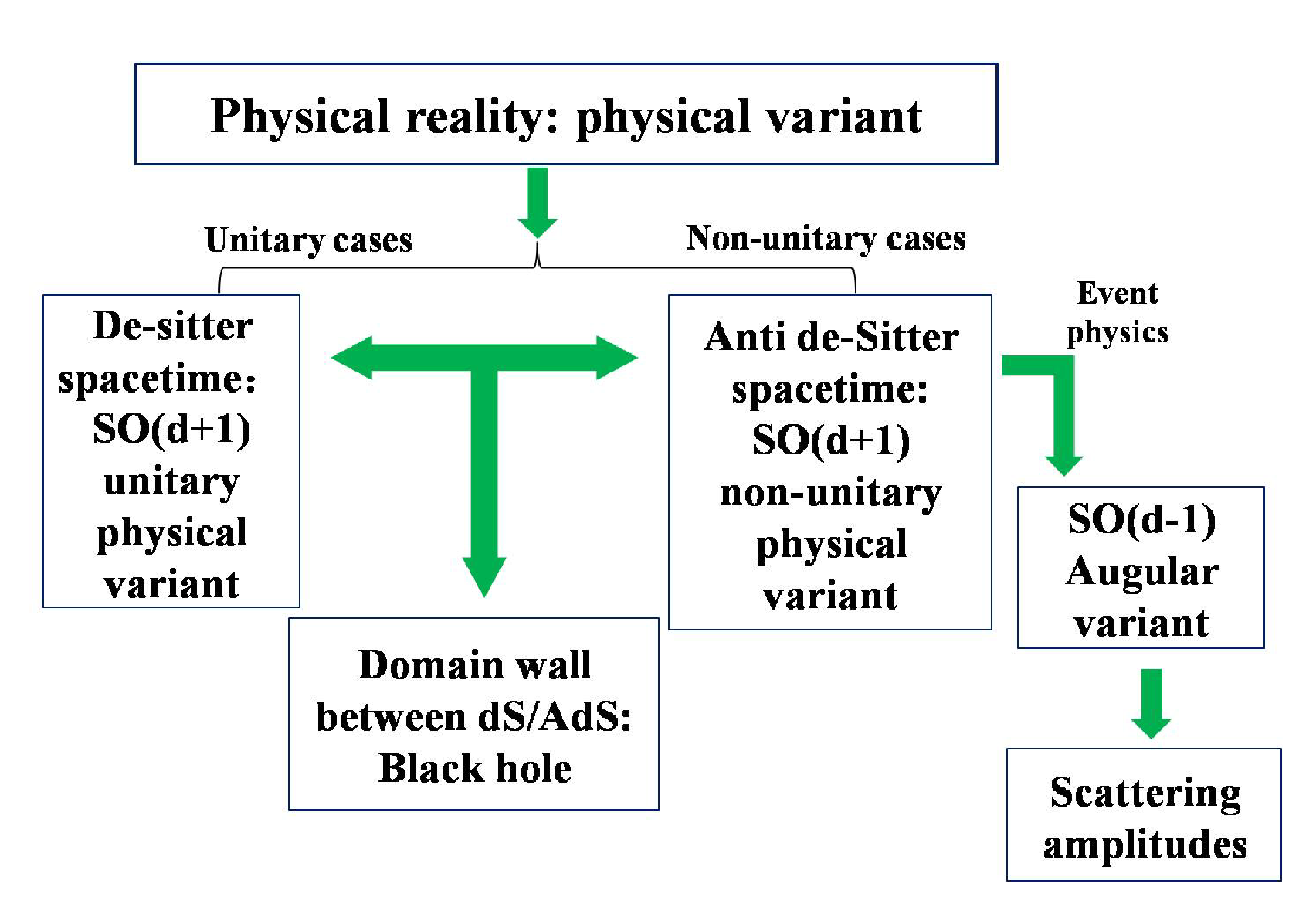}\caption{Logical
structure of the paper}%
\end{figure}

I answer the five unsolve problems for quantum gravity:

1) What's the exact \emph{microstructure} of spacetime near Planck length
$l_{p}$? Does \emph{geometric} structure have quantization characteristics,
and what are the quantization rules?

\textbf{The answer: }

The microstructure of flat spacetime near Planck length is a physical variant
with 1-th order variability. Under geometry representation, the microstructure
of our spacetime is a (uniform/non-uniform) topological lattice; under matrix
representation, the microstructure of our spacetime is a (uniform/deformed)
matrix network. Under matrix representation, the Hilbert space $\mathcal{E}$
of quantum spacetime consists of all four-by-four matrices on links $\{N^{\mu
},M^{\mu}\}$ of the uniform topological lattice, $\mathcal{E}:\mathcal{H}%
_{QST}=\mathcal{H}_{\{(0,0,0,0),(1,0,0,0)\}}\otimes...\mathcal{H}_{\{N^{\mu
},M^{\mu}\}}.$ The states of quantum spacetime are characterized by different
matrix networks $\{ \Gamma_{\mathrm{curved}}^{\{N^{\mu},M^{\mu}\}}%
(x),\mu=x,y,z,t\}.$

In particular, an unexpected result of this paper is obtained -- a Dirac
particle has fixed size rather than a point on spacetime! The volume of a
Dirac particle is obtained to be $4\pi(l_{p})^{3}$ where $l_{p}$ is Planck
constant. The result leads to a great unification of matter and spacetime
--\ the particles constitute the basic blocks of spacetime and spacetime is
really a multi-particle system that is made of matter.

The quantum flat/curved spacetime is uniquely characterized by the coordinates
total size $\Delta x^{\mu}$ and the local vector's unit $\Gamma^{\mu}(x)$.
Now, the changes of a quantum spacetime can be divided into two types, one is
longitudinal about $\Delta x^{\mu}$ (or the contraction/expansion processes
with finite volume changing), and the other is transverse changings about
$\Gamma^{\mu}(x)$ (or shape changings without 3-volume changing). The
transverse changings -- shape changings is just the processes for curving
spacetime that is characterized by a matrix network; the longitudinal
changings -- contraction/expansion changings is just the processes for single
particle annihilation/generation that is characterized by quantum mechanics.
As a result, this leads to the unification of quantum mechanics and gravity.

2) What's the \emph{exact} rule of AdS/CFT correspondence within the framework
of quantum gravity rather than just a conjecture?

\textbf{The answer:}

We found that AdS is (d+1)-dimensional \textrm{\~{S}\~{O}(d+1)} non-unitary
physical variant\textit{ }$V_{\mathrm{\tilde{S}\tilde{O}(d+1)},d+1}$ that is
characterized by 1-th order non-unitary spatial variability along the d-th
direction $\mathcal{T}(\delta x^{d})\leftrightarrow \hat{U}(\delta \phi^{\mu
})=e^{k_{0}x^{d}\Gamma^{d}}$. Then, we develop a microscopic theory for
AdS/CFT correspondence and its updated version -- AdS/NGT equivalence. Here,
NGT is abbreviation of non-Hermitian gauge theory. Based on gravity/N-gauge
equivalence, the quantum fluctuations from gravitational waves both in bulk
and on boundary of AdS can all be characterized by non-Hermitian
\textrm{U(0,1)}$\times$\textrm{SU(0,N)} gauge fields. When we only consider
unitary physical processes on the boundary of AdS, the AdS/NGT equivalence is
reduced to usual AdS/CFT correspondence. See the logical structure of the
paper in Fig.27.

We found that due to spacetime skin effect from non-unitary variability,
AdS/CFT correspondence characterizes the equivalence for the slow motion in
CFT and that on the boundary of AdS.

It was known that the perturbative metric fluctuations $g_{\mu \nu}$ of AdS
correspond to a boundary stress tensor $T_{\mu \nu}$ in CFT within the
framework of quantum gravity. We found that this is really a correspondence
between shape changing of boundary in AdS and expansion/contraction in CFT.
The exact correspondence between metric fluctuations in AdS and the motion
tensor $M_{\mu \nu}$ are given by $g_{\mu \nu}=(l_{0})^{2}M_{\mu \nu}.$ It is the
changing of motion tensor $M_{\mu \nu}$ is equal to energy-momentum tensor
$T_{\mu \nu}$ rather than $M_{\mu \nu}$ itself.

According to the dictionary from AdS/CFT correspondence, the particle's mass
$m$ in AdS plays the role of anomalous dimension $\nu$ in correlation
functions. Why? We indeed have a correspondence between particle's mass $m$ of
AdS and anomalous dimension $\nu$ of correlation functions in CFT. So, it is
correct. The underlying mechanism of this correspondence is the re-definition
the elementary particles in both sides. The anomalous dimension plays the role
of the ratio of the volume of elementary particle in AdS and that in CFT.

According to AdS/CFT correspondence, the gauge fields $A_{\mu}$ in AdS
correspond to usual current in CFT $J^{\mu}$. What does it mean within the
framework of quantum gravity? Abelian/non-Abelian gauge fields characterize
the dynamics of global/relative loop currents on spacetime. In AdS, due to
spacetime skin effect, the loop currents for the gauge fields is naturally
reduced to the current of CFT on the boundary of the AdS, i.e., Loop currents
in AdS $\leftrightarrow$ Currents in CFT.

Another important feature of AdS/CFT correspondence is Ryu-Takayanagi's
formula for the holographic entangled entropy. We derive the same results that
are same to Ryu-Takayanagi's formula.\textbf{ }The underlying mechanism of
holographic entangled entropy in AdS/CFT correspondence really comes from the
geometry quantized for quantum flat spacetime. Each unit cell of quantum flat
spacetime in CFT carry area $l_{0}^{2}$. When one smears out the information
of the unit cells, the entropy is just the RT formula of the holographic
entangled entropy.

3) What's the exact \emph{microstructure} of spacetime around black hole near
Planck length? What's the exact \emph{microstructure} of spacetime inside
black hole? And, how to characterize it?

\textbf{The answer: }

In this paper, we found that black hole becomes really a physical variant with
topological defects. The key point is%
\begin{align*}
&  \text{Black hole (a phenomenological theory)}\\
&  \Longrightarrow \text{Physical variant with topological defect}\\
&  \text{ (a microscopic theory).}%
\end{align*}
Now, the event horizon of a black hole becomes a topological domain wall
between a unitary physical variant (or a dS) and a non-unitary physical
variant (or an AdS).

Firstly, we developed the microscopic theory to learn the nature of the region
inside a black hole. Because the spacetime inside black hole is AdS, we use
Gravity/N-gauge equivalence to characterize its dynamics. Now, the physical
processes for slow motion come from non-Hermitian \textrm{U(0,1)}$\times
$\textrm{SU(0,N)} gauge fields. Near the singularity, the curvature of
spacetime becomes imaginary. So, by using non-Hermitian quantum mechanics, the
trouble about singularity doesn't exist at all.

Next, we developed the microscopic theory to learn the nature of the physical
structure of event horizon. By integrating fast variables, we get effective
model for slow variables. The effective model has three equivalent forms: one
is effective Jackiw-Teitelboim gravity under geometric representation, second
is effective SYK model under matrix representation, third is effective 1D
gauge theory under kinetic representation. The formula can be applied to all
kinds of black hole rather than only extremal one with its fine-tuned magnetic charge.

In particular, we developed the thermodynamics and quantum statistical theory
for a black hole.

In variant theory, the black hole is a topological defect between a unitary
physical variant (or a dS) and a non-unitary physical variant (or an AdS). Due
to the "non-changing" structure along tempo direction, the event horizon of
the black hole becomes a stochastic variant with a random distribution of unit
cells. Under an assumption of Principle of equal probability and the
constraint of energy (or particle number), we have a new statistics of
spacetime $\Omega=\frac{(N_{U})^{N_{U}}}{(N_{U})!}$ where $N_{U}$ is the
number of unit cells. As a result, in thermodynamic limit, a black hole
becomes a classical object with finite temperature. From the statistics of
spacetime, the Hawking entropy, Hawking temperature are exactly derived.

According to above discussion, the randomness from non-variability of event
horizon leads to thermalization and decoherence of the quantum states near
event horizon. Therefore, the quantum information disappear and a pure quantum
state evolves to a mixed state. This indicates usual quantum mechanics becomes
invalid near event horizon! Hence, the \textquotedblleft \textit{black hole
information paradox}\textquotedblright \ is solved. Our results indicate that
Page curve cannot characterize the information process for Hawking evaporation
of black hole.

4) How \emph{quantize} gravitational waves correctly?

\textbf{The answer: }

For a (3+1)D quantum curved spacetime, we have a deformed (3+1)D topological
lattice with fluctuated lattices in geometry representation and a non-uniform
(3+1)D matrix network with fluctuated Gamma matrix on its links in matrix
representation. Under Lorentz covariance, we use $\gamma$-matrix/gauge
representation to characterize the changings of \textrm{SO(4)} matrix network
$\Gamma^{\mu}(x,t)$. This leads to an \textrm{SO(3)}$^{\mathrm{SO(4)}}$ gauge
structure,\ of which each group element of $\mathrm{SO(4)}$ group for a 3D
sub-manifold \textrm{M}$_{3}^{\mu}$ corresponds to an \textrm{SO(3)} gauge
theory. By using the \textrm{SO(3)}$^{\mathrm{SO(4)}}$ gauge theory, we have a
local field description for curved spacetime. This plays important role in the
quantization of spacetime and gravity.

Now, elementary particles become topological defects of quantum spacetime. To
characterize the topological constraint, we introduce topological BF term that
is just the famous Einstein-Hilbert term. The situation is similar to the
Chern-Simons terms in (2+1)D topological field theory. Under the Chern-Simons
term, the local constraint from flux-charge binding is guaranteed. However,
according to the existence of \textrm{SO(3)}$^{\mathrm{SO(4)}}$ gauge
structure, the situation here is more complex. For different 3D sub-manifolds
of the 4D topological lattice, we must define different gauge fields. It is
round-robin of generalized gamma matrices that changes one gauge class to another.

Because the Einstein-Hilbert action $S_{\mathrm{EH}}$ is only a pure
topological constraint term, the Hamiltonian for quantum spacetime themselves
(without considering matter) becomes zero. Therefore, the evolution of quantum
spacetime can not satisfy Schrodinger equation! Instead, the time evolution in
quantum spacetime is determined spacetime Gaussian theorem. Therefore, the
evolution of quantum spacetime is \emph{self-induced} and does not satisfy the
Schrodinger equation. This leads to time evolution in quantum spacetime itself.

5) What's the exact \emph{microstructure} of the scattering amplitudes for
different particles? How to calculate \emph{loop} amplitudes? Why
\emph{amplituhedron}?

\textbf{The answer: }

In this paper, based on angular variant, we develop a new theory beyond
"quantum field theory" to calculate the scattering amplitudes. Now, scattering
process for quantum states is regarded as an event process from initial
quantum states to final quantum states.

The angular variant $V_{\mathrm{\tilde{S}\tilde{O}(d-1)},d-1}%
^{\text{\textrm{Angular}}}$ is defined by a mapping between the
$\mathrm{\tilde{S}\tilde{O}(d-1)}$ group-changing space and the angular space
of the original Cartesian space $\mathrm{S}_{d-1}^{\text{\textrm{Angular}}}$,
i.e.,%
\begin{equation}
V_{\mathrm{\tilde{S}\tilde{O}(d-1)},d-1}^{\text{\textrm{Angular}}}%
:\mathrm{C}_{\mathrm{\tilde{S}\tilde{O}(d-1),}d-1}\Longleftrightarrow
\mathrm{S}_{d-1}^{\text{\textrm{Angular}}}%
\end{equation}
\textit{ w}here the d-1 dimensional angular space $\mathrm{S}_{d-1}%
^{\text{\textrm{Angular}}}$ is sphere in d dimensional Cartesian space with a
radius $R$ (or $\mathrm{S}_{d-1}^{\text{\textrm{Angular}}}$ manifold). A
group-changing space $\mathrm{C}_{\mathrm{\tilde{S}\tilde{O}(d-1)},d-1}$ is a
group-changing space of non-compact $\mathrm{\tilde{S}\tilde{O}(d-1)}$ Lie
group. The angular variant provides a solid physical foundation on ambitwistor
space and the celestial sphere. In general, the angular variant is
characterized by 1-th order variability,%
\begin{equation}
\mathcal{T}(\delta \theta^{\mu})\leftrightarrow \hat{U}(\delta \phi^{\mu
})=e^{i\Gamma^{\mu}\sqrt{N_{\mathrm{tot}}^{F}}\delta \theta^{\mu}}%
\end{equation}
where $\hat{U}(\delta \phi^{\mu})=e^{i\Gamma^{\mu}\delta \phi^{\mu}}$ with
$\delta \varphi^{\mu}=\sqrt{N_{\mathrm{tot}}^{F}}\delta \theta^{\mu}$.

Based on the framework of angular variants, the scattering amplitudes are
obtained, including tree diagrams and loop diagrams. The key point for
calculating loop diagram is to split the single loop scattering amplitude with
$n$ nodes into $n$ tree scattering amplitudes. The final result is
\[
\mathcal{M=}%
{\displaystyle \prod \limits_{l}}
{\displaystyle \int \limits_{-\Lambda}^{\Lambda}}
dp_{l}%
{\displaystyle \prod \limits_{a}}
\frac{1}{(k_{i}^{ll^{\prime}})^{2}}\mathcal{M}_{l}(1^{l},\ldots,n^{l}).
\]
Here, in $\mathcal{M}$,\ each integral comes from an uncertain momenta around
a loop. There are $N$ factors of $\frac{1}{(k_{i}^{ll^{\prime}})^{2}}.$ Each
factor comes from an internal line. Because on angular space the momenta for
different lines (either external lines or internal lines) are real number,
rather than a three dimensional vector, we can easily determine the momentum
of all internal lines and get the loop scattering amplitudes.

We then explored the nature of Amplituhedron. To characterize the geometric
property of tree-level scattering amplitudes, we use the bosonic
representation. The geometric structure of projected external lines become
ribbons with fixed width. After projected on an angular space with finite
radius, the common intersection region of several external lines with common
center becomes Amplituhedron. After determining the phase factors of different
geometric objects, including areas, boundaries, and points, amplituhedron
differential form turns into the scattering amplitude of Parte-Taylor formula.

In addition, we found that string theory become a correct framework for event
physics on angular space rather than dynamical physics on usual spacetime.
Now, supersymmetry and string structure become \emph{emergent} phenomena.

\end{document}